\documentclass[11pt,twoside,openright]{report}
\usepackage[utf8]{inputenc}
\usepackage{graphicx}
\usepackage{color}
\usepackage{geometry}
\usepackage{amsmath}
\usepackage{perpage}
\usepackage{mathtools}
\usepackage{slashed}
\usepackage{bbold}
\usepackage{axodraw2}
\usepackage{cite}
\usepackage{bibentry}
\usepackage{nameref}
\nobibliography*
\usepackage{fancyhdr}
\usepackage[colorlinks=true,urlcolor=blue,anchorcolor=blue,citecolor=blue,filecolor=blue,linkcolor=blue,menucolor=blue,pagecolor=blue,linktocpage=true,pdfproducer=medialab,pdfa=true]{hyperref}

\fancypagestyle{fancychapter}{
\fancyhf{}

}

\graphicspath{{Figures/}}
\MakePerPage{footnote}

\def\Eq#1{Eq.~(#1)}
\def\Eqs#1#2{Eqs.~(#1) and (#2)}
\def\Eqss#1#2#3{Eqs.~(#1), (#2) and (#3)}
\def\Fig#1{Fig.~#1}
\def\Table#1{Table~#1}
\def\Chapter#1{Chapter~#1}
\def\Section#1{Section~#1}
\def\Sections#1#2{Sections~#1 and #2}
\def\Chapters#1#2{Chapters~#1 and #2}
\def\Appendix#1{Appendix~#1}
\def\s{s_{12}}
\def\Sup#1{#1^{(+)}}
\def\deltaplus#1{\delta_+(#1)}
\def\deltatilde#1{\widetilde{\delta}(#1)}
\def\sigmatilde{\widetilde{\sigma}}
\def\sigmabar{\overline{\sigma}}
\def\yp#1{y_{#1}'}
\def\ct{{\rm CT}}
\def\uv{{\rm UV}}
\def\ir{{\rm IR}}
\def\V{{\rm V}}
\def\r{{\rm R}}
\def\nn{\nonumber}
\def\dxi#1{d[\xi_{#1,0}]}
\def\dv#1{d[v_#1]}
\def\cGamma{c_\Gamma}
\def\cGammaTilde{\widetilde{c}_\Gamma}
\def\eiPerp#1{\mathbf{e}_{#1,\perp}}
\def\qbar{\bar{q}}
\def\Oep#1{\mathcal{O}(\epsilon^{#1})}
\def\MSbar{\overline{\text{MS}}}
\def\vev{\langle v\rangle}
\def\Se{\widetilde{S}_{\epsilon}}

\def\mathematica{{{\sc Mathematica}}}
\def\BIGG#1{\left#1\rule{0cm}{1cm}\right.}

\title{Four-dimensional representation of scattering amplitudes and physical observables through the application of the Loop-Tree Duality theorem}
\author{F\'elix Driencourt-Mangin}
\date{Soon!}

\DeclareMathOperator{\Tr}{Tr}
\DeclareMathOperator{\Res}{Res}
\DeclareMathOperator{\ReText}{Re}
\DeclareMathOperator{\ImText}{Im}
\DeclareMathOperator{\Li}{Li}

\begin{document}
\pagenumbering{roman}
\pagestyle{empty}
\begin{titlepage}
	\begin{center}
	\vspace{25pt}
		{\LARGE \bfseries Four-dimensional representation of scattering amplitudes and physical observables through the application of the Loop-Tree Duality theorem}\\
		\vspace{1.25cm}
		{\large Thesis submitted for the Degree of Doctor of Philosophy}\\
		\vspace{1.25cm}
		Author:\\
		\vspace{11pt}
		{\bf Félix Driencourt-Mangin}\\
		\vspace{1cm}
		Directors:\\
		\vspace{11pt}
		{\bf Germán Rodrigo}\\
		{\bf German F. R. Sborlini}\\
	\vspace{2cm}
	\textsc{\LARGE Universitat de València}\\
	\vspace{1cm}
		\begin{figure}[h]
			\centering
			\includegraphics[width=5.5cm]{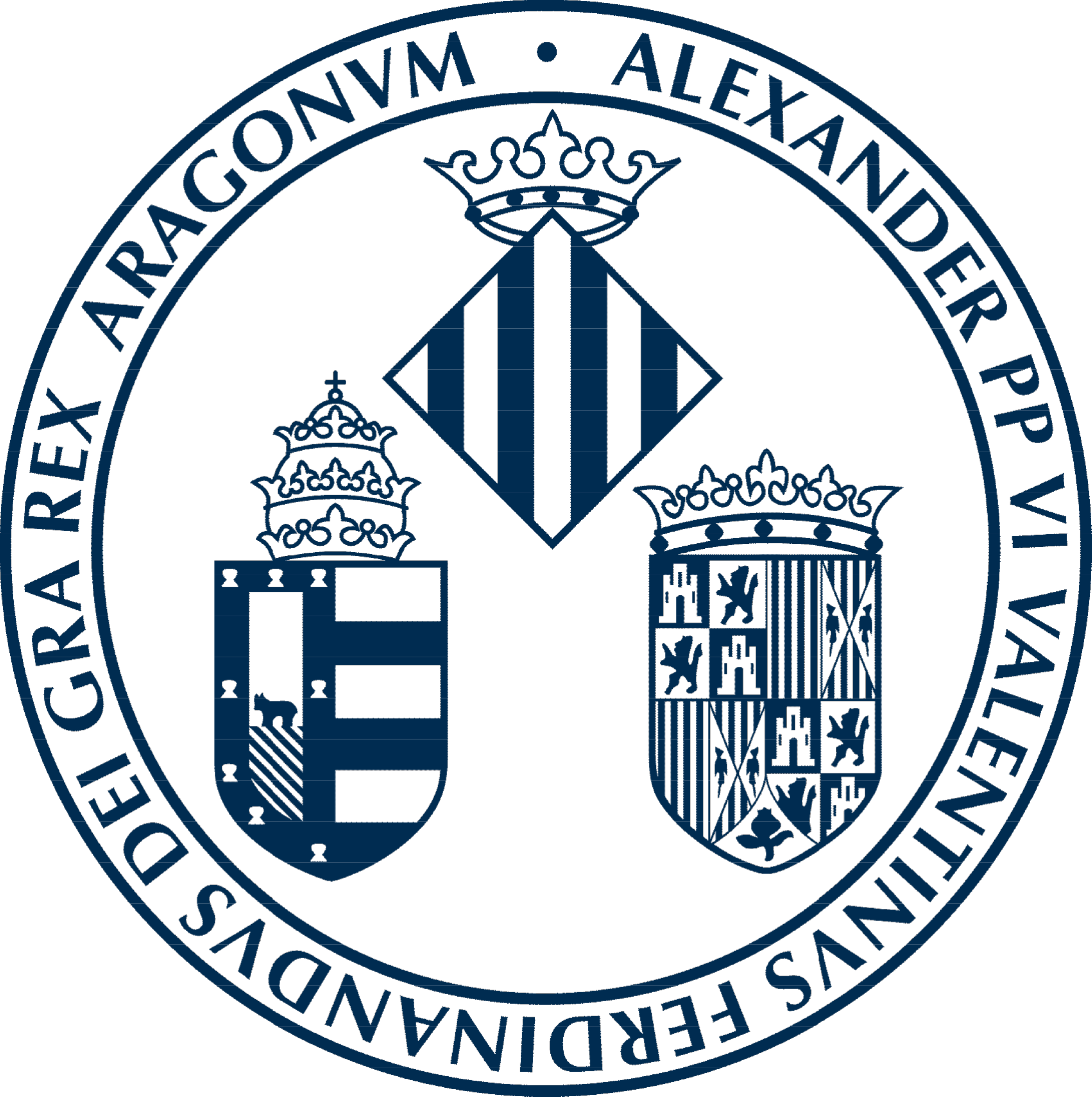}
		\end{figure}
		\vspace{1cm}
		\large \textit{3026 Programa Oficial de Doctorado en F\'isica}\\[0.3cm]
		\vspace{1cm}
		Instituto de Física Corpuscular, Departamento de F\'isica Te\'orica\\
		Universitat de València -- Consejo Superior de Investigaciones Científicas\\
		\vspace{2cm}
		{\large April 2019}\\
		\vfill
	\end{center}
	
\end{titlepage}

\chapter*{Declaration of authorship}
\thispagestyle{empty}
I, Félix Driencourt-Mangin, hereby certify that this thesis, titled \textit{Four-dimensional representation of scattering amplitudes and physical observables through the application of the Loop-Tree Duality theorem}, has been composed by myself. All the work contained herein is my own except where explicitly stated otherwise, and all main sources of help have been properly referenced and acknowledged.\\
\\
The work presented in this thesis is based on original results previously published by myself, in collaboration with Germ\'an Rodrigo, German F. R. Sborlini, Roger Hern\'andez-Pinto and William J. Torres Bobadilla. Chapters~\ref{Chapter:FDU}, \ref{Chapter:FDUM}, \ref{Chapter:HOL} and \ref{Chapter:HTL}, in particular, are based on the following publications:

\section*{Articles}

\begin{itemize}
	\item[\cite{Sborlini:2016gbr}]\bibentry{Sborlini:2016gbr}\\
	\item[\cite{Sborlini:2016hat}]\bibentry{Sborlini:2016hat}\\
	\item[\cite{Driencourt-Mangin:2017gop}]\bibentry{Driencourt-Mangin:2017gop}\\
	\item[\cite{Gnendiger:2017pys}]\bibentry{Gnendiger:2017pys}\\
	\item[\cite{Driencourt-Mangin:2019aix}]\bibentry{Driencourt-Mangin:2019aix}
\end{itemize}

\section*{Proceedings}

\begin{itemize}
	\item[\cite{Rodrigo:2016hqc}]\bibentry{Rodrigo:2016hqc}
	\item[\cite{Hernandez-Pinto:2016uwx}]\bibentry{Hernandez-Pinto:2016uwx}
	\item[\cite{Sborlini:2016bod}]\bibentry{Sborlini:2016bod}
	\item[\cite{Driencourt-Mangin:2016dpf}]\bibentry{Driencourt-Mangin:2016dpf}
	\item[\cite{Sborlini:2017mhc}]\bibentry{Sborlini:2017mhc}
	\item[\cite{Sborlini:2017nee}]\bibentry{Sborlini:2017nee}
	\item[\cite{Rodrigo:2018jme}]\bibentry{Rodrigo:2018jme}
	\item[\cite{Bendavid:2018nar}]\bibentry{Bendavid:2018nar}
\end{itemize}
\vspace{50pt}
Different talks about the topic were also given at the following conferences:
\vspace{20pt}
\begin{itemize}
	\item\href{http://www.cs.infn.it/diff2016}{DIFFRACTION 2016}, \emph{International Workshop on Diffraction in High-Energy Physics}, Acireale (Italy), September 2016
	\item\href{https://www.i-cpan.es/jornadas8/}{VIII CPAN DAYS}, Zaragoza (Spain), November 2016
	\item\href{https://phystev.cnrs.fr/2017-les-houches}{Les Houches Workshop Series}, \emph{Physics at TeV colliders}, Les Houches (France), June 2017
	\item\href{https://www.i-cpan.es/jornadas9/}{IX CPAN DAYS}, Santander (Spain), October 2017
	\item\href{https://indico.ific.uv.es/event/3187/}{LHCPHENO 2017}, \emph{Workshop on High-Energy Physics Phenomenology in the LHC era}, Valencia (Spain), December 2017
	\item\href{http://www.tep.physik.uni-freiburg.de/workshops/hp2-2018}{HP2 2018}, \emph{High Precision for Hard Processes at the LHC}, Freiburg (Germany), October 2018
	\item\href{https://indico.cern.ch/event/761296/}{PARTICLEFACE 2019}, \emph{Working Group Meeting and MC Meeting of the COST Action CA16201}, Coimbra (Portugal), February 2019
\end{itemize}

\newpage
\hspace{0pt}
\thispagestyle{empty}
\clearpage

\newpage
\hspace{0pt}
\thispagestyle{empty}
\clearpage

\pagebreak
\newpage
\hspace{0pt}
\thispagestyle{empty}
\vspace{75mm}
\begin{flushright}
À mon arrière grand-père du Sahara, que je n'ai jamais connu,\linebreak mais à qui probablement je dois beaucoup
\end{flushright}
\vfill

\newpage
\hspace{0pt}
\thispagestyle{empty}
\clearpage

\newpage
\pagebreak
\hspace{0pt}
\vfill
\begin{center}
\textbf{\huge Abstract}\\
\end{center}
\vspace{25pt}
The outstanding quality of the data provided by CERN's Large Hadron Collider (LHC) during the last decade has opened a new thrilling era in particle physics. In the meantime, the excellent agreement between the experimental observations and predictions of the Standard Model of particles has demanded the development of new tools to increase the precision of theoretical predictions. In this thesis we propose a novel method to compute higher-order corrections to physical cross sections, bypassing more traditional approaches. This technique, the Four-Dimensional Unsubtraction (FDU), is based on the Loop-Tree Duality (LTD) theorem, and aims at building pure four-dimensional representations of scattering amplitudes at higher orders in perturbation theory. This is done by locally renormalising the virtual contribution at high energies, and then mapping the kinematics of the corresponding real contribution so it matches the virtual one in the soft and collinear limits. This makes the ultraviolet and infrared singularities vanish at the integrand level when taking the sum of both contributions, meaning that the four-dimensional limit can be taken before integrating, thus allowing for a straightforward numerical implementation. We apply this method for the first time to calculate the decay of a virtual photon to two massless quarks, and then generalise the method to massive particles. We also show new advantages provided by the LTD formalism. One of them is the possibility to write one- and two-loop scattering amplitudes in a universal form at the integrand level, regardless of the nature of the particle running inside the loop. The other is to perform integrand-level asymptotic expansions in the internal mass --~without having to rely on expansion-by-region techniques~-- by taking advantage of the Euclidean nature of the integration domain. Finally, we present a new algorithm to locally renormalise two-loop amplitudes, where the renormalisation scheme can be easily fixed through subleading contributions in the ultraviolet region.
\vfill
\hspace{0pt}

\newpage
\pagebreak
\hspace{0pt}
\vfill
\begin{center}
\textbf{\huge Resumen}\\
\end{center}
\vspace{25pt}
L’excepcional qualitat  de les dades proporcionades per el Gran Col·lisionador d'Hadrons (LHC per les seues sigles en anglès) durant la última dècada ha donat lloc a una nova i emocionant era en la física de partícules. Mentrestant, l’excel·lent compatibilitat entre les observacions experimentals i les prediccions del Model Estàndard de física de partícules ha exigit el desenvolupament de noves eines per tal d’augmentar la precisió de les prediccions teòriques. En aquesta tesi es proposa un nou mètode per calcular correccions d’ordre superior a seccions transversals físiques, evitant enfocaments més tradicionals. Aquesta tècnica, la No-substracció Quatre-Dimensional (FDU en anglès), es basa en el teorema de la Dualitat Bucle-Arbre (LTD en anglès) i té com a objectiu construir representacions d'exactament quatre dimensions de les amplituds de dispersió a ordres superiors en teoria de pertorbacions. Això es fa renormalitzant localment la contribució virtual a altes energies, i després mapejant la cinemàtica de la contribució real corresponent, de manera que coincidisca amb la virtual en els límits colineal i suau. Això fa que les singularitats ultraviolades i infrarojes es cancel·len a nivell de l’integrand quan prenem la suma d’ambdues contribucions. Açò significa que el límit en quatre dimensions es pot prendre abans d’integrar, permetent així una implementació numèrica senzilla. Apliquem aquest mètode per primera vegada per calcular la desintegració d'un fotó virtual a dos quarks sense massa i després generalitzar el mètode a partícules massives. També mostrem nous avantatges proporcionats pel formalisme LTD. Un d’ells és la possibilitat d’escriure amplituds de dispersió d’un i dos bucles d’una manera universal a nivell de l’integrand, independentment de la naturalesa de la partícula que hi haja dins del bucle. L’altre és realitzar expansions asimptòtiques a nivell de l’integrand -- sense haver de dependre de tècniques d’expansió per regió -- aprofitant la naturalesa Euclidiana del domini d’integració. Finalment, es presenta un nou algorisme per renormalitzar localment les amplituds de dos bucles, on l’esquema de renormalització es pot fixar fàcilment mitjançant contribucions subdominants a la regió ultraviolada. 
\vfill
\hspace{0pt}

\chapter*{Acknowledgements}
\thispagestyle{empty}

Writing this PhD thesis is by far the most difficult work I have ever had to do. After all these efforts, I was really looking forward to the day I would hand in, and now that I am about to, it almost feels unreal. Until quite recently, I was not sure I would be up to the task, and sometimes, looking back, I am still surprised by how far I have come. This thesis, though, is not only my work, but also the one of all the people who have helped and supported me during these last three years and a half. I could never express enough gratitude to them with just a few lines, but I will try my best.\\
\\
First of all, I am extremely grateful to my thesis director, Germán Rodrigo, who would always find time for me and for answering the many, many questions I had. His insight, experience and advices were immensely valuable throughout all my time at the IFIC. I am proud and consider myself very lucky to have been his PhD student.\\
\\
I am equally grateful to my co-director and friend, German Sborlini, whose immeasurable patience and great didactic skills played a very important role in my understanding of the many subtleties the world of particle physics offers.\\
\\
I also want to thank Roger who very kindly welcomed me at the IFIC and helped me to integrate in this then completely unknown environment, William for sharing his tricks in \textsc{Mathematica}, and Tomá\v{s} for saving me from some headaches related with polylogarithms. I am also grateful to Sam, who took the time to correct the unenglishnesses of my thesis.\\
\\
Muchas gracias a mis compañeros de despacho, Ana, Fernando y Víctor, por la genial atmósfera, y por hacerme hablar castellano, a pesar de mi fuerte resistencia a hacerlo. Gracias también a Clara, por haber aceptado de utilizar sus talentos para dibujar la cobertura de mi tesis.\\
\\
Many thanks to all the researchers with whom I had very interesting discussions during workshops, conferences and informal meetings in many different institutes. I really enjoyed exchanging about my projects, and their constructive feedback and different perspective were very helpful for the work I carried out during my PhD.\\
\\
I am of course very grateful to the University of Valencia and the CSIC for providing me with financial support during my PhD, and for giving me the opportunity to travel to many great places in Europe. I am also glad to have worked in a country that can appreciate the importance of fundamental research, and that is willing to spend efforts and resources to promote and spread science (and knowledge in general), which is today probably more important than ever.\\
\\
Et grand merci à ma famille, bien entendu. À ma mère pour m'avoir supporté tout ce temps (ce que je vous le garantis n'est pas une mince affaire), à mon père pour m'avoir donné goût aux mathématiques et à la physique dès tout petit, et à mon petit frère pour tous ces moments de complicité que je ne peux avoir qu'avec lui.\\
\\
And finally a huge thanks to all my friends and companions, Louis, Otto, Claudius, Carlos, Christoph, Ryan, Judith, Luismi, Rozenn... and of course to all the others, in or outside the IFIC, for playing volley-ball on the beach, for coming to my (many, too many...) parties, and for sharing the stories and experiences they gathered from all around the world. All of you made my stay in Valencia an unforgettable experience!

\newpage
\thispagestyle{empty}
\pagebreak
\hspace{0pt}
\vfill
\begin{center}
	\textit{``We are very lucky to be living in an age in which we are still making discoveries. (\dots) The age in which we live is the age in which we are discovering\linebreak the fundamental laws of nature.''}\\
	\vspace{11pt}
	—Richard Feynman
\end{center}
\vfill
\hspace{0pt}
\pagebreak

\newpage
\hspace{0pt}
\thispagestyle{empty}
\clearpage

\chapter*{Foreword}
\thispagestyle{empty}

Coincidence has it that I started writing this thesis the year of the hundredth anniversary of Richard Feynman. Without his contribution, and also the ones of many other great scientists, we would not be where we are today. It is very important to always remember that in order to make progress in physics (and in science in general) we have to rely upon the work achieved by other people before us. This means that we must keep in mind that everything we write today will eventually be read and used by other people in the future! My objective when writing this thesis was shaped by this very thought. ``What kind of manuscript do I want to produce?'' Well, very ideally, something that the me of three and a half years ago would be able to read, more or less understand and use as a basis for his future work. Of course, unless we invent time travel, I will never know if I succeeded in doing so, but if at least one fellow student one day tells me they found my thesis useful for their work, then I would think that I did succeed in a way.\\
\\
This thesis represents my tiny contribution to the field of particle physics, and, hopefully, future young (and less young of course) particle physicists will find these pages instructive.

\newpage
\hspace{0pt}
\thispagestyle{empty}
\clearpage

\addtocontents{toc}{\protect\thispagestyle{empty}}
\thispagestyle{empty}
\tableofcontents
\thispagestyle{empty}
\vfill
\hspace{0pt}
\clearpage
\pagebreak

\newpage
\hspace{0pt}
\thispagestyle{empty}
\clearpage

\pagestyle{fancy}
\fancyhead{}
\fancyhead[LE,RO]{\thepage}
\fancyfoot{}

\pagenumbering{arabic}
\chapter{Introduction}\label{Chapter:INTRO}
\thispagestyle{fancychapter}
\fancyhead[RE]{\nameref*{Chapter:INTRO}}

\begin{center}
\textit{``If you want to make an apple pie from scratch, you must first create the Universe.''}\\
\vspace{11pt}
—Carl Sagan
\end{center}
\vspace{20pt}
Scientists from all eras have always been fascinated by the elementary structure of the Universe. From the philosophical theorisation of a discrete unit of matter over two and a half thousands years ago, to the very recent discovery of the Higgs boson, every key milestone mankind reached has raised many questions and opened new passionating possibilities. Various models and theories have emerged and been developed in order to explain results obtained by experiments, and one in particular, known as the Standard Model of particles, has more than proven its efficiency. Its huge success at predicting known results and anticipating new discoveries has made it the almost--indisputable reference model for particle physics nowadays.\\
\\
However, despite its achievements, it is still unable to explain different known and well-established phenomena such as neutrino oscillation and the existence of Dark Matter and Dark Energy, or even include a quantum description of gravity. Therefore, in order to test and challenge even further the validity of the Standard Model, physicists are relying on the biggest machine man has ever built: the Large Hadron Collider (LHC). Its main goals, among many others, are to investigate whether the Standard Model still holds at higher energies, and to probe for Beyond Standard Model physics such as Super Symmetry, Dark Matter particles, gravity mediators, and other exotic particles. More than ever before, it has become essential to develop powerful computational techniques to keep up with the enormous amount of data the LHC will provide, as well as the ever-increasing experimental precision.\\
\\

\section{A short history of the Standard Model}\label{Section:INTROHistory}
\fancyhead[LO]{\ref*{Section:INTROHistory}~~\nameref*{Section:INTROHistory}}

The discovery of the electron by J. J. Thomson in 1897 can be considered as the first crucial step among many of the journey towards an exhaustive representation of the elementary structure of matter. Back then, very little of the very small was known, but as experiments and theories became more sophisticated, the mystery started to unravel itself. The second major breakthrough occurred less than twenty years later, in 1911, when E. Rutherford showed that the positively charged part of the hydrogen atom was limited to a very small and relatively heavy nucleus. The proton had been discovered. A bit later, in 1914, N. Bohr proposed a representation for the hydrogen atom, where the electron revolves around the proton, like the Earth does around the Sun. Although a priori possibly simplistic, this model was the very base on which modern particle physics built itself.\\
\\
Around 1930, P. Dirac exploited Bohr's idea and formulated the foundation of a relativistic model describing the \emph{electromagnetic force}, i.e. the interaction of photons --~the ``particles of light''~-- with charged particles, which would ultimately lead to what we call today Quantum Electrodynamics (QED). Probably his biggest achievement was to introduce the concept of antimatter and antiparticles. Then, after the postulation of the neutrino by W. Pauli in 1930 and the discovery of the neutron by  J. Chadwick in 1932, E. Fermi generalised and extended the work of Dirac by including radioactive $\beta$ decay. Yet, even though the theory was really effective at explaining all kind of electromagnetic interactions, it was still not understood why multiple positively charged protons were holding together in a nucleus without repelling each other.\\
\\
In order to solve this issue, H. Yukawa proposed in 1935 the existence of another fundamental force of nature he called the \emph{strong force}. As its name suggests, this force would be powerful enough to negate the electromagnetic repulsion between two protons and even make them bind together. However, because this interaction had to be of short range\linebreak--~otherwise its effects would be much more dramatic~-- its mediator was thought to be quite heavy, with a mass probably between the one of the electron and the one of the proton. For this reason, Yukawa logically gave the name meson to this hypothetical elementary particle. In 1937, the studies of cosmic rays by two separate groups (C. D. Anderson and S. Neddermeyer, and J. C. Street and E. C. Stevenson) lead to the discovery of a new massive elementary particle matching Yukawa's description, but that interacted too weakly with nuclei to be the strong force mediator. Even more problematic, experiments were showing worrying discrepancies. The puzzle was solved when, in 1947, C. F. Powell showed there were actually two different particles of similar mass in cosmic rays. One had the same properties as the electron, but was much heavier, and the other was characterised, in particular, by its very short lifetime. The former was called the muon and was the first particle of the second generation --~even though it had not yet been established at that time~-- to be discovered, and the latter was called the pion, and, although not an elementary particle, was indeed the meson Yukawa had postulated.\\
\\
On the theory side, in the late 1940's R. Feynman, J. Schwinger and S. Tomonaga made huge progress by finding the means to deal with the divergences appearing when computing perturbative corrections in QED. This new method was called \textit{renormalisation} and was generalised by F. Dyson in 1949. Their work gained a lot of credibility when the measurements of the anomalous magnetic moment of the electron and of the Lamb shift were matching their theoretical predictions.\\
\\
After these breakthroughs and until the end of 1947, it was thought that the puzzle of elementary particle physics was almost resolved, and that only the neutrino was yet to be seen. This belief was soon buried when in December of the same year the kaon --~a heavier version of the pion decaying into two of them~-- was discovered by C. Butler and G. Rochester. In the following years, many other particles, all belonging to the meson family, were also found. One of them in particular --~called the lambda meson and discovered in 1950~-- showed strange properties. First, experiments showed that it was produced much faster than is was decaying, suggesting that the two mechanisms behind each process were completely different. This was one of the first evidences (one of the other being the $\beta$ decay) of a new type of interaction, called today the \textit{weak force} and that is mainly responsible for radioactive decays for instance. Then, as it was also the case for the kaon, the lambda meson was always produced in pairs. In order to explain this property, M. Gell-Mann proposed the introduction of a new quantum number he called strangeness.\\
\\
In 1956, the (electronic) neutrino was finally detected by F. Reines and C. Cowan, confirming Pauli's postulate. Later in the decade, many experiments to test and understand the weak force lead theorists such as Schwinger to suggest the existence of a new hypothetical massive vector particle, called the $W$ boson, that was believed to be the mediator of this interaction. Its relatively high mass and short life-time, however, would make it very difficult to observe with the experimental tools available at that time.\\
\\
The next twenty years were going to be very exciting for particle physicists, as the first sizeable accelerator was built at the Stanford Linear Accelerator Center (SLAC). It was also a very challenging period for theorists as they attempted to classify and organise the ``zoo'' of already-discovered particles. Probably the most notable and elegant contribution in this regard was made by Gell-Mann (and independently by Y. Ne'eman) in 1961. He arranged the known hadrons --~that is, particles governed by the strong force~-- into what he called the ``Eightfold Way'', different geometric shapes classifying hadrons according to their quantum numbers. In order to explain the very curious patterns those particles were fitting into, Gell-Mann and G. Zweig each independently suggested in 1964 that hadrons were in fact composed of smaller elementary blocks, called quarks. There would be three of them --~the up, down and strange quarks~-- and all the possible combinations would result in different particles. However, despite the theoretical beauty of this quark model, the lack of experimental evidence --~no free quark had never been observed~-- coupled with the fact it seemed to violate the Pauli exclusion principle made the particle physicist community quite sceptical.\\
\\
In parallel, S. Glashow proposed in 1961 that the weak and electromagnetic interactions were actually the manifestation of a single force, called the \emph{electroweak force}. This formulation was carried out by A. Salam and J. C. Ward in 1964, and S. Weinberg in 1967, and lead to the prediction of a new mediating, neutral particle, called the $Z$ boson. This theory, called the Glashow-Salam-Weinberg (GSW) model, was very elegant but was failing to explain the non-zero mass of the mediating vectors. A major theoretical breakthrough was made in 1964, when P. Higgs, R. Brout and F. Englert showed how those particles could acquire their mass through a mechanism called symmetry breaking, conjecturing at the same time the existence of a spin-0, neutral boson (today known as the Higgs boson). Another remarkable theoretical consequence of this model was made by Glashow, J. Iliopoulos and L. Maiani in 1964, when they predicted the existence of a fourth quark. Besides, O. W. Greenberg suggested the same year that quarks were carrying an additional quantum number called colour, that would come as a combination of red, green and blue. He proposed this as a possible explanation to how seemingly identical quarks could occupy the same quantum state. These two ideas gave the quark model a second breath, but it is only after the discovery ten years later at SLAC of the $J/\Psi$ meson, a bound state of what appeared to be a new quark, called the charm quark, and its antiparticle, that the quark model was universally accepted, as it was able to perfectly explain the relative long life of the psi meson, as well as predicting many new mesons that were detected not so long after.\\
\\
The beginning of the seventies was a turning point for theoretical particle physics. First, M. Veltam and G. 't Hooft showed that, like QED, the GSW model was renormalisable, opening many new mathematical possibilities. Then, M. Kobayashi and T. Maskawa, as a consequence of their attempt to explain a phenomenon called Charge-Parity violation, postulated the existence of a third generation of quarks. Finally, H. Fritzsch and H. Leutwyler, with the help of Gell-Mann, developed the colour model into what is known today as Quantum Chromodynamics (QCD), the last missing piece of the theory that would ultimately be known as the Standard Model.\\
\\
However, as it is typical in physics, whenever everything finally seemed to be in order, a new discovery would take the community by surprise. This time it was the detection, in 1975 at SLAC again, of what seemed to be a new lepton --~later called the tau lepton~-- and supposedly its corresponding neutrino. Glashow's flavour symmetry was therefore not fulfilled any more, but hope was restored when, in 1977, a fifth quark, called the bottom quark, was detected at Fermilab. Now, if there is one thing we should know, it is that nature works so that the most naive and appealing predictions are also often the correct ones. So undoubtedly, regardless of the prediction of Kobayashi and Maskawa, there had to be a sixth quark out there, but whether it exists or not is something that would have to wait twenty years to be established.\\
\\
At that time particle physicists had actually predicted all elementary particles discovered to date --~even though they were not aware of it~-- and had developed extremely powerful tools in Quantum Field Theory (QFT). And probably for the first time of the century, theory was ahead of experiments. Discoveries were actually expected, instead of taking the scientific community by surprise, as it had often been the case during the previous forty years. Now the only goal that remained was to prove the existence of these hypothetical particles. The first step was to build powerful particle accelerators, such as the Super Proton Synchrotron, built in 1976 at the Organisation europ\'eenne pour la recherche nucl\'eaire (CERN). The ancestor of nowadays proton colliders lead to the discovery in 1983 of the long-awaited $W$ and $Z$ bosons. During the next ten years, the SM passed all the experimental tests with flying colours, and by providing extremely accurate predictions. Besides, although three particles --~the top quark, the tau neutrino and the Higgs boson~-- remained to be discovered, there were already many hints suggesting their existence.\\
\\
And indeed in 1995, the top quark was finally discovered at Fermilab, with the help of the Tevatron, the most powerful particle accelerator at that time. And only five years later, in 2000, the tau neutrino was detected, again at Fermilab. In order to verify even further the validity of the SM and to begin the hunt for the Higgs boson, the Large Hadron Collider (LHC), the world largest and most powerful particle accelerator to date, was built at CERN, and put into service in 2009. The accelerator itself as well as the four main detectors --~ATLAS, CMS, LHCb and Alice~-- are wonders of engineering. They involve collaborations of thousands of physicists whose work contributed, in 2012, to the detection of what is probably considered as one of the most important discovery of mankind, and the last missing piece of the SM: the Higgs boson.\\
\\
With this very last achievement, and the extremely high theoretical precisions it has provided, the SM and its performances are undisputed.\\

\section{Current paradigm in particle physics}\label{Section:INTROMotivations}
\fancyhead[LO]{\ref*{Section:INTROMotivations}~~\nameref*{Section:INTROMotivations}}

Yet, it is also undeniable that the Standard Model bears several major fundamental flaws. It is mostly accepted today that the SM is most likely but a low-energy approximation of an even bigger theory of fields, in which all the three forces it would describe -- as well as possibly gravity -- are unified. Therefore, the main goal of the particle physicists community today is to probe for Beyond Standard Model (BSM) physics, i.e. physics that is not -- or not entirely -- described or explained by the SM.\\
\\
One way of doing so is by increasing the energy and the luminosity at which colliders operate. The LHC, in particular, is currently being upgraded, and its luminosity will increase tenfold by 2026. Several other projects are under consideration, such as the International Linear Collider (Japan)~\cite{Bambade:2019fyw} and the Compact Linear Collider (CERN)~\cite{deBlas:2018mhx,Charles:2018vfv}, two linear electron-positron colliders, and the Future Circular Collider (CERN)~\cite{Mangano:2018mur,Benedikt:2018qee,Benedikt:2018csr,Zimmermann:2018wdi} and the Circular Electron Positron Collider (China)~\cite{CEPC-SPPCStudyGroup:2015csa,CEPC-SPPCStudyGroup:2015esa}, whose precise goals are still to be defined, but that would eventually be much bigger versions of the LHC.\\
\\
Another possibility is to rely on very high precision, which means looking for BSM physics not necessarily at higher energies, but by finding tiny discrepancies between theoretical and experimental results, which could give hints about unknown processes or exotic particles. From the experimental point of view, this is where electron-positron colliders come into play, as they are able to provide very precise measurements (albeit at relatively lower energies). From the theoretical point of view however, increasing the precision of the predictions is far from being a trivial task, as even though the mathematical formulation of the SM is quite simple, it is not possible to solve the equations of motion analytically. Instead, computations rely on what is called perturbation theory. Indeed, because the couplings between interacting particles at colliders' energies are usually much smaller than 1, it is possible to expand a given process into an infinite sum of ``orders'', each consecutive order involving more interactions between particles, but also contributing in general to a lesser extent to the final result. In general, the Leading Order (LO) of the expansion, i.e. its first term, gives a good approximation of the result, but it is necessary to go beyond, e.g. at Next-to-Leading Order (NLO) or Next-to-Next to Leading Order (NNLO), and so on, to achieve the desired precision. The tools available to us nowadays are extremely effective at computing relatively simple processes. As of today, the vast majority of NLO computations can be done by programs such as \textsc{Madgraph}~\cite{Alwall:2014hca} and \textsc{MCFM}~\cite{Campbell:2011bn}, in a completely automated way. Yet, the current experimental precision provided by the LHC requires most of the time to compute processes at least up to NNLO, and/or with more external legs. This can be very challenging from the theoretical point of view, as this means dealing with more scales due to the presence of more particles and internal masses. In the traditional method, the presence of additional scales involve elliptical functions, which can increase by a huge margin the complexity of the computations. So far, most of the $2\longrightarrow2$ processes at NNLO have been computed in the massless limit, but it has become necessary to develop new tools to bypass the difficulties encountered when computing massive processes.\\
\\
New methods have recently been developed in order to overcome the challenges that arise when performing such computations. In this thesis we propose a novel approach to higher-order computations in perturbative theory, called the Four-Dimensional Unsubtraction \cite{Hernandez-Pinto:2015ysa,Sborlini:2016gbr,Sborlini:2016hat,Rodrigo:2016hqc,Hernandez-Pinto:2016uwx,Sborlini:2016bod,Driencourt-Mangin:2016dpf}. It is based on the Loop-Tree Duality theorem \cite{Catani:2008xa,Rodrigo:2008fp,Bierenbaum:2010cy,Bierenbaum:2012th,Buchta:2014dfa,Buchta:2015xda,Buchta:2015wna}, and aims at building quantities that can be evaluated numerically, completely circumventing most complications analytic approaches tend to encounter.

\section{Outline}\label{Section:INTROOutline}
\fancyhead[LO]{\ref*{Section:INTROOutline}~~\nameref*{Section:INTROOutline}}

This thesis is organised as follows. In \Chapter{\ref{Chapter:TBPP}}, we briefly recall the fundamentals of particle physics, from the mathematical formulation of the Standard Model to the evaluation of cross sections. We explain how physical observables in quantum field theory can be constructed from the Lagrangian, through the use of perturbation theory and Feynman diagrams. We study the divergent structure of scattering amplitudes in \Chapter{\ref{Chapter:SAR}}, where we give a succinct overview of regularisation in perturbation theory. There we introduce the traditional method known as Dimensional Regularisation, and explain how it can be used to remove the infinities appearing at intermediate steps of the calculation. In \Chapter{\ref{Chapter:LTD}}, we introduce the Loop-Tree Duality (LTD) theorem and formalism, which will serve as a foundation for the rest of this thesis. We start by deriving the duality relation at one-loop level, and then extend it to multi-loop diagrams, and diagrams involving multiple identical propagators. In \Chapter{\ref{Chapter:FDU}}, we present the Four-Dimensional Unsubtraction (FDU) method, and its very first application to a physical process. We start by illustrating, using a toy model as an example, the fundamental concepts of FDU, which will then be put into practice to calculate the decay rate of a virtual photon into a pair of massless quarks, at NLO. The FDU framework is extended to massive particles in \Chapter{\ref{Chapter:FDUM}}. After explaining how to solve the difficulties introduced by the presence of masses, we compute numerically the decay rate of scalar and vector bosons into a pair of massive quarks, again at NLO. In \Chapter{\ref{Chapter:HOL}}, we present new advantages of the LTD formalism, by considering the Higgs boson production through gluon fusion and decay to two photons amplitudes, at one-loop level. We show how LTD allows for the amplitude to be written in a universal form at the integrand level, regardless of the nature of the particle running inside the loop. We also perform straightforward integrand-level asymptotic expansions in the internal mass, without having to rely on the expansion-by-region approach. We apply for the first time the LTD theorem at two-loop level in \Chapter{\ref{Chapter:HTL}}, by considering the Higgs boson decay to two photons amplitude, and show that the universality of the functional form still holds at this order. There, we present a new algorithm to locally regularise ultraviolet divergences for two-loop processes, allowing for a direct numerical implementation of the computation. We summarise our work in \Chapter{\ref{Chapter:CONCLUSION}}, and discuss future potential directions of research.\\
\\
Most algebraic manipulations and computations related to this thesis have been carried out using \textsc{Mathematica}, and especially the \textsc{FeynArts}~\cite{Hahn:2000kx} and \textsc{FeynCalc}~\cite{Mertig:1990an, Shtabovenko:2016sxi} packages. The Feynman diagrams in this thesis were drawn using graphical tools provided by the Axodraw 2 \LaTeX~package~\cite{Collins:2016aya}.

\chapter{Theoretical basis of particle physics}\label{Chapter:TBPP}
\thispagestyle{fancychapter}
\fancyhead[RE]{\nameref*{Chapter:TBPP}}

In this chapter we briefly introduce the mathematical formalism behind the Standard Model. We start by formulating QED and QCD through their Lagrangians, and from them we explain how to compute scattering amplitudes and ultimately cross sections. We also give some details about the generic strategies to calculate these amplitudes through the use of perturbation theory and Feynman diagrams.

\section{The Standard Model}\label{Section:TBPPSM}
\fancyhead[LO]{\ref*{Section:TBPPSM}~~\nameref*{Section:TBPPSM}}

The Standard Model (SM) is a wonderful theory that has been developed, extended and refined by many physicists over the last half century. It is now able to successfully describe three of the four interactions of nature, which are the electromagnetic force, the strong force, and the weak force. It is able to explain the vast majority of the experimental results in particle physics and has successfully predicted, before they were even discovered, the existence of several elementary particles, namely the charm and top quarks, the gluon, the $W^\pm$ and $Z$ bosons, and the Higgs boson. Furthermore, the theoretical predictions are consistent with observations to a precision far exceeding all other scientific fields, in certain circumstances showing agreement to one part in $10^{-13}$.\\
\\
The SM is a relativistic Quantum Field Theory (QFT) defined by the local
\begin{equation}\label{Equation:TBPPGauge}
SU(3)_C\,\otimes\,SU(2)_L\,\otimes\,U(1)_Y
\end{equation}
gauge symmetry, where $SU(3)_C$ is the gauge group of the strong interaction, and $SU(2)_L\,\otimes\,U(1)_Y$, the gauge group of the electroweak interaction. A very remarkable feature of the SM is that all the interactions it describes can be mathematically defined by imposing its Lagrangian to be locally symmetric, i.e. invariant under any $SU(3)_C\,\otimes\,SU(2)_L\,\otimes\,U(1)_Y$ local transformation\footnote{This means one can choose a different transformation parameter at each space-time point.}. As it is going to be explained in the following, this automatically leads to the emergence of what are called gauge fields --~bosons that are carriers of the distinct forces~-- and in the meantime provides the mathematical rules describing these interactions.

\subsection{Quantum Electrodynamics}{\label{Subsection:TBPPQED}}

QED is an abelian $U(1)$ gauge theory describing the electromagnetic interaction between fermions, particles of spin $\frac{1}{2}$, and photons. It is the relativistic counterpart of the classical theory of electromagnetism developed by J. C. Maxwell during the nineteenth century, and is the first theory to successfully combine quantum mechanics and special relativity.\\
\\
The QED Lagrangian, obtained by imposing the Lagrangian of a given free Dirac fermion to be invariant under a local $U(1)$ symmetry, is written
\begin{equation}\label{Equation:TBPPQEDLagrangian}
\mathcal{L}_{QED}=\bar{\psi}(i\,\gamma^\mu D_\mu-m)\psi-\frac{1}{4}F_{\mu\nu}\,F^{\mu\nu}\;,
\end{equation}
where $\psi$ is the fermion field, $D_\mu$ is the covariant derivative, defined as
\begin{equation}\label{Equation:TBPPCovarQED}
D_\mu=\partial_\mu-i\,e\,A_\mu\;,
\end{equation}
with $e$ the electromagnetic charge of the fermion field, and $F_{\mu\nu}$ is the electromagnetic field strength tensor, written
\begin{equation}\label{Equation:TBPPFieldTensorQED}
F_{\mu\nu}=\partial_\mu A_\nu-\partial_\nu A_\mu\,.
\end{equation}
The vector $A_\mu$ is interpreted as the electronic field generated by the fermion, i.e. the photon field, and naturally appears in the Lagrangian by imposing gauge invariance.\\
\\
In QED, it is easy to show that the photon has to be massless -- which is indeed what is to be expected -- as a photon mass term inside the Lagrangian would not be invariant under a local transformation.

\subsection{Quantum Chromodynamics}\label{Section:TBPPQCD}

QCD is a non-abelian $SU(3)$ gauge theory describing the strong interaction between quarks and gluons. QCD is very similar to QED in the sense that the steps to achieve a mathematical representation of the strong interaction are exactly the same as with the electromagnetic interaction. The main difference however, is that, as said above, QCD is non abelian. Although seemingly only a mathematical peculiarity, it gives very interesting and exclusive physical properties to the strong force, as we will see in the following.\\
\\
Derived in a similar way as the QED Lagrangian, the QCD Lagrangian reads
\begin{equation}\label{Equation:TBPPQCDLagrangian}
\mathcal{L}_{QCD}=\bar{\psi}(i\,\gamma^\mu D_\mu-m)\psi-\frac{1}{4}G_{\mu\nu}^a\,G^{\mu\nu}_a\;,
\end{equation}
where $\psi$ is the quark field, the covariant derivative $D_\mu$ is defined as
\begin{equation}\label{Equation:TBPPCovarQCD}
D_\mu=\partial_\mu-i\,g_S\,A_\mu^a\,T_a\;,
\end{equation}
with $g_S$ being the strong coupling, and the strong field strength tensor is written
\begin{equation}\label{Equation:TBPPFieldTensorQCD}
G_{\mu\nu}^a=\partial_\mu A_\nu^a-\partial_\nu A_\mu^a+g_S\,f_{~bc}^a\,A_\mu^b\,A_\nu^c\,.
\end{equation}
The tensors $A_\mu^a$ are the gluon fields in the adjoint representation of the $SU(3)$ gauge group, and are labelled with the index $a$. Unlike for QED -- for which there is only one mediating field -- there are eight different gluon fields in QCD. This is simply due to the fact $SU(3)$ possesses eight generators\footnote{One can indeed show that the group $SU(N)$ has $N^2-1$ generators, for $N\geq2$.}, written $\mathbf{T}^a$, and fulfilling
\begin{equation}\label{Equation:TBPPSU3Prop}
\Tr(\mathbf{T}^a\,\mathbf{T}^b)=\frac{1}{2}\,\delta^{ab}\qquad\text{and}\qquad[\mathbf{T}^a,\mathbf{T}^b]=i\,f^{abc}\,\mathbf{T}_c\;,
\end{equation}
with $\delta^{ab}$ being the Kronecker delta, and $f^{abc}$ the structure constants of $SU(3)$. This feature, coming from the non-abelian characteristic of QCD, gives birth to the third term in the right-hand side of \Eq{\ref{Equation:TBPPFieldTensorQCD}}, which is particular and essential to QCD as a theory, as it is at the origin of the gluon self-interaction.\\
\\
As a consequence of this, QCD exhibits two remarkable properties known as \emph{confinement} and \emph{asymptotic freedom}. Both phenomena are explained by what is called the running of the coupling $\alpha_S=g_S^2/(4\pi)$, whose value differs depending on the energy scale $\mu$ of the process. One can show for instance that, for $N_C$ colours and $n_f$ different flavours of quarks, $\alpha_S(\mu)$ is given by
\begin{equation}\label{Equation:TBPPRunningQCD}
\alpha_S(\mu)\approx\frac{\alpha_S(\mu_0)}{1+\alpha_S(\mu_0)(11N_C-2n_f)\log(\mu^2/\mu_0^2)/(12\pi)}\;,
\end{equation}
where $\mu_0$ is a reference constant. This means that, given $N_C=3$ and $n_f=6$ in the SM, the higher the energy of the process -- or, equivalently, the lower the distance between two interacting particles -- the lower $\alpha_S$, and vice versa.\\
\\
On one hand, confinement takes place at low energy, or high distances. Although there is for the moment no analytical proof of this phenomenon, is it still possible to give a qualitative explanation. When two colour-charged particles are far enough from one another, the potential energy between them is such that it leads to the creation of a new pair of particles with each one of them biding with one of the original particles. This means that quarks and gluons can never be completely separated and isolated from another hadron, and therefore that in nature, only colourless particles can be seen and detected.\\
\\
On the other hand, asymptotic freedom occurs at high energies. When colour-charged particles are close to each other, the coupling becomes very small, which makes those particles behave as if they were free. This is a very important feature from the theory point of view as it allows perturbative computations, as it will be seen in \Section{\ref{Section:TBPPPTFD}}.

\section{Building a physical observable}\label{Section:TBPPPhysObs}
\fancyhead[LO]{\ref*{Section:TBPPPhysObs}~~\nameref*{Section:TBPPPhysObs}}

The non-deterministic aspect of quantum mechanics -- and consequently of any quantum field theory -- means that ultimately, we can only hope to compute the \emph{probability} of a given outcome. In the context of particle physics, what is usually worth measuring is a \emph{cross section}, which is a quantity expressing the probability two particles (or more) interact. When they do interact, these particles are transformed through a process called scattering, whose mathematical description is achieved through the use of the so-called Scattering matrix, or $S$-matrix. The elements composing this matrix are \emph{scattering amplitudes} describing all possible interactions that are allowed in a given theory.\\
\\
In the following, we will briefly enumerate the successive steps needed to construct relevant mathematical quantities starting from a ``raw'' theory, usually described by a Lagrangian such as the ones we wrote before in this chapter. We will try to stick to the bare minimum, but for a much more exhaustive and detailed discussion of this subject, we refer the reader to e.g.~\cite{Peskin:1995ev,Ryder:1985wq}.\\

\subsection{Scattering amplitudes and cross sections}\label{Section:TBPPSACS}

Say we want to find the probability of transition between an initial state $|\,i\,\rangle=|\{p_{i_1},p_{i_2},\dots\}\rangle$ and a final state $|\,f\,\rangle=|\{p_{f_1},p_{f_2},\dots\}\rangle$, both defined by given fields as well as a given configuration of the momenta of those fields. These states must be assumed to be asymptotically free --~i.e. they must correspond to states of the theory which are not yet or not any more being affected by the interaction. The \emph{probability amplitude} describing the evolution of this system is given by the $S$-matrix element
\begin{equation}\label{Equation:TBBPSfi}
S_{fi}=\langle\,f\,|\,S\,|\,i\,\rangle=\delta_{fi}+i\,T_{fi}\;,
\end{equation}
where the term $\delta_{fi}$ describes the unaltered propagation of the two states, while the interaction is described by the term
\begin{equation}\label{Equation:TBBPTfi}
i\,T_{fi}=i(2\pi)^4\delta\left(\,\sum\limits_j\,p_{f_j}-\,\sum_k\,p_{i_k}\right)\mathcal{M}(i\to f)\;,
\end{equation}
where the four-momentum conservation is enforced by the presence of the $\delta$ function. Inside $T_{fi}$ are encoded all possible paths (some more likely than others) that can follow the interacting particles to go from state $|\,i\,\rangle$ to state $|\,f\,\rangle$, whose probabilities are given by the matrix element $\mathcal{M}(i\to f)$.\\
\\
Generally in physics, theories are confronted with experiments via the use of measurable physical quantities called \emph{observables}. In particle physics, the total cross section, usually written $\sigma$, acts as such an observable. The differential cross section $d\sigma$ of a given process can be obtained from the corresponding $S$-matrix elements $S_{fi}$ through the relation
\begin{equation}
d\sigma=\frac{|\mathcal{M}(i\to f)|^2}{F}\,d\Omega
\end{equation}
where $F$ is a quantity called \emph{flux factor} that only depends on the kinematics of the process, and $d\Omega$ is the solid angle element. The total cross section is then obtained by integrating over the desired final phase space, depending on the observable that has been chosen beforehand. The biggest obstacle is that, as already said before, it is impossible (rather, we do not know how) to calculate an exact analytic expression for $|T_{fi}|^2$ and therefore for the cross section $\sigma$, in the general case. Yet, and luckily for us, it is still possible to obtain an \emph{approximation} of $\sigma$. The approach used to do so is called \emph{perturbation theory}, whose principles will be explained in the following.

\subsection{Perturbation theory and Feynman diagrams}\label{Section:TBPPPTFD}

Perturbation theory is a method used to obtain an approximate solution to a mathematical problem whose formulation is in general too complex, or for which the tools needed to obtain an analytic solution simply do not exist. Its principle is to consider a known solution for a given --~usually simpler~-- configuration of a problem, and to study how a small perturbation affects the solution. This leads to an expression in terms of a power series in a relatively small parameter of the problem.\\
\\
In the simplest case of a generic QFT with a single interaction whose corresponding coupling is $g$, for instance, a probability amplitude $\mathcal{A}$ can be decomposed as
\begin{equation}\label{Equation:TBPPPowerSeriesOfA}
\mathcal{A}=\mathcal{A}^{(0)}+g^2\,\mathcal{A}^{(1)}+g^4\,\mathcal{A}^{(2)}+\cdots\;,
\end{equation}
where $\mathcal{A}^{(n)}$ is the $n$-th order of the perturbative expansion of $\mathcal{A}$. If $g$ is small enough, each consecutive order of the expansion contributes less and less to the total amplitude, meaning that the more orders one considers, the closer one gets to the actual result\footnote{While it is what one should indeed expect from the physical point of view, the rigorous demonstration of this statement in the framework of QFT is highly non-trivial, and requires very sophisticated mathematical techniques. The reader can find more details in e.g. \cite{Peskin:1995ev}.}.\\
\\
One remarkable characteristic of the SM Lagrangian is that under certain conditions\footnote{When it comes to perturbation theory in the SM, QCD is usually the main bottleneck, since $\alpha_S$ is quite big compared to the other coupling constants. For hadron colliders like the Tevatron or the LHC, it is only for energies around 5-10GeV that we start obtaining a good approximation.}, the interactions between particles and gauge bosons are relatively weak, legitimising the use of perturbation theory. The formalism involved in the enumeration and computation of all the configurations that must be taken into account at a given order in the SM, though, is quite complex mathematically speaking. Nevertheless, the integrals that appear in such computations have particular structures that allow them to be easily represented by what are called Feynman diagrams, invented by the eponymous physicist. They are pictorial descriptions of the different terms that appear in a perturbative expansion such as the one in \Eq{\ref{Equation:TBPPPowerSeriesOfA}}. Note that these diagrams do not actually illustrate what happens in nature, but are merely representations of mathematical equations describing the interaction under consideration.
\begin{figure}[h]
	\centering
	\begin{equation*}
	\mathcal{A}=
	\underbrace{\makebox[0pt][l]{\phantom{$\frac{1}{1+\frac{1}{1+\frac{1}{1}}}$}}
		\begin{picture}(64,20)(0,0)
		\ArrowLine[arrowscale=0.75](2,3)(62,3)
		\end{picture}}_{\mathcal{A}^{(0)}}
	+
	\underbrace{\makebox[0pt][l]{\phantom{$\frac{1}{1+\frac{1}{1+\frac{1}{1}}}$}}
		\begin{picture}(64,20)(0,0)
		\ArrowLine[arrowscale=0.75](2,3)(62,3)
		\PhotonArc(32,3)(14,0,180){3}{5.5}
		{\SetColor{red}
			\Vertex(18,3){1.5}
			\Vertex(46,3){1.5}
		}
		\end{picture}}_{e^2\,\mathcal{A}^{(1)}}
	+
	\underbrace{\makebox[0pt][l]{\phantom{$\frac{1}{1+\frac{1}{1+\frac{1}{1}}}$}}
		\begin{picture}(64,20)(0,0)
		\ArrowLine[arrowscale=0.75](2,3)(62,3)
		\PhotonArc(32,3)(23,0,180){3}{7.5}
		\PhotonArc(32,3)(11,0,180){2}{5.5}
		{\SetColor{red}
			\Vertex(9,3){1.5}
			\Vertex(21,3){1.5}
			\Vertex(43,3){1.5}
			\Vertex(55,3){1.5}
		}
		\end{picture}
		+
		\begin{picture}(64,20)(0,0)
		\ArrowLine[arrowscale=0.75](2,3)(62,3)
		\PhotonArc(27,3)(11,0,180){2}{5.5}
		\PhotonArc(37,3)(11,180,0){2}{5.5}
		{\SetColor{red}
			\Vertex(16,3){1.5}
			\Vertex(26,3){1.5}
			\Vertex(38,3){1.5}
			\Vertex(48,3){1.5}
		}
		\end{picture}
		+
		\begin{picture}(64,20)(0,0)
		\ArrowLine[arrowscale=0.75](2,3)(62,3)
		\PhotonArc(37,3)(14,0,78){2.5}{3}
		\PhotonArc[clock](27,3)(14,180,102){2.5}{3}
		\Arc[arrow,arrowpos=0.25,arrowscale=0.75](32,15)(8,0,360)
		{\SetColor{red}
			\Vertex(13,3){1.5}
			\Vertex(24,15){1.5}
			\Vertex(40,15){1.5}
			\Vertex(51,3){1.5}
		}
		\end{picture}}_{e^4\,\mathcal{A}^{(2)}}
	+\ldots
	\end{equation*}
	\caption{First three orders of the perturbative expansion of the fermionic propagator in QED. Each red dot represents an interaction, and therefore introduces an additional power in the coupling constant $e$.}
	\label{Figure:TBPPFeynmanDiagramsExample}
\end{figure}\\
In Fig.~\ref{Figure:TBPPFeynmanDiagramsExample}, we draw the Feynman diagrams contributing to the first three orders of the fermionic propagator. The very first order, called \emph{Leading order} (LO) or \emph{tree-level}\footnote{To be more precise, tree-level means \emph{without loops}. For most processes, leading order and tree-level are equivalent. There are cases though, such as the one studied in \Chapter{\ref{Chapter:HOL}}, for which there are no tree-level contributions as the leading order already involves at least one loop.} contains all the diagrams with the minimum possible amount of vertices (a vertex representing a point-like interaction between three or more particles) for the process under consideration. In the particular case of Fig.~\ref{Figure:TBPPFeynmanDiagramsExample}, there is only one diagram and the amount of vertices is zero. The second order is the \emph{Next-to-Leading order} (NLO), and is the first order \emph{correcting} the LO. In our example, again, there is also only one contributing diagram, containing this time two vertices. The third order is the \emph{Next-to-Next-to-Leading order} (NNLO), and so on... Even though higher orders involve more and more diagrams, these diagrams contain more vertices, and therefore more powers of the coupling -- that is much smaller than one. Consequently, they are expected to contribute less to the amplitude $\mathcal{A}$ than the previous orders.\\
\\
While the mathematical formulation behind Feynman diagrams is quite complex, constructing them is actually fairly easy. They are drawn using a set of fundamental building blocks, each of them associated to different mathematical quantities through what we call \emph{Feynman rules}. For a given theory, these rules can be computed from the corresponding Lagrangian. In QED for instance, we can compute the Feynman rules of the fermion propagator, the photon propagator and the electromagnetic interaction. The complete SM has much more of them, but only a handful (found in \Appendix{\ref{Section:APPFeynmanRules}}) will be used in this thesis.\\
\\
Even though the tools at our disposal to generate higher orders in perturbative QFT are very elegant, there is one major issue. When constructing higher-order diagrams using Feynman rules, we start encountering \emph{loops} that contain internal particles with arbitrary four-momenta. Because we must take into account all possible configurations, integrals over these loop momenta will appear. Performing such computations naively, however, can lead to infinities. Of course, as physical observables must have finite values, it is essential to get rid of these infinities. The different steps and methods needed to achieve this will be discussed in the next chapter.

\chapter{Singularities and regularisation}\label{Chapter:SAR}
\thispagestyle{fancychapter}
\fancyhead[RE]{\nameref*{Chapter:SAR}}

Higher-order computations in perturbative gauge theories such as the SM involve most of the time dealing with ill-defined objects, i.e. mathematical quantities that generate infinities when evaluated in four space-time dimensions. In physics, these infinities are often called \emph{singularities}, or \emph{divergences}. The procedure of removing singularities is known as \emph{regularisation}, and has been a very active field of research for the last half-century. In this chapter, we will cover the different types of divergences one can encounter at intermediate steps when dealing with higher-order computations. We will also introduce the traditional regularisation approach known as Dimensional Regularisation (DREG), and will briefly enumerate the general methods used to deal with infinities in this framework.

\section{The different types of divergences}\label{Section:SARDivergences}
\fancyhead[LO]{\ref*{Section:SARDivergences}~~\nameref*{Section:SARDivergences}}

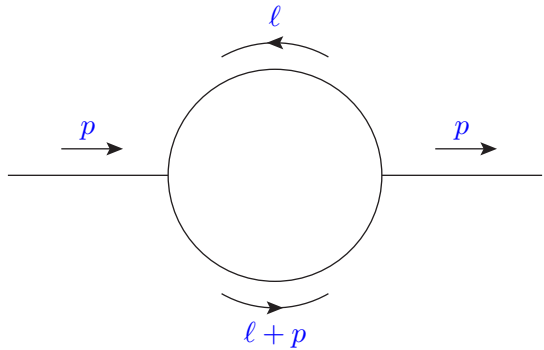
\begin{figure}[h]
	\centering
	\begin{picture}(200,150)(0,-75)
	\Line(0,0)(60,0)
	\LongArrow(20,10)(40,10)
	\Line(140,0)(200,0)
	\LongArrow(160,10)(180,10)
	\Arc(100,0)(40,0,360)
	\Arc[arrow,arrowpos=0.5](100,10)(40,60,120)
	\Arc[arrow,arrowpos=0.5](100,-10)(40,240,300)
	\color{blue}
	\Text(30,17){$p$}
	\Text(170,17){$p$}
	\Text(100,60){$\ell$}
	\Text(100,-60){$\ell+p$}
	\end{picture}
	\caption{The Feynman diagram of the massless scalar two-point function, with external momentum $p$ and loop momentum $\ell$.}
	\label{Fig:SARS2PFDiagram}
\end{figure}

\begin{figure}[h]
	\centering
	\begin{picture}(210,130)(-25,0)
	\Gluon(80,72)(15,72){5}{4}
	\Gluon(80,72)(135,127){5}{5}
	\Gluon(80,72)(135,17){-5}{5}
	\Gluon(98,47)(135,84){5}{3}
	\color{blue}
	\Text(145,17){$p$}
	\Text(145,85){$p_r$}
	\end{picture}
	\hfill
	\begin{picture}(210,130)(0,0)
	\Photon(15,72)(80,72){5}{4}
	\ArrowLine(80,72)(135,127)
	\ArrowLine(135,17)(80,72)
	\Photon(101,51)(135,85){5}{3}
	\color{blue}
	\Text(145,17){$p$}
	\Text(145,85){$p_r$}
	\end{picture}
	\caption{Feynman diagrams illustrating the emission of a massless particle (here a gluon or a photon). If the integration is performed over the low-energy region of the phase space, the left diagram will generate both soft and collinear singularities, while the right diagram will only generate soft singularities, assuming the fermion is massive.}
	\label{Figure:SARMasslessEmission}
\end{figure}
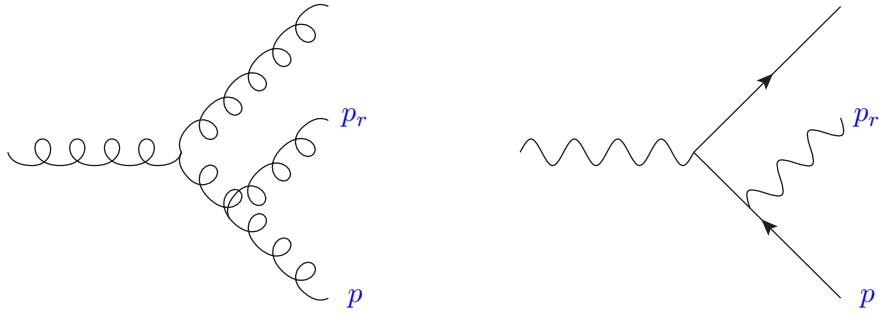

When a loop appears in a Feynman diagram, there is an implicit additional integration that must be performed over the loop four-momentum. Since the integration domain is unrestricted, this momentum can have arbitrary large energies, including ones at which we know the SM is not a valid theory any more\footnote{For instance, our description of fundamental interactions cannot be any more valid above the Planck scale, since quantum gravitational effects would have to be taken into account.}. Depending on the process under consideration, this can lead to infinities when trying to evaluate the integral in the high-energy region. In this case, and for this very reason, we refer to these divergences as \emph{ultraviolet} (UV) singularities. As an explicit example, let's consider the most basic one-loop divergent diagram, namely the massless scalar two-point function $L^{(1)}_{Bubble}$, shown in \Fig{\ref{Fig:SARS2PFDiagram}}.\\
\\
Explicitly, we have
\begin{equation}\label{Equation:SARLBubble}
L_{Bubble}^{(1)}(p,-p)=i\int\,\frac{d^4\ell}{(2\pi)^4}\frac{1}{(\ell^2+i0)((\ell+p)^2+i0)}\;.
\end{equation}
Simply by counting the powers of $\ell$ coming from the measure and the denominators, we can see that the integral appearing in \Eq{\ref{Equation:SARLBubble}} does not converge for large\footnote{Note that $\ell$ being a four-vector in a Minkowski space of signature (1,-1,-1,-1), associating a magnitude to it does not make much sense a priori. It is understood here that ``large $\ell$'' implies ``large energy component for $\ell$''.} $\ell$. In this particular case, it is logarithmically divergent, which means that if we were to integrate for energies only up to a cutoff $\Lambda$, the integrated results would exhibit a term proportional to $\log(\Lambda)$, which indeed goes to infinity when the limit $\Lambda\to\infty$ is taken. A logarithmic divergence is the ``smallest'' UV divergence one can encounter; anything less divergent would automatically lead to a finite result. On the other hand, there exists much more divergent amplitudes, exhibiting for instance linear and quadratic divergences (that would go as $\Lambda$ and $\Lambda^2$, respectively). In any case, removing UV singularities in practice is roughly equivalent to removing any divergent term in this limit. The use of a cutoff to achieve this, however, is not very practical for several reasons; we will see in the following -- and in many places in this thesis -- that there are more advanced and efficient techniques to deal with UV singularities.\\
\\
Another type of infinities, called \emph{infrared} (IR) singularities, are associated this time with the low-energy region of the integration domain. These divergences are related with the degeneration of observable states and can only emerge in theories with massless fields. In order to understand how such infinities are generated in practice, we will work with an actual example. Let's consider the situation in which a massless particle with four-momentum $p_r$ is emitted from another particle with four-momentum $p$, as shown in \Fig{\ref{Figure:SARMasslessEmission}}. A first issue arises from the fact that the integration domain of the external particles (i.e. the phase space) includes regions in which the radiated particle can have zero energy -- and therefore zero four-momentum (i.e. $p_r=0$), since it is on shell. While from the experimental point of view, there is no difference between no emission and the emission of a particle with zero energy, the limit in the theoretical point of view is not mathematically well-defined and gives rise to infinities at intermediate steps of the calculation. In this particular case, we refer to these divergences as \emph{soft} singularities. A very similar issue can occur if the radiating field is also massless. In this case, infinities will be generated when the radiated particle is emitted in the same direction as the radiating one (i.e. $p_r\parallel p$ or, equivalently, $p_r\cdot p=0$). As for the first case, there is no experimental difference between two massless particles with four-momenta $p$ and $p_r$ travelling very close from each other, or a single one whose momenta is $p+p_r$. But once again, the limit is not mathematically well-defined and leads to infinities when studying this kind of configurations. This second type of IR divergence is called \emph{collinear singularities} (sometimes also referred to as \emph{mass singularities}).\\
\\
Additionally, it is very important to note that loop diagrams can also generate IR singularities. They come in general from the low-energy region of the loop four-momentum ($\ell\to0$), when one or more internal lines are massless. We will see in \Section{\ref{Section:SARIR}} that there is a fundamental connection between IR divergences coming from real emissions, and the ones coming from virtual (i.e. loop) amplitudes.\\
\\
Finally, a third type of singularities can manifest themselves, when computing loop diagrams involving at least two scales. In this case, and since the virtuality of the internal particle is arbitrary, some denominators can vanish in points of the integration space that are not end points. This generates what are called \emph{threshold singularities}. Unlike UV or IR singularities though, the threshold singularities are actually integrable.\\
\\
The three types of divergence we just discussed about are the most common one will encounter when computing higher-order contributions in perturbation theory, and loop diagrams in general. Other types of divergences also exist, but usually only appear under certain conditions, or when working with modified frameworks. In \Chapter{\ref{Chapter:LTD}} for instance, we will encounter unphysical spurious threshold singularities inside intermediate expressions, and will see that they naturally cancel themselves at the end of the computation. There are also singularities arising from a particular choice of gauge for instance, such as the light-cone gauge that introduces an additional denominator that is linear in the momentum of the propagating particle.\\
\\
To compute observables that have a physical relevance, UV and IR singularities must be removed before the final step of the computation. In the following, we will see the most common methods used to deal with divergences in general.

\section{Regularisation schemes}\label{Section:SARRegSch}
\fancyhead[LO]{\ref*{Section:SARRegSch}~~\nameref*{Section:SARRegSch}}

A regularisation scheme is a framework used to deal with singularities of different types, and ultimately to avoid or remove them. The regularisation process will most of the time depend on which singularities are to be removed. For instance, the introduction of a cutoff $\Lambda$ to limit the integration space, as briefly explained earlier, is one possibility to deal with UV singularities. Another possibility -- known as the Pauli-Villars regularisation -- is to introduce and associate to each particle of the theory an unphysical \emph{ghost} particle, with arbitrary large mass (also usually written $\Lambda$), designed so that their loop contributions cancel the UV behaviour of physical loop amplitudes. Note that in both these regularisation methods, one has to keep $\Lambda$ finite at intermediate steps, and make every dependence in this regulator vanish at the end of the computation to obtain a meaningful physical result. There also exists other completely different approaches, such as for instance lattice regularisation. The idea of this method is to discretise the space-time onto a lattice of a given grid size, which acts as a natural cutoff for the momentum of the propagating particles. This modifies the definition of the propagator, and allows for the computation of otherwise divergent probability amplitudes, as a function of the grid spacing. The continuous (i.e. physical) limit is recovered by extrapolating the grid spacing to 0.\\
\\
For IR singularities, similar ideas can be developed. As we saw earlier in \Section{\ref{Section:SARDivergences}}, IR singularities appear at intermediate steps when dealing with massless particles. One possibility to completely avoid this kind of divergences is simply to give an infinitesimally small mass to these particles. This mass would act as a lower regulator, preventing altogether the emergence of infinities in soft or collinear configurations.\\
\\
However, the methods we just described usually introduce several problematic issues. They may for instance violate fundamental symmetries of the theory such as gauge invariance or Lorentz invariance. In some cases, this can mean the necessity of performing additional complex steps in order to recover the expected result. There exists an alternative and very effective approach, tough, that does not encounter such issues. Its principles will be introduced in the next section.

\subsection{Dimensional regularisation}\label{Section:SARDREG}

Dimensional regularisation (DREG) is a regularisation scheme that was introduced independently by Giambiagi and Bollini~\cite{Bollini:1972ui}, and by `t Hooft and Veltman \cite{tHooft:1972tcz} in the seventies. It is without any doubt the most popular regularisation method used by particle physicists nowadays, and has so far proven to be a very powerful tool for computing higher-order amplitudes in perturbation theory.\\
\\
The principle of DREG is to alter the dimension of space-time and momentum space, from 4 to an arbitrary $d\in\mathbb{C}$, and to modify loop and phase-space measures accordingly. For loop integrals, for instance,
\begin{equation}\label{Equation:SARdDimensionalMeasure}
\int\,\frac{d^4\ell}{(2\pi)^4}~~\longrightarrow~~\mu^{4-d}\int\,\frac{d^d\ell}{(2\pi)^d}\;.
\end{equation}	
The arbitrary mass $\mu$ is introduced to compensate the extra dimensions generated by the $d$-dimensional integration measure. Note that while the dimension $d$ can a priori take any value, a space (or space-time) of a finite non-integer dimension would not make much sense. This is why it is important to remember that this ``trick'' is just a purely formal way of expressing mathematical quantities. The main idea behind DREG is that for any loop or phase-space integral, there exists a value of $d$ for which this integral converges (for example, the integral appearing in \Eq{\ref{Equation:SARLBubble}} is convergent for $\ReText(d)<4$). This means that one can work with mathematically well-defined quantities by keeping the dependence in $d$ at intermediate steps of the computation, and only take the limit $d\to4$ after having regularised the divergent terms.\\
\\
In order to get a grasp on the method, we will apply it to two very basic examples. Let's first consider the function
\begin{equation}\label{Equation:SARExample1Functionf}
f(x)=\frac{1}{x+1}\;,\quad\text{for}\quad x\in\mathbb{R^+}\;.
\end{equation}
The function $f$ is clearly not integrable around infinity (we can say this is the one-dimensional equivalent of a UV singularity). Now if we assume $x$ to instead live in a $d$-dimensional space, with $d\in\mathbb{C}$, we can examine the integral
\begin{equation}\label{Equation:SARExample1Integral}
I_d=\int_{(\mathbb{R}^+)^d}\,f(x)\,d^dx=c_d\int_0^\infty\,\frac{x^{d-1}}{x+1}\,dx\;,
\end{equation}
with $c_d$ being some volume factor depending on the dimension $d$. This quantity is only well-defined if $\ReText(d)<1$, and leads after integration to
\begin{equation}\label{Equation:SARExample1Integrated}
I_d=\left.c_d\,\frac{\pi}{\sin(\pi d)}\right|_{\ReText(d)<1}\;.
\end{equation}
By expanding \Eq{\ref{Equation:SARExample1Integrated}} around $d=1$, we have
\begin{equation}\label{Equation:SARExample1Expanded}
I_d=-\frac{1}{d-1}+\mathcal{O}\left((d-1)^0\right)\;.
\end{equation}
where the divergence for $d\to1$ explicitly appears in the first term, which we call a \emph{pole}. It is important to notice that while we had to assume $\ReText(d)<1$ to integrate \Eq{\ref{Equation:SARExample1Integral}}, there is nothing actually preventing $\ReText(d)$ in \Eqs{\ref{Equation:SARExample1Integrated}}{\ref{Equation:SARExample1Expanded}} from taking values greater than 1. As a matter of fact, by carrying out an analytical continuation of the integral\footnote{We would like to point out, however, that showing that Feynman integrals written in terms of the parameter $d$ can be analytically continued is far from being an easy task. We refer the interested reader to \cite{Collins:1986a} where very complete demonstrations can be found.} in \Eq{\ref{Equation:SARExample1Integral}}, it is possible to extend the results in \Eq{\ref{Equation:SARExample1Integrated}} to any value of $d\in\mathbb{C}$. This is a very important property of DREG, because it allows in practice for the calculation of any integral in $d$ dimensions without worrying about the potential conditions on the values of $d$, which can therefore remain completely arbitrary.\\
\\
Now let's work with a very similar example, and consider
\begin{equation}\label{Equation:SARExample2Functiong}
g(x)=\frac{1}{x}\;,\quad\text{for}\quad \{x\in\mathbb{R^+}~|~x\neq0\}\;,
\end{equation}
and
\begin{equation}\label{Equation:SARExample2Integral}
J_d=\int_{(\mathbb{R}^+)^d}\,g(x)\,d^dx=c_d\int_0^\infty\,\frac{dx}{x^{2-d}}\;.
\end{equation}
Note that the quantity in \Eq{\ref{Equation:SARExample2Integral}} is \emph{a priori} not well-defined, since there is no value of $\ReText(d)$ for which we have convergence both around 0 and infinity. Yet, we will see that within DREG, we can consider this integral to be \emph{a posteriori} well-defined. By splitting the integrand into two pieces, namely
\begin{equation}\label{Equation:SARExampleSplitted}
J_d=c_d\int_0^1\,\frac{dx}{x^{2-d}}+c_d\int_1^\infty\,\frac{dx}{x^{2-d}}=J_d^{[0,1]}+J_d^{[1,\infty)}\;,
\end{equation}
and by assuming $\ReText(d)>1$ for the first integral, and $\ReText(d)<1$ for the second, we have convergence for both of them independently, with
\begin{align}\label{Equation:SARExampleSplittedIntegrated}
J_d^{[0,1]}=&~\left.\frac{c_d}{d-1}\right|_{\ReText(d)<1}\;,\nn\\
J_d^{[1,\infty)}=&~\left.\frac{c_d}{1-d}\right|_{\ReText(d)>1}\;.
\end{align}
In this case, the integral $J_d^{[0,1]}$ exhibits a IR pole, and $J_d^{[1,\infty)}$ exhibits a UV pole. Now, by using the arbitrariness of $d$ in \Eq{\ref{Equation:SARExampleSplitted}}, we have\footnote{While this reasoning may seem a bit dubious and questionable since we needed to make incompatible assumptions on $d$, the main argument is that we can perform an analytical continuation of each integral in \Eq{\ref{Equation:SARExampleSplittedIntegrated}}, making them valid for any value of $d\in\mathbb{C}$. This is something specific to DREG, where essentially IR and UV poles can cancel each other.} 
\begin{equation}\label{Equation:SARExampleIntegrated}
J_d=J_d^{[0,1]}+J_d^{[1,\infty)}=0\;,
\end{equation}
for any $d\in\mathbb{C}$, and in particular
\begin{equation}\label{Equation:SARExample0}
\int_{\mathbb{R}^+}\,g(x)\,dx=0\;.
\end{equation}
In general, DREG does not discriminate between IR and UV poles. They can partially or even completely cancel each other, as we just saw in this last example. In other formalisms though, it may be necessary to keep track of the origin of each pole, since they have different physical meanings.\\
\\
Of course, calculating actual loop integrals in DREG is much more complicated, and involve several tricks and mathematical techniques that will not be enumerated here. Detailed and exhaustive step-by-step calculations are, once again, available in standard textbooks. The integral appearing in \Eq{\ref{Equation:SARLBubble}} for instance, leads after integration to
\begin{align}\label{Equation:SARBubbleDREG}
L_{Bubble}^{(1)}(p,-p)=&~i\,\mu^{4-d}\int\,\frac{d^d\ell}{(2\pi)^d}\frac{1}{(\ell^2+i0)((\ell+p)^2+i0)}\nn\\
=&~\frac{\Gamma(1+\epsilon)\Gamma^2(1-\epsilon)}{(4\pi)^{2-\epsilon}\Gamma(1-2\epsilon)}\,\frac{\mu^{2\epsilon}}{\epsilon(1-2\epsilon)}(-p^2-i0)^{-\epsilon}\;,
\end{align}
with $i0$ being an infinitely small quantity with positive imaginary part, and where we rewrote the dimensions $d$ in terms of the parameter $\epsilon$, as
\begin{equation}\label{Equation:SARd}
d=4-2\epsilon\;.
\end{equation}
This way, taking the limit $d\to4$ is equivalent to taking the limit $\epsilon\to0$, and the poles will simply be negative powers of $\epsilon$. Note that the convention chosen in \Eq{\ref{Equation:SARd}} is one possibility. Others exist, such as for example $d=4-\epsilon$ or $d=4+\epsilon$.\\
\\
One of the biggest advantage of DREG is that fundamental symmetries of the theories are conserved, which leads to more compact results compared to other approaches, and reduces the risk of encountering quantities at intermediate steps that may have no physical meaning. By working with well-defined objects throughout the whole calculation, there is usually no need to carry out extra, usually quite heavy steps to obtain a relevant result.

\subsection{A brief review of other regularisation methods}\label{Section:SARReview}

A regularisation method in general must establish a framework in which consistent guidelines are given to deal with all the different types of singularities. Furthermore, it also needs to specify how to handle different quantities appearing in Feynman amplitudes, such as metric tensors, momenta, and gamma matrices. In this section, we will write a brief summary of a recent review on the subject, found in~\cite{Gnendiger:2017pys}.\\
\\
The most commonly used regularisation schemes are the traditional dimensional schemes, where the dimension $d$ of the space-time is altered. Among them are the Conventional Dimensional Regularisation (CDR, introduced in \Section{\ref{Section:SARDREG}}), the `t Hooft and Veltman (HV) scheme~\cite{tHooft:1972tcz}, the Four-Dimensional Helicity scheme (FDH)~\cite{Bern:1991aq,Bern:2002zk} and Dimensional Reduction (DRED)~\cite{Stockinger:2005gx,Signer:2008va}. Depending on the scheme, different vector spaces will be used for different objects. From the original four-dimensional vector space can be constructed\footnote{More details about the mathematical formulation of these extensions can be found in~\cite{Wilson:1972cf,Collins:1986a,Stockinger:2005gx}.} the $d$-dimensional space and the $d_s$-dimensional space. The former is the natural space-time used in DREG, and the one that is used for momentum integration in all the aforementioned schemes, while the latter is the direct sum of the former with an extra orthogonal space of dimension $n_\epsilon$, with $d_s=d+n_\epsilon$. The dimension $n_\epsilon$ is most of the time taken to be the difference between four and the dimension $d$, meaning that we essentially have $d_s=4$, even though $d_s$ is usually kept as a free parameter until the end of the computation. In \Table{\ref{Table:SARDimensions}}, we show the differences between the schemes listed above by specifying how each of them treats singular and regular vector fields.\\
\begin{table}
	\begin{center}
		\begin{tabular}{l|cccc}
			&\hphantom{i}CDR\hphantom{i}&\hphantom{~}HV\hphantom{~}&\hphantom{i}FDH\hphantom{i}&DRED\\
			\hline
			&&&&\\
			Singular vector fields&$d$&$d$&$d_s$&$d_s$\\
			&&&&\\
			Regular vector fields&$d$&$4$&$4$&$d_s$\\
			&&&&\\
		\end{tabular}
		\caption{Value of the dimensions used in the treatment of singular and regular vector fields, depending on the regularisation scheme. The singular vector fields are the ones associated with parts of diagrams that can exhibit UV or IR singularities. The regular vectors fields are all the other ones.}
		\label{Table:SARDimensions}
	\end{center}
\end{table}\\
While these dimensional schemes have been proven to be very effective for any type of calculations in gauge theories, it can be in some cases more practical to work in a space-time of fixed finite integer dimension. For instance the Four-Dimensional Formulation (FDF) of FDH \cite{Fazio:2014xea} adapts the FDH scheme so it can be used in a four-dimensional framework. In brief, FDF allows for integrands to be rewritten in terms of tree-level amplitudes, by taking advantage of unitarity-based methods. The divergences are regulated by representing particles propagating inside loops as massive internal states. The Six-Dimensional Formalism (SDF) \cite{Bern:2002zk} is very similar to FDF but, as its name suggests, sets the number of dimensions to 6. It is based on the six-dimensional spinor-helicity formalism \cite{Cheung:2009a}, and is optimised for the analytic calculation of two-loop amplitudes.\\
\\
Other schemes that do not modify at all the dimension of space-time have also been developed. They aim at building pure four-dimensional representations of loop and phase-space integrals, usually by removing the singularities at the integrand level. Implicit Regularisation (IREG) \cite{BaetaScarpelli:2000zs,BaetaScarpelli:2001ix}, for instance, isolate the UV singular part of divergent loop integrals and express it in terms of implicit integrals and other boundary terms. The Four-Dimensional Regularisation (FDR) scheme \cite{Pittau:2012zd,Donati:2013iya} is very similar to IREG, but avoids renormalisation altogether by simply setting to zero the UV part of the integral, which renders the introduction of UV counterterms at the Lagrangian level unnecessary. In both IREG and FDR, the IR singularities are regularised by restricting the phase-space integration around the soft and collinear regions, by using an infinitesimal shift of momenta. Finally, another non-dimensional scheme, the Four-Dimensional Unsubtraction (FDU) \cite{Hernandez-Pinto:2015ysa,Sborlini:2016gbr,Sborlini:2016hat,Rodrigo:2016hqc,Hernandez-Pinto:2016uwx,Sborlini:2016bod,Driencourt-Mangin:2016dpf,Sborlini:2017nee}, will be covered in its entirety in this thesis and will be the main subject of \Chapters{\ref{Chapter:FDU}}{\ref{Chapter:FDUM}}.\\
\\
In the rest of this chapter, we will explain the main principles to perform explicit regularisations of divergences. To simplify the discussion, we will stick to DREG.

\section{Renormalisation of UV singularities in gauge theories}\label{Section:SARUV}
\fancyhead[LO]{\ref*{Section:SARUV}~~\nameref*{Section:SARUV}}

Because of the infinities that are generated at very high energies in perturbation theory, the physical quantities that we measure --~such as the electric charge and the mass~-- are not the ones that directly appear in the original, unrenormalised theory. This is due to the fact the \emph{bare} parameters of the Lagrangian (that is to say, the parameters of the Lagrangian before renormalisation) do not take into account the effects that quantum corrections have on the associated physical quantities. The main idea of renormalisation is to rewrite the Lagrangian by introducing small shifts in these parameters, in order to generate counterterms at the Lagrangian level, allowing for a direct cancellation of infinities while simultaneously promoting the bare parameters to physical ones.\\
\\
In this section, we will detail how to systematise the renormalisation procedure for perturbation theories, and show how to obtain a finite theory formulated in terms of physical parameters that can be measured experimentally. We will first introduce the (modified) minimal subtraction renormalisation scheme, and then illustrate the procedure by using as an example the QED Lagrangian, found in \Eq{\ref{Equation:TBPPQEDLagrangian}}. We will try to avoid most explicit calculations here, and will mostly focus on the more conceptual aspects.

\subsection{The minimal subtraction scheme}\label{Section:SARMSbar}

Essentially and as said above, renormalising a quantity means removing (UV) infinities generated by the theory at a given order in perturbation theory. In practice, this is done by introducing a counterterm at a given step of the calculation in order to cancel the poles. For instance, let's take $I_d$ in \Eq{\ref{Equation:SARExample1Expanded}}. The counterterm
\begin{equation}\label{Equation:SARIdCT}
	I_d^\ct(A)=\frac{1}{d-1}+A\;,\qquad A~\text{finite}
\end{equation}
will do the required job, since
\begin{equation}\label{Equation:SARIdRegularised}
	I_d-I_d^\ct(A)
\end{equation}
is regular when the limit $d\to1$ is taken. The finite part $A$ appearing in \Eq{\ref{Equation:SARIdCT}}, however, is completely arbitrary. Since the final result must of course be independent of $A$, its value must be fixed at some point of the computation, but this has to be done in a consistent manner. The rules to fix $A$ are given by a renormalisation scheme that must be chosen beforehand. Probably the simplest of them is the Minimal Subtraction (MS) scheme, introduced independently by `t Hooft \cite{tHooft:1973a} and Weinberg \cite{Weinberg:1973a}. In this scheme, only the pole is removed, meaning that in our previous example, $A$ is simply set to 0. Now let's do the same exercise with the integral appearing in \Eq{\ref{Equation:SARBubbleDREG}}. By expanding around $\epsilon=0$, we have
\begin{equation}\label{Equation:SARBubbleDREGExpanded}
	L_{Bubble}^{(1)}(p,-p)=\frac{1}{16\pi^2}\left(\frac{1}{\epsilon}-\log\left(\frac{-p^2-i0}{\mu^2}\right)+2-\gamma_E+\log(4\pi)+\Oep{1}\right)\;,
\end{equation}
where $\gamma_E$ is the Euler-Mascheroni constant. In the MS scheme, only the pole is removed, meaning that the integrated counterterm would be of the form
\begin{equation}\label{Equation:SARBubbleCT}
	L^{(1,\textrm{CT},\textrm{MS})}_{Bubble}=\frac{1}{16\pi^2}\frac{1}{\epsilon}\;.
\end{equation}
It would be more practical, though, to also include in the counterterm the very commonly encountered $\gamma_E$ and $\log(4\pi)$ (also appearing in \Eq{\ref{Equation:SARBubbleDREGExpanded}}). To do so, we define the Modified Minimal Subtraction scheme, written $\MSbar$, where in addition to removing the pole, we factorise in the counterterm the quantity
\begin{equation}\label{Equation:SARSMSBar}
	S_\epsilon^{\MSbar}=(4\pi)^\epsilon e^{-\epsilon\,\gamma_E}\;.
\end{equation}
Note that, up to $\Oep{2}$, this is equivalent to applying the shift
\begin{equation}\label{Equation:SARMSbarPoleShift}
	\frac{1}{\epsilon}\longrightarrow\frac{1}{\epsilon}-\gamma_E+\log(4\pi)
\end{equation}
inside the counterterm. We therefore have
\begin{equation}
	L^{(1,\textrm{CT},\MSbar)}_{Bubble}=\frac{S_\epsilon^{\MSbar}}{16\pi^2}\frac{1}{\epsilon}\;,
\end{equation}
which we use to obtain the renormalised amplitude,
\begin{equation}\label{Equation:SARRenormalisedBubble}
	L_{Bubble}^{(1,\r)}(p,-p)=L_{Bubble}^{(1)}(p,-p)-L_{Bubble}^{(1,\textrm{CT},\MSbar)}=\frac{1}{16\pi^2}\left(-\log\left(\frac{-p^2-i0}{\mu^2}\right)+2\right)\;,
\end{equation}
in the $\MSbar$ scheme and in the limit $\epsilon\to0$.

\subsection{Generating the counterterms from the Lagrangian}

In terms of the bare parameters, written with an index 0, the (unrenormalised) QED Lagrangian reads
\begin{equation}\label{Equation:SARQEDLagrangianBare}
\mathcal{L}_0=\bar{\psi}_0(i\,\gamma^\mu(\partial_\mu+i\,e_0\,{A_0}_\mu)-m_0)\psi_0-\frac{1}{4}(\partial_\mu {A_0}_\nu-\partial_\nu {A_0}_\mu)(\partial^\mu A_0^\nu-\partial_\nu A_0^\mu)\;,
\end{equation}
Since there are four such parameters, we logically expect in the end the same number of counterterms. By using \Eq{\ref{Equation:SARQEDLagrangianBare}}, it is possible to show that computing the electron and photon propagator leads to
\begin{equation*}
\centering
\begin{picture}(100,25)(-50,-3)
\PhotonArc(0,0)(15,0,180){3}{4.5}
\ArrowLine(-45,0)(-15,0)
\ArrowLine(-15,0)(15,0)
\ArrowLine(15,0)(45,0)
\end{picture}
=\frac{i\,Z_2}{\slashed{p}-m}+\cdots\;,
\qquad
\begin{picture}(100,25)(-50,-3)
\Arc[arrow,arrowpos=0.25](0,0)(15,0,360)
\Photon(-45,0)(-15,0){3}{3}
\Photon(15,0)(45,0){3}{3}
\end{picture}
=-\frac{i\,Z_3\,g_{\mu\nu}}{q^2}+\cdots\;,
\end{equation*}
\\
with $Z_2$ and $Z_3$ being the wave-function renormalisation factors of the electron and the photon, respectively. The first step is therefore to rescale the bare fermion field $\psi_0$ and photon field $A_0^\mu$ in order to absorb $Z_2$ and $Z_3$ into the Lagrangian. This is done by considering the new fields $\psi=Z_2^{-1/2}\psi_0$ and $A^\mu=Z_3^{-1/2}\,A_0^\mu$, leading to
\begin{equation}\label{Equation:SARQEDLagrangianRescaled}
\mathcal{L}_0=Z_2\,\bar{\psi}(i\,\gamma^\mu\partial_\mu-m_0)\psi-e\,Z_1\,\bar{\psi}\,\gamma^\mu\,\psi\,A_\mu-\frac{Z_3}{4}(\partial_\mu A_\nu-\partial_\nu A_\mu)(\partial^\mu A^\nu-\partial^\nu A^\mu)\;,
\end{equation}
where we also rescaled the bare electric charge $e_0$ as $e_0\,Z_2\,Z_3^{1/2}=e\,Z_1$, with $e$ being interpreted as the \emph{physical} electric charge. By defining
\begin{align}\label{Equation:SARdeltas}
\delta_2=&~Z_2-1\;,&\quad\delta_3=&~Z_3-1\;,\nn\\ \delta_m=&~Z_2\,m_0-m&\quad\delta_1=&~Z_1-1=(e_0/e)Z_2\,Z_3^{1/2}-1
\end{align}
and expanding accordingly \Eq{\ref{Equation:SARQEDLagrangianRescaled}}, we find
\begin{equation}
\mathcal{L}_{QED}=~\mathcal{L}_0-\mathcal{L}_{CT}
\end{equation}
where the second piece $\mathcal{L}_{CT}$ contains all four counterterms. Explicitly,
\begin{equation}\label{Equation:SARLCT}
\mathcal{L}_{CT}=\bar{\psi}(i\,\delta_2\,\gamma^\mu\partial_\mu-\delta_m)\psi-e\,\delta_1\,\bar{\psi}\,\gamma^\mu\,\psi\,A_\mu-\frac{\delta_3}{4}(\partial_\mu A_\nu-\partial_\nu A_\mu)(\partial^\mu A^\nu-\partial_\nu A^\mu)\;.
\end{equation}
Now that we have built the renormalised Lagrangian, the next step consists in finding the values of the counterterms so they effectively cancel the infinities coming from the original theory. In general, they are fixed through what are called \emph{renormalisation conditions}. In QED, there are four of them, reading
\begin{align}\label{Equation:SARLCTRenCond}
\Sigma(\slashed{p}=m)=&~0\;,\nn\\
\left.\frac{\partial}{\partial\slashed{p}}\Sigma(\slashed{p})\right|_{\slashed{p}=m}=&~0\;,\nn\\
\Pi(q^2=0)=&~0\;,\nn\\
-i\,e\,\Gamma^\mu(q=0)=&~-i\,e\,\gamma^\mu\;.
\end{align}
where $i\,\Sigma$ (resp. $\Pi$) is the coefficient appearing in the evaluation of the corrections of the fermion (resp. photon) propagator, and $-i\,e\,\Gamma^\mu$ is the term involved in the computation of the corrections of the three-point function. The first condition sets the fermion mass to the physical value $m$, and when combined with the second, it forces the cancellation of on-shell singularities. The third condition forces the on-shell photonic propagator to remain unmodified by the renormalisation procedure. Finally, the fourth condition fixes the three-point coupling to the physical one, which sets $e$ to be the physical electric charge.\\
\\
In the $\MSbar$ scheme (see \Section{\ref{Section:SARMSbar}}), with dimension $d=4-2\epsilon$, the QED counterterms (at one-loop order) read
\begin{align}
\delta_1=\delta_2=-e^2\,\frac{S_\epsilon^{\MSbar}}{16\pi^2}\frac{1}{\epsilon}\;,\qquad\delta_3=-e^2\,\frac{S_\epsilon^{\MSbar}}{16\pi^2}\frac{4}{3\epsilon}\;,\qquad\delta_m=-e^2\,\frac{S_\epsilon^{\MSbar}}{16\pi^2}\frac{3}{\epsilon}\;,
\end{align}
and their associated Feynman rules are
\vspace{25pt}
\\
\begin{minipage}{0.5\textwidth}
	\centering
	\begin{picture}(0,25)(75,27)
		\Arc(75,35)(8,0,360)
		\Line(69.5,29.5)(80.5,40.5)
		\Line(69.5,40.5)(80.5,29.5)
		\Photon(25,35)(67,35){3}{3}
		\Photon(83,35)(125,35){3}{3}
		\LongArrow(60,20)(90,20)
		\color{blue}
		\Text(17,35){$\mu$}
		\Text(133,35){$\nu$}
		\Text(75,5){$q$}
	\end{picture}
\end{minipage}
\begin{minipage}{0.5\textwidth}
	\begin{equation*}
		-i(g^{\mu\nu}\,q^2-q^\mu\,q^\nu)\delta_3
	\end{equation*}
\end{minipage}
\vspace{25pt}
\\
\begin{minipage}{0.5\textwidth}
	\centering
	\begin{picture}(0,25)(75,27)
		\Arc(75,35)(8,0,360)
		\Line(69.5,29.5)(80.5,40.5)
		\Line(69.5,40.5)(80.5,29.5)
		\ArrowLine(25,35)(67,35)
		\ArrowLine(83,35)(125,35)
		\LongArrow(60,20)(90,20)
		\color{blue}
		\Text(75,5){$p$}
	\end{picture}
\end{minipage}
\begin{minipage}{0.5\textwidth}
	\begin{equation*}
		i(\slashed{p}\,\delta_2-\delta_m)
	\end{equation*}
\end{minipage}
\vspace{45pt}
\\
\begin{minipage}{0.5\textwidth}
	\centering
	\begin{picture}(0,72)(75,38)
		\Photon(25,72)(72,72){3}{3.5}
		\Arc(80,72)(8,0,360)
		\Line(74.5,66.5)(85.5,77.5)
		\Line(74.5,77.5)(85.5,66.5)
		\ArrowLine(85.5,77.5)(125,117)
		\ArrowLine(125,27)(85.5,66.5)
		\color{blue}
		\Text(15,72){$\mu$}
	\end{picture}
\end{minipage}
\begin{minipage}{0.5\textwidth}
	\begin{equation*}
		-i\,e\,\gamma^\mu\,\delta_1
	\end{equation*}
\end{minipage}
\vspace{45pt}
\\
Removing the infinities is therefore simply done by taking into account and calculating all the extra diagrams generated from these new Feynman rules. For instance, the singularities generated by the one-loop self-energy contribution of the photon propagator are cancelled by the corresponding counterterm, i.e.
\begin{equation}
\centering
\begin{picture}(110,25)(-55,-3)
\Arc[arrow,arrowpos=0.25](0,0)(15,0,360)
\Photon(-50,0)(-15,0){3}{2.5}
\Photon(15,0)(50,0){3}{2.5}
\end{picture}
\qquad
-
\qquad
\begin{picture}(110,25)(-55,-3)
\Arc(0,0)(8,0,360)
\Line(-5.5,-5.5)(5.5,5.5)
\Line(-5.5,5.5)(5.5,-5.5)
\Photon(-50,0)(-8,0){3}{3}
\Photon(8,0)(50,0){3}{3}
\end{picture}
\qquad
=\qquad\Oep{0}\;.
\end{equation}
\\
This procedure is completely equivalent to simply removing the infinities by hand, but makes things easier in general when dealing with more loops, thanks to its systematic aspect.

\section{IR singularities and the KLN theorem}\label{Section:SARIR}
\fancyhead[LO]{\ref*{Section:SARIR}~~\nameref*{Section:SARIR}}

While the treatment of UV divergences can be done systematically and order by order in perturbation theory by introducing counterterms at the Lagrangian level, cancelling IR divergences requires a very different approach. As we briefly saw before, they can arise from both virtual contributions (i.e. diagrams involving loops) and real contributions (i.e. diagram with the emission of an extra physical particle). We will see that, by properly handling each of these quantities, IR singularities will actually vanish by themselves at the end of the computation.\\
\\
In order to illustrate this idea, we will consider the very simple $1\longrightarrow2$ process at NLO in QED of a virtual photon $\gamma^*$ decaying into two (massless) electrons. In order to obtain the complete\footnote{Here, we consider the sum of both the LO and NLO, instead of only the NLO corrections.} cross section $\sigma_{\gamma^*\to e^+e^-(\gamma)}^\textrm{NLO}$, we need to calculate two quantities. The first, written $\sigma_{\V(\gamma^*\to e^+e^-)}^\textrm{NLO}$, is evaluated by taking into account all possible one-loop corrections, and considering the interference between them and the tree-level diagram. The second, written $\sigma_{\r(\gamma^*\to e^+e^-\gamma)}^\textrm{NLO}$, is obtained by squaring the sum of all diagrams where an additional physical particle is emitted. Essentially and in our example, we want to compute
\begin{align*}
&~\hspace{110pt}\sigma_{\V(\gamma^*\to e^+e^-)}^\textrm{NLO}\hspace{84pt}~\hspace{57pt}\sigma_{\r(\gamma^*\to e^+e^-\gamma)}^\textrm{NLO}\\
&~
\overbrace{\left(
\begin{picture}(40,30)(0,-3)
\Photon(0,0)(20,0){3}{2}
\ArrowLine[arrowscale=0.75](20,0)(40,20)
\ArrowLine[flip,arrowscale=0.75](20,0)(40,-20)
\end{picture}
~+~
\begin{picture}(40,30)(0,-3)
\Photon(0,0)(20,0){3}{2}
\ArrowLine[arrowpos=0.40,arrowscale=0.75](20,0)(40,20)
\ArrowLine[arrowpos=0.40,flip,arrowscale=0.75](20,0)(40,-20)
\PhotonArc(20,0)(17,-45,45){2}{3.5}
\end{picture}
~+~
\begin{picture}(40,30)(0,-3)
\Photon(0,0)(20,0){3}{2}
\ArrowLine[arrowscale=0.75](20,0)(40,20)
\ArrowLine[flip,arrowscale=0.75](20,0)(40,-20)
\PhotonArc(30,10)(9,45,225){2}{3.5}
\end{picture}
~+~
\begin{picture}(40,30)(0,-3)
\Photon(0,0)(20,0){3}{2}
\ArrowLine[arrowscale=0.75](20,0)(40,20)
\ArrowLine[flip,arrowscale=0.75](20,0)(40,-20)
\PhotonArc(30,-10)(9,135,315){2}{3.5}
\end{picture}
\right)^2}
+
\overbrace{\left(
\begin{picture}(40,30)(0,-3)
\Photon(0,0)(20,0){3}{2}
\ArrowLine[arrowpos=0.25,arrowscale=0.75](20,0)(40,20)
\ArrowLine[flip,arrowscale=0.75](20,0)(40,-20)
\Photon(30,10)(40,0){3}{1.5}
\end{picture}
~+~
\begin{picture}(40,30)(0,-3)
\Photon(0,0)(20,0){3}{2}
\ArrowLine[arrowscale=0.75](20,0)(40,20)
\ArrowLine[arrowpos=0.25,flip,arrowscale=0.75](20,0)(40,-20)
\Photon(30,-10)(40,0){-3}{1.5}
\end{picture}
\right)^2}\;,\\
\end{align*}
where we must ignore the terms given by the products of two one-loop diagrams, since they are of a higher order than the one we consider here. Indeed, in this case, we restrict ourselves to terms in the perturbative expansion that are at most proportional to $\alpha^2$, with
\begin{equation}\label{SARAlpha}
\alpha=\frac{e^2}{4\pi}\;.
\end{equation}
Performing the explicit calculation\footnote{We refer the reader to e.g. \cite{Field:1989uq} for the complete derivation.} leads to
\begin{equation}\label{SARSigmaVirtual}
\sigma_{\V(\gamma^*\to e^+e^-)}^\textrm{NLO}=\sigma^\textrm{LO}\left(1+\frac{2\alpha}{3\pi}\left(\frac{\s}{\mu}\right)^{-\epsilon}\frac{\Gamma(1+\epsilon)\Gamma^2(1-\epsilon)}{(4\pi)^{-\epsilon}\Gamma(1-2\epsilon)}\left(-\frac{2}{\epsilon^2}-\frac{2}{\epsilon}-8+\Oep{1}\right)\right)\;,
\end{equation}
with $\sigma^\textrm{LO}$ the value of the cross section at LO, obtained when computing the square of the tree-level diagram, $\s$ the virtuality of the incoming photon, and $\mu$ an arbitrary mass scale. The $\epsilon$-poles appearing in \Eq{\ref{SARSigmaVirtual}} only come from the low-energy region of the loop integration domain, since in this particular case the UV singularities cancel between the different virtual contributions\footnote{In the most general case, the amplitude must be renormalised beforehand, so only IR singularities remain.}.\\
\\
On the other hand, the real contribution reads
\begin{equation}\label{SARSigmaReal}
\sigma_{\r(\gamma^*\to e^+e^-\gamma)}^\textrm{NLO}=\sigma^\textrm{LO}\left(1+\,\frac{2\alpha}{3\pi}\left(\frac{\s}{\mu}\right)^{-\epsilon}\frac{\Gamma(1+\epsilon)\Gamma^2(1-\epsilon)}{(4\pi)^{-\epsilon}\Gamma(1-2\epsilon)}\left(\frac{2}{\epsilon^2}+\frac{2}{\epsilon}+\frac{19}{2}+\Oep{1}\right)\right)\;.
\end{equation}
The double pole (i.e. the term proportional to $1/\epsilon^2$) in \Eq{\ref{SARSigmaReal}} comes from the region of the phase space where soft and collinear singularities manifest themselves simultaneously, whereas the single pole (i.e. the term proportional to $1/\epsilon$) is generated when these singularities occur separately. One can immediately notice that not only the poles of $\sigma_{\V(\gamma^*\to e^+e^-)}^\textrm{NLO}$ and $\sigma_{\r(\gamma^*\to e^+e^-\gamma)}^\textrm{NLO}$ have the same absolute value, they are also of \emph{opposite signs}. As a consequence of this, $\sigma_{\gamma^*\to e^+e^-(\gamma)}^\textrm{NLO}$ is completely free of $\epsilon$-poles, and the limit $\epsilon\to0$ can be taken after summing, i.e.
\begin{equation}\label{SARSigmaSum}
\sigma_{\gamma^*\to e^+e^-(\gamma)}^\textrm{NLO}=\sigma_0\left(1+\frac{\alpha}{\pi}\right)\;.
\end{equation}
The two quantities $\sigma_{\V(\gamma^*\to e^+e^-)}^\textrm{NLO}$ and $\sigma_{\r(\gamma^*\to e^+e^-\gamma)}^\textrm{NLO}$ having opposite $\epsilon$-poles is of course not a coincidence. As a matter of fact, the Block-Nordsieck theorem states that in perturbative QED, the cancellation between IR poles generated by virtual contributions and IR poles generated by real contributions is verified at all orders. In non-abelian theories, the IR divergences exhibit a more complex structure because of the existence of self-interaction vertices between gauge bosons for instance. In the case of QCD, collinear singularities can manifest themselves even when considering massive quarks, since a final-state gluon can be emitted from another final-state gluon. However and fortunately, the cancellation is guaranteed by the more general Kinoshita-Lee-Nauenberg (KLN) theorem~\cite{Kinoshita:1962ur,Lee:1964is}. This theorem states that any quantum field theory with massless fields is free of IR singularities, which disappear after performing the sum over virtual and degenerate initial- and final-state real contributions. This property is a crucial aspect of perturbation theory, since it allows one to build infrared-safe observables at any order.\\
\\
The traditional approach to regularise IR divergences is known as the subtraction formalism \cite{Kunszt:1992tn,Frixione:1995ms,Catani:1996jh,Catani:1996vz,GehrmannDeRidder:2005cm,Seth:2016hmv,Catani:2007vq,Catani:2009sm} (and its variants \cite{Czakon:2010td,Bolzoni:2010bt,DelDuca:2015zqa,Boughezal:2015dva,Gaunt:2015pea,DelDuca:2016ily,DelDuca:2016csb}). It exploits the KLN theorem by building counterterms which locally mimic the real radiation contribution. The resulting integrals can then be computed, and reproduce the divergences present in the virtual part.

\section{Threshold singularities and the optical theorem}\label{Section:SARTSOT}
\fancyhead[LO]{\ref*{Section:SARTSOT}~~\nameref*{Section:SARTSOT}}

One can show~\cite{Cutkosky:1960sp} that a given loop diagram contributing to a given transition probability amplitude $\mathcal{M}(i\to f)$ will always be real, unless there exists non-trivial points of the loop integration domain for which one or more internal propagators vanish. This can only happen if intermediate states can go on shell, which usually gives a condition on the lightest internal mass. In this case, the discontinuity is called a threshold singularity; it is integrable, and will contribute to the imaginary part of the scattering amplitude. As we will see, techniques exist to isolate and calculate this imaginary component.\\
\\
But first, let's go back to the definition of the $S$-matrix given in \Section{\ref{Section:TBPPSACS}}. Knowing that $S$ is Hermitian, we can write
\begin{equation}\label{Equation:SARSdaggerSDecomposition}
S^\dagger S=1=(1-i\,T^\dagger)(1+i\,T^\dagger)
\end{equation}
which leads to
\begin{equation}\label{Equation:SARTRelation}
i(T-T^\dagger)=T^\dagger T\;.
\end{equation}
For given initial and final states $|\,i\,\rangle$ and $|\,f\,\rangle$, we can rewrite the right-hand side as
\begin{equation}\label{Equation:SARTRelationRHS}
\langle\,f\,|\,T^\dagger T\,|\,i\,\rangle=\,\sum_x\,\int\,d\Omega_x\langle\,f\,|\,T\,|\,x\,\rangle\langle\,x\,|\,T^\dagger\,|\,i\,\rangle
\end{equation}
where the sum is performed over all possible intermediate states $|\,x\,\rangle=|\{p_{x_1},p_{x_2},\dots\}\rangle$, whose corresponding phase-space factor is $\Omega_x$. From \Eq{\ref{Equation:SARTRelation}}, we therefore have
\begin{equation}\label{Equation:SARGOT}
\mathcal{M}(i\to f)-\mathcal{M}^*(f\to i)=i\,\sum_x\,\int\,d\Omega_\textrm{LIPS}\,\mathcal{M}^*(f\to x)\mathcal{M}(i\to x)\;,
\end{equation}
with
\begin{equation}\label{Equation:SAROmegaLIPS}
 d\Omega_\textrm{LIPS}=(2\pi)^4\delta\left(\,\sum\limits_j\,p_{f_j}-\,\sum_k\,p_{x_k}\right)d\Omega_x\;.
\end{equation}
The relation obtained in \Eq{\ref{Equation:SARGOT}} is called the \emph{generalised optical theorem}, and holds order by order in perturbation theory. A more interesting case of the generalised optical theorem is when $|\,i\,\rangle=|\,f\,\rangle=|\,k\,\rangle$, which reduces \Eq{\ref{Equation:SARGOT}} to
\begin{equation}\label{Equation:SAROT}
2i\,\ImText\mathcal{M}(k\to k)=i\,\sum_x\,\int\,d\Omega_\textrm{LIPS}\,|\mathcal{M}(k\to x)|^2\;.
\end{equation}
For instance if $|\,k\,\rangle$ is a one-particle state, \Eq{\ref{Equation:SAROT}} tells us that the imaginary part of the amplitude $\mathcal{M}(k\to k)$ is equal to the total decay rate of this particle times its mass. On the other hand, if $|\,k\,\rangle$ is a two-particle state, then \Eq{\ref{Equation:SAROT}} tells us that the total scattering cross section can be deduced from the imaginary part of the probability amplitude $\mathcal{M}(k\to k)$ and vice-versa.\\
\\
This theorem is very convenient, since it is in practice quite easy to evaluate the imaginary part of a loop diagram. Indeed, it is possible to show that for an internal particle of momentum $q$,
\begin{equation}\label{Equation:SARImProp}
\ImText\frac{1}{q^2-m^2+i0}=-\pi\,\delta(q^2-m^2)\;,
\end{equation}
which means that the imaginary part of a loop amplitude can be directly obtained from the discontinuity arising when internal lines go on shell. A systematic approach to evaluate the imaginary part of an amplitude involves a three-step algorithm known as \emph{Cutkosky's cutting rules} \cite{Cutkosky:1960sp}. First, take a loop diagram and cut through it in all possible ways such that the cut propagators can simultaneously be put on shell (momentum conservation must hold). Then, for each cut, perform the replacement $(q^2-m^2+i0)^{-1}\to-2i\,\pi\,\delta(q^2-m^2)$. Finally, the integrable discontinuity is given by the sum of all cuts, and is equal to minus two times the imaginary part of the original amplitude.\\
\\
Although very useful, this method has its limits. If there are more than one internal scale for instance, additional discontinuities -- known as \emph{anomalous thresholds} -- may manifest themselves \cite{Cutkosky:1961a,Mandelstam:1960zz}. They still generate an imaginary component, but are much less simple to evaluate. This is even more challenging at two-loop level, where having more degrees of freedom inside the integrand means that more internal lines can go on shell simultaneously.

\newpage
\hspace{0pt}
\thispagestyle{empty}
\clearpage

\chapter{The Loop-Tree Duality theorem}\label{Chapter:LTD}
\thispagestyle{fancychapter}
\fancyhead[RE]{\nameref*{Chapter:LTD}}

The Loop-Tree Duality (LTD) theorem \cite{Catani:2008xa,Rodrigo:2008fp,Bierenbaum:2010cy,Bierenbaum:2012th,Buchta:2014dfa,Buchta:2015xda,Buchta:2015wna} is a mathematical method based on\linebreak Cauchy's residue theorem that can be applied to any QFT in Minkowski space with an arbitrary number $d$ of space-time dimensions. In short, and as its name suggests, it allows the loop scattering amplitudes to be rewritten as a sum of tree-level-like objects, thus demonstrating the existence of an underlying and formal connection among loops and phase-space integrals. As explained in more details in this chapter, its main strength lies in the fact it reduces the integration space from a $(L\,d)$-dimensional Minkowski space to a $(L(d-1))$-dimensional Euclidean space with $L$ being the number of loops, which allows one to circumvent many of the difficulties that may arise in the traditional approach.\\
\\
In the following, we will establish the theoretical grounds on which LTD is built, and will introduce most of the notations that will be used throughout this thesis. Without any loss of generality and in order to simplify all the demonstrations, the particles considered in this chapter will all be scalars.

\section{Loop-Tree Duality at one loop}\label{Section:LTDOneLoop}
\fancyhead[LO]{\ref*{Section:LTDOneLoop}~~\nameref*{Section:LTDOneLoop}}

\subsection{The Loop-Tree Duality theorem}\label{Section:LTDTheorem}

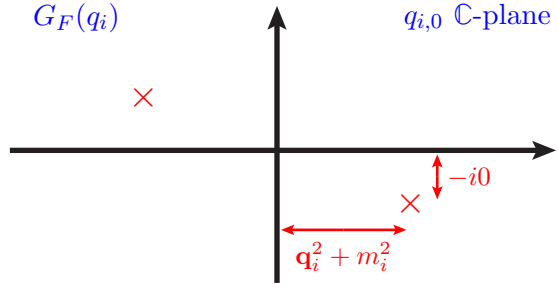
\begin{figure}[h]
	\centering
	\begin{picture}(300,150)(0,0)
	\SetWidth{2}
	\LongArrow(150,25)(150,125)
	\LongArrow(50,75)(250,75)
	\color{blue}
	\Text(75,125){$G_F(q_i)$}
	\Text(225,125){$q_{i,0}$ $\mathbb{C}$-plane}
	\color{red}
	\Text(200,55){\Large$\times$}
	\Text(100,95){\Large$\times$}
	\SetWidth{1}
	\LongArrow[arrowwidth=2](210,65)(210,59)
	\LongArrow[arrowwidth=2](210,65)(210,71)
	\Text(222,65){\small$-i0$}
	\LongArrow[arrowwidth=2](175,45)(196,45)
	\LongArrow[arrowwidth=2](175,45)(154,45)
	\Text(175,35){\small$\mathbf{q}_i^2+m_i^2$}
	\end{picture}
	\caption{Position of the poles of $G_F(q_i)$ in the $q_{i,0}$ complex plane.}
	\label{Figure:LTDPolesGF}
\end{figure}

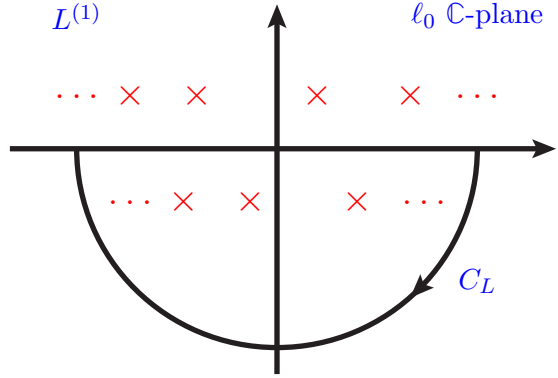
\begin{figure}[h]
	\centering
	\begin{picture}(300,150)(0,-20)
	\SetWidth{2}
	\LongArrow(150,-10)(150,125)
	\LongArrow(50,75)(250,75)
	\Arc[arrow,arrowpos=0.75,flip](150,75)(75,180,0)
	\color{blue}
	\Text(75,125){$L^{(1)}$}
	\Text(225,125){$\ell_0$ $\mathbb{C}$-plane}
	\Text(225,25){$C_L$}
	\color{red}
	\Text(95,55){\Large$\ldots$}
	\Text(115,55){\Large$\times$}
	\Text(140,55){\Large$\times$}
	\Text(180,55){\Large$\times$}
	\Text(205,55){\Large$\ldots$}
	\Text(75,95){\Large$\ldots$}
	\Text(95,95){\Large$\times$}
	\Text(120,95){\Large$\times$}
	\Text(165,95){\Large$\times$}
	\Text(200,95){\Large$\times$}
	\Text(225,95){\Large$\ldots$}
	\end{picture}
	\caption{Position of the poles of $L^{(1)}$ in the $\ell_0$ complex plane. The closed contour $C_L$ includes all poles with negative imaginary part.}
	\label{Figure:LTDPolesL1}
\end{figure}

In this section we introduce the key concepts of LTD \cite{Catani:2008xa} by applying the theorem to a generic one-loop scalar integral we will call $L^{(1)}$, and deriving the associated duality relation.
	\begin{figure}[h]
	\centering
	\begin{picture}(200,200)(-100,-100)
	\SetWidth{2}
	\Arc[arrow](0,0)(50,300,30)
	\Arc[arrow](0,0)(50,30,120)
	\Arc[arrow](0,0)(50,120,210)
	\Arc[arrow](0,0)(50,210,300)
	\Arc[arrow,arrowpos=0.5](0,0)(35,60,180)
	\ArrowLine(-16,47)(-30,91)
	\ArrowLine(44,24)(85,45)
	\ArrowLine(35,-35)(68,-68)
	\ArrowLine(-49,-10)(-94,-20)
	\Vertex(-19,-65){3}
	\Vertex(17,-66){3}
	\Vertex(-50,-46){3}
	\color{blue}
	\large
	\Text(-33,105){$p_1$}
	\Text(103,45){$p_2$}
	\Text(83,-73){$p_3$}
	\Text(-110,-23){$p_N$}
	\Text(25,63){$q_1$}
	\Text(67,-10){$q_2$}
	\Text(-63,34){$q_N$}
	\Text(-9,16){$\ell$}
	\end{picture}
	\caption{Momentum configuration of the one-loop $N$-point scalar integral.}
	\label{Figure:LTDLabelOneLoop}
\end{figure}
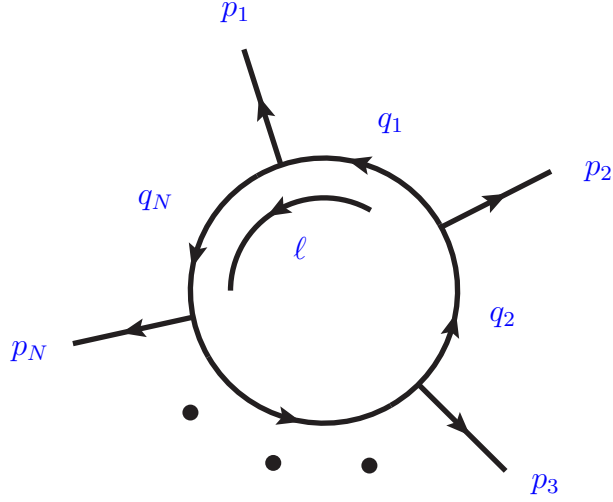
As shown in \Fig{\ref{Figure:LTDLabelOneLoop}}, the external momenta are labelled as $p_i$ with $i\in\alpha=\{1,2,\dots,N\}$. They also are clockwise ordered, and are taken as outgoing. The internal momenta are written\footnote{Throughout this thesis, we will label the internal momenta more adequately, depending on the process under consideration. The notations used in this chapter are more suited for a formal and general demonstration.}
\begin{equation}\label{Equation:LTDOneLoopqi}
q_i=\ell+k_i\quad\text{with}\quad k_i=\sum\limits_{j=1}^{i}\,p_j\;,
\end{equation}
and with $\ell$ being the loop momenta, which flows anti-clockwise. Consequently, and because of the momentum conservation relation between the external particles, namely
\begin{equation}\label{Equation:LTDMomentumConservation}
\,\sum\limits_{i=1}^N\,p_i=0\;,
\end{equation}
we have $q_N=\ell$.\\
\\
With all this in mind, $L^{(1)}$ can be written in a straightforward manner
\begin{equation}\label{Equation:LTDL1}
L^{(1)}(p_1,\dots,p_N)=\int_\ell\,\,\prod\limits_{i\in\alpha}\,G_F(q_i)\quad\text{where}\quad\int_\ell\,~\bullet~=-i\,\mu^{4-d}\,\int\,\frac{d^d\ell}{(2\pi)^d}\,~\bullet~\;,
\end{equation}
and with $d$ the dimension of the Minkowski space-time, on which the metric tensor is the usual $g^{\mu\nu}$ metric of signature $(+1,-1,\dots,-1)$. The Feynman propagator $G_F$ is written
\begin{equation}\label{Equation:LTDGF}
G_F(q_i)=\frac{1}{q_i^2-m_i^2+i0}\;,
\end{equation}
with $m_i$ being the mass of the internal particle with momentum $q_i$.\\
\\
The propagator $G_F(q_i)$ has two complex poles, and if we write $q_{i,0}$ the energy component of $q_i$, and $\mathbf{q}_i$ its space component, we have
\begin{equation}\label{Equation:LTDGF0}
\big(G_F(q_i)\big)^{-1}=0\quad\Longleftrightarrow\quad q_{i,0}=q_{i,0}^{(\pm)}=\pm\,\sqrt{\mathbf{q}_i^2+m_i^2-i0}\;.
\end{equation}
This means that, in the complex plane of the variable $q_{i,0}$, the positive energy solution has negative imaginary part, and vice-versa, as shown in \Fig{\ref{Figure:LTDPolesGF}}.
\\
\\
Deriving the Loop-Tree Duality theorem is done by directly applying Cauchy's residue theorem on $L^{(1)}$, with a well-chosen contour $C_L$ that selects all the poles with negative imaginary part, i.e. all the poles with positive energy of each individual Feynman propagator $G_F(q_i)$, as shown in \Fig{\ref{Figure:LTDPolesL1}}.
\\
\\
Assuming there are only single poles\footnote{At one-loop level, this is always the case, assuming a suitable choice of gauge \cite{Catani:2008xa}.}, we have
\begin{align}\label{Equation:LTDCauchy}
L^{(1)}(p_1,p_2,\dots,p_N)=&~\int_{\boldsymbol{\ell}}\,\int\,d\ell_0\,\prod\limits_{i\in\alpha}\,G_F(q_i)=\int_{\boldsymbol{\ell}}\,\int_{C_L}\,d\ell_0\,\prod\limits_{i\in\alpha}\,G_F(q_i)\nn\\
=&~-2\pi i \int_{\boldsymbol{\ell}}\,\,\sum\,\Res_{\{\ImText(\ell_0)<0\}}\left(\,\prod\limits_{i\in\alpha}G_F(q_i)\right)\;,
\end{align}
where integrating for all values of $\ell_0$ is equivalent to integrating over the closed contour $C_L$, as the integrand is convergent when $\ell_0\to\infty$. The next step consists in computing the $N$ residues that appear in \Eq{\ref{Equation:LTDCauchy}}. While it seems quite straightforward, it involves several subtleties that should not be overlooked.\\
\\
Let's first focus solely on the residue at the $i$-th pole for instance, and write
\begin{equation}\label{Equation:LTDResidueithPole}
\Res_{\{i-\text{th pole}\}}\left(\,\prod\limits_{j\in\alpha}\,G_F(q_j)\right)=\Res_{\{i-\text{th pole}\}}\big(G_F(q_i)\big)\times\hspace{-18pt}\underbracket[0.5pt][5pt]{\left[\,\prod\limits_{\substack{j\in\alpha\\j\neq i}}\,G_F(q_j)\right]}_{\{\text{Evaluated at the }i\text{-th pole}\}}\;,
\end{equation}
where none of the propagators $G_F(q_j)$ with $j\neq i$ are singular at the value of the pole of $G_F(q_i)$, because of the fact there are no multi-poles. The product appearing in the right-hand-side of \Eq{\ref{Equation:LTDResidueithPole}} can therefore directly be evaluated at this particular value. Furthermore, the contour $C_L$ only selects poles with negative imaginary part, which means that in practice the residues at the $i$-th pole are to be evaluated when $q_{i,0}=\Sup{q_{i,0}}$, as a consequence of \Eq{\ref{Equation:LTDGF0}}.\\
 \\
 In order to simplify the notations, and because of the fact the rest of the computation is independent of the index $i$, we will write
 \begin{equation}\label{Equation:LTDGFqj}
 G_F(q_j)=G_F(q_i+(q_j-q_i))=G_F(q_i+k_{ji})\;,\quad\text{for }j\neq i\;.
 \end{equation}
 One can observe that $k_{ji}=q_j-q_i$ does not depend on the loop momentum $\ell$, and is simply a linear combination of the external momenta. We will also remove any dependence in $i$ and relabel the momenta and the mass
 \begin{equation}\label{Equation:LTDqiToq}
 q_i\to q\;,\quad q_j\to q+k_j\quad\text{and}\quad m_i\to m\;.
 \end{equation}
 The right-hand side of \Eq{\ref{Equation:LTDResidueithPole}} is therefore rewritten
 \begin{equation}\label{Equation:LTDResidueq0Pole}
\Res_{\{q_0=\Sup{q_0}\}}\big(G_F(q)\big)\times\left[\,\prod\limits_{j}\,G_F(q+k_j)\right]_{q_0=\Sup{q_0}}\;,
 \end{equation}
 where $\Sup{q_0}=\sqrt{\mathbf{q}^2+m^2-i0}$.\\
 \\
The residue itself is very easily computed, and gives
\begin{align}\label{Equation:LTDResidueValue}
\Res_{\{q_0=\Sup{q_0}\}}\big(G_F(q)\big)=&~\lim\limits_{~~q_0\to\Sup{q_0}}\left((q_0-\Sup{q_0})\frac{1}{q_0^2-\mathbf{q}^2-m^2+i0}\right)\nn\\
=&~\frac{1}{2\Sup{q_0}}=\frac{1}{2\sqrt{\mathbf{q^2}+m^2}}=\int\,d\ell_0\,\deltaplus{q^2-m^2}\;,
\end{align}
where we used the short hand notation
\begin{equation}\label{Equation:LTDdeltaplus}
\deltaplus{q^2-m^2}=\theta(q_0)\,\delta(q^2-m^2)\;,
\end{equation}
which sets the internal particle of momentum $q$ on shell, while selecting the positive energy mode thanks to the presence of the Heaviside function $\theta$. Note that the prescription in \Eq{\ref{Equation:LTDResidueValue}} has been dropped inside the definition of $\Sup{q_0}$. This is justified by the fact that in the massive case ($m\neq0$), $\Sup{q_0}$ is always strictly positive and thus cannot vanish, and in the massless case ($m=0$), the singularity appearing when $\mathbf{q}^2\to 0$ corresponds to an end-point singularity in the integration over $\mathbf{q}$, for which the $i0$ prescription has no regularisation effect whatsoever. The last equality of the second line of \Eq{\ref{Equation:LTDResidueValue}} is simply obtained from the mathematical definition of the delta function.\\
\\
Thus, the calculation of the residue finally gives
\begin{equation}\label{Equation:LTDResidueFinal}
\Res_{\{i-\text{th pole}\}}\big(G_F(q_i)\big)=\int\,d\ell_0\,\deltaplus{q^2-m^2}\;.
\end{equation}
This equality shows that after applying Cauchy's residue theorem to the loop integral, the residue of the Feynman propagator of the internal line with momentum $q_i$ can be substituted with the corresponding on-shell propagator $\deltaplus{q_i^2-m_i^2}$. This is equivalent to cutting -- i.e. putting on shell -- this particular line in the appropriate term of the sum in \Eq{\ref{Equation:LTDCauchy}}. Consequently, and by inserting \Eq{\ref{Equation:LTDResidueFinal}} into \Eq{\ref{Equation:LTDCauchy}}, we achieve a representation of the one-loop integral as a linear combination of $N$ phase-space integrals.\\
\\
The only thing that remains to be done is to evaluate the residue pre-factor (the product in \Eq{\ref{Equation:LTDResidueq0Pole}}). But before moving forward, it is worth mentioning that even though we removed the prescription coming from the Feynman propagator in \Eq{\ref{Equation:LTDResidueValue}}, it still played an important role in the application of Cauchy's Residue Theorem, as it is the reason why we selected the pole with positive energy. Now, the computation of the pre-factor is a bit more subtle, as the prescription plays an even bigger role, and its careful and rigorous treatment is critical for the consistency of the method.\\
\\
Explicitly, we have
\begin{align}\label{Equation:LTDPrefactorExplicit}
\left[\,\prod\limits_j\,G_F(q+k_j)\,\right]_{q_0=\Sup{q_0}}=&~\left[\,\prod\limits_j\,\frac{1}{(q+k_j)^2-m_j^2+i0}\right]_{q_0=\Sup{q_0}}\nn\\
=&~\prod\limits_j\,\frac{1}{m^2+2\Sup{q_0}k_{j,0}-2\mathbf{q}\cdot\mathbf{k}_j+k_j^2-m_j^2}\;,\nn\\
\end{align}
where we replaced $q_0$ by $\Sup{q_0}=\sqrt{\mathbf{q}^2+m^2-i0}$ in the last equality. Then, by noting that $2\Sup{q_0}\approx2\sqrt{\mathbf{q}^2+m^2}-i0/\sqrt{\mathbf{q}^2+m^2}$, we can write
\begin{align}\label{Equation:LTDPrefactorExplicit2}
\left[\,\prod\limits_j\,G_F(q+k_j)\,\right]_{q_0=\Sup{q_0}}=&~\prod\limits_j\,\frac{1}{m^2+2\sqrt{\mathbf{q}^2+m^2}\,k_{j,0}-2\mathbf{q}\cdot\mathbf{k}_j+k_j^2-m_j^2-i0\,k_{j,0}/\sqrt{\mathbf{q}^2+m^2}}\;.\nn\\
=&~\left[\,\prod\limits_j\,\frac{1}{q^2+2q\cdot k_j+k_j^2-m_j^2-i0\,k_{j,0}/q_0}\right]_{q_0=\sqrt{\mathbf{q}^2+m^2}}\;.
\end{align}
The potential singularities appearing in \Eq{\ref{Equation:LTDPrefactorExplicit2}} are regularised thanks to the shift from the real axis produced by the imaginary term $i0\,k_{j,0}/q_0$. In addition, it is worth mentioning that in the limit of the infinitesimal prescription, only the sign in front of $i0$ matters\footnote{In general, the sign in front of a given prescription is always unambiguously defined at one-loop level. This may not be true any more, though, when dealing with two-loop amplitudes.}. Therefore, knowing that $q_0>0$, we can perform the replacement $i0\,k_{j,0}/q_0\to i0\,\eta\cdot k_j$, where $\eta$ can be any future-like vector, namely fulfilling
\begin{equation}\label{Equation:LTDeta}
\eta^\mu=(\eta_0,\boldsymbol{\eta})\quad\text{with}\quad
\begin{cases}
\eta_0>0\\
\eta^2\geq0
\end{cases}\;.
\end{equation}
Thus, we finally obtain
\begin{equation}\label{Equation:LTDPrefactorFinal}
\left[\,\prod\limits_j\,G_F(q+k_j)\,\right]_{q_0=\Sup{q_0}}=\left[\,\prod\limits_j\,\frac{1}{(q+k_j)^2-m_j^2-i0\,\eta\cdot k_j}\right]_{q_0=\sqrt{\mathbf{q}^2+m^2}}\;,
\end{equation}
which, after reintroducing the index $i$ we dropped earlier, gives us
\begin{equation}\label{Equation:LTDPrefactorithpole}
\underbracket[0.5pt][5pt]{\left[\,\prod\limits_{\substack{j\in\alpha\\j\neq i}}\,G_F(q_j)\right]}_{\{\text{Evaluated at the }i\text{-th pole}\}}=\left[\,\prod\limits_{\substack{j\in\alpha\\j\neq i}}\,\frac{1}{q_j^2-m_j^2-i0\,\eta\cdot(q_j-q_i)}\right]_{q_{i,0}=\Sup{q_{i,0}}}\;.
\end{equation}
As we can see, setting on shell the internal line with momentum $q_i$ also affects the remaining propagators. Indeed, the singularity appearing when $q_j^2=m_j^2$ for $i\neq j$ is no longer regularised by the customary $+i0$ Feynman prescription, but instead by a new prescription, $-i0\,\eta\cdot(q_j-q_i)$, which we call the \emph{dual prescription}. This prescription arises from the fact the original Feynman propagator $G_F(q_j)$ is evaluated at the complex value of the loop momentum $q_i$, which is determined by the location of the pole at $\big(G_F(q_i)\big)^{-1}$. The $i0$ dependence from the pole has to be combined with the $i0$ dependence as given by the dual prescription.\\
\\
Finally, inserting \Eq{\ref{Equation:LTDResidueFinal}} and \Eq{\ref{Equation:LTDPrefactorithpole}} into \Eq{\ref{Equation:LTDCauchy}} gives the duality relation between one-loop integrals and phase-space integrals
\begin{equation}\label{Equation:LTDDualityRelation}
L^{(1)}(p_1,p_2,\dots,p_N)=-\widetilde{L}^{(1)}(p_1,p_2,\dots,p_N)\;,
\end{equation}
where the expression of the phase-space integral $\widetilde{L}^{(1)}$ is given by
\begin{equation}\label{Equation:LTDL1tilde}
\widetilde{L}^{(1)}(p_1,p_2,\dots,p_N)=\int_\ell\,\,\sum\limits_{i\in\alpha}\,\deltatilde{q_i}\,\prod\limits_{{\substack{j\in\alpha\\j\neq i}}}\,G_D(q_i;q_j)\;,
\end{equation}
with
\begin{equation}\label{Equation:LTDDeltaTilde}
\deltatilde{q_i}=2\pi i\,\deltaplus{q_i^2-m_i^2}=2\pi i\,\theta(q_{i,0})\,\delta(q_i^2-m_i^2)\;,\nn\\
\end{equation}
and
\begin{equation}\label{Equation:LTDDualPropagator}
G_D(q_i;q_j)=\frac{1}{q_j^2-m_j^2-i0\,\eta\cdot(q_j-q_i)}\;.
\end{equation}
In a given dual contribution, the $N-1$ Feynman propagators $G_F(q_j)$ are substituted by the corresponding \emph{dual propagators} $G_D(q_i;q_j)$. Only their prescriptions have been modified, and these are completely independent of the internal momentum $\ell$. They are therefore fixed\footnote{When dealing with several loops, this statement is not true any more. See Section \ref{Section:LTDMultiLoop}.}. As said above, the value of $\eta$ is arbitrary, as long it is taken to be a future-like vector. One important point to keep into consideration, however, is that in order to cancel the dependence in $\eta$ in $\widetilde{L}^{(1)}$, one has to take the same $\eta$ across all contributions. For simplicity, we will take
\begin{equation}\label{Equation:Eta1000}
\eta^\mu=(1,0,0,0)
\end{equation}
for any subsequent computations in this thesis.\\
\\
The presence of $\eta$ is a consequence of using Cauchy's residue theorem, and is essential to the consistency of the LTD theorem. In addition, the fact that the prescription depends on $\eta$ indicates that the residues at each of the poles are not Lorentz-invariant quantities. It is only when summing over all the residues that one recovers Lorentz invariance. Indeed, the one-loop integral $L^{(1)}$ is a function of the $N^2$ Lorentz invariants $(p_i\cdot p_j)$, and has a complex analytic structure involving poles and branch-cut singularities in the multi-dimensional space of these variables. The usual $+i0$ prescription of the Feynman propagators select a Riemann sheet in this multi-dimensional space and therefore unambiguously defines $L^{(1)}$ as a single-valued function. However, after applying the LTD theorem, each contribution to $\widetilde{L}^{(1)}$ has additional -- unphysical -- singularities in the multidimensional complex space. The auxiliary vector $\eta$, however, fixes the position of those singularities by correlating the various single-cut contributions, so that they are evaluated on the same Riemann sheet, which leads to the cancellation of the unphysical singularities we just mentioned. More details about this subject are given in the following section.

\subsection{Cancellation of singularities among dual integrals}\label{Section:LTDCancellationOfThreshold}

In this section, we explicitly show how the cancellation of unphysical spurious singularities takes place within the LTD formalism, by carefully studying the structure of the pole of the dual propagators \cite{Buchta:2014dfa}.\\
\\
Let's start by considering the dual contribution
\begin{equation}\label{Equation:LTDIi}
I_i=-\int_\ell\,\deltatilde{q_i}\,\prod\limits_{j\neq i}\,G_D(q_i;q_j)=-\int_\ell\,\deltatilde{q_i}\,\prod\limits_{j\neq i}\,\frac{1}{q_j^2-m_j^2-i0\,\eta\cdot k_{ji}}\;.
\end{equation}
A crucial point of our discussion is to notice that it is possible to rewrite a given dual propagator as
\begin{equation}\label{Equation:LTDDualPropagatorRewritten}
\deltatilde{q_i}\,G_D(q_i;q_j)=2\pi i\,\frac{\delta(q_{i,0}-\Sup{q_{i,0}})}{2\Sup{q_{i,0}}}\,\frac{1}{(\Sup{q_{i,0}}+k_{ji,0})^2-(\Sup{q_{j,0}})^2}\;,
\end{equation}
where we recall (\Eq{\ref{Equation:LTDGF0}})
\begin{equation}\label{Equation:LTDqi0p}
\Sup{q_{i,0}}=\sqrt{\mathbf{q}_i^2+m_i^2-i0}
\end{equation}
is the loop energy measured along the on-shell hyperboloid whose origin is at $-k_i$. While the factor $1/\Sup{q_{i,0}}$ can become singular if $m_i=0$, the integral
\begin{equation}\label{Equation:LTDIntegrateNoIR}
\int_\ell\,\frac{\delta(q_{i,0}-\Sup{q_{i,0}})}{2\Sup{q_{i,0}}}
\end{equation}
is still convergent by two powers in the IR region, thanks to the presence of the integration measure. Soft singularities indeed require at least two dual propagators to vanish simultaneously. If we define
\begin{equation}\label{Equation:LTDLambdapm}
\lambda_{ij}^{\pm\pm}=\pm \Sup{q_{i,0}}\pm \Sup{q_{j,0}}+k_{ji,0}\;,
\end{equation}
we can see from \Eq{\ref{Equation:LTDDualPropagatorRewritten}} that the inverse dual propagator $\big(G_D(q_i;q_j)\big)^{-1}$ vanishes if either one of
\begin{equation}\label{Equation:LTDPoleConditions}
\lambda_{ij}^{++}=0\qquad\text{or}\qquad\lambda_{ij}^{+-}=0
\end{equation}
is fulfilled. The first condition of \Eq{\ref{Equation:LTDPoleConditions}} is satisfied if the forward hyperboloid of $-k_i$ intersects with the backward hyperboloid of $-k_j$. The second condition is verified if the two forward hyperboloids intersect.\\
\\
In the massless case, the hyperboloids reduce to light-cones in the loop three-momentum space, and $\Sup{q_{i,0}}$ and $\Sup{q_{j,0}}$ are the distance from the foci located at $-\mathbf{k}_i$ and $-\mathbf{k}_j$, respectively, with the distance between these two foci being $\sqrt{k_{ji}^2}$. In the massive case, \Eq{\ref{Equation:LTDqi0p}} can be reinterpreted as the distance associated to a four-dimensional space with one ``massive'' dimension and the foci now located at $(-\mathbf{k}_i,-m_i)$ and $(-\mathbf{k}_j,-m_j)$, respectively. Then, the singularity arises at the intersection of the conic section given by \Eq{\ref{Equation:LTDPoleConditions}} in this generalised space with the zero mass plane. This picture is useful to identify the singular regions of the loop integrand in the loop three-momentum space.\\
\\
The solution to the first condition of \Eq{\ref{Equation:LTDPoleConditions}} is an ellipsoid and clearly requires $k_{ji,0}<0$. Moreover, since it is the result of the intersection of a forward with a backward hyperboloid, the distance between the two propagators has to be future-like, i.e. we must have $k_{ji}^2>0$. In fact, internal masses further restrict this condition. Bearing in mind the image of the conic sections in the generalised massive space, we can intuitively deduce\footnote{Of course, a more mathematical approach would lead to the same conclusion.} that this first condition has a solution if and only if
\begin{equation}\label{Equation:LTDlambdappCondition}
\begin{cases}
k_{ji,0}<0\\
k_{ji}^2-(m_j+m_i)^2\geq0
\end{cases}\;.
\end{equation}
On the other hand, the solution of the second condition of \Eq{\ref{Equation:LTDPoleConditions}} is an hyperboloid in the generalised space, and this time there can be solutions for $k_{ji}$ either positive or negative, i.e. when either of the two momenta are set on shell. However, by interpreting the result in the generalised space, it is clear that the intersection with the zero mass plane does not always exist, and if it does, it can either be an ellipsoid or a hyperboloid in the loop three-momentum space. Here, the distance between the momenta of the propagators has to be space-like, although time-like configurations can fulfil this second condition as well, as long as the time-like distance is small, or close to light-like. We can show that the condition
\begin{equation}\label{Equation:LTDlambdapmCondition}
k_{ji}^2-(m_j-m_i)^2\leq0\;.
\end{equation}
is necessary for the existence of a solution. In any other configuration, the singularity only appears for loop-three momenta with imaginary components.\\
\begin{figure}[t]
	\centering
	\includegraphics[width=7cm]{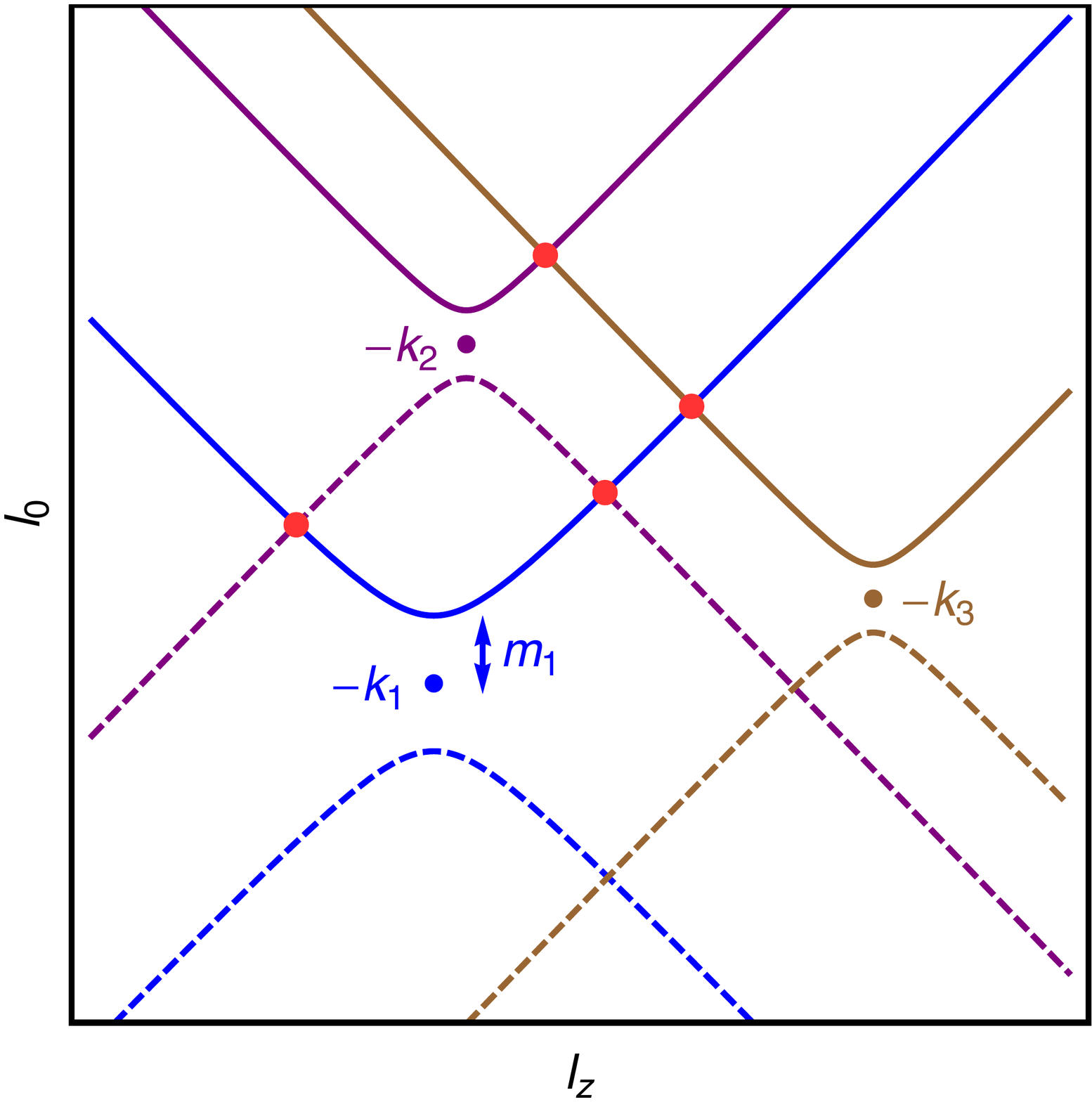}
	\includegraphics[width=7cm]{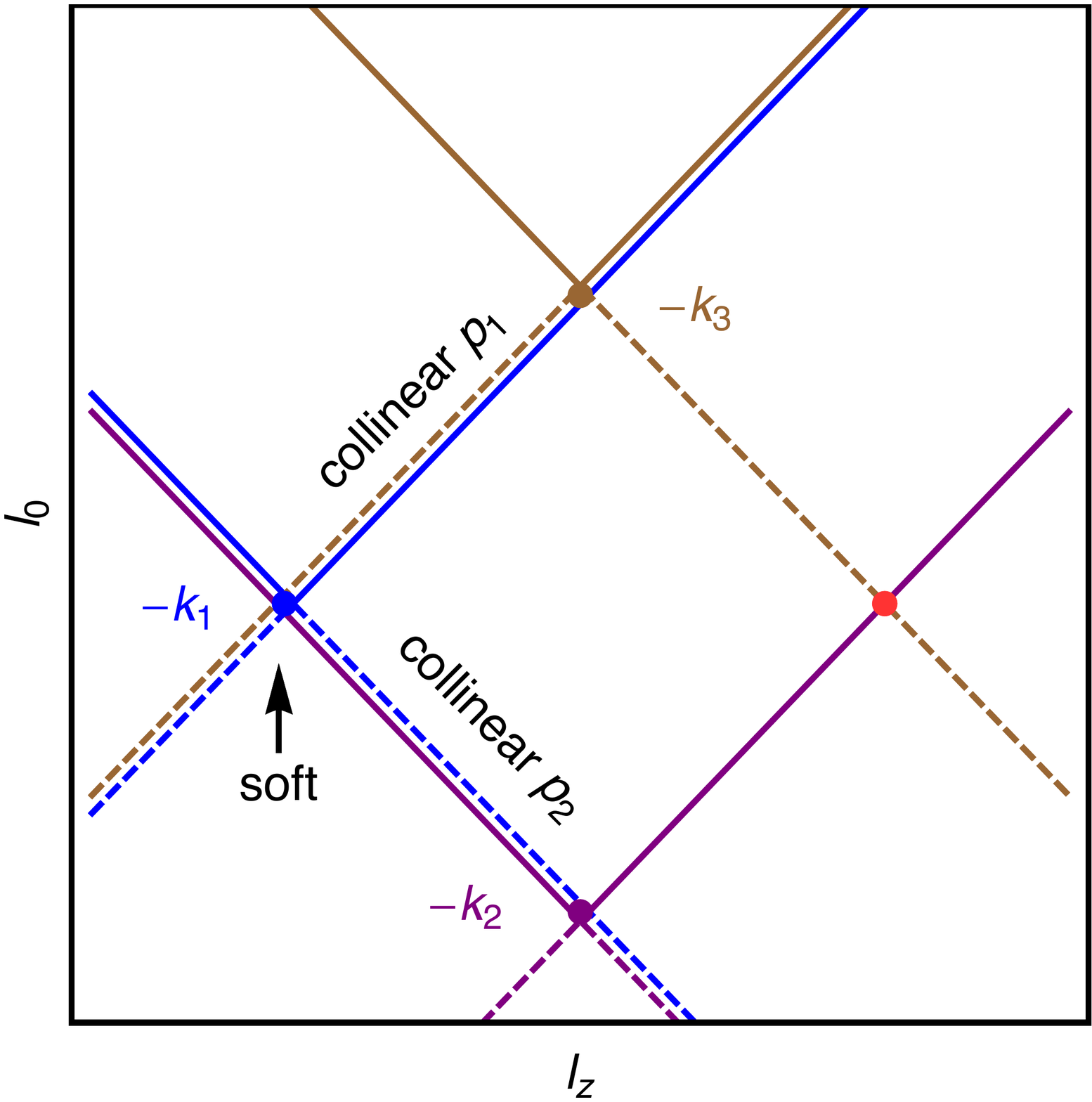}
	\caption{On-shell hyperboloids for three arbitrary propagators in Cartesian coordinates in the ($\ell_0$,$\boldsymbol{\ell}_z$) space (left), and kinematical configurations with infrared singularities (right). In the latter case, the on-shell hyperboloids degenerate to light-cones.}
	\label{Figure:LTDCartesean}
\end{figure}
\\
In the following, we will prove the partial cancellation of singularities among dual integrands that has been briefly discussed in the previous subsection. Let's first consider two Feynman propagators separated by a space-like distance $k_{ji}^2<0$ (or more generally, fulfilling \Eq{\ref{Equation:LTDlambdapmCondition}}). In the corresponding dual representations, one of these propagators is cut and the corresponding particle is set on shell, the other becomes dual, and the integration occurs along the on-shell hyperboloid of the former. A graphical representation of this configuration is shown in \Fig{\ref{Figure:LTDCartesean}} (left). There, the two forward hyperboloids of $-k_1$ and $-k_3$ intersect at a single point. Integrating over $\boldsymbol{\ell}_z$ along the forward hyperboloid of $-k_1$, we find that the dual propagator $G_D(q_1;q_3)$, which is negative below the intersection point where the integrand becomes singular, changes sign above this particular point as we move from outside to inside the on-shell hyperboloid of $-k_3$. The opposite occurs in the contribution where $q_3$ is set on shell; $G_D(q_3;q_1)$ is positive below the intersection point, and negative above. The change of sign leads to the cancellation of the singularity they have in common. Notice also that the dual $i0$ prescription changes sign. Analytically, we have
\begin{equation}\label{Equation:LTDThresholdCancellation1}
\lim\limits_{\lambda_{ij}^{+-}\to0}\left(\frac{\deltatilde{q_i}\,G_D(q_i;q_j)}{2\pi i}+\{i\leftrightarrow j\}\right)=\left(\frac{1}{\lambda_{ij}^{+-}}-\frac{1}{\lambda_{ij}^{+-}}\right)\frac{\delta(q_{i,0}-\Sup{q_{i,0}})}{(2\Sup{q_{i,0}})(2\Sup{q_{j,0}})}+\mathcal{O}\big((\lambda_{ij}^{+-})^0\big)\;,
\end{equation}
demonstrating that the leading behaviour indeed cancels among the two dual contributions. The cancellation of these singularities is neither altered by the presence of other non-vanishing dual propagators nor by a numerator, as
\begin{equation}\label{Equation:LTDThresholdOtherProp}
\lim\limits_{\lambda_{ij}^{+-}\to0}G_D(q_j;q_k)=\lim\limits_{\lambda_{ij}^{+-}\to0}\frac{1}{(\Sup{q_{j,0}}+k_{ki,0}-k_{ji,0})^2-(\Sup{q_{k,0}})^2}=\lim\limits_{\lambda_{ij}^{+-}\to0}G_D(q_i;q_k)\;,
\end{equation}
where we have used $k_{kj}=k_{ki}-k_{ji}$. If instead, the separation is time-like (in the sense of the second condition of \Eq{\ref{Equation:LTDPoleConditions}}, we find
\begin{equation}\label{Equation:LTDThresholdNoCancellation1}
\lim\limits_{\lambda_{ij}^{++}\to0}\left(\frac{\deltatilde{q_i}\,G_D(q_i;q_j)}{2\pi i}+\{i\leftrightarrow j\}\right)=-\theta(-k_{ji,0})\frac{1}{\lambda_{ij}^{++}}\,\frac{\delta(q_{i,0}-\Sup{q_{i,0}})}{(2\Sup{q_{i,0}})(2\Sup{q_{j,0}})}+\{i\leftrightarrow j\}+\mathcal{O}\big((\lambda_{ij}^{++})^0\big)\;.
\end{equation}
In this case, the singularity of the integrand persists because of the presence of the Heaviside step function.\\
\\
In the case where more than two propagators become simultaneously singular, we need to define
\begin{equation}\label{Equation:LTDThreePropLimitParam}
\lambda\,\lambda_{ab}^{\pm\pm}=\pm\Sup{q_{a,0}}\pm\Sup{q_{b,0}}+k_{ba,0}\;,
\end{equation}
with $a,b\in\{i,j,k\}$ and $a\neq b$. We also added a weight $\lambda$ to the parameters so we can take simultaneous limits. When three propagators become simultaneously singular at the intersection of three forward hyperboloids, we have
\begin{align}\label{Equation:LTDThreePropSingularFFF}
&\lim\limits_{\lambda\to0}\left(\frac{\deltatilde{q_i}\,G_D(q_i;q_j)\,G_D(q_i,q_k)}{2\pi i}+\{(i,j,k)\to(j,k,i)\}+\{(i,j,k)\to(k,i,j)\}\right)\nn\\
&=\frac{1}{\lambda^2}\left(\frac{1}{\lambda_{ij}^{+-}\,\lambda_{ik}^{+-}}+\frac{1}{\lambda_{ij}^{+-}(\lambda_{ij}^{+-}-\lambda_{ik}^{+-})}+\frac{1}{\lambda_{ik}^{+-}(\lambda_{ik}^{+-}-\lambda_{ij}^{+-})}\right)\frac{\delta(q_{i,0}-\Sup{q_{i,0}})}{(2\Sup{q_{i,0}})(2\Sup{q_{j,0}})(2\Sup{q_{k,0}})}+\mathcal{O}\left(\lambda^{-1}\right)\;,
\end{align}
where the leading singular behaviour vanishes when taking the sum of all three contributions. Although not shown since its expression is quite complex, the $\mathcal{O}(\lambda^{-1})$ term also vanishes, thus rendering the full integrand finite in the limit $\lambda\to0$. When two forward hyperboloids intersect simultaneously with a backward hyperboloid, we have
\begin{align}\label{Equation:LTDThreePropSingularFFB}
&\lim\limits_{\lambda\to0}\left(\frac{\deltatilde{q_i}\,G_D(q_i;q_j)\,G_D(q_i,q_k)}{2\pi i}+\{(i,j,k)\to(j,k,i)\}+\{(i,j,k)\to(k,i,j)\}\right)\nn\\
&=\theta(-k_{ki,0})\theta(-k_{kj,0})\frac{1}{\lambda^2}\left(\frac{1}{\lambda_{ik}^{++}(\lambda_{jk}^{++}-\lambda_{ik}^{++})}+\frac{1}{\lambda_{jk}^{++}(\lambda_{ik}^{++}-\lambda_{jk}^{++})}\right)\frac{\delta(q_{i,0}-\Sup{q_{i,0}})}{(2\Sup{q_{i,0}})(2\Sup{q_{j,0}})(2\Sup{q_{k,0}})}+\mathcal{O}\left(\lambda^{-1}\right)\;,
\end{align}
where the singularity in $1/(\lambda_{jk}^{++}-\lambda_{ik}^{++})$ cancels (it is not shown but it is once again also the case at $\mathcal{O}\left(\lambda^{-1}\right)$). Finally, when two backward hyperboloids intersect with a forward hyperboloid,
\begin{align}
&\lim\limits_{\lambda\to0}\left(\frac{\deltatilde{q_i}\,G_D(q_i;q_j)\,G_D(q_i,q_k)}{2\pi i}+\{(i,j,k)\to(j,k,i)\}+\{(i,j,k)\to(k,i,j)\}\right)\nn\\
&=\theta(-k_{ki,0})\theta(-k_{ji,0})\frac{1}{\lambda^2\,\lambda_{ij}^{++}\,\lambda_{ik}^{++}}\frac{\delta(q_{i,0}-\Sup{q_{i,0}})}{(2\Sup{q_{i,0}})(2\Sup{q_{j,0}})(2\Sup{q_{k,0}})}+\mathcal{O}\left(\lambda^{-1}\right)\;.
\end{align}
The same exercise can be carried out for the case where four hyperboloids intersect, for which it is straightforward to prove that it does not lead to any common singularities.\\
\\
By studying all the different possible scenarios, we showed that singularities of space-like separated propagators, taking place where forward on-shell hyperboloids intersect, vanish in the dual representation of loop integrand. This cancellation of singularities represents a big advantage of the LTD formalism with respect to the direct integration in the four-dimensional loop space; it makes unnecessary the use of contour deformation to deal numerically with the integrable singularities of these configurations.\\
\\
The remaining singularities (coming from forward-backward intersections) are IR singularities and are inherent to the process under consideration; they will therefore remain in the sum of the dual contributions. Collinear singularities occur when two massless propagators are separated by a light-like distance, $k_{ji}^2=0$. In that case, the corresponding light-cones overlap along an infinite interval. If we assume $k_{i,0}>k_{j,0}$ for instance, the collinear singularity for $\ell_0>-k_{j,0}$ appears at the intersection of the two forward light-cones, with the forward light-cone of $-k_j$ located inside the forward light-cone of $-k_i$, or equivalently with the forward light-cone of $-k_i$ located outside the forward light-cone of $-k_j$. As a consequence of this, the singular behaviours of the two dual contributions cancel one another. If instead we have $-k_{i,0}<\ell_0<-k_{j,0}$, then it is the forward light-cone of $-k_i$ that intersects tangentially with the backward light-cone of $-k_j$, according to the first condition of \Eq{\ref{Equation:LTDPoleConditions}}. In this region, collinear singularities remain, but the range of the loop three-momentum is limited. If there are more than one momenta separated by light-like distances, the region in which we have IR singularities is this time limited by the minimal and maximal energies of the external momenta. The singularity of the integrand at $\Sup{q_{i,0}}$ leads to an actual soft divergence only if two other propagators --~each one contributing one power in the infrared~-- are light-like separated from $-k_i$. In \Fig{\ref{Figure:LTDCartesean}}, this condition is fulfilled only for $\Sup{q_{1,0}}=0$, but not for $\Sup{q_{2,0}}=0$ or $\Sup{q_{3,0}}=0$.\\
\\
In summary, both threshold and IR singularities are limited to a compact region of the loop three-momentum, which is of the order of the external momenta. Outside this region, singularities can only take place at forward-forward intersection, and thus vanish in the sum of all dual contributions.

\subsection{Similarities with the Feynman Tree Theorem}

Since it shares many aspects with the LTD theorem, it is worth discussing a similar relation known as Feynman's Tree Theorem (FTT)~\cite{Feynman:1963ax}. In this subsection we recall the FTT and examine how it compares to the LTD theorem\footnote{For a more thorough comparison between FTT and the LTD theorem, see~\cite{Catani:2008xa}.}.\\
\\
We start by defining, for an internal particle with four-momentum $q$ and mass $m$, the \emph{advanced propagator} $G_A(q)$ as
\begin{equation}\label{Equation:LTDGA}
G_A(q)=\frac{1}{q^2-m^2-i0\,q_0}\;.
\end{equation}
By applying Cauchy's residue theorem on
\begin{equation}\label{Equation:LTDLA}
L_A^{(1)}(p_1,p_2,\dots\,p_N)=\int_\ell\,\,\prod\limits_{i\in\alpha}\,G_A(q_i)\;,
\end{equation}
using the same contour $C_L$ as when we derived the LTD theorem, we remark that
\begin{equation}\label{Equation:LTDLA0}
L_A^{(1)}(p_1,p_2,\dots\,p_N)=0\;,
\end{equation}
since $G_A$ only has poles with positive imaginary part. By noting that
\begin{equation}\label{Equation:LTDGARelation}
G_A(q)=G_F(q)+\deltatilde{q}\;,
\end{equation}
with $G_F(q)$ and $\deltatilde{q}$ given in \Eqs{\ref{Equation:LTDGF}}{\ref{Equation:LTDDeltaTilde}}, respectively, we have
\begin{align}\label{Equation:LTDLADecomposed}
L_A^{(1)}(p_1,p_2,\dots\,p_N)=&~\int_\ell\,\,\prod\limits_{i\in\alpha}\,\left(G_F(q_i)+\deltatilde{q_i}\right)\nn\\
=&~L^{(1)}(p_1,p_2,\dots\,p_N)+\,\sum_{i=1}^N\,L_{i-\textrm{cut}}^{(1)}(p_1,p_2,\dots\,p_N)\;,
\end{align}
where $L_{m-\textrm{cut}}^{(1)}(p_1,p_2,\dots\,p_N)$ is the sum of all contributions where $m$ propagators $G_F(q_i)$ have been cut, i.e. replaced by a $\deltatilde{q_i}$. FTT is obtained by combining \Eqs{\ref{Equation:LTDLA0}}{\ref{Equation:LTDLADecomposed}}, and relates the one-loop amplitude to all\footnote{Note that for $m>d$, $L_{m-\textrm{cut}}^{(1)}(p_1,p_2,\dots\,p_N)=0$ since the delta functions give more conditions than the number of available integration variables.} the $m$-cut contributions, namely
\begin{equation}\label{Equation:LTDFTT}
L^{(1)}(p_1,p_2,\dots\,p_N)=-\,\sum_{i=1}^N\,L_{i-\textrm{cut}}^{(1)}(p_1,p_2,\dots\,p_N)\;.
\end{equation}
The first contribution, which reads
\begin{equation}\label{Equation:LTDL1cut}
L_{1-\textrm{cut}}^{(1)}(p_1,p_2,\dots\,p_N)=\int_\ell\,\,\sum\limits_{i\in\alpha}\,\deltatilde{q_i}\,\prod\limits_{{\substack{j\in\alpha\\j\neq i}}}\,G_F(q_i)\;,
\end{equation}
is very similar to $\widetilde{L}^{(1)}(p_1,p_2,\dots,p_N)$ given in \Eq{\ref{Equation:LTDL1tilde}}, the only difference being that it involves Feynman propagators instead of dual ones. But since in the end, calculating the one-loop amplitude must lead to the same result regardless of how it is decomposed, we deduce that the modified prescriptions that appear in the dual propagators in the LTD relation account for every remaining multi-cuts that appear in FTT. This is a very strong advantage of LTD over FTT, since only single-cut contributions have to be taken into account, removing completely the need to evaluate multi-cut contributions. A careful bookkeeping of the dual prescriptions throughout the entire calculation, though, is the price to pay.

\section{Loop-Tree Duality beyond one loop}\label{Section:LTDMultiLoop}
\fancyhead[LO]{\ref*{Section:LTDMultiLoop}~~\nameref*{Section:LTDMultiLoop}}

\begin{figure}[h]
	\centering
	\begin{picture}(350,320)(-175,-160)
	\SetWidth{1.5}
	\Arc[arrow](0,0)(100,-30,30)
	\Arc[arrow](0,0)(100,30,60)
	\Arc[arrow](0,0)(100,60,90)
	\Arc[arrow,flip](0,0)(100,90,120)
	\Arc[arrow,flip](0,0)(100,90,120)
	\Arc[arrow,flip](0,0)(100,120,150)
	\Arc[arrow,flip](0,0)(100,150,210)
	\Arc[arrow,flip](0,0)(100,210,240)
	\Arc[arrow,flip](0,0)(100,240,270)
	\Arc[arrow](0,0)(100,-60,-30)
	\Arc[arrow](0,0)(100,-90,-60)
	\Line[arrow](0,100)(0,60)
	\Line[arrow](0,60)(0,20)
	\Line[arrow](0,20)(0,-60)
	\Line[arrow](0,-60)(0,-100)
	\Line[arrow](87,50)(130,75)
	\Line[arrow](87,-50)(130,-75)
	\Line[arrow](-87,-50)(-130,-75)
	\Line[arrow](50,87)(75,130)
	\Line[arrow](0,100)(0,150)
	\Line[arrow](-50,87)(-75,130)
	\Line[arrow](-87,50)(-130,75)
	\Line[arrow](-50,-87)(-75,-130)
	\Line[arrow](50,-87)(75,-130)
	\Line[arrow](0,-100)(0,-150)
	\Line[arrow](0,60)(44,60)
	\Line[arrow](0,20)(44,20)
	\Line[arrow](0,-60)(44,-60)
	\Arc[arrow](0,0)(80,325,5)
	\Arc[arrow,flip](0,0)(80,175,215)
	\Vertex(-125,0){2}
	\Vertex(125,0){2}
	\Vertex(-121,-32){2}
	\Vertex(-121,32){2}
	\Vertex(121,32){2}
	\Vertex(121,-32){2}
	\Vertex(25,0){2}
	\Vertex(25,-20){2}
	\Vertex(25,-40){2}
	\color{blue}
	\large
	\Text(85,140){$p_1$}
	\Text(140,85){$p_2$}
	\Text(140,-85){$p_{r-1}$}
	\Text(85,-140){$p_r$}
	\Text(0,-160){$p_N$}
	\Text(-85,-140){$p_{r+1}$}
	\Text(-140,-85){$p_{r+2}$}
	\Text(-140,85){$p_{s-2}$}
	\Text(-85,140){$p_{s-1}$}
	\Text(0,160){$p_s$}
	\Text(60,60){$p_{s+1}$}
	\Text(60,20){$p_{s+2}$}
	\Text(62,-60){$p_{N-1}$}
	\Text(81,81){$q_1$}
	\Text(30,111){$q_0$}
	\Text(86,-81){$q_{r-1}$}
	\Text(-86,-81){$q_{r+2}$}
	\Text(-30,111){$q_s$}
	\Text(-81,81){$q_{s-1}$}
	\Text(-30,-111){$q_{r+1}$}
	\Text(30,-111){$q_r$}
	\Text(18,80){$q_{s+1}$}
	\Text(18,40){$q_{s+2}$}
	\Text(15,-80){$q_N$}
	\Text(63,-17){$\ell_1$}
	\Text(-63,-17){$\ell_2$}
	\end{picture}
	\caption{Assignment of momenta in the two-loop the master diagram.}
	\label{Figure:LTDTwoLoopMasterDiagram}
\end{figure}
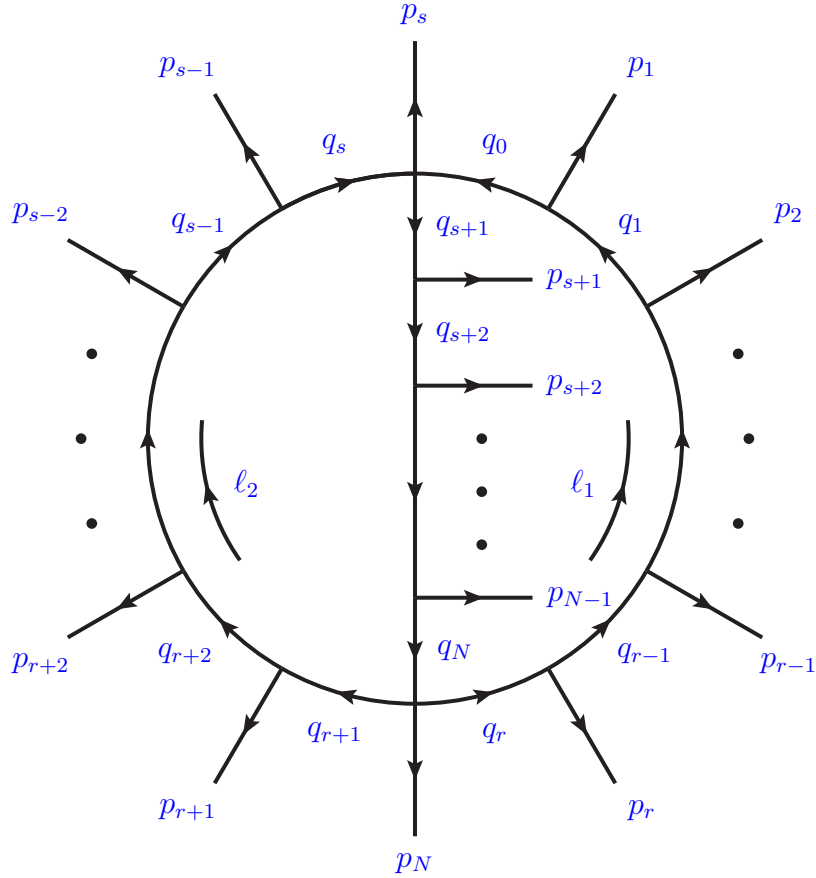

In the previous section, we established the LTD framework, and derived the duality relation at one-loop level. Here, we iteratively extend the LTD formalism so it can deal with an arbitrary number of loops. We will mostly focus on the two-loop case, which is sufficient for the work carried out later in this thesis. For a more detailed and exhaustive discussion, see \cite{Bierenbaum:2010cy}. In the following, we will start by introducing notations that are more suited for multi-loop computations within our framework. We will then describe the general iterative procedure to obtain a LTD representation of multi-loop integrals. And finally, we will explicitly apply this procedure to the two-loop case to derive the corresponding duality relation.\\
\\
First, we need to extend the definition of the Feynman and dual propagators -- whose argument in the one-loop case was a single internal momentum $q_i$ -- to sets of internal momenta. Let $\alpha_k$ be a set of indices, namely $\alpha_k=\{i_1,i_2,\dots\}$, where $q_{i_j}$ are four-momenta of internal propagators. From now on and all throughout this thesis, we will use $\alpha_k$ to denote both the set of indices and the corresponding set of internal momenta, as there is an unambiguous correspondence between the two. For instance, we do not differentiate $\alpha_k=\{1,2,3\}$ and $\alpha_k=\{q_1,q_2,q_3\}$. With these notations, we generalise the definition of the Feynman and dual propagators to a set of internal lines $\alpha_k$ by defining
\begin{equation}\label{LTDPropagatorsOfSets}
G_F(\alpha_k)=\,\prod\limits_{i\in\alpha_k}\,G_F(q_i)\;,\qquad G_D(\alpha_k)=\,\sum\limits_{i\in\alpha_k}\,\deltatilde{q_i}\,\prod\limits_{\substack{j\in\alpha_k\\j\neq i}}\,G_D(q_i;q_j)\;.
\end{equation}
In the case where $\alpha_k=\{i\}$, then $G_D(\alpha_k)=\deltatilde{q_i}$ by definition. Although individual terms inside $G_D(\alpha_k)$ depend on the dual vector $\eta$, the dependence vanishes after considering the sum over all terms appearing in $G_D(\alpha_k)$, as it was already the case at one loop. For further use, we also define $-\alpha_k$ to be the set $\alpha_k$ in which all internal lines have had their flow reversed. Accordingly, we have
\begin{equation}\label{Equation:LTDGDMinusAlphak}
G_D(-\alpha_k)=\,\sum\limits_{i\in\alpha_k}\,\deltatilde{-q_i}\,\prod\limits_{\substack{j\in\alpha_k\\j\neq i}}\,G_D(-q_i;-q_j)\;.
\end{equation}
 In practice, this means that instead of selecting the positive energy modes as we would usually do, we select the negative ones, i.e. for a given internal momenta $q_i$ belonging to $\alpha_k$,
\begin{equation}\label{LTDDeltaTildeMinus}
\deltatilde{-q_i}=\frac{i\pi}{\Sup{q_{i,0}}}\,\delta(q_{i,0}+\Sup{q_{i,0}})\;.
\end{equation}
Note that $G_F(-\alpha_k)=G_F(\alpha_k)$, since the four-momentum is squared, and the $+i0$ prescription is independent of the direction of the flow.\\
\\
With these notations, the one-loop duality relation in \Eq{\ref{Equation:LTDDualityRelation}} simply reads
\begin{equation}\label{Equation:LTDOneLoopNewNotations}
\int_\ell\,G_F(\alpha)=-\int_\ell\,G_D(\alpha)\;,
\end{equation}
with $\alpha=\{1,2,\dots,N\}$, where we reused the notations of Section \ref{Section:LTDOneLoop}. Moreover, while for two sets $\alpha_i$ and $\alpha_j$, we always have $G_F(\alpha_i\cup\alpha_j)=G_F(\alpha_i)\,G_F(\alpha_j)$, this is not true for dual propagators in general. Instead, it is possible to show that we have the very interesting identity\footnote{For a complete proof of \Eq{\ref{Equation:LTDRelationGDGF}}, as well as the general relation for an arbitrary number of sets, see \cite{Bierenbaum:2010cy}.}
\begin{equation}\label{Equation:LTDRelationGDGF}
G_D(\alpha_i\cup\alpha_j)=G_D(\alpha_i)\,G_D(\alpha_j)+G_F(\alpha_i)\,G_D(\alpha_j)+G_D(\alpha_i)\,G_F(\alpha_j)\;,
\end{equation}
which is essential to the iterative procedure we are about to describe. Besides, if we have several sets $\alpha_1,\alpha_2,\dots,\alpha_N$ whose momenta depend on the same integration variable $\ell_i$, then
\begin{equation}\label{Equation:LTDGFtoGDli}
\int_{\ell_i}\,G_F(\alpha_1\cup\alpha_2\cup\dots\cup\alpha_N)=-\int_{\ell_i}\,G_D(\alpha_1\cup\alpha_2\cup\dots\cup\alpha_N)\;,
\end{equation}
which is a direct consequence of \Eq{\ref{Equation:LTDOneLoopNewNotations}}, and is simply the generalisation of the LTD theorem to a set of internal lines belonging to a single given loop of a multi-loop diagram. The main idea is to subsequently apply \Eq{\ref{Equation:LTDGFtoGDli}} to all loops one after the other, introducing an extra single-cut at each step of the process. One important thing to remember however, is that the one-loop LTD theorem only applies to Feynman propagators, and some loops that have not yet been cut may have had their Feynman propagators previously transformed into dual propagators (if they depend on more than one loop momenta, for instance). For that reason, we need to cleverly apply \Eq{\ref{Equation:LTDRelationGDGF}} (or rather, its general form for an arbitrary number of sets) whenever it is needed, in order to transform some of the dual propagators back into Feynman ones. Our goal in the end is to obtain an expression that has as many cuts (i.e. dual propagators) as loops involved.\\
\\
As an explicit example, let's now consider a general two-loop diagram, as shown in \Fig{\ref{Figure:LTDTwoLoopMasterDiagram}}. The two loop momenta are denoted $\ell_1$ and $\ell_2$, and they flow anti-clockwise and clockwise, respectively. As for the one-loop case, all external momenta are outgoing. We also write
\begin{align}\label{Equation:LTDpij}
p_{i,j}=&~p_{i+1}+p_{i+2}+\dots+p_{j-1}+p_j&\quad&\text{if}\quad i<j\;,\nn\\ p_{i,j}=&~-p_{i-1}-p_{i-2}-\dots-p_{j+1}-p_j&\quad&\text{if}\quad i>j\;,\nn\\
p_{i,j}=&~0&\qquad&\text{if}\quad i=j\;,
\end{align}
where $p_{0,N}=0$ because of momentum conservation. Note that unlike for the one-loop case, the number of external momenta and internal momenta may not be the same, hence the need for a different, more elaborated labelling.\\
\\
At two loops, according to \Fig{\ref{Figure:LTDTwoLoopMasterDiagram}}, we write
\begin{equation}
q_i=
\begin{cases}
\ell_1+p_{0,i}\;,\qquad&i\in\alpha_1\\
\ell_2+p_{i-1,s-1}\;,\qquad&i\in\alpha_2\\
\ell_1+\ell_2+p_{i,s}\;,\qquad&i\in\alpha_3\\
\end{cases}
\end{equation}
where $\alpha_k$, $k=1,2,3$, are sets of internal lines, defined by
\begin{equation}\label{Equation:LTDAlphaSets}
\alpha_1=\{0,1,\dots,r\}\;,\qquad\alpha_2=\{r+1,r+2,\dots,s\}\;,\qquad\alpha_3=\{s+1,s+2,\dots,N\}\;.
\end{equation}
The corresponding two-loop scalar integral $L^{(2)}$ reads
\begin{equation}\label{Equation:LTDL2}
L^{(2)}(p_1,p_2,\dots,p_N)=\int_{\ell_1}\,\int_{\ell_2}\,G_F(\alpha_1\cup\alpha_2\cup\alpha_3)=\int_{\ell_1}\,\int_{\ell_2}\,G_F(\alpha_1)\,G_F(\alpha_2\cup\alpha_3)\;,
\end{equation}
where we explicitly isolated one of the two loops, namely the loop depending on $\ell_2$ and $\ell_1+\ell_2$. This loop is the one that is on the left of \Fig{\ref{Figure:LTDTwoLoopMasterDiagram}}. The first step is to apply the LTD theorem at one loop, i.e. \Eq{\ref{Equation:LTDOneLoopNewNotations}}, to $\alpha_2\cup\alpha_3$. We obtain
\begin{equation}\label{Equation:LTDL2FirstStep}
L^{(2)}(p_1,p_2,\dots,p_N)=-\int_{\ell_1}\,\int_{\ell_2}\,G_F(\alpha_1)\,G_D(\alpha_2\cup\alpha_3)\;.
\end{equation}
As one can see, and as it was explained previously, it is not possible to directly apply the LTD theorem a second time, since neither $G_F(\alpha_1\cup\alpha_2)$ nor $G_F(\alpha_1\cup\alpha_3)$ appears in \Eq{\ref{Equation:LTDL2FirstStep}}. This is where we need to take advantage of the identity in \Eq{\ref{Equation:LTDRelationGDGF}}, allowing us to write
\begin{align}\label{Equation:LTDL2FirstStepBis}
L^{(2)}(p_1,p_2,\dots,p_N)=&~-\int_{\ell_1}\,\int_{\ell_2}\,G_F(\alpha_1)\Big[G_D(\alpha_2)\,G_D(\alpha_3)\nn\\
&~+G_F(\alpha_2)\,G_D(\alpha_3)+G_D(\alpha_2)\,G_F(\alpha_3)\Big]\;,
\end{align}
where $G_F(\alpha_1\cup\alpha_2)$ and $G_F(\alpha_1\cup\alpha_3)$ appear in the second and third term, respectively. The first term already exhibits two dual propagators, we can therefore leave it as it is. Applying \Eq{\ref{Equation:LTDOneLoopNewNotations}} to the remaining two terms leads to the final dual representation of the two-loop amplitude,
\begin{align}\label{Equation:LTDL2SecondStep}
L^{(2)}(p_1,p_2,\dots,p_N)=&~\int_{\ell_1}\,\int_{\ell_2}\,\Big[G_D(\alpha_2)\,G_D(\alpha_1\cup\alpha_3)\nn\\
&~+G_D(-\alpha_2\cup\alpha_1)\,G_D(\alpha_3)-G_F(\alpha_1)\,G_D(\alpha_2)\,G_D(\alpha_3)\Big]\;,
\end{align}
where each of the three terms now exhibits two dual cuts, as desired. Note that the indices of the sets appearing in \Eq{\ref{Equation:LTDL2SecondStep}} are completely interchangeable. It is important to take into account that the momenta in $\alpha_1$ and the ones in $\alpha_2$ have opposite flow when considering the loop formed by the union of these two sets. Therefore, in order to be able to apply the LTD theorem to this loop, we had to reverse the flow of $\alpha_2$ (for instance), which was done by adding a minus sign in front of this set in \Eq{\ref{Equation:LTDL2SecondStep}}, as illustrated in \Eq{\ref{Equation:LTDGDMinusAlphak}}. This duality relation at two-loop level can be, under certain conditions, reduced to a simpler version. We will see how in \Chapter{\ref{Chapter:HTL}}.\\
\\
While a rigorous demonstration of the cancellation of unphysical singularities among dual integrand at two-loop level within the LTD formalism has not yet been carried out, it is done in \Chapter{\ref{Chapter:HTL}} in the particular case of a below-threshold amplitude.

\section{Loop-Tree Duality beyond simple poles}\label{Section:LTDMultiPole}
\fancyhead[LO]{\ref*{Section:LTDMultiPole}~~\nameref*{Section:LTDMultiPole}}

\begin{figure}[t]
	\centering
	\begin{picture}(350,320)(-175,-160)
	\SetWidth{1.5}
	\Arc[arrow](0,0)(100,-30,30)
	\Arc[arrow](0,0)(100,30,60)
	\Arc[arrow,flip](0,0)(100,90,270)
	\Arc[arrow](0,0)(100,-60,-30)
	\Line[arrow](0,100)(0,-100)
	\Line[arrow](87,50)(130,75)
	\Line[arrow](50,87)(75,130)
	\Line[arrow](50,-87)(75,-130)
	\Line[arrow](87,-50)(130,-75)
	\Arc[arrow](0,0)(80,325,5)
	\Arc[arrow,flip](0,0)(80,175,215)
	\Vertex(125,0){2}
	\Vertex(121,32){2}
	\Vertex(121,-32){2}
	\color{blue}
	\large
	\Text(85,140){$p_1$}
	\Text(140,85){$p_2$}
	\Text(85,-140){$p_r$}
	\Text(140,-85){$p_{r-1}$}
	\Text(81,81){$q_1$}
	\Text(87,-81){$q_{r-1}$}
	\Text(63,-17){$\ell_1$}
	\Text(-63,-17){$\ell_2$}
	\color{red}
	\Arc[arrow](0,0)(100,60,90)
	\Arc[arrow](0,0)(100,270,300)
	\Text(30,111){$q_0$}
	\Text(30,-111){$q_r$}
	\end{picture}
	\caption{Example of a two-loop diagram involving a squared propagator, which here has momentum $q_0=q_r$.}
	\label{Figure:LTDDoublePole}
\end{figure}
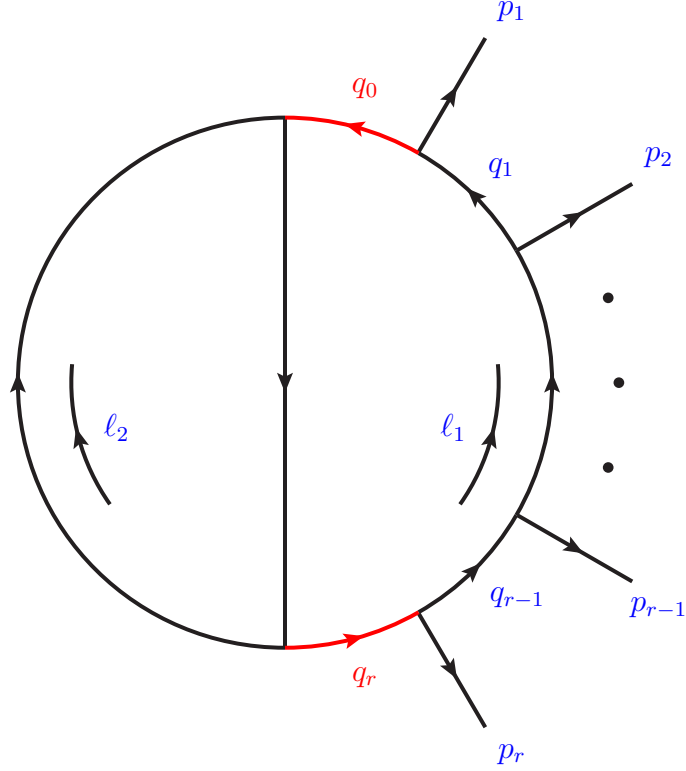

So far in our derivations of duality relations, we only considered diagrams and expressions with single poles, i.e. that only involve single-power propagators. Assuming a suitable choice of gauge, Feynman amplitudes at one-loop level will never exhibit multi-pole propagators\footnote{Note however that the local UV counterterms we will build in the following chapters will involve squared and cubed UV propagators, even at one-loop level.} \cite{Catani:2008xa}. At two-loop level and beyond, though, higher powers of the propagators can appear, from self-energy insertions on internal lines for instance, as seen in \Fig{\ref{Figure:LTDDoublePole}}. To compute such amplitudes using LTD, we must extend its formalism so it is able to deal with expressions involving multi-pole propagators. We will try to keep the following analysis as brief as possible, and refer the reader to \cite{Bierenbaum:2012th} for more details.\\
\\
Starting from the formulation of Cauchy's residue theorem for multiple poles, namely
\begin{equation}\label{Equation:LTDCauchyMultiPole}
\Res_{\{z=z_0\}}\,f(z)=\frac{1}{(k-1)!}\left[\frac{d^{k-1}}{dz^{k-1}}(z-z_0)^kf(z)\right]_{z=z_0}\;,
\end{equation}
for a function $f$ having a pole at $z_0$, of multiplicity at most $k$, we can generalise the duality relation. Let's consider the case where we have a Feynman propagator $G_F(q)$ raised at the $n$-th power. By explicitly writing
\begin{equation}\label{Equation:LTDGFn}
\big(G_F(q)\big)^n=\frac{1}{(q_0-\Sup{q_0})^n(q_0+\Sup{q_0})^n}\;,
\end{equation}
where we recall $\Sup{q_0}=\sqrt{\mathbf{q}^2+m^2-i0}$, we have
\begin{align}\label{Equation:LTDGFnResidue}
\Res_{\{q_0=\Sup{q_0}\}}\big(G_F(q)\big)^n=&~\lim\limits_{~~q_0\to\Sup{q_0}}\frac{1}{(n-1)!}\left(\frac{d^{n-1}}{dq_0^{n-1}}\frac{1}{(q_0+\Sup{q_0})^n}\right)\nn\\
=&~(-1)^{n-1}\frac{(2n-2)!}{((n-1)!)^2}\frac{1}{(2\Sup{q_0})^{2n-1}}\;,
\end{align}
and in particular
\begin{align}\label{Equation:LTDGF2GF3Residue}
\Res_{\{q_0=\Sup{q_0}\}}\big(G_F(q)\big)^2=&~-\frac{2}{(2\Sup{q_0})^3}\;,\nn\\
\Res_{\{q_0=\Sup{q_0}\}}\big(G_F(q)\big)^3=&~\frac{6}{(2\Sup{q_0})^5}\;.
\end{align}
Note that we can completely ignore the prescription here, since the denominator appearing in the second line of \Eq{\ref{Equation:LTDGFnResidue}} is always positive.\\
\\
Now we need to evaluate the residue of the complete amplitude. In the single-pole case, the numerator does not play a crucial role, since we only need to evaluate it at the pole value. However in the multi-pole case, since it can depend on $q_0$, the numerator may be affected by the presence of the derivative in \Eq{\ref{Equation:LTDGFnResidue}}. Moreover, this derivative also acts on the remaining Feynman propagators, as
\begin{align}\label{Equation:LTDGDDerivated}
\left[\frac{d}{dq_0}G_F(q;q_j)\right]_{q_0=\Sup{q_0}}=&~\left[\frac{d}{dq_0}\frac{1}{(q+k_j)^2-m_j^2}\right]_{q_0=\Sup{q_0}}\nn\\
=&~\left[-\frac{2q_0+2k_{j,0}}{((q+k_j)^2-m_j^2)^2}\right]_{q_0=\Sup{q_0}}=-2q_{j,0}\,\big(G_D(q;q_j)\big)^2
\end{align}
where we assumed that each of these propagators only has a single pole. Now let consider a set $\alpha=\{q\}\cup\{q_j\}_j$, with $q_j\neq q$ for all $j$. If $G_F(q)$ exhibits a double pole, then we have  
\begin{align}\label{Equation:LTDMultiPoleFormula}
\Res_{\{q_0=\Sup{q_0}\}}&\left((G_F(q)\big)^2\left(\,\prod_j\,G_F(q_j)\right)\mathcal{N}(\ell,\{p_i\}_N)\right)\nn\\
=&~\left[\frac{d}{dq_0}\frac{1}{(q_0+\Sup{q_0})^2}\left(\,\prod_j\,G_F(q_j)\right)\mathcal{N}(\ell,\{p_i\}_N)\right]_{q_0=\Sup{q_0}}\nn\\
=&~\left(\,\prod_j\,G_D(q;q_j)\right)\frac{1}{(2\Sup{q_0})^2}\left[-\frac{1}{\Sup{q_0}}-\,\sum_j\,(2q_{j,0})G_D(q;q_j)+\frac{d}{dq_0}\right]\mathcal{N}(\ell,\{p_i\}_N)\;.
\end{align}
This result can be easily extended for poles with higher multiplicities.

\chapter{The Four-Dimensional Unsubtraction from the Loop-Tree Duality}\label{Chapter:FDU}
\thispagestyle{fancychapter}
\fancyhead[RE]{\nameref*{Chapter:FDU}}

In this chapter, we present a new algorithm~\cite{Sborlini:2016gbr} to construct a purely four-dimensional representation of higher-order perturbative corrections to physical cross sections at NLO. The algorithm is based on the LTD theorem, and is implemented by introducing a suitable mapping between the external and loop momenta of the virtual scattering amplitudes, and the external momenta of the real emission corrections. In this way, the sum over degenerate states is performed at the integrand level, and the cancellation of infrared divergences occurs locally without introducing subtraction counterterms to deal with soft and final-state collinear singularities. The calculation of UV counterterms -- and in particular of self-energies -- within the LTD formalism is also discussed in detail. The method is first illustrated with the scalar three-point function, before being applied to the physical cross section for the $\gamma^*\to q\qbar(g)$ process at NLO. We also discuss about the possibility to generalise the method to multi-leg processes, and briefly comment about the extension to NNLO.\\

\section{Introduction}\label{Section:FDUIntro}
\fancyhead[LO]{\ref*{Section:FDUIntro}~~\nameref*{Section:FDUIntro}}

As a consequence of the KLN theorem~\cite{Kinoshita:1962ur,Lee:1964is}, theoretical predictions in theories with massless particles can only be obtained after defining infrared-safe physical observables. These involve performing a sum over all degenerate states, which means adding together real and virtual contributions. After UV renormalisation of virtual scattering amplitudes, the remaining contributions develop IR singularities that vanish when putting all the terms together, implying that the IR divergent structure of real and virtual corrections are closely related. The standard approach to calculating higher order corrections in perturbative QCD relies on the application of the subtraction formalism \cite{Kunszt:1992tn,Frixione:1995ms,Catani:1996jh,Catani:1996vz,GehrmannDeRidder:2005cm,Seth:2016hmv,Catani:2007vq,Catani:2009sm,Czakon:2010td,Bolzoni:2010bt,DelDuca:2015zqa,Boughezal:2015dva,Gaunt:2015pea,DelDuca:2016ily,DelDuca:2016csb}, where the real and virtual contributions are treated separately. This formalism exploits factorisation properties of QCD~\cite{Collins:1989gx,Catani:2011st} to define suitable subtraction counterterms mimicking the IR singular behaviour.\\
\\
Nowadays, several variants of the subtraction method at NLO and beyond have been developed~\cite{Kunszt:1992tn,Frixione:1995ms,Catani:1996jh,Catani:1996vz,GehrmannDeRidder:2005cm,Seth:2016hmv,Catani:2007vq,Catani:2009sm,Czakon:2010td,Bolzoni:2010bt,DelDuca:2015zqa,Boughezal:2015dva,Gaunt:2015pea}. However, these methods might not be efficient enough, from a computational point of view, for multi-particle processes. The main reason is that final-state phase space (PS) of the different contributions involves different numbers of particles. For instance at NLO, virtual corrections with Born kinematics have to be combined with real contributions involving an additional final-state particle. The IR counterterms have to be local in the real PS, as well as analytically integrable over the extra-radiation factorised PS, in order to properly cancel the divergent structure present in the virtual corrections. Building these counterterms represents a significant challenge and introduces a potential bottleneck when it comes to efficiently carrying out the IR subtraction for multi-leg multi-loop processes.\\
\\
With the aim of avoiding the introduction of IR counterterms, we explore an alternative idea based on the application of the LTD theorem \cite{Catani:2008xa,Rodrigo:2008fp,Bierenbaum:2010cy,Bierenbaum:2012th,Buchta:2014dfa,Buchta:2015xda,Buchta:2015wna}, whose formalism has been introduced in \Chapter{\ref{Chapter:LTD}}. As we saw, it establishes that loop scattering amplitudes can be expressed as a sum of PS integrals (or dual integrals), with an additional on-shell particle. Since dual integrands and real-radiation contributions exhibit a similar structure, they can be combined at the integrand level. As shown in~\cite{Hernandez-Pinto:2015ysa}, the divergent behaviour of both contributions match, and the expression is finite. In other words, working in the context of DREG, with $d=4-2\epsilon$ the number of space-time dimensions, the mapped real-virtual contributions do not lead to $\epsilon$-poles, meaning that the limit $\epsilon\to0$ can be safely considered. The possibility of carrying out pure four-dimensional implementations for any observables at NLO and beyond is a strong implication of this last fact. In this chapter, we develop a novel algorithm to perform a four-dimensional regularisation of multi-leg physical cross sections at NLO, that does not require the use of soft and final-state collinear subtraction.\\
\\
It is worth mentioning that the idea of obtaining purely four-dimensional expressions to compute higher-order observables has been previously studied. For instance, it was proposed to apply momentum smearing~\cite{Soper:1999rd,Soper:2000a,Soper:2001hu,Kramer:2002cd} to combine real and virtual contributions, thus achieving a local cancellation of singularities. Other methods consist in rewriting the standard UV/IR subtraction counterterms in local form, as discussed in~\cite{Becker:2010ng,Becker:2012aqa}, or in modifying the structure of the propagators (and the associated Feynman rules) to regularise the singularities~\cite{Pittau:2012zd,Donati:2013iya,Fazio:2014xea}. Besides that, the numerical computation of virtual corrections has received a lot of attention in recent years~\cite{Passarino:2001a,Ferroglia:2002mz,Nagy:2003qn,Nagy:2006a,Anastasiou:2007qb,Moretti:2008jj,Gong:2008ww,Kilian:2009wy,Becker:2012nk,Freitas:2016sty}. For these reasons, through the application of LTD, we will tackle both problems simultaneously; we will express virtual amplitudes as phase-space integrals and combine them with the real contributions, working directly at the \emph{integrand level}. Moreover, physically interpretable results will emerge in a natural way.\\
\\
The outline of this chapter is the following. First, we describe in detail the implementation of the Four-Dimensional Unsubtraction (FDU) algorithm at NLO with a scalar toy example. We start by commenting on the IR singular structure of the scalar three-point function and its associated dual integrands in Section \ref{Section:FDUS3PF}. We then define the mapping of momenta between real and virtual corrections for this toy example in \Section{\ref{Section:FDUUnsubtraction}}, where our strongly physically motivated four-dimensional regularisation of soft and collinear singularities is presented. In \Section{\ref{Section:FDUUV}}, we study the renormalisation of UV divergences at the integrand level in the LTD framework. The discussion is focused on the treatment of scalar two-point functions, where we properly rewrite unintegrated dual counterterms in a fully local way. After that, we carefully analyse the implementation of these techniques to the process $\gamma\to q\qbar(g)$ in \Section{\ref{Section:FDUNLO}}. We put special emphasis on the algorithmic construction of the integrands, and on the numerical implementation of the purely four-dimensional representation. In \Section{\ref{Section:FDUGeneralisation}}, we generalise the unsubtraction algorithm to multi-leg processes, and briefly comment about the extension of the algorithm to NNLO. Finally, we present our conclusions and discuss future research directions in \Section{\ref{Section:FDUConclusion}}.

\section{Singularities of the scalar three-point function}\label{Section:FDUS3PF}
\fancyhead[LO]{\ref*{Section:FDUS3PF}~~\nameref*{Section:FDUS3PF}}

In this section, we show a detailed derivation of the results presented in~\cite{Hernandez-Pinto:2015ysa}, namely those concerning the scalar three-point function with massless internal particles. This discussion is useful to analyse and understand the application of LTD to the realistic case presented in \Section{\ref{Section:FDUNLO}} and the subsequent generalisation to multi-leg processes in \Section{\ref{Section:FDUGeneralisation}}.\\
\\
We consider final-state massless and on-shell particles with momenta $p_1$ and $p_2$, with the incoming momentum being $p_3=p_1+p_2=p_{12}$ by momentum conservation, with virtuality $p_3^2=\s>0$. The internal momenta are written $q_1=\ell+p_1$, $q_2=\ell+p_{12}$ and $q_3=\ell$, where $\ell$ is the loop four-momentum. The scalar three-point function at one-loop is given by the well-known result~\cite{Ellis:2007qk,Sborlini:2013jba}
\begin{equation}\label{Equation:FDUS3PFLiterature}
L^{(1)}(p_1,p_2,-p_3)=\int_\ell\,\,\prod\limits_{i=1}^3\,G_F(q_i)=-\cGamma\,\frac{\mu^{2\epsilon}}{\epsilon^2}(-\s-i0)^{-1-\epsilon}\;,
\end{equation}
where
\begin{equation}\label{Equation:FDUcGamma}
\cGamma=\frac{\Gamma(1+\epsilon)\Gamma^2(1-\epsilon)}{(4\pi)^{2-\epsilon}\Gamma(1-2\epsilon)}
\end{equation}
is the usual one-loop volume factor in $d$ dimensions.\\
\\
As for this particular example there are three internal particles, $L^{(1)}$ will be split into three dual contributions. Applying the LTD theorem yields
\begin{equation}\label{Equation:FDUS3PF}
L^{(1)}=\,\sum\limits_{i=1}^3\,I_i=-\,\sum\limits_{i=1}^3\,\int_\ell\,\deltatilde{q_i}\,G_D(q_i,q_j)\,G_D(q_i,q_k)\;,
\end{equation}
with $i$, $j$ and $k$ all different. The dual contributions are written $I_i$, for $i\in\{1,2,3\}$, and in each of them a different internal line is cut. In $I_1$ for instance, we put on shell the internal particle with momentum $q_1$ by promoting its corresponding propagator to the dual delta function $\deltatilde{q_1}$. We also replace the Feynman propagators $G_F(q_2)$ and $G_F(q_3)$ by their dual counterparts $G_D(q_1;q_2)$ and $G_D(q_1;q_3)$ (see \Eq{\ref{Equation:LTDL1tilde}}). Explicitly, we have 
\begin{align}\label{Equation:FDUI1S3PF}
I_1=&~-\int_\ell\,\deltatilde{q_1}\,G_D(q_1;q_2)\,G_D(q_1;q_3)\nn\\
=&~-\int_\ell\,\frac{\deltatilde{q_1}}{((q_1-p_2)^2-i0\,\eta\cdot p_2)((q_1-p_1)^2+i0\,\eta\cdot p_1)}\nn\\
=&~-\int_\ell\,\frac{\deltatilde{q_1}}{(2q_1\cdot p_2-i0)(-2q_1\cdot p_1+i0)}\:,
\end{align}
where we wrote both $q_2$ and $q_3$ in terms of $q_1$ to take advantage of the fact $q_1^2=0$, on account of the presence of the $\deltatilde{q_1}$. Furthermore, because of the fact the external momenta have positive energy, $\eta\cdot p_1$ and $\eta\cdot p_2$ are also both positive, meaning they can be taken out as only the sign in front of the prescription is relevant.\\
\\
Likewise,
\begin{align}\label{Equation:FDUI2I3S3PF}
I_2=&~-\int_\ell\,\frac{\deltatilde{q_2}}{(-2q_2\cdot p_2+i0)(-2q_2\cdot p_{12}+\s+i0)}\nn\\
I_3=&~-\int_\ell\,\frac{\deltatilde{q_3}}{(2q_3\cdot p_1-i0)(2q_3\cdot p_{12}+\s-i0)}\;.
\end{align}
In order to simplify the computation of these three integrals, we work in the centre-of-mass frame of $p_1$ and $p_2$, and parametrise all the momenta as
\begin{align}\label{Equation:FDUParam}
p_1^\mu=&~\frac{\sqrt{\s}}{2}(1,\mathbf{0}_\perp,1)\;,\quad p_2^\mu=\frac{\sqrt{\s}}{2}(1,\mathbf{0}_\perp,-1)\;,\nn\\
q_i^\mu=&~\frac{\sqrt{\s}}{2}\xi_{i,0}\left(1,2\sqrt{v_i(1-v_i)}\,\eiPerp{i},1-2v_i\right)\;,
\end{align}
with $\xi_{i,0}\in[0,\infty)$ and $v_i\in[0,1]$ being the integration variables describing the energy and polar angle of the loop momenta, respectively. The integration over the transverse plane $\text{Vect}[\mathbf{e}_{i,\perp}]$ is trivial in this case.\\
\\
The scalar products involving internal and external momenta are therefore given by
\begin{align}\label{Equation:FDUScalarProductsS3PF}
2\frac{q_i\cdot p_1}{\s}=&~\xi_{i,0}\,v_i\;,\nn\\
2\frac{q_i\cdot p_2}{\s}=&~\xi_{i,0}(1-v_i)\;,
\end{align}
which allows us to rewrite the dual integrals of \Eqs{\ref{Equation:FDUI1S3PF}}{\ref{Equation:FDUI2I3S3PF}} as
\begin{align}\label{Equation:FDUS3PFDualsBis}
I_1=&~\frac{1}{\s}\int\,\frac{\dxi{1}\,\dv{1}}{\xi_{1,0}(v_1(1-v_1))}\;,\nn\\
I_2=&~\frac{1}{\s}\int\,\frac{\dxi{2}\,\dv{2}}{(1-\xi_{2,0}+i0)(1-v_2)}\;,\nn\\
I_3=&~-\frac{1}{\s}\int\,\frac{\dxi{3}\,\dv{3}}{(1+\xi_{3,0})v_3}\;,
\end{align}
with the integration measure in $d$-dimension given by the direct product of (see \Eq{\ref{Equation:APPIntegrationMeasure}})
\begin{equation}\label{Equation:FDUIntegrationMeasure}
\dxi{i}=\frac{(4\pi)^{\epsilon-2}}{\Gamma(1-\epsilon)}\left(\frac{\s}{\mu^2}\right)^{-\epsilon}\xi_{i,0}^{-2\epsilon}\,d\xi_{i,0}\quad\text{and}\quad \dv{i}=(v_i(1-v_i))^{-\epsilon}dv_i\,.
\end{equation}
It is possible to perform the integrations analytically, leading to
\begin{align}\label{Equation:FDUIntegratedS3PF}
I_1=&~0\;,\nn\\
I_2=&~\cGammaTilde\,\frac{\mu^{2\epsilon}}{\epsilon^2}\,\s^{-1-\epsilon}\,e^{i2\pi\epsilon}\;,\nn\\
I_3=&~\cGammaTilde\,\frac{\mu^{2\epsilon}}{\epsilon^2}\,\s^{-1-\epsilon}\;,
\end{align}
where
\begin{equation}\label{Equation:FDUPSVolumeFactor}
\cGammaTilde=\frac{\Gamma(1-\epsilon)\Gamma(1+2\epsilon)}{(4\pi)^{2-\epsilon}}=\frac{\cGamma}{\cos(\pi\epsilon)}
\end{equation}
is the $d$-dimensional phase-space volume factor. As expected, the sum of the three dual integrands
\begin{equation}\label{Equation:FDUSumS3PF}
\,\sum\limits_{i=1}^3\,I_i=-\cGamma\,\frac{\mu^{2\epsilon}}{\epsilon^2}(-\s-i0)^{-1-\epsilon}
\end{equation}
agrees with the literature result from \Eq{\ref{Equation:FDUS3PFLiterature}}. One can observe that even though $I_1$ vanishes because of the absence of scale, it still contains IR and UV singularities. They do not manifest after integration, however, because they lead to two $\epsilon$-poles of opposite signs that cancel each other. Notice also that in \Eq{\ref{Equation:FDUIntegratedS3PF}} the dual $+i0$ prescription is crucial for computing $I_2$, because of the fact $1-\xi_{2,0}$ changes sign inside the integration region, leading to a threshold singularity.\\
\\
For a later use, it is necessary to obtain an explicit expression of the imaginary part of $I_2$, which is done by setting $G_F(q_3)$ on shell, with negative energy mode\footnote{This is an explicit example of a forward-backward intersection, as discussed in \Section{\ref{Section:LTDCancellationOfThreshold}}.}, inside the expression of $I_2$,
\begin{align}\label{Equation:FDUImI2S3PF}
&~i\ImText\,L^{(1)}(p_1,p_2,-p_3)=i\ImText\,I_2=\frac{1}{2}\int_\ell\,\deltatilde{q_2}\,\deltatilde{-q_3}\,G_D(q_2;q_1)\nn\\
&~=-\frac{i\pi}{\s}\int\,\dxi{2}\,\dv{2}(1-v_2)^{-1}\theta(2-\xi_{2,0})\delta(1-\xi_{2,0})=i\,\cGamma\,\frac{\mu^{2\epsilon}}{2\epsilon^2}\,\s^{-1-\epsilon}\sin(2\pi\epsilon)\;.
\end{align}
We can remark that $G_D(q_2;q_1)=G_F(q_1)$ because of the fact $\eta\cdot k_{12}<0$. \Eq{\ref{Equation:FDUImI2S3PF}} is therefore consistent with Cutkosky's rule~\cite{Catani:2008xa}. This is the causality connection mentioned in \Section{\ref{Section:LTDOneLoop}}, and it becomes relevant in our computation because the $\epsilon$-expansion in \Eq{\ref{Equation:FDUImI2S3PF}} reveals the presence of a purely imaginary single-pole in $I_2$ that will not be cancelled by real corrections. At the integrand level, this means that the real part of $I_2$ exhibits an integrable singularity in the neighbourhood of $\xi_{2,0}=1$, but also a non-integrable one that must be cancelled by properly removing its imaginary component before performing a four-dimensional numerical implementation. Thus, the real part of $I_2$ is defined as
\begin{equation}\label{Equation:FDUReI2}
\ReText I_2=I_2-i\,\ImText I_2=\frac{1}{\s}\int\,\dxi{2}\dv{2}(1-v_2)^{-1}\left(\frac{1}{1-\xi_{2,0}+i0}+i\,\pi\,\delta(1-\xi_{2,0})\right)\;,
\end{equation}
and, by virtue of the Sokhotski-Plemelj theorem,
\begin{equation}\label{Equation:FDUReI2Final}
\ReText I_2=\frac{1}{\s}\int\,\dxi{2}\dv{2}(1-v_2)^{-1}\textrm{PV}\left(\frac{1}{1-\xi_{2,0}}\right)\;,
\end{equation}
where we made use of Cauchy's principal value (here written $\textrm{PV}$) to get rid of the $+i0$ prescription and the imaginary pole. From the formal point of view, we could have performed this computation by simply working with the real part of the integrand (and neglecting the prescription). This would however introduce numerical instabilities, making the application of PV prescription a more efficient implementation.\\
\begin{figure}[ht]
	\begin{center}
		\includegraphics[width=0.3\textwidth]{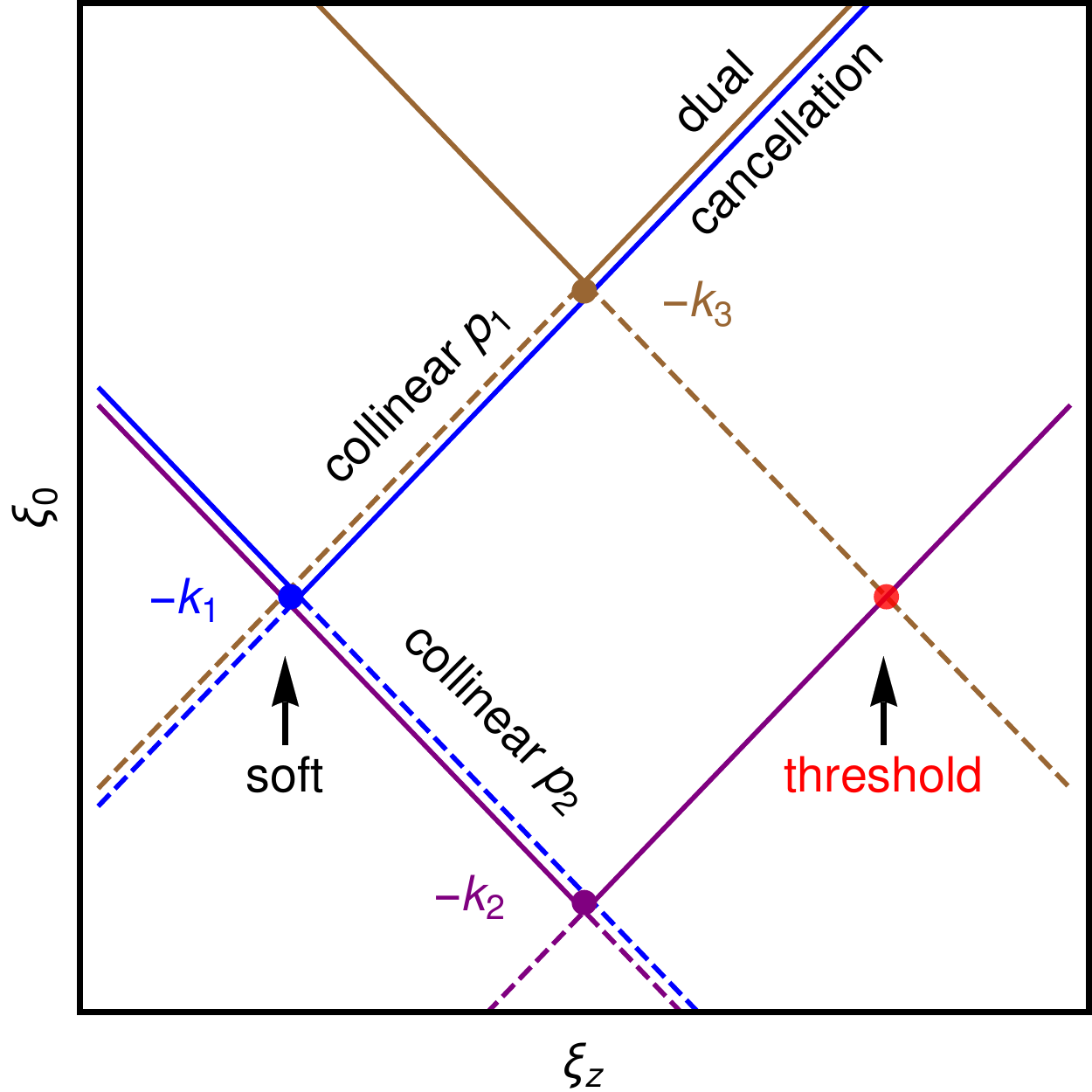} $\quad$
		\includegraphics[width=0.3\textwidth]{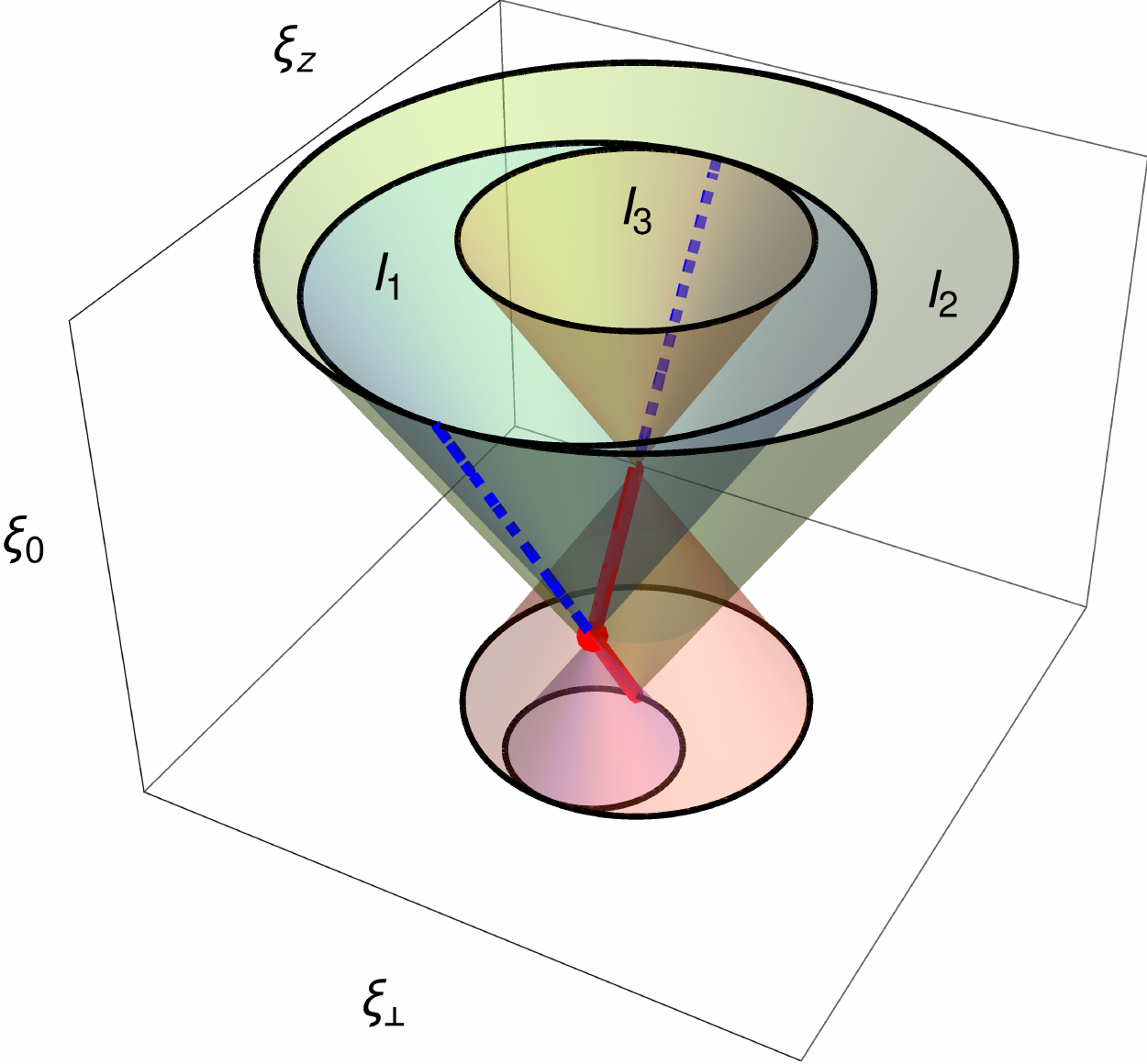} $\quad$
		\includegraphics[width=0.3\textwidth]{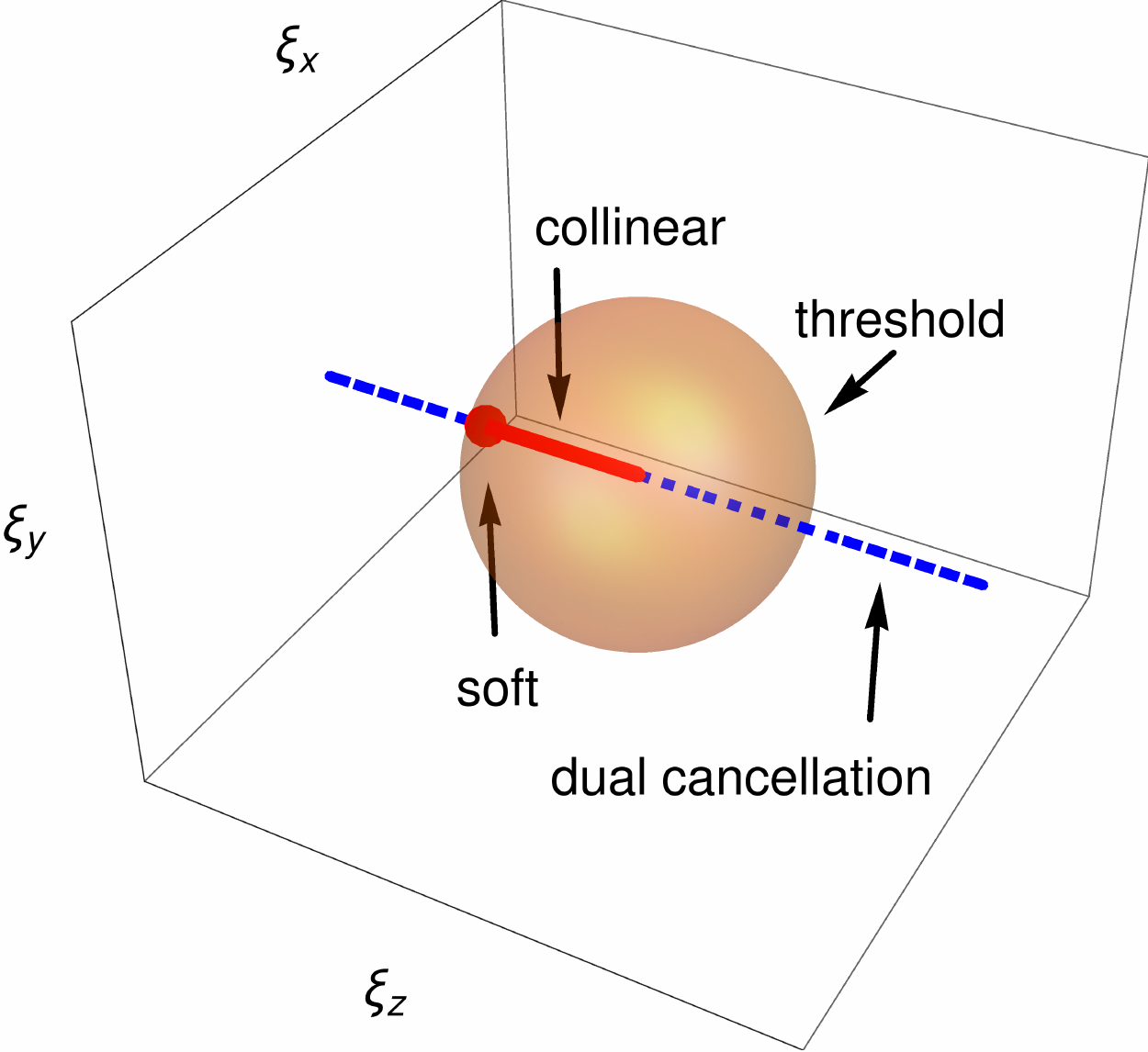}
		\caption{Light-cones of the three-point scalar function in the loop coordinates $\ell^{\mu}=\sqrt{s_{12}}/2\,(\xi_0,\xi_\perp,\xi_z)$, with $\xi_\perp=\sqrt{\xi_x^2+\xi_y^2}$; two dimensions (left) and three-dimensions (middle). LTD is equivalent to integrating along the forward light-cones (solid lines in the left plot). Backward light-cones are represented by dashed lines. The intersection of light-cones leads to soft, collinear, and threshold singularities in the loop three-momentum space (right plot), or to the cancellation of singularities among dual contributions.}
		\label{Figure:FDURegionLightcone}
	\end{center}
\end{figure}
\\
LTD can further be exploited to have a deeper and detailed understanding of the origin of the singularities of the loop integral under consideration. As commented before, the origin of the singularities can be underlined by analysing the relative position and intersections of the on-shell hyperboloids or light-cones of the propagators of the loop integrand~\cite{Buchta:2014dfa}. In \Fig{\ref{Figure:FDURegionLightcone}}, we plot the light-cones that support each of the dual integrals of the dual representation of the three-point function (\Eq{\ref{Equation:FDUS3PFLiterature}}). Although the scalar three-point function is UV finite, the individual dual integrands in \Eqs{\ref{Equation:FDUI1S3PF}}{\ref{Equation:FDUI2I3S3PF}} diverge in the UV region as the propagators are linear in the loop momentum. However, by taking their sum we recover the same UV structure as the original integral (in this case, it is UV safe), as expected by the LTD theorem. We can thus focus on its IR behaviour, which we do in the following; renormalisation of UV divergences in general will be considered later in \Section{\ref{Section:FDUUV}}.\\
\\
Collinear divergences are associated with regions of the phase space where light-cones intersect or overlap, as shown in \Fig{\ref{Figure:FDURegionLightcone}}. At large loop momentum, the intersections occur among forward light-cones which ensures the cancellation of the associated collinear singularities in the sum of dual integrals. Though, there are still collinear divergences that originate from compact regions defined by the intersection of forward and backward light-cones. Soft divergences arise at $\Sup{q_{i,0}}=0$ (which is a point-like solution), but it leads to actual singularities only if two other propagators -- each of them contributing one power in the infrared -- are light-like separated from the $i$-th propagator. We can see in \Fig{\ref{Figure:FDURegionLightcone}} that this condition is fulfilled only with $\Sup{q_{1,0}}=0$. Finally, a threshold singularity appears inside the dual integrand $I_2$ through the intersection of the backward light-cone of $G_F(q_3)$ with the forward light-cone of $G_F(q_2)$. The imaginary part of $I_2$ is singular but the singularity can be removed using \Eq{\ref{Equation:FDUImI2S3PF}}. The singularity of the real part of $I_2$ is integrable but generates numerical instabilities. In that case, a contour deformation must be employed to achieve a stable numerical implementation~\cite{Buchta:2015xda,Buchta:2015wna}.\\
\\
Motivated by \Fig{\ref{Figure:FDURegionLightcone}} and in order to isolate the IR divergences of the scalar three-point function, we define the soft and collinear components of the dual integrals in well-defined compact regions of the loop three-momentum, i.e.
\begin{align}\label{Equation:FDUSoftCollinearRegions}
I_1^{(s)}=&~I_1(\xi_{1,0}\leq w)\;,\nn\\
I_1^{(c)}=&~I_1(w\leq\xi_{1,0}\leq1\,;\,v_1\leq1/2)\;,\nn\\
I_2^{(c)}=&~I_2(\xi_{2,0}\leq1+w\,;\,v_2\geq1/2)\;,
\end{align}
where $0<w<1$ is a cut in the energy of the internal on-shell particles. The collinear singularity of the dual integral $I_2$ appears at $v_2=1$ with $\xi_{2,0}\in[0,1]$, but $I_2$ develops also a threshold singularity at $\xi_{2,0}=1$. For that reason, we have imposed a finite $w$-cut to include the threshold region in the definition of $I_2^{(c)}$. The integral $I_1^{(s)}$ includes the soft singularity of the dual integral $I_1$ at $\xi_{1,0}=0$, and the collinear singularities in the neighbourhood of $\xi_{1,0}=0$ at $v_1=0$. The $\epsilon$-poles present in the integral $I_1^{(c)}$ are due to collinear singularities only. Note that there is some arbitrariness in the definition of the integration regions of these integrals. Indeed, provided that we always include the soft and collinear singularities, different definitions will result in different finite contributions. We will redefine them later with a better motivated physical target, but for the current illustrative purpose, we use the simplest choice of \Eq{\ref{Equation:FDUSoftCollinearRegions}}.\\
\\
Analytically integrating the dual contributions, we get
\begin{align}\label{Equation:FDUDualsIR}
I_1^{(s)}=&~\cGammaTilde\,\frac{w^{-2\epsilon}}{\epsilon^2}\,\mu^{2\epsilon}\,\s^{-1-\epsilon}\,\frac{\sin(2\pi\epsilon)}{2\pi\epsilon}\;,\nn\\
I_1^{(c)}=&~\cGammaTilde\,\frac{1-w^{-2\epsilon}}{2\epsilon^2}\,\mu^{2\epsilon}\,\s^{-1-\epsilon}\,\frac{\sin(2\pi\epsilon)}{2\pi\epsilon}\;,\nn\\
I_2^{(c)}=&~-\cGammaTilde\,\frac{(1+w)^{1-2\epsilon}}{2\epsilon(1-2\epsilon)}\,\mu^{2\epsilon}\,\s^{-1-\epsilon}\,\frac{\sin(2\pi\epsilon)}{2\pi\epsilon}\nn\\
&~\times\,_2F_1\left(1,1-2\epsilon,2-2\epsilon;\frac{\s(1+w)}{\s+i0}\right)\left(1+\frac{4^\epsilon\,\Gamma(1-2\epsilon)}{\Gamma(1-\epsilon)}\right)\,.
\end{align}
Using Pfaff and shift identities, the hypergeometric function in $I_2^{(c)}$ can be written in the physical region with $w>0$ and $\s>0$. This leads to
\begin{align}\label{Equation:FDUI2cPfaff}
I_2^{(c)}=&~\cGammaTilde\,\frac{\mu^{2\epsilon}}{4\epsilon^2}\,\s^{-1-\epsilon}\,\frac{\sin(2\pi\epsilon)}{2\pi\epsilon}\nn\\
&~\times\left(e^{i2\pi\epsilon}-w^{-2\epsilon}\,_2F_1\left(2\epsilon,2\epsilon,1+2\epsilon;-\frac{1}{w}\right)\right)\left(1+\frac{4^\epsilon\,\Gamma(1-2\epsilon)}{\Gamma(1-\epsilon)}\right)\;,
\end{align}
with
\begin{equation}\label{Equation:FDU2F1}
_2F_1\left(2\epsilon,2\epsilon,1+2\epsilon;z\right)=1+4\epsilon^2\Li_2(z)+\mathcal{O}(\epsilon^3)\;.
\end{equation}
As expected, the soft integral $I_1^{(s)}$ in \Eq{\ref{Equation:FDUDualsIR}} contains double poles, while the collinear integrals develop single poles only. Although each individual integral depends on the cut $w$, the poles of the sum are independent of $w$ and agree with the total divergences of the full integral. We indeed obtain
\begin{equation}\label{Equation:FDUFullIntIR}
L^{(1)}(p_1,p_2,-p_3)=I^\ir+\Oep{0}\;,
\end{equation}
with
\begin{align}\label{Equation:FDUIRResult}
I^\ir=&~I_1^{(s)}+I_1^{(c)}+I_2^{(c)}=\frac{\cGamma}{\s}\left(\frac{-\s-i0}{\mu^2}\right)^{-\epsilon}\nn\\
&~\times\left(\frac{1}{\epsilon^2}+\log(2)\log(w)-\frac{\pi^2}{3}-2\Li_2\left(-\frac{1}{w}\right)+i\,\pi\log(2)\right)+\Oep{1}\,.
\end{align}
Outside the region that contains the IR poles, the sum of the dual integrands is finite, although they are separately divergent. A suitable combination is required to obtain a finite result. We consequently define the forward and backward regions as those delimited by $v_i\leq 1/2$ and $v_i\geq 1/2$, respectively. This separation does not have any physical meaning; it is just more convenient for the analytical computation. Explicitly, we define
\begin{align}\label{Equation:FDUIRForward}
I^{(f)}=&~I_1(\xi_{1,0}\geq1\,;\,v_1\leq1/2)+I_2(v_2\leq1/2)+I_3(v_3\leq1/2)\nn\\
=&~\frac{\cGamma}{\s}\int_{0}^{\infty}\,d\xi_0\int_0^{1/2}\,dv\left[\frac{1}{1+\xi_0}\left(\frac{1}{1-v}+2\log(2)\left(\frac{1+\xi_0}{\xi_0}\right)\delta(v)\right)\right.\nn\\
&~~~~~~~~~~~~~~~~~~~~~~~~~~~~~+\left.\frac{1}{(1-v)(1-\xi_0+i0)}\right]+\Oep{1}\;,
\end{align}
where we have performed the trivial change of the integration variables
\begin{equation}\label{Equation:FDUCOVForward}
\xi_{1,0}=1+\xi_0\;,\quad\xi_{2,0}=\xi_{3,0}=\xi_0\;,\quad v_i=v\;,
\end{equation}
and have taken the limit $\epsilon\to0$ at the integrand level. Notice that each dual integrand is still individually singular. For instance, $I_1$ and $I_3$ are divergent for $v_1=v_3=0$, but their sum is finite in the IR limit, although UV divergences are still present. This divergent behaviour at high energies is cancelled once we add $I_2$. These cross-cancellations of singularities allow to perform the integral of the forward contribution after setting $\epsilon=0$. It is worth noting that the logarithmic term in \Eq{\ref{Equation:FDUIRForward}} is originated from the fact we are using different coordinate systems for each dual integral. This produces a mismatch of the integration measure that is of $\Oep{1}$. And since the integral behaves as $\mathcal{O}(\epsilon^{-1})$ in the collinear limit, a non-vanishing finite contribution will arise from the collinear region. Explicitly, the expansion of $v^{-1-\epsilon}$ by using \Eq{\ref{Equation:APPSIntegrationBasicFormula}} leads to
\begin{equation}\label{Equation:FDUExpansionLogForward}
\left(-\frac{\delta(v)}{\epsilon}+\left(\frac{1}{v}\right)_C+\Oep{1}\right)\left(\frac{(1+\xi_0)^{-2\epsilon}}{1-v}-\xi_0^{-2\epsilon}\right)=\frac{1}{1-v}+2\log\left(\frac{1+\xi_0}{\xi_0}\right)\delta(v)+\Oep{1}\;,
\end{equation}
where we removed the $C$-distribution. These logarithms are avoidable by a proper reparametrisation of the integration variables, which we will do in details \Section{\ref{Section:FDUUCS}}.\\
\\
Integrating \Eq{\ref{Equation:FDUIRForward}} gives
\begin{equation}\label{Equation:FDUIRForwardResult}
I^{(f)}=\frac{\cGamma}{\s}\left(\frac{\pi^2}{3}-i\,\pi\log(2)\right)+\Oep{1}\,.
\end{equation}
In a analogous way, we can calculate the finite contribution originated in the backward region $(v_i\geq1/2)$, given by
\begin{align}\label{Equation:FDUIRBackward}
I^{(b)}=&~I_1(\xi_{1,0}\geq w\,;\,v_1\geq1/2)+I_2(\xi_{2,0}\geq1+w\,;,v_2\geq1/2)+I_3(v_3\geq1/2)\nn\\
=&~\frac{\cGamma}{\s}\int_0^\infty\,d\xi_0\int_{1/2}^1\,dv\nn\\
&\left[\frac{1}{\xi_0+w}\left(\frac{1}{v}+2\log\left(\frac{w+\xi_0}{1+w+\xi_0}\right)\delta(1-v)\right)-\frac{1}{v(1+\xi_0)}\right]+\Oep{1}\;,
\end{align}
where we have abandoned the $i0$ prescription because $w>0$ excludes the threshold singularity from the integration region. Additionally, to obtain \Eq{\ref{Equation:FDUIRBackward}} from \Eq{\ref{Equation:FDUS3PFDualsBis}} we used the change of variables
\begin{equation}\label{Equation:FDUCOVBackward}
\xi_{1,0}=w+\xi_0\;,\quad\xi_{2,0}=1+w+\xi_0\;,\quad\xi_{3,0}=\xi_0\;,\quad v_i=v\;,
\end{equation}
and then took again the limit $\epsilon\to0$ at the integrand level. Similarly as for the forward integral, there is a cancellation of collinear singularities among $I_1$ and $I_2$, which takes place at $v_1=1=v_2$ in this case. $I_1+I_2$ is therefore IR-finite but still UV divergent. And -- as it was also the case for the forward contribution --, the high-energy singular behaviour will be regularised after we add the $I_3$ contribution. Again, notice that a logarithmic term appears in \Eq{\ref{Equation:FDUIRBackward}}. It is this time due to the mismatch in the collinear behaviour of $I_1$ and $I_2$ at $\Oep{1}$. As we did for the forward case, we derived the logarithmic corrections by expanding the collinear factor $(1-v)^{(-1-\epsilon)}$, leading to
\begin{align}\label{Equation:FDUExpansionLogBackward}
&~\left(-\frac{\delta(1-v)}{\epsilon}+\left(\frac{1}{1-v}\right)_C+\Oep{1}\right)\left(\frac{(w+\xi_0)^{-2\epsilon}}{v}-(1+w+\xi_0)^{-2\epsilon}\right)\nn\\
=&~\frac{1}{v}+2\log\left(\frac{w+\xi_0}{1+w+\xi_0}\right)\delta(1-v)+\Oep{1}\;,
\end{align}
where, again, the $C$-distribution can be removed because the integrand is regular for $v=1$.\\
\\
Once again, the integral of the sum of the three dual integrands can be performed in the limit $\epsilon\to0$. We obtain
\begin{equation}\label{Equation:FDUIRBackwardResult}
I^{(b)}=\frac{\cGamma}{\s}\left(\Li_2\left(-\frac{1}{w}\right)-2\log(2)\log(w)\right)+\Oep{1}\\
\end{equation}
The sum of \Eq{\ref{Equation:FDUIRResult}}, \Eq{\ref{Equation:FDUIRForwardResult}} and \Eq{\ref{Equation:FDUIRBackwardResult}} leads to the correct result up to $\Oep{1}$,
\begin{equation}\label{Equation:FDUIRFullResult}
L^{(1)}(p_1,p_2,-p_3)=I^\ir+I^{(f)}+I^{(b)}+\Oep{1}\;.
\end{equation}
This result is independent of $w$, which is expected because $w$ is a non-physical cut. It is important to note that only $I^\ir$ contains $\epsilon$-poles, whereas the remaining contributions have been computed directly with $\epsilon=0$. Moreover, through the application of \Eqs{\ref{Equation:FDUExpansionLogForward}}{\ref{Equation:FDUExpansionLogBackward}}, the integrand can be easily expressed as the $\epsilon\to0$ limit of the original DREG expression plus some logarithmic corrections, which leads to the right result.

\subsection{Unification of the coordinate system}\label{Section:FDUUCS}

In this section, we show that it is possible to avoid potential extra logarithmic terms to appear when performing the sum and the integration over the different dual contributions, by using as example $I^{(f)}$ and $I^{(b)}$. These logarithmic terms are originated from the fact each dual integrand has been expressed in a different coordinate system, which makes them approach the collinear limit in a slightly different way at $\Oep{1}$. The solution therefore consists in using the same coordinate system for all the dual integrals, where the loop three-momenta $\mathbf{q}_i$ are mapped exactly. Although for analytic calculations this leads to more complex intermediate expressions, it is the natural choice for numerical computations.\\
\\
The idea is to rewrite all integration variables in terms of $\boldsymbol{\ell}$. This means that in our example (since $q_3=\ell$) we need to rewrite $(\xi_{1,0},v_1)$ in terms of $(\xi_{3,0},v_3)$ inside $I^{(f)}$. Notice that $q_3=\ell$ is set on shell in $I_3$, but not in $I_1$ where $q_1^2=0$. According to \Eq{\ref{Equation:FDUParam}}, the spatial components of the internal loop momenta are parametrised as
\begin{align}
\mathbf{q}_1=&~\frac{\sqrt{\s}}{2}\,\xi_{1,0}\left(2\sqrt{v_1(1-v_1)}\,\eiPerp{1},1-2v_1\right)\nn\\
=&~\mathbf{q}_3+\mathbf{p}_1=\frac{\sqrt{\s}}{2}\left(2\xi_{3,0}\sqrt{v_3(1-v_3)}\,\eiPerp{3},\xi_{3,0}(1-2v_3)+1\right)\;,
\end{align}
with $\Sup{q_{1,0}}=\sqrt{(\mathbf{q}_3+\mathbf{p}_1)^2-i0}$ when $q_1$ is on shell. Solving the system for $(\xi_{1,0},v_1)$ gives
\begin{align}\label{Equation:FDUCOVUnification}
\xi_{1,0}&=~\sqrt{(1+\xi_{3,0})^2-4v_3\,\xi_{3,0}}\;,\nn\\
v_1&=~\dfrac{1}{2}\left(1-\dfrac{1+(1-2v_3)\xi_{3,0}}{\sqrt{(1+\xi_{3,0})^2-4v_3\,\xi_{3,0}}}\right)\;,
\end{align}
with the associated Jacobian reading
\begin{equation}\label{Equation:FDUUCSJacobian}
\mathcal{J}(\xi_3,v)=\frac{\xi_3^2}{(1+\xi_3)^2-4v_3\,\xi_3}\;.
\end{equation}
With this change of variables, the forward integral can be rewritten (with $\epsilon=0$) as
\begin{align}\label{Equation:FDUIRForwardWithCOV}
I^{(f)}=&~\frac{\cGamma}{\s}\left[\int_0^{1/2}\,dv\int_0^\infty\,d\xi_0\left(v^{-1}\left(\frac{(1-v)^{-1}}{\sqrt{(1+\xi_0)^2-4v\,\xi_0}}-\frac{1}{1+\xi_0}\right)+\frac{(1-v)^{-1}}{(1-\xi_0+i0)}\right)\right.\nn\\
&~+\left.\int_0^{1/2}\,dv_1\int_1^{1/(1-2v_1)}d\xi_{1,0}\frac{(v_1(1-v_1))^{-1}}{\xi_{1,0}}\right]+\Oep{1}\;,
\end{align}
which is free of the logarithmic contributions that appear in \Eq{\ref{Equation:FDUIRForward}}, and leads to the same result as in \Eq{\ref{Equation:FDUIRForwardResult}}. Notice that in \Eq{\ref{Equation:FDUIRForwardWithCOV}} we used $\xi_{2,0}=\xi_{3,0}=\xi_0$ and $v_2=v_3=v$. A similar representation is available for the backward integral, where this time we must combine $I_1$ and $I_2$ by expressing $(\xi_{1,0},v_1)$ in terms of $(\xi_{2,0},v_2)$. By noticing that $\mathbf{q}_2=\mathbf{q}_3+\mathbf{p}_1+\mathbf{p}_2=\mathbf{q}_3$ in the centre-of-mass frame, we can use the change of variable in \Eq{\ref{Equation:FDUCOVUnification}} where we simply replace $(\xi_{3,0},v_3)$ by $(\xi_{2,0},v_2)$. We obtain
\begin{align}\label{Equation:FDUIRBackwardWithCOV}
I^{(b)}=&~\frac{\cGamma}{\s}\int_{1/2}^{0}\,dv\int_0^\infty\,dx_0\Bigg((1-v_1)^{-1}\Bigg(v^{-1}\frac{\theta\left(\sqrt{(1+\xi_0)^2-4v\,\xi_0}-w\right)}{\sqrt{(1+\xi_0)^2-4v\,\xi_0}}\theta\left(\xi_0-\frac{1}{2v-1}\right)\nn\\
&~+\frac{\theta(\xi_0-1-w)}{1-\xi_0+i0}\Bigg)-\frac{v^{-1}}{1+\xi_0}\Bigg)+\Oep{1}
\end{align}
where we also applied $\xi_{2,0}=\xi_{3,0}=\xi_0$ and $v_2=v_3=v$. The integration limits in \Eq{\ref{Equation:FDUIRBackwardWithCOV}}, codified through Heaviside theta functions, are more cumbersome than in \Eq{\ref{Equation:FDUExpansionLogBackward}}, but the result of both expressions is the same, given by \Eq{\ref{Equation:FDUIRBackwardResult}}.\\
\\
As a final note, we would like to emphasise on the fact that this step can be completely avoided if the dual integrals are not to be analytically calculated individually, when aiming for a full numerical implementation for instance. In this case, it is only necessary to parametrise the loop three-momentum $\boldsymbol{\ell}$ itself as well as the external momenta, and parametrise accordingly each internal line by simply summing the parametrisations.\\
\\
To summarise, since the IR singularities of loop integrals are restricted to a compact area of the integration domain, the finite remnants are expressible in terms of pure four-dimensional functions, which implies that DREG could be avoided. Still, we need to keep $d\neq4$ to deal with $I^\ir$. The next section is dedicated to the discussion of how to overcome this issue, in order to achieve a full four-dimensional implementation.

\section{Unsubtraction of soft and collinear divergences}\label{Section:FDUUnsubtraction}
\fancyhead[LO]{\ref*{Section:FDUUnsubtraction}~~\nameref*{Section:FDUUnsubtraction}}

\begin{figure}[t]
	\begin{center}
		\includegraphics[width=7cm]{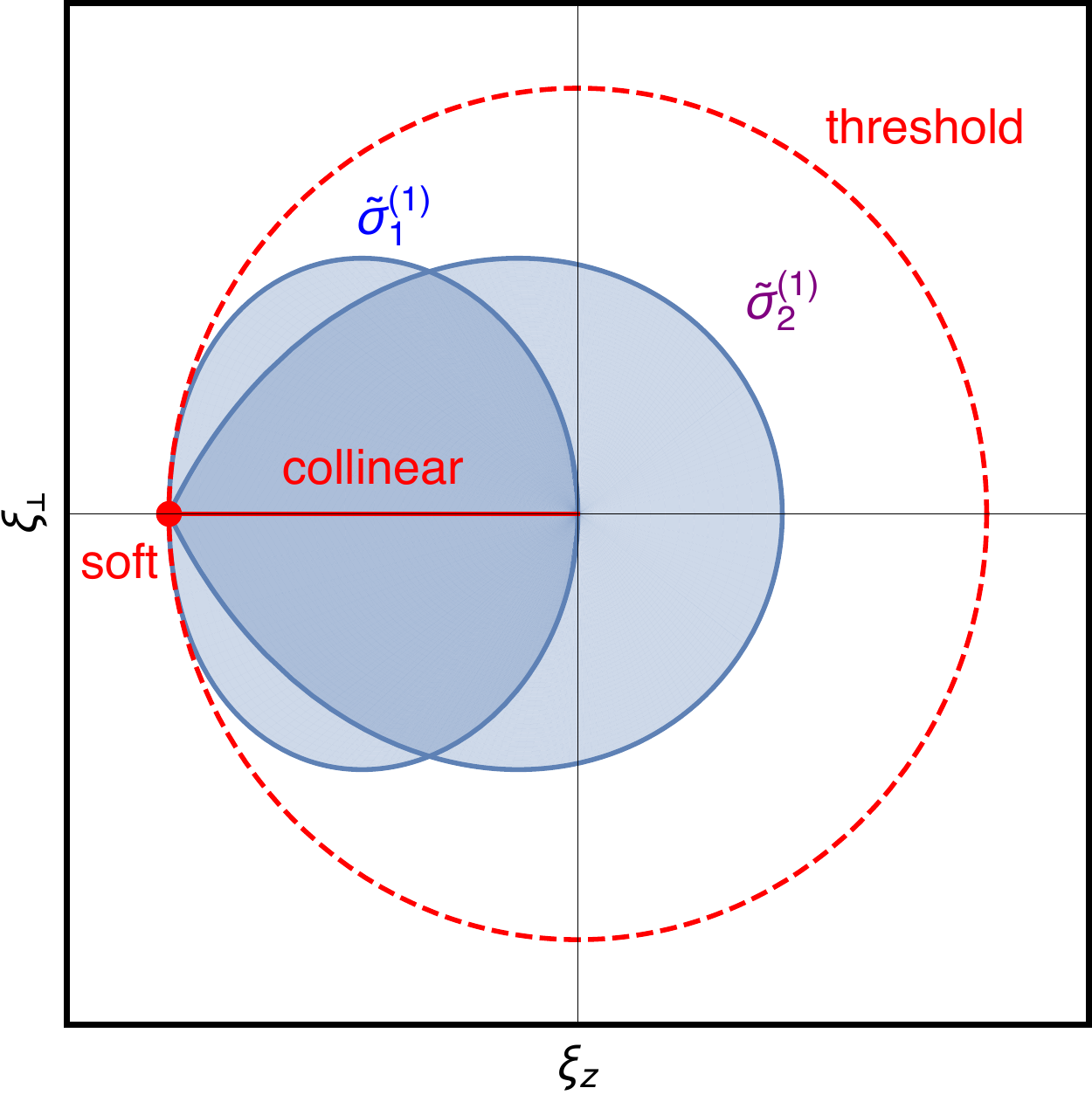}
		\caption{The dual integration regions in the loop three-momentum space.}
		\label{Figure:FDUDualRegions} 
	\end{center}
\end{figure}

In \Section{\ref{Section:FDUS3PF}}, we have illustrated in detail the application of the LTD theorem to a scalar one-loop Feynman integral, and we have isolated its infrared divergences in the function $I^{\rm IR}$ (see \Eq{\ref{Equation:FDUIRResult}}), which is obtained from a compact region of the loop momenta. In the framework of LTD, a suitable mapping of external and loop momenta between virtual and real corrections allows for the cancellation of  the IR singularities at the integrand level, such that a full four-dimensional implementation is achieved without the need to introduce soft and collinear subtraction terms \cite{Hernandez-Pinto:2015ysa}. We illustrate the method with a simplified toy scalar example before performing a complete calculation in a realistic physical process in \Section{\ref{Section:FDUNLO}}.\\
\\
We consider the one-loop virtual corrections to the cross section, which are proportional to the scalar three-point function
\begin{equation}\label{Equation:FDUVirtualCorrection}
\sigma_\V^{(1)}=\frac{1}{2\s}\int\, d\Phi_{1\to2}\,2\ReText\,\langle\mathcal{M}^{(0)}|\mathcal{M}^{(1)}\rangle=-\sigma^{(0)}2g^2\,\s\,\ReText\,L^{(1)}(p_1,p_2,-p_3)\;,
\end{equation}
where
\begin{equation}\label{Equation:FDUSigma0}
\sigma^{(0)}=\frac{g^2}{2\s}\int d\Phi_{1\to2}
\end{equation}
is the Born cross section, $\int\,d\Phi_{1\to2}$ is the integrated phase-space volume, given by \Eq{\ref{Equation:APPPhaseSpace1to2}}, and $g$ is a generic coupling.\\
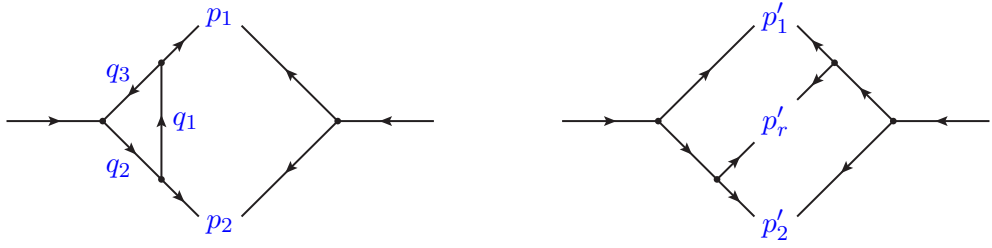
\begin{figure}
	\centering
\begin{picture}(400,100)(-200,-50)
\SetWidth{0.75}
\ArrowLine[arrowscale=0.75](-184,0)(-148,0)
\ArrowLine[arrowscale=0.75](-126,22)(-148,0)
\ArrowLine[arrowscale=0.75](-126,22)(-112,36)
\ArrowLine[arrowscale=0.75](-148,0)(-126,-22)
\ArrowLine[arrowscale=0.75](-126,-22)(-112,-36)
\ArrowLine[arrowscale=0.75](-126,-22)(-126,22)
\Vertex(-148,0){1.25}
\Vertex(-126,22){1.25}
\Vertex(-126,-22){1.25}
\ArrowLine[arrowscale=0.75](-60,0)(-96,36)
\ArrowLine[arrowscale=0.75](-60,0)(-96,-36)
\ArrowLine[arrowscale=0.75](-24,0)(-60,0)
\Vertex(-60,0){1.25}
\ArrowLine[arrowscale=0.75](24,0)(60,0)
\ArrowLine[arrowscale=0.75](60,0)(96,36)
\ArrowLine[arrowscale=0.75](60,0)(82,-22)
\ArrowLine[arrowscale=0.75](82,-22)(96,-8)
\ArrowLine[arrowscale=0.75](82,-22)(96,-36)
\Vertex(60,0){1.25}
\Vertex(82,-22){1.25}
\ArrowLine[arrowscale=0.75](184,0)(148,0)
\ArrowLine[arrowscale=0.75](148,0)(126,22)
\ArrowLine[arrowscale=0.75](126,22)(112,36)
\ArrowLine[arrowscale=0.75](126,22)(112,8)
\ArrowLine[arrowscale=0.75](148,0)(112,-36)
\Vertex(148,0){1.25}
\Vertex(126,22){1.25}
\color{blue}
\Text(-104,39){$p_1$}
\Text(-104,-39){$p_2$}
\Text(-117,0){$q_1$}
\Text(-142,18){$q_3$}
\Text(-142,-18){$q_2$}
\Text(104,39){$p'_1$}
\Text(104,-39){$p'_2$}
\Text(104,0){$p'_r$}
\end{picture}
	\caption{Momentum configuration of the virtual and real contributions to the process $\phi\to\phi\phi$ at NLO.}
	\label{Figure:FDUNLOConfiguration}
\end{figure}
\\
To obtain the full NLO correction to the cross section, it is also necessary to include the real radiation given by $1\longrightarrow3$ processes. As shown in \Fig{\ref{Figure:FDUNLOConfiguration}}, the momentum configuration is $p_3\to p_1'+p_2'+p_r'$, where we keep the same incoming momentum as in the $1\longrightarrow2$ contribution, namely $p_3=p_1+p_2$ with $p_3^2=\s$. The real radiation correction to the cross section is given by
\begin{equation}\label{Equation:FDURealRadiation}
\sigma_\r^{(1)}=\frac{1}{2\s}\int\, d\Phi_{1\to3}\,2\ReText\,\langle\mathcal{M}^{(0)}_{2r}|\mathcal{M}^{(0)}_{1r}\rangle=\frac{g^4}{\s}\int\, d\Phi_{1\to3}\frac{2\s}{s_{1r}'\,s_{2r}'}\;,
\end{equation}
with $s_{ir}'=(p_i'+p_r')^2$. The real corrections included in \Eq{\ref{Equation:FDURealRadiation}} can be understood as the interference of the two scattering amplitudes corresponding to the emission of the real radiation from each of the outgoing particles. We do not take into account for the moment the squares of these amplitudes, which are proportional to $(1/s_{ir}')^2$. They are topologically related to self-energy diagrams and will be considered explicitly in \Section{\ref{Section:FDUNLO}} for the physical process $\gamma^*\to q\bar{q}(g)$. Thus, for the current illustrative purpose, we will only consider the above-mentioned interference.\\
\\
Then, we need to split the three-body phase space to isolate the different IR singular regions. This strategy is a common practice in the context of subtraction methods \cite{Kunszt:1992tn,Frixione:1995ms}, because it allows for the optimisation of the local cancellation of the collinear singularities at the integrand level. Since there are three particles in the final state, and the incoming one is off shell, it is enough to separate the three-body phase space into two pieces by making use of the identity
\begin{equation}\label{Equation:FDUThetaIdentity}
1=\theta(\yp{2r}-\yp{1r})+\theta(\yp{1r}-\yp{2r})\;,
\end{equation}
which leads to the definitions
\begin{equation}\label{Equation:FDUSigmaTildeR}
\sigmatilde_{\r,i}^{(1)}=\frac{1}{2\s}\int\, d\Phi_{1\to3}\,2\ReText\langle\mathcal{M}_{2r}^{(0)}|\mathcal{M}_{1r}^{(0)}\rangle\theta(\yp{jr}-\yp{ir})\;,\quad i,j\in\{1,2\}\;,
\end{equation}
where the dimensionless scalar products $\yp{ir}$ are defined as $\yp{ir}=s_{ir}'/\s$. Analogously, we define the corresponding dual contribution to the virtual cross section as
\begin{equation}\label{Equation:FDUSigmaTildeV}
\sigmatilde_{\V,i}^{(1)}=\frac{1}{2\s}\int\, d\Phi_{1\to2}\,2\ReText\langle\mathcal{M}^{(0)}|\mathcal{M}_i^{(1)}\rangle\theta(\yp{jr}-\yp{ir})\;,\quad i,j\in\{1,2\}\;,
\end{equation}
with
\begin{equation}\label{Equation:FDUDualComponent}
\langle\mathcal{M}^{(0)}|\mathcal{M}_i^{(1)}\rangle=-g^4\,\s\,I_i
\end{equation}
being the $i$-th dual contribution of the one-loop scattering amplitude, according to the decomposition written in \Eq{\ref{Equation:FDUS3PF}}. We claim that the quantity
\begin{equation}\label{Equation:FDUSigmaTilde}
\sigmatilde_i^{(1)}=\sigmatilde_{\V,i}^{(1)}+\sigmatilde_{\r,i}^{(1)}\;,
\end{equation}
with $i\in\{1,2\}$, is finite in the limit $\epsilon\to0$ and can be expressed using a purely four-dimensional representation. It is worth appreciating that the dual integral $I_3$ is not necessary to cancel the IR singularities present in the real corrections. This behaviour was expected from the analysis shown in \Section{\ref{Section:FDUS3PF}}, explicitly from \Eqs{\ref{Equation:FDUDualsIR}}{\ref{Equation:FDUIRResult}}, where we showed that $I_3$ was not leading to collinear divergences that are not cancelled by the other dual contributions. Defining the quantity $\sigmatilde_3^{(1)}$ is therefore not needed. In fact, $I_3$ will solely contribute to the definition of the IR finite virtual remnant, formerly described in terms of the backward and forward integrals.\\
\\
In the following we implement a mapping between the final-state momenta of the loop amplitudes, $(p_1,p_2)$, the loop three-momentum $\boldsymbol{\ell}$, and the final-state momenta of the real amplitudes, $(p_1',p_2',p_r')$. Momentum conservation and on-shell constraints must be fulfilled by $p_i$ and $p_i'$ simultaneously. Hence, assuming $q_1$ is on shell, we propose\footnote{The notations have been slightly modified compared to~\cite{Sborlini:2016gbr}, in order to match the ones used in \Chapter{\ref{Chapter:FDUM}} and in~\cite{Sborlini:2016hat}.}
\begin{align}\label{Equation:FDUMappingR1}
p_r'^\mu=&~q_1^\mu\;,\quad&p_1'^\mu=&~p_1^\mu-q_1^\mu+(1-\gamma_1)p_2^\mu\;,\nn\\
p_2'^\mu=&~\gamma_1\,p_2^\mu\;,\quad&\gamma_1=&~1-\frac{(q_1-p_1)^2}{2(q_1-p_1)\cdot p_2}\;,
\end{align}
to perform the evaluation of the dual cross section in \Eq{\ref{Equation:FDUSigmaTilde}}. The parameter $\gamma_1$ has been determined by solving the on-shell condition $(p'_1)^2=0$. This mapping has many interesting properties that deserved to be discussed. First, momentum conservation, $p_1'+p_2'+p_r'=p_1+p_2$, is automatically fulfilled. Then, all the final-state momenta in \Eq{\ref{Equation:FDUMappingR1}} are on shell. Finally, the mapping has been designed to describe collinear configurations -- when $p_1\parallel q_1$, reached for $\gamma_1\to1$ -- which is crucial to properly combine the divergent regions of the virtual and real contributions, in order to achieve a fully local regularisation. Although this mapping is only suitable for $1\longrightarrow2$ and $1\longrightarrow3$ kinematics, it can easily be extended to processes with an arbitrary number of external particles (see \Section{\ref{Section:FDUGeneralisation}}).\\
\\
The next step consists in using the parametrisation of $q_i$ and $p_i$ from \Eq{\ref{Equation:FDUParam}}, together with the mapping in \Eq{\ref{Equation:FDUMappingR1}}, to rewrite the two-body kinematic invariants $\yp{ij}$ in terms of the integrations variables $(\xi_{1,0},v_1)$. Expressing the scalar products $p_i'\cdot p_j'$ with both sets of variables, we obtain
\begin{equation}\label{Equation:FDUyijpR1}
\yp{1r}=\frac{\xi_{1,0}\,v_1}{1-(1-v_1)\xi_{1,0}}\;,\quad\yp{2r}=\frac{\xi_{1,0}(1-\xi_{1,0})(1-v_1)}{1-(1-v_1)\xi_{1,0}}\;,\quad\yp{12}=1-\xi_{1,0}\;.
\end{equation}
Since this mapping is optimised for the description of the collinear limit $p_1\parallel q_1$, it must be used\footnote{To be mathematically rigorous, the transformation proposed in \Eq{\ref{Equation:FDUMappingR1}} is a diffeomorphism connecting the physical three-body phase space and its image in the integration domain of the dual contributions. In that respect, in principle, it would not be necessary to define a second mapping, since the entire phase space is covered by the first. However, the collinear limit $p_2\parallel q_1$ would not be dealt with in an optimal way.} in the region of the two-body and the three-body phase spaces where $\yp{1r}<\yp{2r}$. By giving a lower limit to the value of $\yp{2r}$, we avoid dealing, for the moment, with the other collinear singularity, manifesting when $p_2\parallel q_1$. A second mapping is indeed necessary to treat this singularity -- occurring when $\yp{2r}\to0$ -- that can be isolated in the complementary regions where $\yp{2r}<\yp{1r}$. Similarly, for $q_2$ on shell, we propose
\begin{align}\label{Equation:FDUMappingR2}
p_2'^\mu=&~q_2^\mu\;,\quad&p_r'^\mu=&~p_2^\mu-q_2^\mu+(1-\gamma_2)p_1^\mu\;,\nn\\
p_1'^\mu=&~\gamma_2\,p_1^\mu\;,\quad&\gamma_2=&~1-\frac{(q_2-p_2)^2}{2(q_2-p_2)\cdot p_1}\;,
\end{align}
where this time $\gamma_2$ has been determined by solving the on shell condition $(p'_r)^2=0$. With this mapping, we obtain
\begin{equation}\label{Equation:FDUyijpR2}
\yp{1r}=1-\xi_{2,0}\;,\quad\yp{2r}=\frac{\xi_{2,0}(1-v_2)}{1-v_2\,\xi_{2,0}}\;,\quad\yp{12}=\frac{\xi_{2,0}(1-\xi_{2,0})v_2}{1-v_2\,\xi_{2,0}}\;,
\end{equation}
for the corresponding two-body invariants. By virtue of \Eq{\ref{Equation:FDUThetaIdentity}}, the complete three-body phase space for the real radiation can be parametrised by applying each mapping in their respective region. It is in fact useful to define
\begin{align}
\label{Equation:FDURegion1}\mathcal{R}_1(\xi_{1,0},v_1)=&~\theta(\yp{2r}-\yp{1r})=\theta(1-2v_1)\,\theta\left(\frac{1-2v_1}{1-v_1}-\xi_{1,0}\right)\;,\\
\label{Equation:FDURegion2}\mathcal{R}_2(\xi_{2,0},v_2)=&~\theta(\yp{1r}-\yp{2r})=\theta\left(\frac{1}{1+\sqrt{1-v_2}}-\xi_{2,0}\right)\;,
\end{align}
to explicitly parametrise the integration regions for $\sigmatilde_1^{(1)}$ and $\sigmatilde_2^{(1)}$. A graphical representation of integration regions defined by \Eqs{\ref{Equation:FDURegion1}}{\ref{Equation:FDURegion2}} is shown in \Fig{\ref{Figure:FDUDualRegions}}.\\
\\
Once the momenta are properly parametrised, we proceed to evaluate together the real and virtual contributions at the integrand level. For $\yp{1r}<\yp{2r}$, we obtain
\begin{align}
\label{Equation:FDUSigmaTilde1}\sigmatilde_1^{(1)}=\sigmatilde_{\V,1}^{(1)}+\sigmatilde_{\r,1}^{(1)}=&~\sigma^{(0)}\,2g^2\int\,d[\xi_{1,0}]\,d[v_1]\,\mathcal{R}_1(\xi_{1,0},v_1)\nn\\
&\times\xi_{1,0}^{-1}(v_1(1-v_1))^{-1}\left(\left(\frac{1-\xi_{1,0}}{1-(1-v_1)\xi_{1,0}}\right)^{-2\epsilon}-1\right)\;,
\end{align}
while
\begin{align}
\label{Equation:FDUSigmaTilde2}\sigmatilde_2^{(1)}=\sigmatilde_{\V,2}^{(1)}+\sigmatilde_{\r,2}^{(1)}=&~\sigma^{(0)}\,2g^2\int\,d[\xi_{2,0}]\,d[v_2]\,\mathcal{R}_2(\xi_{2,0},v_2)\nn\\
&\times(1-v_2)^{-1}\left(\frac{(1-\xi_{1,0})^{-2\epsilon}}{(1-v_2\,\xi_{2,0})^{1-2\epsilon}}-\frac{1}{1-\xi_{2,0}+i0}-i\pi\,\delta(1-\xi_{2,0})\right)\;,
\end{align}
represents the analogous expression for $\yp{2r}<\yp{1r}$. The integrand in \Eq{\ref{Equation:FDUSigmaTilde1}} has the form
\begin{equation}\label{Equation:FDUSigmaTilde1Form}
\xi_{1,0}^{-1-2\epsilon}\,v_1^{-1-\epsilon}\,f(\xi_{1,0},v_1)\;,
\end{equation}
where the function $f(\xi_{1,0},v_1)$ vanishes in the soft and collinear regions ($\xi_{1,0}=0$ and/or $v_1=0$). Moreover, $f(\xi_{1,0},v_1)=\Oep{1}$, which implies that
\begin{equation}\label{Equation:FDUSigmaTilde1Result}
\sigmatilde_1^{(1)}=\Oep{1}\;.
\end{equation}
On the other hand, the integrand $\sigmatilde_2^{(1)}$ in \Eq{\ref{Equation:FDUSigmaTilde2}} behaves as $(1-v_2)^{-1-\epsilon}f(\xi_{2,0},v_2)$ where $f(\xi_{2,0},v_2)$ vanishes for $v_2=1$. The delta function in \Eq{\ref{Equation:FDUSigmaTilde2}} cancels the imaginary part of the $I_2$ dual integral, given by \Eq{\ref{Equation:FDUImI2S3PF}}. However, the condition $\yp{2r}<\yp{1r}$ excludes the threshold singularity of $I_2$ from the integration region, with the exception of the single point $(\xi_{2,0},v_2)=(1,1)$. This allows for the calculation of \Eq{\ref{Equation:FDUSigmaTilde2}} after removing the $+i0$ prescription as well as the delta function, leading to
\begin{equation}\label{Equation:FDUSigmaTilde2Result}
\sigmatilde_2^{(1)}=-\sigma^{(0)}\,a\,\frac{\pi^2}{6}+\Oep{1}\;,
\end{equation}
where $a=g^2/(4\pi)^2$. It is important to notice that this result can be reached following two different paths. The first one consists in using DREG and integrating in $d=4-2\epsilon$ dimensions, and once an analytic expression is obtained, we verify that no $\epsilon$-poles are present and we take the limit $\epsilon\to0$. The other possibility is to consider the limit $\epsilon\to0$ at the integrand level, leading directly to integrable expressions. The results agree in both cases.\\
\\
After combining virtual and real corrections, we define what we call the virtual remnant $\sigmabar_\V^{(1)}$, as the sum of the three dual integrals, excluding the regions of the loop three-momentum already included in \Eq{\ref{Equation:FDUSigmaTilde1}} and \Eq{\ref{Equation:FDUSigmaTilde2}}. We have
\begin{align}\label{Equation:FDUSigmaBarV}
\sigmabar_\V^{(1)}=&~\sigma^{(0)}2g^2\int\, d[\xi_0]d[v]\left[-(1-\mathcal{R}_1(\xi_0,v))\frac{v^{-1}(1-v)^{-1}}{\sqrt{(1+\xi_0)^2-4v\,\xi_0}}\right.\nn\\
&~\left.-(1-\mathcal{R}_2(\xi_{2,0},v))(1-v)^{-1}\left(\frac{1}{1-\xi_{2,0}+i0}+i\pi\delta(1-\xi_0)\right)+\frac{v^{-1}}{1+\xi_0}\right]\;.
\end{align}
This expression is analogous to the sum of the forward and backward contributions defined in \Section{\ref{Section:FDUS3PF}}, but does not require any unphysical cut $w$ to properly deal with the threshold singularity. In \Eq{\ref{Equation:FDUSigmaBarV}}, we have identified all the integration variables,\linebreak $\xi_{2,0}=\xi_{3,0}=\xi_0$ and $v_2=v_3=v$, while $\xi_{1,0}$ and $v_1$ are expressed in terms of $\xi_{3,0}$ and $v_3$ by using the change of variables in \Eq{\ref{Equation:FDUCOVUnification}} to directly avoid the appearance of logarithmic contributions coming from the expansion of the integration measure. The integration regions are defined as
\begin{align}
\label{Equation:FDUR1Dual}\mathcal{R}_1(\xi_0,v)=&~\theta(1-2v_1)\left.\theta\left(\frac{1-2v_1}{1-v_1}-\xi_{1,0}\right)\right|_{(\xi_{1,0},v_1)\to(\xi_{3,0},v_3)=(\xi_0,v)}\;,\\
\label{Equation:FDUR2Dual}\mathcal{R}_2(\xi_0,v)=&~\theta\left(\frac{1}{1+\sqrt{1-v}-\xi_0}\right)\;.
\end{align}
The explicit expression for $\mathcal{R}_1(\xi_0,v)$ is rather cumbersome (though this is not an issue for numerical computations), so for the analytic integration, we use the expansion
\begin{align}\label{Equation:FDUR1Expansion}
1-\mathcal{R}_1(\xi_0,v)=&~\theta(1-2v)+\theta(\xi_0-1)\theta(2v-1)+\theta\left(\xi_{1,0}-\frac{1-2v_1}{1-v_1}\right)\nn\\
&~\times\left[\theta\left(\frac{1}{1-2v_1}-\xi_{1,0}\right)\theta\left(\frac{2-\sqrt{2}}{4}-v_1\right)\right.\nn\\
&~~~~~~+\left.\theta(2-4v_1-\xi_{1,0})\theta\left(v_1-\frac{2-\sqrt{2}}{4}\right)\theta(1-2v_1)\right]
\end{align}
that exploits both reference systems. The first two terms in the right-hand side of \Eq{\ref{Equation:FDUR1Expansion}} contribute at large loop three-momenta making the integral defined by \Eq{\ref{Equation:FDUSigmaBarV}} finite in the UV limit. The next two terms provide a finite contribution. The virtual remnant in \Eq{\ref{Equation:FDUSigmaBarV}} is also IR finite, and therefore can be calculated in the limit $\epsilon=0$. In particular, we get
\begin{equation}\label{Equation:FDUSigmaBarVResult}
\sigmabar_\V^{(1)}=\sigma^{(0)}a\,\frac{\pi^2}{6}+\Oep{1}\;.
\end{equation}
The sum of all the contributions, namely \Eq{\ref{Equation:FDUSigmaTilde1Result}}, \Eq{\ref{Equation:FDUSigmaTilde2Result}} and \Eq{\ref{Equation:FDUSigmaBarVResult}}, gives the total cross section up to $\Oep{1}$, which is in agreement with the result that is obtained from the standard calculation in DREG.

\section{Ultraviolet renormalisation}\label{Section:FDUUV}
\fancyhead[LO]{\ref*{Section:FDUUV}~~\nameref*{Section:FDUUV}}

In the previous section we have shown how to avoid the introduction of subtraction counterterms to cancel soft and collinear singularities by applying a suitable mapping of momenta between virtual and real corrections. In any practical computation in QFT, UV divergences must also be taken into account. Another advantage of LTD is to illustrate the physical aspects of renormalisation. In order to explain the proposed approach, we first consider the simple example of the scalar two-points function, with massless internal particles.

\subsection{Dual representation of the renormalised scalar two-point function}\label{Section:FDUS2PFRenormalisation}

\begin{figure}[t]
	\begin{center}
		\includegraphics[width=6.5cm]{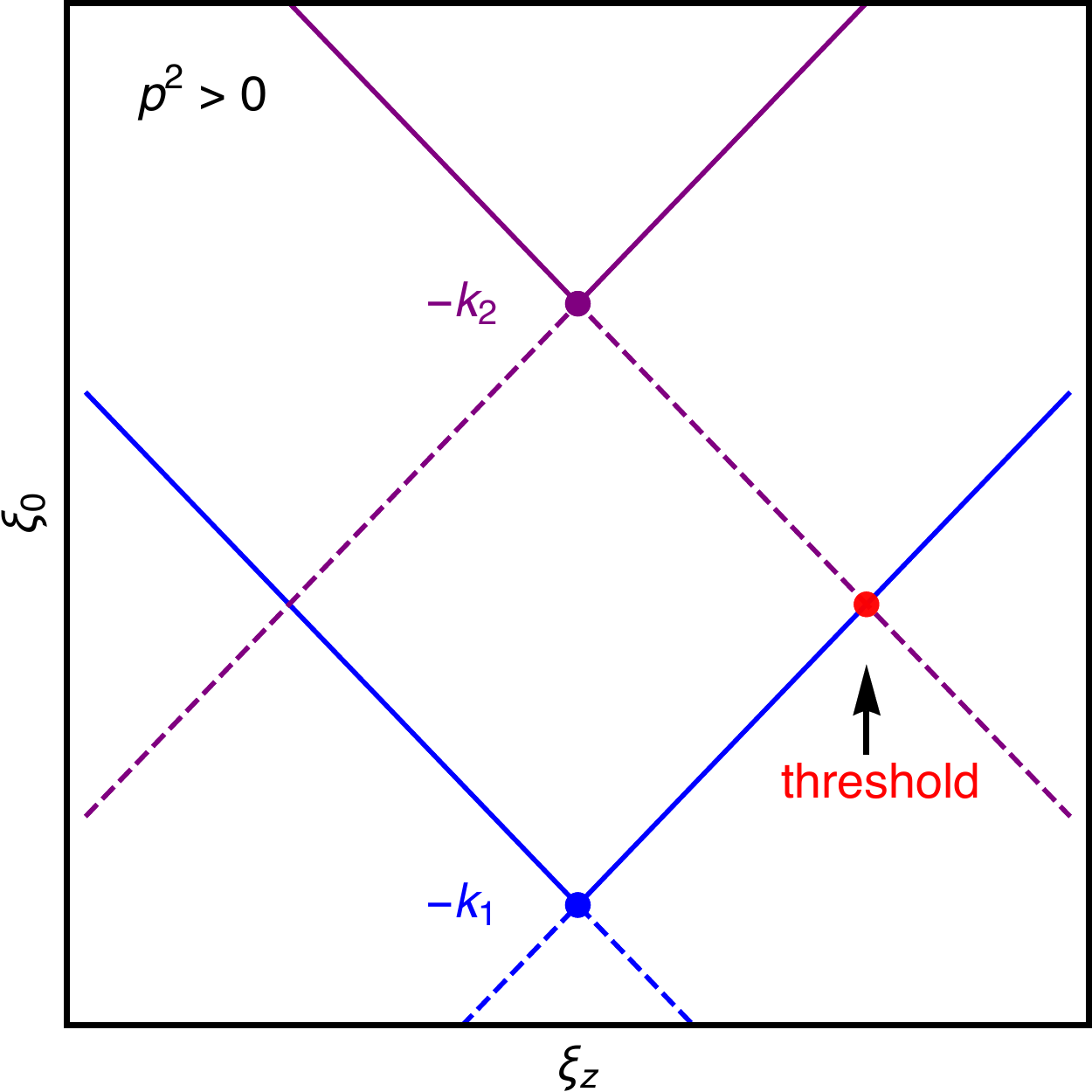} $\qquad$
		\includegraphics[width=6.5cm]{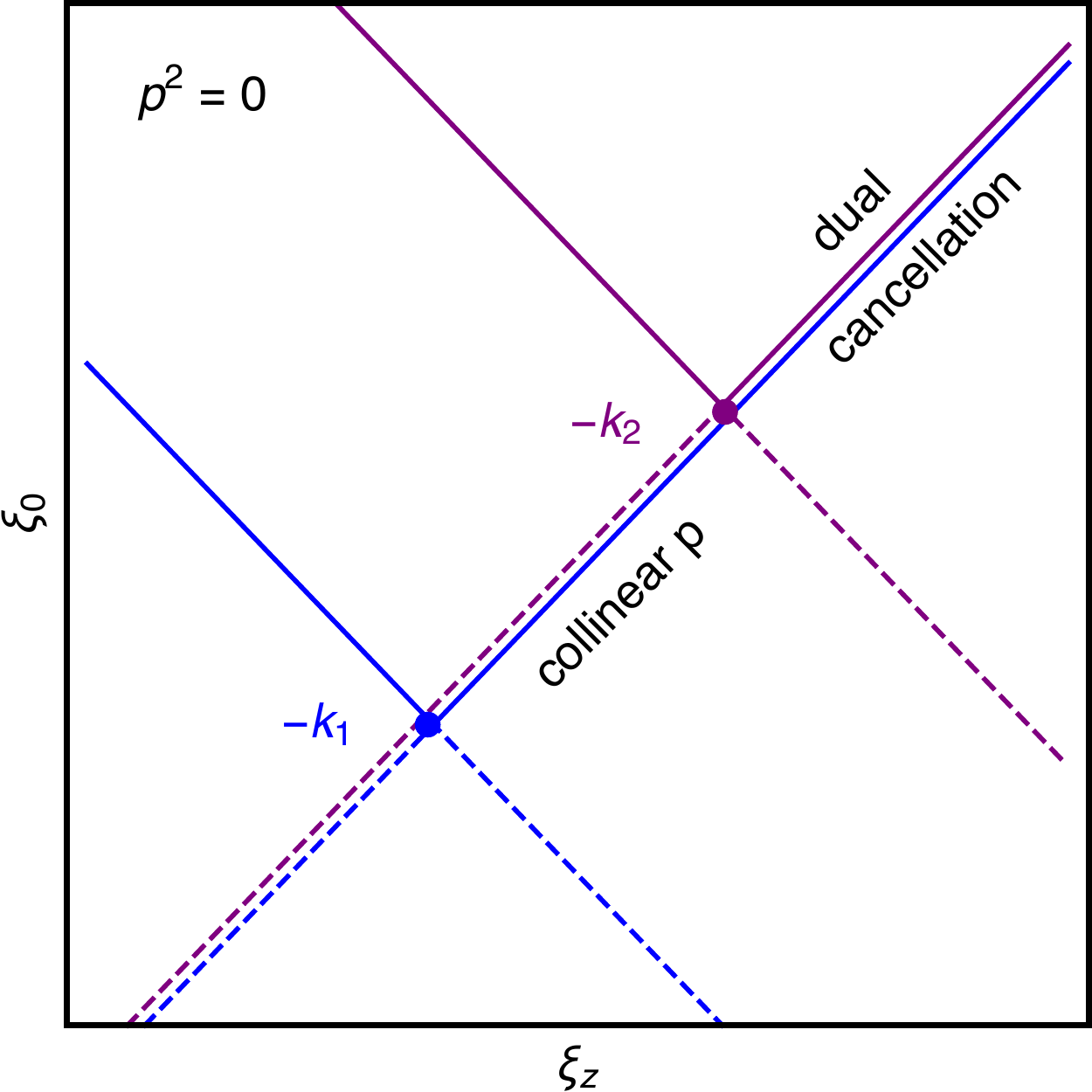}
		\caption{Light-cones of the two-point function in the loop coordinates $\ell^{\mu}=p_0\,(\xi_0, \xi_{\perp},\xi_z)$ for a time-like, $p^2>0$ (left), and a light-like, $p^2 = 0$ (right), configuration. The internal momenta are $q_1=\ell+p$ and $q_2=\ell$. In the time-like case, a threshold singularity appears when the backward light-cone (dashed) of $G_F(q_2)$ intersects with the forward light-cone (solid) of $G_F(q_1)$. In the light-like case, the IR singularities are restricted to the compact region $\xi_{1,0}\leq1$.}
		\label{Figure:FDUBubbles} 
	\end{center}
\end{figure}

There is only one external momentum which we denote $p_\mu$, and the amplitude is free of IR singularities if this momentum is not light-like. Due to the fact that the virtuality of the incoming particle is the unique physical scale involved in the process, the integral vanishes if we set $p^2=0$. The non-trivial massless scalar two-point function therefore requires $p^2\neq0$, and in this case is only UV divergent. Labelling the internal momenta as $q_1=\ell+p$ and $q_2=\ell$, with $\ell$ the loop four-momentum, we have
\begin{equation}\label{Equation:FDUS2PFLiterature}
L_{Bubble}^{(1)}(p,-p)=\int_\ell\,\,\prod\limits_{i=1}^2\,G_F(q_i)=\cGamma\,\frac{\mu^{2\epsilon}}{\epsilon(1-2\epsilon)}(-p^2-i0)^{-\epsilon}\;,
\end{equation}
as shown in the literature (e.g.~\cite{Catani:2008xa,Ellis:2007qk,Sborlini:2013jba}). The LTD representation of the scalar two-point function reads
\begin{equation}\label{Equation:FDUS2PFLTDRep}
L_{Bubble}^{(1)}(p,-p)=\,\sum\limits_{i=1}^2\,I_i\;,
\end{equation}
with the dual integrals
\begin{align}\label{Equation:FDUS2PFI1I2}
I_1=&~-\int_\ell\,\frac{\deltatilde{q_1}}{-2q_1\cdot p+p^2+i0}\;,\nn\\
I_2=&~-\int_\ell\,\frac{\deltatilde{q_2}}{2q_2\cdot p+p^2-i0}\;,
\end{align}
where for simplicity we consider $p_0>0$ and $p^2>0$ (i.e. the incoming particle has positive energy and we work in the time-like region). Following the discussion presented in \Section{\ref{Section:FDUS3PF}}, we parametrise the momenta using
\begin{equation}
p^\mu=(p_0,\mathbf{0})\;,\qquad q_i^\mu=p_0\,\xi_{i,0}\left(1,2\sqrt{v_i(1-v_i)}\,\eiPerp{i},1-2v_i\right)\;,
\end{equation}
which is equivalent to working in the rest frame of the incoming particle. With this choice, the dual integrals are rewritten as
\begin{align}
I_1=&~-\int\,\dxi{1}\dv{1}\frac{4\xi_{1,0}}{1-2\xi_{1,0}+i0}\;,\label{Equation:FDUS2PFDual1}\\
I_2=&~-\int\,\dxi{2}\dv{2}\frac{4\xi_{2,0}}{1+2\xi_{2,0}}\;,\label{Equation:FDUS2PFDual2}
\end{align}
where, instead of using $\s$ as the normalisation scale, we use $p^2$ inside $\dxi{i}$, namely
\begin{equation}
\dxi{i}=\frac{(4\pi)^{\epsilon-2}}{\Gamma(1-\epsilon)}\left(\frac{4p^2}{\mu^2}\right)^{-\epsilon}\xi_{i,0}^{-2\epsilon}d\xi_{i,0}\;.
\end{equation}
 The integration can be performed analytically, and yields
\begin{align}
\label{Equation:FDUS2PFI1}I_1=&~\frac{\cGammaTilde}{2\epsilon(1-2\epsilon)}\left(\frac{p^2}{\mu^2}\right)^{-\epsilon}e^{i2\pi\epsilon}\;,\\
\label{Equation:FDUS2PFI2}I_2=&~\frac{\cGammaTilde}{2\epsilon(1-2\epsilon)}\left(\frac{p^2}{\mu^2}\right)^{-\epsilon}\;.
\end{align}
The sum of both contributions gives the standard DREG result given in \Eq{\ref{Equation:FDUS2PFLiterature}}. The imaginary part of the scalar two-point function can be calculated and gives
\begin{align}
i\ImText L_{Bubble}^{(1)}(p,-p)=&~i\ImText I_1=\frac{1}{2}\int_\ell\,\deltatilde{q_1}\deltatilde{-q_2}=i\,\pi\int\,\deltatilde{q_1}\,\theta(p_0-q_{1,0})\,\delta(p^2-2q_1\cdot p)\nn\\
=&~i\,\frac{\cGammaTilde}{2\epsilon(1-2\epsilon)}\left(\frac{p^2}{\mu^2}\right)^{-\epsilon}\sin(2\pi\epsilon)\;,
\end{align}
which also agrees with \Eq{\ref{Equation:FDUS2PFLiterature}}. This imaginary component is related to $I_1$ and is generated by an integrable threshold singularity. Indeed, as showed in \Fig{\ref{Figure:FDUBubbles}}, the forward region of the light-cone associated with $I_1$ intersects the backward region of the light-cone associated with $I_2$. It is in this case necessary to keep the explicit $i0$ prescription, as we did in \Eq{\ref{Equation:FDUS2PFDual1}}.\\
\\
In order to build a suitable local UV counterterm of the two-point function, we follow the ideas presented in~\cite{Becker:2010ng} and we consider
\begin{equation}\label{Equation:FDUS2PFUV}
L_{Bubble}^{(1,\textrm{CT})}=\int_\ell\,\frac{1}{(q_\uv^2-\mu_\uv^2+i0)^2}\;,
\end{equation}
where $\mu_\uv$ is an arbitrary scale, and $q_\uv=\ell+k_\uv$, with $k_\uv$ an arbitrary four-momentum. Notice that the counterterm is expressed as a Feynman integral using the customary $+i0$ prescription. The next step is to calculate the corresponding dual representation and combine it with \Eqs{\ref{Equation:FDUS2PFDual1}}{\ref{Equation:FDUS2PFDual2}}. The counterterm given in \Eq{\ref{Equation:FDUS2PFUV}} exhibits a double pole in the complex plane of the loop energy component, which requires to apply the more general duality relation, valid for integrals involving multi-pole propagators and given in \Section{\ref{Section:LTDMultiLoop}}. The location of the double pole of the UV propagator in \Eq{\ref{Equation:FDUS2PFUV}} is obtained from the on-shell condition, i.e.
\begin{equation}
\big(G_F(q_\uv)\big)^{-1}=q_\uv^2-\mu_\uv^2+i0=0\quad\Leftrightarrow\quad q_{\uv,0}=q_{\uv,0}^{(\pm)}=\pm\sqrt{\mathbf{q}_\uv^2+\mu_\uv^2-i0}\;,
\end{equation}
where only $\Sup{q_{\uv,0}}$ is relevant. According to \Eq{\ref{Equation:LTDGF2GF3Residue}},
\begin{equation}
\Res_{\{q_{\uv,0}=\Sup{q_{\uv,0}}\}}\big(G_F(q_\uv)\big)^2=-\frac{2}{(2\Sup{q_0})^3}\;,
\end{equation}
which means that after applying LTD, we have
\begin{equation}\label{Equation:FDUS2PFIUV}
L_{Bubble}^{(1,\textrm{CT})}=\int_\ell\,\frac{\deltatilde{q_\uv}}{2(\Sup{q_\uv})^2}\;.
\end{equation}
By parametrising as $\xi_\uv=\sqrt{\mathbf{q}_\uv^2}/p_0$ and $m_\uv=\mu_\uv/p_0$, the UV counterterm becomes
\begin{equation}\label{Equation:FDUS2PFIUVDual}
L_{Bubble}^{(1,\textrm{CT})}=\int\,d[\xi_\uv]\,d[v_\uv]\frac{2\xi_\uv^2}{(\xi_\uv^2+m_\uv^2)^{3/2}}\;,
\end{equation}
where we dropped the $+i0$ prescription because it is not needed here. After integration, \Eq{\ref{Equation:FDUS2PFIUVDual}} leads to
\begin{equation}\label{Equation:FDUS2PFIUVIntegrated}
L_{Bubble}^{(1,\textrm{CT})}=\frac{\Se}{16\pi^2}\left(\frac{\mu_\uv^2}{\mu^2}\right)^{-\epsilon}\frac{1}{\epsilon}=\left(\frac{\mu_\uv^2}{\mu^2}\right)^{-\epsilon}\frac{\cGamma}{\epsilon}+\Oep{2}\;,
\end{equation}
which indeed cancels the single $\epsilon$-pole exhibited by the two-point scalar function. The prefactor is defined\footnote{$\Se$ is slightly different from the prefactor we used in \Section{\ref{Section:SARMSbar}}, i.e. $S_\epsilon^{\MSbar}$ defined in \Eq{\ref{Equation:SARSMSBar}}. At the considered order they both lead to the same expressions, but at NNLO using one factor over the other may lead to a different bookkeeping of the IR and UV poles at intermediate steps.} as $\Se=(4\pi)^\epsilon\Gamma(1+\epsilon)$.\\
\\
By combining the dual representations of the UV counterterm with the dual integrals $I_1$ and $I_2$, we obtain the locally renormalised scalar two-point function, reading
\begin{align}\label{Equation:FDUS2PFRenormalised}
L_{Bubble}^{(1,\r)}(p,-p)=&~L_{Bubble}^{(1)}(p,-p)-L_{Bubble}^{(1,\textrm{CT})}\nn\\
=&~-4\int\,d[\xi]\,d[v]\left(\frac{\xi}{1-2\xi+i0}+\frac{\xi}{1+2\xi}+\frac{\xi^2}{2(\xi^2+m_\uv^2)^{3/2}}\right)\;,
\end{align}
where we set $\xi_{1,0}=\xi_{2,0}=\xi_\uv=\xi$ and $v_1=v_2=v_\uv=v$. Because it is completely free of local UV singularities, the integral appearing in \Eq{\ref{Equation:FDUS2PFRenormalised}} can be integrated \emph{after} taking the limit $\epsilon\to0$, leading to
\begin{equation}\label{Equation:FDUS2PFRenormalisedIntegrated}
L_{Bubble}^{(1,\r)}(p,-p)=\frac{1}{16\pi^2}\left(-\log\left(\frac{-p^2-i0}{\mu^2}\right)+2\right)\;,
\end{equation}
which agrees with \Eq{\ref{Equation:SARRenormalisedBubble}}. We therefore succeeded in finding a pure four-dimensional representation of the renormalised two-point function. Note that a strong implication of this is that in \Eq{\ref{Equation:FDUS2PFRenormalised}}, the integration and the limit $\epsilon\to0$ commute.

\subsection{Scaleless two-point function}

As mentioned before, if we have $p^2=0$, then $L_{Bubble}^{(1)}(p,-p)=0$ since it does not involve any scale. From a physical point of view, this implies that self-energy corrections to on-shell massless scalar particles vanish. However, this is true \emph{after integration}, meaning that the associated integrand is not necessarily zero. This is particularly an issue within the LTD formalism, where the aim is to cancel the singularities locally, which therefore implies to separate explicitly the IR and UV behaviours.\\
\\
In the light-like case, the external momentum can be parametrised as
\begin{equation}\label{Equation:FDULightLikeParametrisation}
p^\mu=p_0(1,\mathbf{0}_\perp,1)\;.
\end{equation}
In this limit, the dual contributions in \Eq{\ref{Equation:FDUS2PFI1I2}} become
\begin{align}\label{Equation:FDUS2PFI1I2LightLike}
I_1=&~\int\,\dxi{1}\,\dv{1}^{-1}\;,\nn\\
I_2=&~-\int\,\dxi{2}\,\dv{2}^{-1}\;,
\end{align}
whose sum is obviously equal to zero. In the context of DREG, the scaleless two-point function develops both IR and UV divergences that cancel each other because the parameter $\epsilon_\ir$ regularising the IR singularity is identified with the parameter $\epsilon_\uv$ regularising the UV singularity. The important fact is that we can exploit LTD to separate them, proceeding in a similar way as we did in \Section{\ref{Section:FDUS3PF}}. Analysing the integration domain of the dual contributions, as well as the relative position of the light-cone in \Fig{\ref{Figure:FDUBubbles}} (right), we can see that the IR singularity is associated with $I_1$, because its forward light-cone overlaps with the backward light-cone of $G_F(q_2)$ in the region $\xi_{1,0}\leq1$. For this reason we define
\begin{equation}\label{Equation:FDUS2PFLightLikeIR}
\left.L_\ir^{(1)}(p,-p)\right|_{p^2=0}=I_1(\xi_{1,0}\leq1)=-\frac{\cGammaTilde}{\epsilon(1-2\epsilon)}\left(\frac{4p_0^2}{\mu^2}\right)^{-\epsilon}\frac{\sin(2\pi\,\epsilon)}{2\pi\,\epsilon}\;,
\end{equation}
which contains a single $\epsilon$-pole, associated with the IR region. As it was the case for the massless three-point scalar function, the IR singularities are confined in a compact region of the loop three-momentum space. Outside this region, the remnant is given by
\begin{equation}\label{Equation:FDUS2PFLightLikeRemnant}
\left.L_\uv^{(1)}(p,-p)\right|_{p^2=0}=I_1(\xi_{1,0}>1)+I_2=-\frac{\cGammaTilde}{\epsilon(1-2\epsilon)}\left(\frac{4p_0^2}{\mu^2}\right)^{-\epsilon}\frac{\sin(2\pi\,\epsilon)}{2\pi\,\epsilon}\;,
\end{equation}
which we must renormalise with the UV counterterm defined in \Eq{\ref{Equation:FDUS2PFIUV}}. It is worth noting that LTD naturally leads to this separation of IR/UV regions, which is crucial to achieve a local cancellation of singularities in the computation of physical observables at higher orders.

\subsection{Renormalisation of scattering amplitudes and physical interpretation}

\begin{figure}[t]
	\centering
	\includegraphics[width=8cm]{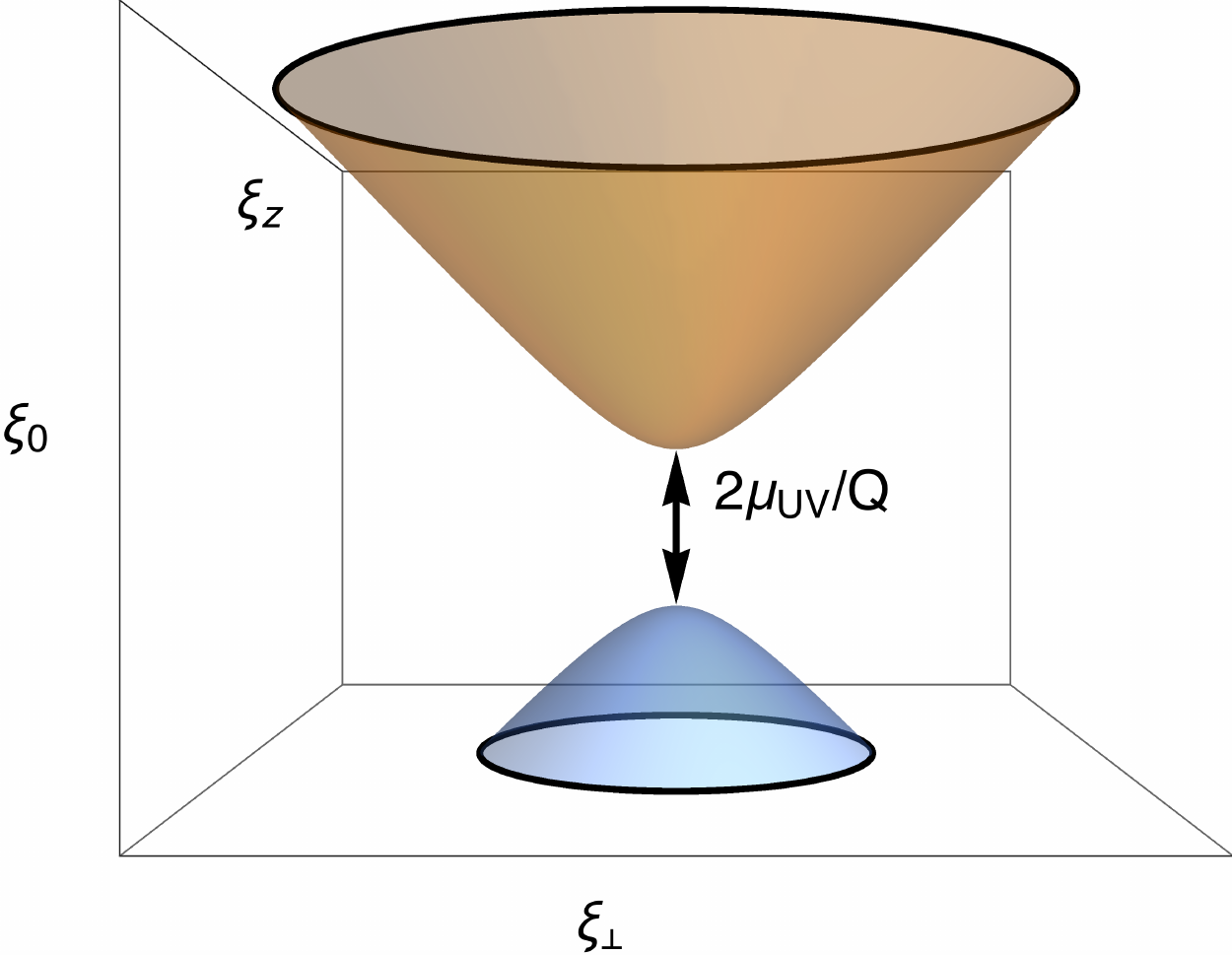}
	\caption{On-shell hyperboloids of the UV counterterm. The forward and backward on-shell hyperboloids are separated by a distance $2\mu_\uv/Q$, where $Q$ is the characteristic hard scale.}
	\label{Figure:FDUCarteseanUV}
\end{figure}

For general processes, the local UV counterterms of scattering amplitudes are derived by expanding the internal propagators around the UV propagator~\cite{Becker:2010ng}. For a single propagator, we can write
\begin{align}
\frac{1}{q_i^2-m_i^2+i0}=&~\frac{1}{q_\uv^2-\mu_\uv^2+i0}\nn\\
&\times\left(1-\frac{2q_\uv\cdot k_{i,\uv}-m_i^2+\mu_\uv^2}{q_\uv^2-\mu_\uv^2+i0}+\frac{(2q_\uv\cdot k_{i,\uv})^2}{(q_\uv^2-\mu_\uv^2+i0)^2}\right)+\mathcal{O}\left((q_\uv^2)^{-5/2}\right)\;,
\end{align}
with $k_{i,\uv}=q_i-q_\uv$. The same kind of expansion must also be performed with numerators. In order to improve the convergence in numerical implementations, it was suggested in~\cite{Becker:2012aqa} to expand to even higher orders of $\big(G_F(q_\uv)\big)^{-1}$.\\
\\
Once the desired UV expansion is obtained, we need to calculate the corresponding dual representation. As explained in \Section{\ref{Section:FDUS2PFRenormalisation}}, it will be necessary to deal with multi-poles and non-trivial numerators that depend on $q_\uv$. From \Eq{\ref{Equation:LTDCauchyMultiPole}}, we can calculate the identities
\begin{align}
\int_\ell\,\big(G_F(q_\uv)\big)^n=&~\frac{(-1)^n(2n-2)}{\big((n-1)!\big)^2}\int_\ell\frac{\deltatilde{q_\uv}}{(2\Sup{q_\uv})^{2n-2}}\;,\nn\\
\int_\ell\,\big(G_F(q_\uv)\big)^nq_\uv=&~0\;,\nn\\
\int_\ell\,\big(G_F(q_\uv)\big)^nq_\uv^2=&~\frac{(-1)^n(2n-4)}{(n-1)!(n-2)!}\int_\ell\frac{\deltatilde{q_\uv}}{(2\Sup{q_\uv})^{2n-4}}\;,
\end{align}
that can be used to construct most of the UV counterterms. Explicit examples will be presented in \Section{\ref{Section:FDUNLO}}. We recall that only the genuine UV singularities of the original scattering amplitudes need to be subtracted with this procedure. The spurious UV singularities of the individual dual integrals vanish when taking the sum of all dual contributions.\\
\\
To conclude this section, it is worth making a comment about the physical interpretation of the energy scale $\mu_\uv$ introduced in the renormalisation procedure~\cite{Hernandez-Pinto:2015ysa}. Since the UV counterterm affects the behaviour of the integrand only in the high-energy region, this arbitrary scale can be interpreted as a renormalisation scale. In fact, as seen in \Fig{\ref{Figure:FDUCarteseanUV}}, the dual representation of the counterterm contributes only for energies larger than $k_{\uv,0}+\mu_\uv$ -- even though it is unconstrained in the loop three-momentum. This means we should choose $\mu_\uv\geq\mathcal{O}(Q)$, with $Q$ the physical hard scale that determines the size of the compact region in the loop momentum space where the IR and threshold singularities are located. In other terms, we must choose $\mu_\uv$ in such a way that the on-shell hyperboloids of the UV propagator do not intersect with any of the on-shell hyperboloids of the original integral. Since the UV forward and backward on-shell hyperboloids are separated by a distance $2\mu_\uv$, it is possible to give a physical motivation for an optimal choice of $\mu_\uv$ and $k_\uv$. If we take $\mu_\uv=Q/2$, and $k_\uv$ to be the centre of the physical compact region, these conditions are fulfilled in a minimal way, i.e. we naturally avoid intersections with the physical on-shell hyperboloids.

\section{NLO corrections to $\gamma^*\to q\qbar(g)$}\label{Section:FDUNLO}
\fancyhead[LO]{\ref*{Section:FDUNLO}~~\nameref*{Section:FDUNLO}}

\begin{figure}[t]
	\centering
	\begin{picture}(425,100)(0,-50)
	\SetWidth{0.75}
	\Photon(0,0)(30,0){3}{3}
	\ArrowLine[arrowpos=0.35,arrowscale=0.75](30,0)(60,30)
	\ArrowLine[arrowpos=0.35,flip,arrowscale=0.75](30,0)(60,-30)
	\GluonArc(30,0)(25,-45,45){3}{5.5}
	\Arc[arrow](20,0)(27,-20,20)
	\Vertex(30,0){1.25}
	\Vertex(48,18){1.25}
	\Vertex(48,-18){1.25}
	\Photon(88,0)(118,0){3}{3}
	\ArrowLine[arrowscale=0.75](118,0)(148,30)
	\ArrowLine[flip,arrowscale=0.75](118,0)(148,-30)
	\GluonArc(133,15)(11,45,225){3}{4.5}
	\Vertex(118,0){1.25}
	\Vertex(125,7){1.25}
	\Vertex(141,23){1.25}
	\Photon(176,0)(206,0){3}{3}
	\ArrowLine[arrowscale=0.75](206,0)(236,30)
	\ArrowLine[flip,arrowscale=0.75](206,0)(236,-30)
	\GluonArc(221,-15)(11,135,315){3}{4.5}
	\Vertex(206,0){1.25}
	\Vertex(213,-7){1.25}
	\Vertex(229,-23){1.25}
	\Photon(264,0)(294,0){3}{3}
	\ArrowLine[arrowpos=0.25,arrowscale=0.75](294,0)(324,30)
	\ArrowLine[flip,arrowscale=0.75](294,0)(324,-30)
	\Gluon(309,15)(324,0){-3}{3}
	\Vertex(294,0){1.25}
	\Vertex(308,14){1.25}
	\Photon(352,0)(382,0){3}{3}
	\ArrowLine[arrowscale=0.75](382,0)(412,30)
	\ArrowLine[arrowpos=0.25,flip,arrowscale=0.75](382,0)(412,-30)
	\Gluon(397,-15)(412,0){3}{3}
	\Vertex(382,0){1.25}
	\Vertex(396,-14){1.25}
	\color{blue}
	\Text(69,34){$p_1$}
	\Text(69,-34){$p_2$}
	\Text(68,0){$q_1$}
	\Text(37,-18){$q_2$}
	\Text(37,18){$q_3$}
	\Text(40,0){$\ell$}
	\Text(157,34){$p_1$}
	\Text(157,-34){$p_2$}
	\Text(120,31){$q_1$}
	\Text(139,9){$q_3$}
	\Text(245,34){$p_1$}
	\Text(245,-34){$p_2$}
	\Text(208,-31){$q_1$}
	\Text(227,-9){$q_2$}
	\Text(333,34){$p'_1$}
	\Text(333,-34){$p'_2$}
	\Text(333,0){$p'_r$}
	\Text(421,34){$p'_1$}
	\Text(421,-34){$p'_2$}
	\Text(421,0){$p'_r$}
	\end{picture}
	\caption{Momentum configuration and Feynman diagrams associated with the process $\gamma^*\to q\qbar(g)$, up to $\mathcal{O}(\alpha\,\alpha_S)$. Notice that we also consider self-energy corrections to the on-shell outgoing particles, even if their total contribution after integration is zero.}
	\label{Figure:FDUVirtualRealDiagrams}
\end{figure}
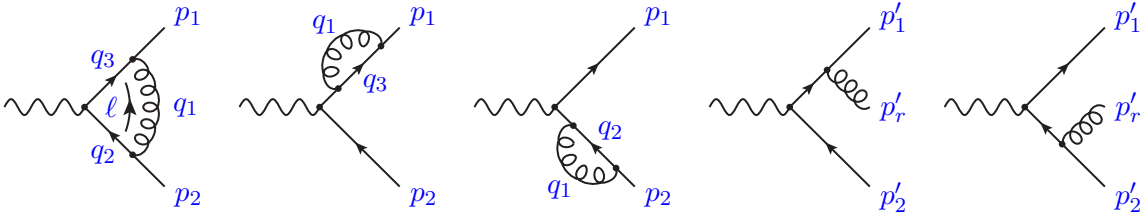

In this section, we discuss in detail the computation of NLO QCD corrections to the total cross section for the process $\gamma^*\to q\qbar(g)$ by using the LTD/FDU approach. We would like to emphasise that this constitutes the first realistic physical application of this method, as already anticipated in~\cite{Sborlini:2016fcj}.\\
\\
The computation is done with massless quarks, up to $\mathcal{O}(\alpha\,\alpha_S)$. The corresponding Feynman diagrams are shown in \Fig{\ref{Figure:FDUVirtualRealDiagrams}}. Starting at LO, we have
\begin{equation}\label{Equation:FDUAmpSquaredMasslessLO}
|\mathcal{M}_{q\qbar}^{(0)}|^2=2C_A\,(e\,e_q)^2\s\,(1-\epsilon)\;,
\end{equation}
for $\gamma^*(p_{12})\to q(p_1)+\qbar(p_2)$ with $p_{12}=p_1+p_2$ and $p_{12}^2=\s>0$,\linebreak where $e$ and $e_q$ denote the electromagnetic coupling and the quark electric charge, respectively\footnote{As usual, the squared matrix elements are averaged over the number of spin degrees of freedom of the incoming particles, which is taking to be $2(1-\epsilon)$ for the photon. In any case, we normalise the NLO results by the LO contributions.}. The corresponding Born level cross section is given by
\begin{equation}\label{Equation:FDUSigma0MasslessBorn}
\sigma^{(0)}=\frac{1}{2\s}\int\, d\Phi_{1\to2}|\mathcal{M}_{q\qbar}^{(0)}|^2=\frac{1}{2}\,\alpha\,e_q^2\,C_A+\Oep{1}\;,
\end{equation}
where the two-body phase-space factor is shown in \Eq{\ref{Equation:APPPhaseSpace1to2}}.\\
\\
Let's now consider the NLO contributions. On the first hand, the real correction from the radiative process $\gamma^*(p_{12})\to q(p_1')+\qbar(p_2')+g(p_r')$ is
\begin{align}\label{Equation:FDUSigmaRMasslessNLO}
\sigma_\r^{(1)}=&~\sigma^{(0)}\frac{(4\pi)^{\epsilon-2}}{\Gamma(1-\epsilon)}\,g_S^2\,C_F\left(\frac{\s}{\mu^2}\right)^{-\epsilon}\int_0^1\,d\yp{1r}\int_0^{1-\yp{1r}}\,d\yp{2r}(\yp{1r}\,\yp{2r}\,\yp{12})^{-\epsilon}\nn\\
&~\times\left[4\left(\frac{\yp{12}}{\yp{1r}\,\yp{2r}}-\epsilon\right)+2(1-\epsilon)\left(\frac{\yp{2r}}{\yp{1r}}+\frac{\yp{1r}}{\yp{2r}}\right)\right]\nn\\
=&~\sigma^{(0)}\,g_S^2\,C_F\,\cGamma\left(\frac{\s}{\mu^2}\right)^{-\epsilon}\frac{4(2-2\epsilon+3\epsilon^2)\Gamma(2-2\epsilon)}{\epsilon^2\Gamma(3-3\epsilon)\Gamma(1+\epsilon)}\;,
\end{align}
where the virtuality of the photon is $\s$. The primed momenta are the real final-state momenta, and we define the two-body invariants as $\yp{ij}=2p_i'\cdot p_j'/\s$. To obtain the expression in \Eq{\ref{Equation:FDUSigmaRMasslessNLO}}, we used the expansion of the three-body phase space shown in \Eq{\ref{Equation:APPPhaseSpace1to3}}. If we expand this result for $\epsilon$ close to 0, we find
\begin{equation}\label{Equation:FDUSigmaRMasslessNLOepsilon}
\sigma_\r^{(1)}=\sigma^{(0)}\,g_S^2\,C_F\,\cGamma\left(\frac{\s}{\mu^2}\right)^{-\epsilon}\left(\frac{4}{\epsilon^2}+\frac{6}{\epsilon}+19-2\pi+\Oep{1}\right)\;,
\end{equation}
which contains both double and single $\epsilon$-poles, associated with soft and collinear singularities.\\
\\
On the other hand, the virtual contribution is generated by the interference of the one-loop vertex correction with the Born amplitude, which is given by
\begin{align}\label{Equation:FDUVirtualInterferenceMassless}
\langle\mathcal{M}_{q\qbar}^{(0)}|\mathcal{M}_{q\qbar}^{(1)}\rangle=&~-\frac{g_S^2\,C_F}{4\s(1-\epsilon)}|\mathcal{M}_{q\qbar}^{(0)}|^2\int_\ell\,\left(\,\prod\limits_{i=1}^{3}\,G_F(q_i)\right)\nn\\
&~\times\Tr\left(\slashed{p}_1\,\gamma^{\sigma_1}\,\slashed{q}_3\,\gamma^{\mu_1}\,\slashed{q}_2\,\gamma^{\sigma_2}\,\slashed{p}_2\,\gamma^{\mu_2}\right)d_{\sigma_1\,\sigma_2}(q_1)\,d_{\mu_1\,\mu_2}(p_{12})\;,
\end{align}
with $d_{\alpha_1\,\alpha_2}$ the gluon (and photon) polarisation tensor. In this expression, we use the same notation for the internal momenta previously used in \Section{\ref{Section:FDUS3PF}} for the scalar three-point function, i.e. $q_1=\ell+p_1$, $q_2=\ell+p_{12}$ and $q_3=\ell$. In the Feynman gauge, \Eq{\ref{Equation:FDUVirtualInterferenceMassless}} takes the form
\begin{align}\label{Equation:FDUVirtualInterferenceMasslessTraced}
\langle\mathcal{M}_{q\qbar}^{(0)}|\mathcal{M}_{q\qbar}^{(1)}\rangle=&~\frac{4g_S^2\,C_F}{\s}|\mathcal{M}_{q\qbar}^{(0)}|^2\int_\ell\,\left(\,\prod\limits_{i=1}^{3}\,G_F(q_i)\right)\nn\\
&~\times\left[(2+\epsilon)(q_2\cdot p_1)(q_3\cdot p_2)-\epsilon\left((q_2\cdot p_2)(q_3\cdot p_1)+\frac{\s}{2}(q_2\cdot q_3)\right)\right]\;,
\end{align}
which leads, after applying LTD, to the three dual contributions
\begin{align}\label{Equation:FDUVirtualInterferenceMasslessDual}
\langle\mathcal{M}_{q\qbar}^{(0)}|\mathcal{M}_{q\qbar,1}^{(1)}\rangle=&~-2g_S^2\,C_F\,|\mathcal{M}_{q\qbar}^{(0)}|^2\int\,\dxi{1}\,\dv{1}\left(\frac{1}{\xi_{1,0}\,v_1}+1\right)\left(\frac{1}{1-v_1}-\xi_{1,0}\right)\;,\nn\\
\langle\mathcal{M}_{q\qbar}^{(0)}|\mathcal{M}_{q\qbar,2}^{(1)}\rangle=&~-2g_S^2\,C_F\,|\mathcal{M}_{q\qbar}^{(0)}|^2\int\,\dxi{2}\,\dv{2}\frac{1}{1-\xi_{2,0}+i0}\left(\left(\frac{1}{1-v_2}-\xi_{2,0}\right)v_2-\epsilon\right)\;,\nn\\
\langle\mathcal{M}_{q\qbar}^{(0)}|\mathcal{M}_{q\qbar,3}^{(1)}\rangle=&~-2g_S^2\,C_F\,|\mathcal{M}_{q\qbar}^{(0)}|^2\int\,\dxi{3}\,\dv{3}\frac{\xi_{3,0}}{1+\xi_{3,0}}\left((1-v_3)\left(\frac{1}{v_3}+\xi_{3,0}\right)-\epsilon\right)\;,
\end{align}
with $|\mathcal{M}_{q\qbar}^{(1)}\rangle=\sum_{i=1}^3|\mathcal{M}_{q\qbar,i}^{(1)}\rangle$. Note that although a simple power counting in the original loop integral in \Eq{\ref{Equation:FDUVirtualInterferenceMassless}} shows it only exhibits logarithmic divergences in the UV, each of the individual dual integrals of \Eq{\ref{Equation:FDUVirtualInterferenceMasslessDual}} contains divergences up to quadratic degree. When summing the three contributions, however, the quadratic divergences cancel. Furthermore, the linear divergences disappear after integration over the polar angle. Nonetheless, applying the change of variables from \Eq{\ref{Equation:FDUCOVUnification}} to the first dual contribution makes the cancellation of the linear divergences explicit at the integrand level. Therefore, and as expected, only the logarithmic UV divergences remain in the sum of the dual contributions. Performing the explicit integration over the loop variables leads to
\begin{align}\label{Equation:FDUVirtualInterferenceMasslessDualResult}
\langle\mathcal{M}_{q\qbar}^{(0)}|\mathcal{M}_{q\qbar,1}^{(1)}\rangle=&~0\;,\nn\\
\langle\mathcal{M}_{q\qbar}^{(0)}|\mathcal{M}_{q\qbar,2}^{(1)}\rangle=&~g_S^2\,C_F\,\cGammaTilde\,|\mathcal{M}_{q\qbar}^{(0)}|^2\left(\frac{\s}{\mu^2}\right)^{-\epsilon}\frac{1}{\epsilon^2}\left(\frac{\epsilon}{2}-\frac{1}{1-2\epsilon}\right)e^{i2\pi\epsilon}\;,\nn\\
\langle\mathcal{M}_{q\qbar}^{(0)}|\mathcal{M}_{q\qbar,3}^{(1)}\rangle=&~g_S^2\,C_F\,\cGammaTilde\,|\mathcal{M}_{q\qbar}^{(0)}|^2\left(\frac{\s}{\mu^2}\right)^{-\epsilon}\frac{1}{\epsilon^2}\left(\frac{\epsilon}{2}-\frac{1}{1-2\epsilon}\right)\;.
\end{align}
Putting together the three dual contributions, we obtain
\begin{align}\label{Equation:FDUSigmaVMasslessNLOepsilon}
\sigma_\V^{(1)}=&~\sigma^{(0)}\,g_S^2\,C_F\,\cGamma\left(\frac{\s}{\mu^2}\right)\frac{2}{\epsilon^2}\left(\epsilon-\frac{1}{1-2\epsilon}\right)\cos(\pi\epsilon)\nn\\
=&~\sigma^{(0)}\,g_S^2\,C_F\,\cGamma\left(\frac{\s}{\mu^2}\right)\left(-\frac{4}{\epsilon^2}-\frac{6}{\epsilon}-16+2\pi+\Oep{1}\right)\;,
\end{align}
recovering the virtual contribution to the total cross section at NLO. Notice that it was unnecessary to introduce any tensor reduction. Indeed, Gram determinants are naturally avoided when using the LTD formalism, as are the spurious singularities that the tensor reduction may introduce and that could lead to numerical instabilities when integrating over the phase space. Finally, if we sum the real and virtual contributions (\Eq{\ref{Equation:FDUSigmaRMasslessNLOepsilon}} and \Eq{\ref{Equation:FDUSigmaVMasslessNLOepsilon}}, respectively), we obtain
\begin{equation}\label{Equation:FDUSigmaMasslessNLOepsilon}
\sigma^{(1)}=\sigma^{(0)}\left(1+3C_F\,\frac{\alpha_S}{4\pi}+\mathcal{O}(\alpha_S^2)\right)\;,
\end{equation}
which agrees with the well-known result available in the literature. The $\epsilon$-poles cancel between real and virtual contributions, as predicted by the KLN theorem. The limit $\epsilon\to0$ can safely be taken \emph{after integration}.\\
\\
The purpose of the rest of this section, however, is to show that the four-dimensional limit can also be considered \emph{before integration}, once a proper combination of the real and virtual terms is done. In the context of LTD, we must also carefully consider the contributions introduced by the self-energy diagrams. On-shell massless quarks do not introduce further corrections to the total cross section in \Eq{\ref{Equation:FDUSigmaMasslessNLOepsilon}} due to the renormalisation of the wave-function because IR and UV divergences are not discriminated in DREG. In the on-shell scheme, the wave function renormalisation factors contain both IR and UV divergences, but they exactly cancel each other, which justifies the exclusion of the corresponding Feynman diagrams when carrying out the computation in the traditional approach. Nevertheless, in order to build a complete LTD representation of the virtual contributions, it is required to include the renormalised self-energy corrections to the external particles, and properly disentangle IR and UV singularities at the integrand level. This step is crucial to achieve a dual representation for which we have a fully local cancellation of singularities, in order to be able to integrate in four dimensions.\\
\\
The quark and antiquark self-energy contributions at one-loop are given by
\begin{align}\label{Equation:FDUSelfEnergiesMassless}
-i\Sigma(p_1)=&~-i\,g_S^2\,C_F\int_\ell\left(\,\prod\limits_{i=1,3}\,G_F(q_i)\right)\gamma^{\sigma_1}\,\slashed{q}_3\,\gamma^{\sigma_2}\,d_{\sigma_1\sigma_2}(q_1)\;,\nn\\
-i\Sigma(-p_2)=&~-i\,g_S^2\,C_F\int_\ell\left(\,\prod\limits_{i=1,2}\,G_F(q_i)\right)\gamma^{\sigma_1}\,\slashed{q}_2\,\gamma^{\sigma_2}\,d_{\sigma_1\sigma_2}(q_1)\;,
\end{align}
where $\Sigma(p_i)=\Sigma_2\,\slashed{p}_i$. In these expressions, we keep the same internal momenta $q_i$ that were used to define the vertex corrections in \Eq{\ref{Equation:FDUVirtualInterferenceMassless}}. According to the usual renormalisation procedure, the self-energy contribution is related to the renormalisation factor $Z_2=1+\Delta Z_2$. Applying on-shell renormalisation conditions to the quark and antiquark self-energies in the Feynman gauge, we obtain the contributions
\begin{align}\label{Equation:FDUSelfEnergiesMasslessTraced}
\langle\mathcal{M}_{q\qbar}^{(0)}|\Sigma(p_1)\rangle=&~-2(1-\epsilon)g_S^2\,C_F|\mathcal{M}_{q\qbar}^{(0)}|^2\int_\ell\left(\,\prod\limits_{i=1,3}\,G_F(q_i)\right)\left(1+\frac{q_3\cdot p_2}{p1\cdot p_2}\right)\;,\nn\\
\langle\mathcal{M}_{q\qbar}^{(0)}|\Sigma(p_2)\rangle=&~-2(1-\epsilon)g_S^2\,C_F|\mathcal{M}_{q\qbar}^{(0)}|^2\int_\ell\left(\,\prod\limits_{i=1,2}\,G_F(q_i)\right)\left(1-\frac{q_2\cdot p_1}{p1\cdot p_2}\right)\;.
\end{align}
Formally, these contributions vanish in DREG. Yet, they feature non-trivial IR and UV behaviours at the integrand level that we need to make explicit in order to achieve a complete local cancellation of all the singularities with those present in the real corrections. After applying LTD to the loop integrals given in \Eq{\ref{Equation:FDUSelfEnergiesMasslessTraced}}, we obtain the two dual representations
\begin{align}\label{Equation:FDUSelfEnergiesMasslessDual}
\langle\mathcal{M}_{q\qbar}^{(0)}|\Sigma(p_1)\rangle=&~-2(1-\epsilon)g_S^2\,C_F|\mathcal{M}_{q\qbar}^{(0)}|^2\left[\int\,\dxi{1}\,\dv{1}\frac{1-v_1}{v_1}\xi_{1,0}\right.\nn\\
&~-\left.\int\,\dxi{3}\,\dv{3}\frac{1+(1+v_3)\xi_{3,0}}{v_3}\right]\;,\nn\\
\langle\mathcal{M}_{q\qbar}^{(0)}|\Sigma(p_2)\rangle=&~-2(1-\epsilon)g_S^2\,C_F|\mathcal{M}_{q\qbar}^{(0)}|^2\left[\int\,\dxi{1}\dv{1}\frac{1-v_1}{v_1}\xi_{1,0}\right.\nn\\
&~+\left.\int\,\dxi{2}\,\dv{2}\frac{1-v_2\,\xi_{2,0}}{1-v_2}\right]\;,\nn\\
\end{align}
where we kept explicit the integration variables associated with each cut.\\
\\
The UV divergences of the wave function cancel exactly with the UV divergences of the vertex corrections, because conserved (or partially conserved) currents, such as the vector and axial ones, do not get renormalised. In order to achieve a local cancellation of the UV divergences, it is relevant to note beforehand that the vertex corrections diverge logarithmically in the UV, while the expressions in \Eq{\ref{Equation:FDUSelfEnergiesMasslessTraced}} behave linearly in the UV. As it was also the case for the virtual contribution, this linear divergence will vanish after angular integration. However, it is still necessary to include an additional UV counterterm if we wish to locally cancel the linear singularities.\\
\\
Assuming $k_\uv=0$ (i.e. $q_\uv=\ell$) and using the notations of \Section{\ref{Section:FDUUV}}, we define
\begin{align}\label{Equation:FDUSelfEnergiesMasslessCTUV}
\langle\mathcal{M}_{q\qbar}^{(0)}|\Sigma_\uv(p_1)\rangle=&~-2(1-\epsilon)g_S^2\,C_F|\mathcal{M}_{q\qbar}^{(0)}|^2\int_\ell\,\big(G_F(q_\uv)\big)^2\left(1+\frac{q_\uv\cdot p_2}{p_1\cdot p_2}\right)\nn\\
&~\times\big(1-G_F(q_\uv)(2q_\uv\cdot p_1+\mu_\uv^2)\big)\;,
\end{align}
whose dual representation is given by
\begin{align}\label{Equation:FDUSelfEnergiesMasslessCTUVDual}
\langle\mathcal{M}_{q\qbar}^{(0)}|\Sigma_\uv(p_1)\rangle=&~-2(1-\epsilon)g_S^2\,C_F|\mathcal{M}_{q\qbar}^{(0)}|^2\int_\ell\,\frac{\deltatilde{q_\uv}}{2\left(\Sup{q_{\uv,0}}\right)^2}\nn\\
&~\times\left[\left(1-\frac{\mathbf{q}_\uv\cdot \mathbf{p}_2}{p_1\cdot p_2}\right)\left(1-\frac{3(2\mathbf{q}_\uv\cdot \mathbf{p}_1-\mu_\uv^2)}{4\left(\Sup{q_{\uv,0}}\right)^2}\right)-\frac{1}{4}\right]\nn\\
=&~-2(1-\epsilon)g_S^2\,C_F|\mathcal{M}_{q\qbar}^{(0)}|^2\int\,d[\xi_\uv]d[v_\uv]\frac{\xi_\uv^2}{\left(\xi_\uv^2+m_\uv^2\right)^{3/2}}\nn\\
&~\times\left[\Big(2+\xi_\uv(1-2v_\uv)\Big)\left(1-\frac{3(2\xi_\uv(1-2v_\uv)-m_\uv^2)}{4(\xi_\uv^2+m_\uv^2)}\right)-\frac{1}{2}\right]\;,
\end{align}
where $m_\uv=2\mu_\uv/\sqrt{\s}$. It is worth noting that the term proportional to $(1-2v_\uv)$ integrates to zero, but locally cancels the linear UV singularity. Integrating \Eq{\ref{Equation:FDUSelfEnergiesMasslessCTUVDual}} leads to
\begin{equation}\label{Equation:FDUSelfEnergiesMasslessCTUVResult}
\langle\mathcal{M}_{q\qbar}^{(0)}|\Sigma_\uv(p_1)\rangle=-g_S^2\,C_F\,\frac{\Se}{16\pi^2}|\mathcal{M}_{q\qbar}^{(0)}|^2\left(\frac{\mu_\uv^2}{\mu^2}\right)^{-\epsilon}\frac{1}{\epsilon}+\Oep{1}\;.
\end{equation}
Thus, $\langle\mathcal{M}_{q\qbar}^{(0)}|\Sigma(p_1)-\Sigma_\uv(p_1)\rangle$ only develops IR singularities. The procedure to compute the counterterm of the antiquark leg is identical, and we obtain the same expression, i.e. $\langle\mathcal{M}_{q\qbar}^{(0)}|\Sigma(p_2)\rangle=\langle\mathcal{M}_{q\qbar}^{(0)}|\Sigma(p_1)\rangle$.\\
\\
Likewise, it is necessary to remove the UV divergences from the vertex correction included in $|\mathcal{M}_{q\qbar}^{(1)}\rangle$. To this end, we use the following counterterm
\begin{align}\label{Equation:FDUVertexMasslessCTUV}
\langle\mathcal{M}_{q\qbar}^{(0)}|\mathcal{M}_{q\qbar}^{(1,\uv)}\rangle=&~g_S^2\,C_F|\mathcal{M}_{q\qbar}^{(0)}|^2\int\,d[\xi_\uv]d[v_\uv]\nn\\
&~\times\frac{\xi_\uv^2}{(\xi_\uv^2+m_\uv^2)^{5/2}}\Big(4(1-3v_\uv(1-v_\uv)-\epsilon)\xi_\uv^2+(7-4\epsilon)m_\uv^2\Big)\;,
\end{align}
where the subleading terms have been fixed such that only the pole is subtracted, i.e.
\begin{equation}\label{Equation:FDUVertexMasslessCTUVResult}
\langle\mathcal{M}_{q\qbar}^{(0)}|\mathcal{M}_{q\qbar}^{(1,\uv)}\rangle=g_S^2\,C_F\,\frac{\Se}{16\pi^2}|\mathcal{M}_{q\qbar}^{(0)}|^2\left(\frac{\mu_\uv^2}{\mu^2}\right)^{-\epsilon}\frac{1}{\epsilon}\;.
\end{equation}
The crucial observation here is that \Eq{\ref{Equation:FDUSelfEnergiesMasslessCTUVResult}} and \Eq{\ref{Equation:FDUVertexMasslessCTUVResult}} share the same divergent structure, allowing for a complete removal of the UV divergences.\\
\\
With all these ingredients, we are now able to define a full four-dimensional representation of the total cross section at NLO, with a local cancellation of the IR divergences of the loop and the real corrections. First, the UV renormalised virtual cross section is given by
\begin{equation}\label{Equation:FDUSigmaRenormalisedMassless}
\sigma_\V^{(1,\r)}=\frac{1}{\s}\int\,d\Phi_{1\to2}\ReText\langle\mathcal{M}_{q\qbar}^{(0)}|\mathcal{M}_{q\qbar}^{(1,\r)}\rangle
\end{equation}
with
\begin{equation}\label{Equation:FDURenormalisedOneLoop}
|\mathcal{M}_{q\qbar}^{(1,\r)}\rangle=|\mathcal{M}_{q\qbar}^{(1)}\rangle-|\mathcal{M}_{q\qbar}^{(1,\uv)}\rangle\;,
\end{equation}
where to lighten the expressions, we included the self-energy counterterms inside $|\mathcal{M}_{q\qbar}^{(1,\uv)}\rangle$, meaning that only the IR part of the self-energy contributions are present inside $|\mathcal{M}_{q\qbar}^{(1,\r)}\rangle$. The renormalised virtual cross section $\sigma_\V^{(1,\r)}$ therefore contains only IR singularities at the integrand level, but the UV counterterms involve a non-trivial integrand level cancellation of UV singularities that must be taken into account in order to find a proper four-dimensional representation of the total cross section. For this reason, we start by splitting $\sigma_\V^{(1,\r)}$ into
\begin{equation}\label{Equation:FDUSigmaRSplitMassless}
\sigma_\V^{(1,\r)}=\sigma_\V^{(1)}-\sigma_\V^{(1,\uv)}\;,
\end{equation}
where
\begin{align}
\sigma_\V^{(1)}=&~\frac{1}{\s}\int\,d\Phi_{1\to2}\ReText\langle\mathcal{M}_{q\qbar}^{(0)}|\mathcal{M}_{q\qbar}^{(1)}\rangle\;,\label{Equation:FDUSigmaVMasslessDef}\\
\sigma_\V^{(1,\uv)}=&~\frac{1}{\s}\int\,d\Phi_{1\to2}\ReText\langle\mathcal{M}_{q\qbar}^{(0)}|\mathcal{M}_{q\qbar}^{(1,\uv)}\rangle\;,\label{Equation:FDUSigmaVUVMasslessDef}
\end{align}
are the original virtual terms (also including self-energy contributions) and the UV counterterms, respectively. From \Eq{\ref{Equation:FDUSigmaVMasslessDef}}, we collect all the dual terms arising when either of the internal momenta $q_1$ or $q_2$ is set on shell, and restrict the loop integration by the dual mapping conditions defined in \Eq{\ref{Equation:FDURegion1}} and \Eq{\ref{Equation:FDURegion2}} respectively. As with the toy model in \Section{\ref{Section:FDUS3PF}}, we define the virtual dual contributions to the cross section as
\begin{align}
\sigmatilde_{\V,1}^{(1)}=&~-\sigma^{(0)}\,g_S^2\,C_F\int\,\dxi{1}\,\dv{1}\mathcal{R}_1(\xi_{1,0},v_1)\nn\\
&~\times\left[4\left(\frac{1}{\xi_{1,0}\,v_1}+1\right)\left(\frac{1}{1-v_1}-\xi_{1,0}\right)+2(1-\epsilon)\xi_{1,0}\left(\frac{1-v_1}{v_1}+\frac{v_1}{1-v_1}\right)\right]\;,\label{Equation:FDUSigmaTildeV1Massless}\\
\sigmatilde_{\V,2}^{(1)}=&~-\sigma^{(0)}\,g_S^2\,C_F\int\,\dxi{2}\,\dv{2}\mathcal{R}_2(\xi_{2,0},v_2)\nn\\
&~\times\left[\frac{4\xi_{2,0}}{1-\xi_{2,0}}\left(v_2\left(\frac{1}{1-v_2}-\xi_{2,0}\right)-\epsilon\right)+2(1-\epsilon)\frac{1-v_2\,\xi_{2,0}}{1-v_2}\right]\;,\label{Equation:FDUSigmaTildeV2Massless}
\end{align}
together with the virtual dual remnants
\begin{align}
\sigmabar_{\V,1}^{(1)}=&~-\sigma^{(0)}\,g_S^2\,C_F\int\,\dxi{1}\,\dv{1}\big(1-\mathcal{R}_1(\xi_{1,0},v_1)\big)\nn\\
&~\times\left[4\left(\frac{1}{\xi_{1,0}\,v_1}+1\right)\left(\frac{1}{1-v_1}-\xi_{1,0}\right)+2(1-\epsilon)\xi_{1,0}\left(\frac{1-v_1}{v_1}+\frac{v_1}{1-v_1}\right)\right]\;,\\
\sigmabar_{\V,2}^{(1)}=&~-\sigma^{(0)}\,g_S^2\,C_F\int\,\dxi{2}\,\dv{2}(1-\mathcal{R}_2(\xi_{2,0},v_2))\nn\\
&~\times\left[4\xi_{2,0}\left(\frac{1}{1-\xi_{2,0}+i0}+i\,\pi\,\delta(1-\xi_{2,0})\right)\left(v_2\left(\frac{1}{1-v_2}-\xi_{2,0}\right)-\epsilon\right)+2(1-\epsilon)\frac{1-v_2\,\xi_{2,0}}{1-v_2}\right]\;,\\
\sigma_{\V,3}^{(1)}=&~-\sigma^{(0)}\,g_S^2\,C_F\int\,\dxi{3}\,\dv{3}\nn\\
&~\times\left[\frac{4\xi_{3,0}}{1-\xi_{3,0}}\left((1-v_3)\left(\frac{1}{v_3}+\xi_{3,0}\right)-\epsilon\right)-2(1-\epsilon)\frac{1+(1-v_3)\xi_{3,0}}{v_3}\right]\;,
\end{align}
which fulfil
\begin{equation}\label{Equation:FDUSigmaVEquality}
\sigma_\V^{(1)}=\,\sum\limits_{i=1,2}\,\left(\sigmatilde_{\V,i}^{(1)}+\sigmabar_{\V,i}^{(1)}\right)+\sigma_{\V,3}^{(1)}\;.
\end{equation}
Notice that in \Eq{\ref{Equation:FDUSigmaTildeV2Massless}}, the $+i0$ prescription has been removed. This is due to the fact that in the considered integration phase space -- delimited by $\mathcal{R}_2$ --, the denominator $(1-\xi_{2,0})^{-1}$ cannot vanish, making the use of a prescription unnecessary.\\
\\
Finally, the UV counterterm is given by
\begin{align}\label{Equation:FDUSigmaVUVMasslessFullDual}
\sigma_\V^{(1,\uv)}=&~\sigma^{(0)}\,g_S^2\,C_F\int\,d[\xi_\uv]d[v_\uv]\left[\frac{(1-\epsilon)(1-2v_\uv)\xi_\uv^3(12-7m_\uv^2+4\xi_\uv^2)}{(\xi_\uv^2+m_\uv^2)^{5/2}}\right.\nn\\
&~+\left.\frac{2\xi_\uv^2\big((1+2\epsilon)m_\uv^2+4\xi_\uv^2(1-\epsilon-3(2-\epsilon)v_\uv(1-v_\uv))\big)}{(\xi_\uv^2+m_\uv^2)^{5/2}}\right]\;.
\end{align}
The next step is to separate the real three-body phase space into two regions to isolate the different collinear configurations. We introduce
\begin{equation}\label{Equation:FDUSigmaTildeRiMassless}
\sigmatilde_{\r,i}^{(1)}=\frac{1}{\s}\int\,d\Phi_{1\to3}|\mathcal{M}_{q\qbar g}|^2\theta(\yp{jr}-\yp{ir})\;,\quad i,j\in\{1,2\}\;,\quad i\neq j\;,
\end{equation}
that share the same form as \Eq{\ref{Equation:FDUSigmaTildeR}}, and fulfil $\sum_i\sigmatilde_{\r,i}^{(1)}=\sigma_\r^{(1)}$. Explicitly,
\begin{align}
\sigmatilde_{\r,1}^{(1)}=&~\sigma^{(0)}g_S^2\,C_f\int\,\dxi{1}\,\dv{1}\mathcal{R}_1(\xi_{1,0},v_1)\frac{2(1-2\xi_{1,0})^{-2\epsilon}}{(1-(1-v_1)\xi_{1,0})^{-2\epsilon}}\nn\\
&~\times\frac{1}{v_1(1-v_1)}\left(\frac{(1-\xi_{1,0})^2+\big((1-\xi_{1,0})^2(1-v_1)+v_1\big)^2}{\xi_{1,0}(1-(1-v_1)\xi_{1,0})^2}-\epsilon\,\xi_{1,0}\right)\;,\label{SigmaTildeR1Massless}\\
\sigmatilde_{\r,2}^{(1)}=&~\sigma^{(0)}g_S^2\,C_f\int\,\dxi{2}\,\dv{2}\mathcal{R}_2(\xi_{2,0},v_2)\frac{2(1-2\xi_{2,0})^{-2\epsilon}}{(1-v_2\,\xi_{2,0})^{1-2\epsilon}}\nn\\
&~\times\frac{1}{1-v_2}\left(\frac{(1-\xi_{2,0})^2-\epsilon\big((1-\xi_{2,0})^2\,v_2+1-v_2\big)^2}{(1-v_2\xi_{2,0})^2}+\xi_{2,0}^2\right)\;,\label{SigmaTildeR2Massless}
\end{align}
which are obtained after applying to \Eq{\ref{Equation:FDUSigmaRMasslessNLO}} the mappings defined in \Eqs{\ref{Equation:FDUMappingR1}}{\ref{Equation:FDUMappingR2}}, respectively. Since the total cross section at NLO is given by
\begin{equation}\label{Equation:FDUSigmaTotalMassless}
\sigma^{(1)}=\sigma_\V^{(1,\r)}+\sigma_\r^{(1)}\;,
\end{equation}
we put together $\sigmatilde_{\V,1}^{(1)}$ and $\sigmatilde_{\r,1}^{(1)}$ to define
\begin{equation}\label{Equation:FDUSigmaTildeiMassless}
\sigmatilde_i^{(1)}=\sigmatilde_{\V,i}^{(1)}+\sigmatilde_{\r,i}^{(1)}\;,\quad i\in\{1,2\}\;,
\end{equation}
that are finite in the limit $\epsilon\to0$. Actually, through the application of the momentum mappings and the separation of the integration region, a local cancellation of singularities takes place, allowing us to take the limit $\epsilon\to0$ at the \emph{integrand level}, i.e. before integration.\\
\\
According to \Eq{\ref{Equation:FDUSigmaTotalMassless}}, we can observe that some contributions are still missing. In fact, we must combine all the virtual terms and UV counterterms that have not yet been included in the virtual dual cross sections of \Eqs{\ref{Equation:FDUSigmaTildeV1Massless}}{\ref{Equation:FDUSigmaTildeV2Massless}}. We therefore define
\begin{equation}\label{Equation:FDUSigmaBarVMasslessNLO}
\sigmabar_\V^{(1)}=\left(\sum\limits_{i=1,2}\sigmabar_{\V,i}^{(1)}\right)+\sigma_{\V,3}^{(1)}-\sigma_\V^{(1,\uv)}\;,
\end{equation}
as the remnant virtual correction. It is worth appreciating that this expression admits a four-dimensional representation, which can be built applying the change of variables shown in \Section{\ref{Section:FDUUCS}}.\\
\\
Finally, the calculation of the integrals gives
\begin{align}\label{Equation:FDUIntegratedCrossSections}
\sigmatilde_1^{(1)}=&~\sigma^{(0)}\frac{\alpha_S}{4\pi}\,C_F\big(19-32\log(2)\big)\nn\;,\\
\sigmatilde_2^{(1)}=&~\sigma^{(0)}\frac{\alpha_S}{4\pi}\,C_F\big(-\frac{11}{2}+8\log(2)-\frac{\pi^2}{3}\big)\nn\;,\\
\sigmabar_\V^{(1)}=&~\sigma^{(0)}\frac{\alpha_S}{4\pi}\,C_F\big(-\frac{21}{2}+24\log(2)+\frac{\pi^2}{3}\big)\;,
\end{align}
whose sum is
\begin{equation}\label{Equation:FDUSumSigmasMasslessNLO}
\sigmatilde_1^{(1)}+\sigmatilde_2^{(1)}+\sigmabar_\V^{(1)}=\sigma^{(0)}3\frac{\alpha_S}{4\pi}\,C_F\;,
\end{equation}
that agrees with the $\epsilon\to0$ limit of the result obtained through DREG. The crucial difference is that we get the result after integrating expressions evaluated in four dimensions, thus avoiding potentially complicated $\epsilon$ expansions. The complete four-dimensional representations of the quantities appearing in \Eq{\ref{Equation:FDUIntegratedCrossSections}} can be found in \Appendix{\ref{Appendix:FourDimensionalRepresentationsMassless}}.\\
\\
To conclude this section, note that the procedure shown here can be significantly simplified. In order to obtain an analytic result, it was necessary to split the integration region and the virtual corrections into several different pieces. In a numerical implementation, this procedure is not needed. The loop integration will occur unrestricted, and the real corrections will be switched on to cancel the collinear and soft divergences in the region of the loop three-momentum where the respective momentum mapping conditions are fulfilled.

\section{Generalisation to multi-leg processes and NNLO}\label{Section:FDUGeneralisation}
\fancyhead[LO]{\ref*{Section:FDUGeneralisation}~~\nameref*{Section:FDUGeneralisation}}

\begin{figure}[t]
	\centering
	\begin{picture}(184,150)(0,-75)
		\SetWidth{1}
		\GOval(32,0)(8,8)(0){0.6}
		\ArrowLine(25,4)(0,16)
		\ArrowLine(25,-4)(0,-16)
		\Arc[arrow](56,0)(24,110,161)
		\Arc[arrow](56,0)(24,199,250)
		\Arc[arrow](56,0)(24,0,70)
		\Arc[arrow](56,0)(24,-70,0)
		\ArrowLine(80,0)(104,0)
		\DashLine(112,50)(112,-50){5}
		\ArrowLine(120,0)(144,0)
		\GOval(152,0)(9,9)(0){0.6}
		\ArrowLine(184,16)(159,4)
		\ArrowLine(184,-16)(159,-4)
		\Vertex(6,7){1.25}
		\Vertex(5,0){1.25}
		\Vertex(6,-7){1.25}
		\Vertex(178,7){1.25}
		\Vertex(179,0){1.25}
		\Vertex(178,-7){1.25}
		\color{blue}
		\Text(56,32){$\widetilde{q}_{i-1}$}
		\Text(56,-32){$\deltatilde{q_i}$}
		\Text(132,8){$p_i$}
	\end{picture}
	\hfill
	\begin{picture}(184,150)(0,-75)
		\SetWidth{1}
		\GOval(32,0)(8,8)(0){0.6}
		\ArrowLine(25,4)(0,16)
		\ArrowLine(25,-4)(0,-16)
		\Arc[arrow](56,0)(24,110,161)
		\Arc[arrow](56,0)(24,199,250)
		\DashLine(56,50)(56,-50){5}
		\Arc[arrow](56,0)(24,0,70)
		\Arc[arrow](56,0)(24,-70,0)
		\ArrowLine(80,0)(104,0)
		\ArrowLine(120,0)(144,0)
		\GOval(152,0)(9,9)(0){0.6}
		\ArrowLine(184,16)(159,4)
		\ArrowLine(184,-16)(159,-4)
		\Vertex(6,7){1.25}
		\Vertex(5,0){1.25}
		\Vertex(6,-7){1.25}
		\Vertex(178,7){1.25}
		\Vertex(179,0){1.25}
		\Vertex(178,-7){1.25}
		\color{blue}
		\Text(33,24){$p'_i$}
		\Text(33,-24){$p'_r$}
		\Text(112,8){$\widetilde{p}'_{ir}$}
	\end{picture}
	\caption{Factorisation of the dual one-loop and tree-level squared amplitudes in the collinear limit. The dashed line represents the momentum conservation cut. Interference of the Born process with the one-loop scattering amplitude with internal momentum $q_i$ on-shell, $\mathcal{M}_{N}^{(1)}(\deltatilde{q_i})\otimes\left(\mathcal{M}_N^{(0)}\right)^\dagger$ (left), and interference of real processes with the parton splitting $p_{ir}'\to p_i'+p_r'$, $\mathcal{M}_{N+1}^{(0)}\otimes \left(\mathcal{M}_{N+1,ir}^{(0)}\right)^\dagger$ (right). In this limit the momenta $q_{i-1}=q_i-p_i$ and $p_{ir}'$ become on-shell and the scattering amplitudes factorise.}
	\label{Figure:FDUFactorisation}
\end{figure}
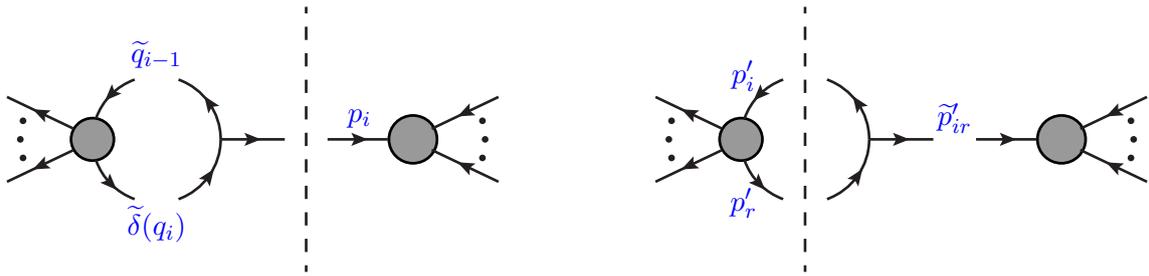

Assuming there are no initial-state partons (for instance in $e^+e^-$ annihilation), the generalisation to multi-leg processes is straightforward, starting from the results presented in the previous sections. As usual, the NLO cross section is constructed from the sum of the one-loop virtual correction with $m$ partons in the final state and the exclusive real cross section with $m+1$ partons in the final state,
\begin{equation}\label{Equation:FDUSigmamNLO}
\sigma^\textrm{NLO}=\int_m\,d\sigma_\V^{(1,\r)}+\int_{m+1}\,d\sigma_\r^{(1)}\;,
\end{equation}
where the virtual contribution is obtained from its dual representation
\begin{equation}\label{Equation:FDUSigmamV}
d\sigma_\V^{(1,\r)}=\,\sum_{i=1}^N\,\int_\ell\,2\ReText\langle\,\mathcal{M}_N^{(0)}|\mathcal{M}_N^{(1,\r)}\big(\deltatilde{q_i}\big)\rangle\mathcal{O}_N(\{p_j\})\;.
\end{equation}
In \Eq{\ref{Equation:FDUSigmamV}}, the term $\mathcal{O}_N(\{p_j\})$ defines a given IR-safe physical observable (e.g. a jet function) by accordingly constraining the integration domain, $\mathcal{M}_N^{(0)}$ denotes the $N$-leg scattering amplitude at LO, with $N>m$, and $\mathcal{M}_N^{(1,\r)}$ is the renormalised one-loop scattering amplitude, which also contains the self-energy corrections of the external legs. By \emph{renormalised}, we mean that appropriate UV counterterms have been subtracted locally, according to the discussion presented in \Section{\ref{Section:FDUUV}}, including UV singularities of degree higher than logarithmic that integrate to zero, and the UV contributions from the wave function of the external particles, in such a way that only IR singularities arise in $d\sigma_\V^{(1,\r)}$. We have also assumed a definite ordering of the external particles that leads a definite set of internal momenta $q_i$. Therefore, the one-loop scattering amplitude $\mathcal{M}_N^{(1,\r)}$ contains not only the contribution from the maximal one-loop $N$-point function, but also all the terms that can be constructed with the same set of internal momenta. Keeping this ordering is necessary to preserve the partial cancellation of singularities among dual contributions at the integrand level. Obviously, all the possible permutations and symmetry factors have to be considered to obtain the physical cross sections.\\
\\
The real cross section is given by
\begin{equation}\label{Equation:FDUSigmamR}
\int_{m+1}\,d\sigma_\r^{(1)}=\,\sum_{i=1}^N\,\int_{m+1}\,|\mathcal{M}_{N+1}^{(0)}(q_i,p_i)|^2\,\mathcal{R}_i(q_i,p_i)\,\mathcal{O}_{N+1}(\{p_j'\})\;,
\end{equation}
where the external momenta $\{p_j'\}$, the phase space and the tree-level scattering amplitude $\mathcal{M}_{N+1}^{(0)}$ have been rewritten in terms of the loop three-momentum (equivalently, the internal loop on-shell momenta) and the external momenta $p_i$ of the tree-level process. The momentum mapping links the soft and collinear states of the real and virtual corrections. Therefore, if the physical observable is IR-safe, then $\mathcal{O}_{N+1}$ reduces to $\mathcal{O}_{N}$ in all the possible IR-degenerate configurations of the $(N+1)$-particle process. In this way, it is guaranteed that the simultaneous implementation of the real-emission terms with the corresponding dual contributions leads to an integrand-level cancellation of IR singularities.\\
\\
Analogously to the dipole method \cite{Catani:1996jh,Catani:1996vz}, in order to construct the momentum mapping between the $m$ and $m+1$ kinematics, we single out two partons for each contribution. The first parton is the \emph{emitter} and the second parton is the \emph{spectator}. The difference with respect to the dipole formalism is that both the emitter and the spectator are initially related to external momenta of the virtual scattering amplitudes, and not to internal or external momenta of the real emission processes. Then, the loop three-momentum and the four-momenta of the emitter and the spectator are used to reconstruct the kinematics of the corresponding real emission cross section in the region of the real phase space where the twin of the emitter decays into two partons in a soft or collinear configuration. Explicitly, if the momentum of the final-state emitter is $p_j$, the internal momentum prior\footnote{We assume that the internal and external momenta are ordered according to \Fig{\ref{Figure:LTDLabelOneLoop}}} to the emitter is $q_i$ and is on shell, and $p_j$ is the momentum of the final-state spectator, then, the multi-leg momentum mapping is given by
\begin{align}\label{Equation:FDUMappingGeneralisation}
p_r'^\mu=&~q_i^\mu\;,\quad&p_i'^\mu=&~p_i^\mu-q_i^\mu+(1-\gamma_i)p_j^\mu\;,\nn\\
p_j'^\mu=&~\gamma_i\,p_j^\mu\;,\quad&\gamma_i=&~1-\frac{(q_i-p_i)^2}{2(q_i-p_i)\cdot p_j}\;,\nn\\
p_k'^\mu=&~p_k^\mu\;,\quad&k\neq&~i,j,r\;,
\end{align}
where the incoming initial-state momenta, $p_a$ and $p_b$, are left unchanged. In \Eq{\ref{Equation:FDUMappingGeneralisation}}, momentum conservation is fulfilled by construction, since
\begin{equation}
p_i+p_j+\,\sum_{k\neq i,j}\,p_k=p_i'+p_r'+p_j'+\,\sum_{k\neq i,j}\,p_k'\;.
\end{equation}
All the primed final-state momenta are massless and on shell if the virtual unprimed momenta are also massless. The momentum mapping in \Eq{\ref{Equation:FDUMappingGeneralisation}} is once again motivated by general factorisation properties in QCD~\cite{Collins:1989gx,Catani:2011st}, and is graphically explained in \Fig{\ref{Figure:FDUFactorisation}}.\\
\\
The emitter $p_i$ has the same flavour as $p_{ir}'$ (see \Fig{\ref{Figure:FDUFactorisation}} (right)), the twin emitter or parent (called emitter in the dipole formalism) of the real splitting configuration that is mapped. The spectator $p_j$ is used to balance momentum conservation, and has the same flavour in the virtual and real contributions. As for the dipoles, there are alternatives to treat the recoiling momentum, but the option with a single spectator is the most suitable. Note that the radiated particles $p_i'$ and $p_r'$ might have a different flavour than the emitter $p_i$. If the emitter is a quark or and antiquark, the role of $p_i'$ and $p_r'$ can be exchanged if the subindex $i$ is used to denote the flavour, as we did in \Eq{\ref{Equation:FDUMappingR2}}. If the emitter is a gluon, the radiated partons are two gluons, or a quark-antiquark pair.\\
\\
The momentum mapping in \Eq{\ref{Equation:FDUMappingGeneralisation}} is suitable for the region of the loop-momentum space where $q_i$ is soft or collinear with $p_i$, and therefore in the region of the real phase space where $p_i'$ and $p_r'$ are collinear or where either one of them is soft. In \Eq{\ref{Equation:FDUSigmamR}}, we have introduced a complete partition of the real phase space
\begin{equation}\label{Equation:FDUMultiLegPartition}
\sum\,\mathcal{R}_i(p_i,q_i)=\sum\,\,\prod_{jk\neq ir}\,\theta(\yp{jk}-\yp{ir})=1\;,
\end{equation}
which is equivalent to dividing the phase space by the minimal two-body invariant $\yp{ir}$. Since the real and virtual kinematics are related, the real phase-space partition defines equivalent regions in the loop three-momentum space. Notice, however, that we have not imposed these constraints in the definition of the virtual cross section in \Eq{\ref{Equation:FDUSigmamR}}, as we did for the analytic applications in \Sections{\ref{Section:FDUUnsubtraction}}{\ref{Section:FDUNLO}}. The actual implementation of the NLO cross section in a Monte Carlo event generator is a single unconstrained integral in the loop three-momentum, and the phase space with $m$ final-state particles. By virtue of the momentum mapping, real corrections are switched on in the region of the loop three-momentum where they map the corresponding soft and collinear divergences. This region is compact, and is of the size of the representative hard scale of the scattering process. At large loop three-momentum, only the virtual corrections contribute, and their UV singularities are subtracted locally by the use of suitable counterterms. These are the only counterterms that are required for the implementation of the method, the IR singularities being unsubtracted as their cancellation is achieved simultaneously. The full calculation is implemented in four dimensions, i.e. with the DREG parameter $\epsilon=0$. Moreover, there is no need to perform any tensor reduction in the calculation of the virtual contributions, hence avoiding the appearance of Gram determinants that usually lead to spurious numerical instabilities. Integrable threshold singularities of the loop contributions are also restricted to the physical compact region, and are treated numerically by contour deformation~\cite{Buchta:2015xda,Buchta:2015wna} in a Monte Carlo implementation.\\
\\
The case of lepton-hadron and hadron-hadron collisions deserves an aside. The cross section is computed by convoluting the corresponding partonic cross section with the process-independent parton distribution functions (PDF) of the incoming hadrons. Since the initial-state partons carry a well-defined momentum, the partonic subprocesses are not collinear safe. By virtue of the universal factorisation properties of QCD for massless incoming partons~\cite{Collins:1989gx}, the initial-state collinear singularities are factorised and reabsorbed into the definition of the non-perturbative PDF, and are removed from the partonic cross section by suitable collinear counterterms that are proportional to the product of $1/\epsilon$ and the Altarelli-Parisi splitting functions~\cite{Sborlini:2013jba,Altarelli:1977zs,deFlorian:2015ujt,Sborlini:2014kla,Sborlini:2014mpa,Catani:2003vu}. The initial-state collinear counterterms are then convoluted with the Born cross section. However, their standard form cannot be used in the LTD approach because the convolution is a single integral in longitudinal momentum with Born kinematics. Its unintegrated form -- which should also depend on the transverse momentum of the real radiation -- is necessary, and will be presented in a future publication.\\
\\
The extension to NNLO, although not straightforward, can instinctively be anticipated and will also be the subject of a future publication. The NNLO cross section consists of three contributions
\begin{equation}\label{Equation:FDUNNLOCrossSection}
\sigma^\textrm{NNLO}=\int_m\,d\sigma_{\V\V}^{(2)}+\int_{m+1}\,d\sigma_{\V\r}^{(2)}+\int_{m+2}\,d\sigma_{\r\r}^{(2)}\;.
\end{equation}
The double virtual cross section $d\sigma_{\V\V}^{(2)}$ receives contributions from the interference of the two-loop with the Born scattering amplitudes, and the square of the one-loop scattering amplitude with $m$ final-state particles. The virtual-real cross section $d\sigma_{\V\r}^{(2)}$ includes the contributions from the interference of the one-loop and tree-level scattering amplitudes with one extra final-state particle. Finally, the 
double real cross section $d\sigma_{\r\r}^{(2)}$ is the sum of all tree-level contributions where two extra particles are emitted. The LTD representation of the two-loop scattering amplitude is obtained by setting two internal lines on shell~\cite{Bierenbaum:2010cy}. It leads to the two-loop dual components $\langle\mathcal{M}_N^{(0)}|\mathcal{M}_N^{(2)}\big(\deltatilde{q_i,q_j}\big)\rangle$, while the two-loop momenta of the squared one-loop amplitudes are independent and generate dual contributions of the type $\langle\mathcal{M}_N^{(1)}\big(\deltatilde{q_i}\big)|\mathcal{M}_N^{(1)}\big(\deltatilde{q_i}\big)\rangle$. In both cases, we have at our disposal two independent loop three-momenta and $m$ final-state momenta, from which we can reconstruct the kinematics of the one-loop corrections inside $d\sigma_{\V\r}^{(2)}$, and the tree-level corrections in $d\sigma_{\r\r}^{(2)}$.

\section{Conclusion}\label{Section:FDUConclusion}
\fancyhead[LO]{\ref*{Section:FDUConclusion}~~\nameref*{Section:FDUConclusion}}

In this chapter, we have carefully discussed the implementation of a novel algorithm to compute higher-order corrections to physical observables. This method is based on the LTD theorem, which states that one-loop virtual contributions can be expressed as the sum over single-cuts, whose structure closely resembles real-emission amplitudes. We have exploited this knowledge to perform an integrand-level combination of real and virtual terms, which has lead to a fully local cancellation of singularities and allows for the implementation of the calculation without making use of DREG.\\
\\
One of the interesting uses of LTD lies in the possibility to explore the causal structure of virtual contributions and to disentangle their singularities. In particular, we have applied this technique to prove that the IR divergences are generated in a compact region of the loop-momentum space. This is a crucial fact in order to achieve the real-virtual cancellation of singularities in IR-safe observables.\\
\\
To illustrate the importance of the compactness of the IR singular regions, we have studied the NLO corrections to a $1\to2$ process in the context of a toy scalar model. By using suitable momentum mappings, we have generated $1\to3$ on-shell massless kinematics from the $1\to2$ process and the loop three-momentum. These mappings relate exactly the integration regions where the singularities are originated. In this example, we have distinguished two regions in the real-emission phase space and defined suitable mappings to cover them in the dual space of the loop three-momentum. In this way, the combination of dual and real contributions led to expressions that are integrable in four dimensions; the remainders of the virtual part were also represented by a four-dimensional integral. The algorithm is named ``unsubtraction'' because the summation over degenerate soft and collinear final states is performed thanks to these momentum mappings, thus making the introduction of IR subtractions unnecessary.\\
\\
On the other hand, we have investigated the cancellation of UV divergences at the integrand level. We have started with the simplest example of a massless scalar two-point function. Using the ideas presented in \cite{Becker:2010ng}, we have obtained the dual representation of the local UV counterterm that exactly cancels the divergences in the high-energy region of the loop momentum, achieving an integrable representation in four dimensions. We have also extended the procedure to deal with arbitrary scattering amplitudes, and provided a subtle physical interpretation of the energy scale appearing in the UV counterterms by identifying it to the renormalisation scale.\\
\\
Then, we have applied the FDU algorithm to the physical process $\gamma^*\to q\qbar(g)$ at NLO in QCD. We have written the dual representation of the virtual contribution and made use of the momentum mappings to perform the real-virtual combination. It is worth appreciating that we have had to take into account self-energy corrections to the external on-shell legs, even though they are usually ignored in the traditional approach given that their integrated form vanishes in DREG due to the lack of physical scale. Since our approach explicitly splits the IR and UV regions of the dual integrals, it was indeed necessary to disentangle their IR/UV behaviour. In this way, we have computed the full NLO corrections to $\gamma^*\to q\qbar(g)$ by making use of pure four-dimensional expressions.\\
\\
The generalisation of the algorithm to deal with multi-particle processes is quite straightforward, at least when only final-state singularities take place. Essentially, the real-emission phase space must be split to isolate the different collinear configurations, so a proper momentum mapping for each of these configurations can then be defined. Combining the dual integrands with the real matrix elements in the corresponding regions leads to integrable expressions in four dimensions. The cancellation of UV divergences is done following the same ideas as in the $1\to2$ case. Finally, we have succinctly sketched the extension of the FDU algorithm to remove initial-state collinear singularities in the multi-leg case, and briefly discussed its generalisation to NNLO.\\
\\
In summary, we have explicitly presented a well-defined algorithm that allows us to bypass DREG by exploiting the LTD formalism. This achievement constitutes a new paradigm in perturbative calculations as it allows for the direct combination of real and virtual corrections in an integrable four-dimensional representation, while providing a simple physical interpretation of the singularities of the scattering amplitudes and unveiling their hidden nature.

\chapter{The Four-Dimensional Unsubtraction with massive particles}\label{Chapter:FDUM}
\thispagestyle{fancychapter}
\fancyhead[RE]{\nameref*{Chapter:FDUM}}

So far, only toy models and processes involving massless particles were considered. Needless to say that in order to compute the vast majority of physical processes of any kind, the FDU formalism must be able to deal with massive particles as well. In this chapter, we extend the algorithm previously presented in \Chapter{\ref{Chapter:FDU}} so it can deal with processes involving massive particles \cite{Sborlini:2016hat}. We first illustrate the method with a scalar toy example, and then analyse the case of the decay of a scalar or vector boson into a pair of massive quarks. The results presented in this chapter are suitable for the application of the method to any multipartonic process.

\section{Introduction}\label{Section:FDUMIntro}
\fancyhead[LO]{\ref*{Section:FDUMIntro}~~\nameref*{Section:FDUMIntro}}

As said before, the standard approach to compute IR-safe observables at higher orders relies on the subtraction formalism~\cite{Kunszt:1992tn,Frixione:1995ms,Catani:1996jh,Catani:1996vz,GehrmannDeRidder:2005cm,Seth:2016hmv,Catani:2007vq,Catani:2009sm,Czakon:2010td,Bolzoni:2010bt,DelDuca:2015zqa,Boughezal:2015dva,Gaunt:2015pea}. The treatment of massive particles within this framework has been considered specifically in~\cite{Catani:2002hc,GehrmannDeRidder:2009fz,Abelof:2012he,Abelof:2014fza,Bonciani:2015sha}. Within FDU, and from the kinematical point of view, having massive particles slightly modifies the momentum mapping used when combining the real and the virtual contribution, and alters the IR-divergent structure. For instance, collinear configurations are not any more singular, strictly speaking. They are rather referred as \emph{quasi-collinear} configurations, where the $\epsilon$-pole is -- roughly -- replaced by a logarithm in the mass of the particle. These configurations will be mapped in such a way that logarithmic contributions are cancelled at the integrand level, allowing for a smooth massless limit. Another major difference is the treatment of the self-energy corrections, since they must now fulfil additional non-trivial constraints both in the IR and UV regions. Something similar also happens with the definition of local vertex renormalisation counterterms. In both cases, the dual representation involves dealing with multi-pole propagators. We restrict the discussion to the treatment of massive partons to the final state because the validity of the QCD factorisation theorem requires that the partons that initiate the hard-scattering subprocess in hadronic collisions have to be massless~\cite{Catani:2002hc}.\\
\\
The outline of this chapter is the following. In \Section{\ref{Section:FDUMS3PF}}, we apply the LTD theorem to the one-loop scalar three-point function with massive particles. In \Section{\ref{Section:FDUMMassiveScalarDecayRate}}, we introduce a scalar toy example and compute NLO corrections through the application of the conventional DREG approach. After describing the real emission phase-space partition and introducing a proper momentum mapping in \Section{\ref{Section:FDUMRealVirtual}}, we calculate the NLO corrections of the scalar toy model within the FDU formalism in \Section{\ref{Section:FDUMDecayRateFDU}}. We use LTD and the momentum mapping to perform the real-virtual combination at the integrand level, and build purely four-dimensional integrable expressions. In \Section{\ref{Section:FDUMRenormalisation}}, we present integrand-level representations of the wave function and mass renormalisation factors for heavy quarks in the on-shell scheme. Renormalisation of UV divergences is then achieved in \Section{\ref{Section:FDUMUV}}. We put into practice the FDU method by computing the NLO QCD corrections to the decay rate $A^*\to q\qbar(g)$ with massless quarks and $A=\phi,\gamma,Z$, and present our results in \Section{\ref{Section:FDUMNLO}}. Finally, in \Section{\ref{Section:FDUMConclusion}}, we conclude and discuss future implications of this work.

\section{Massive scalar three-point function within LTD}\label{Section:FDUMS3PF}
\fancyhead[LO]{\ref*{Section:FDUMS3PF}~~\nameref*{Section:FDUMS3PF}}

\begin{figure}[t]
	\begin{center}
		\includegraphics[width=0.4\textwidth]{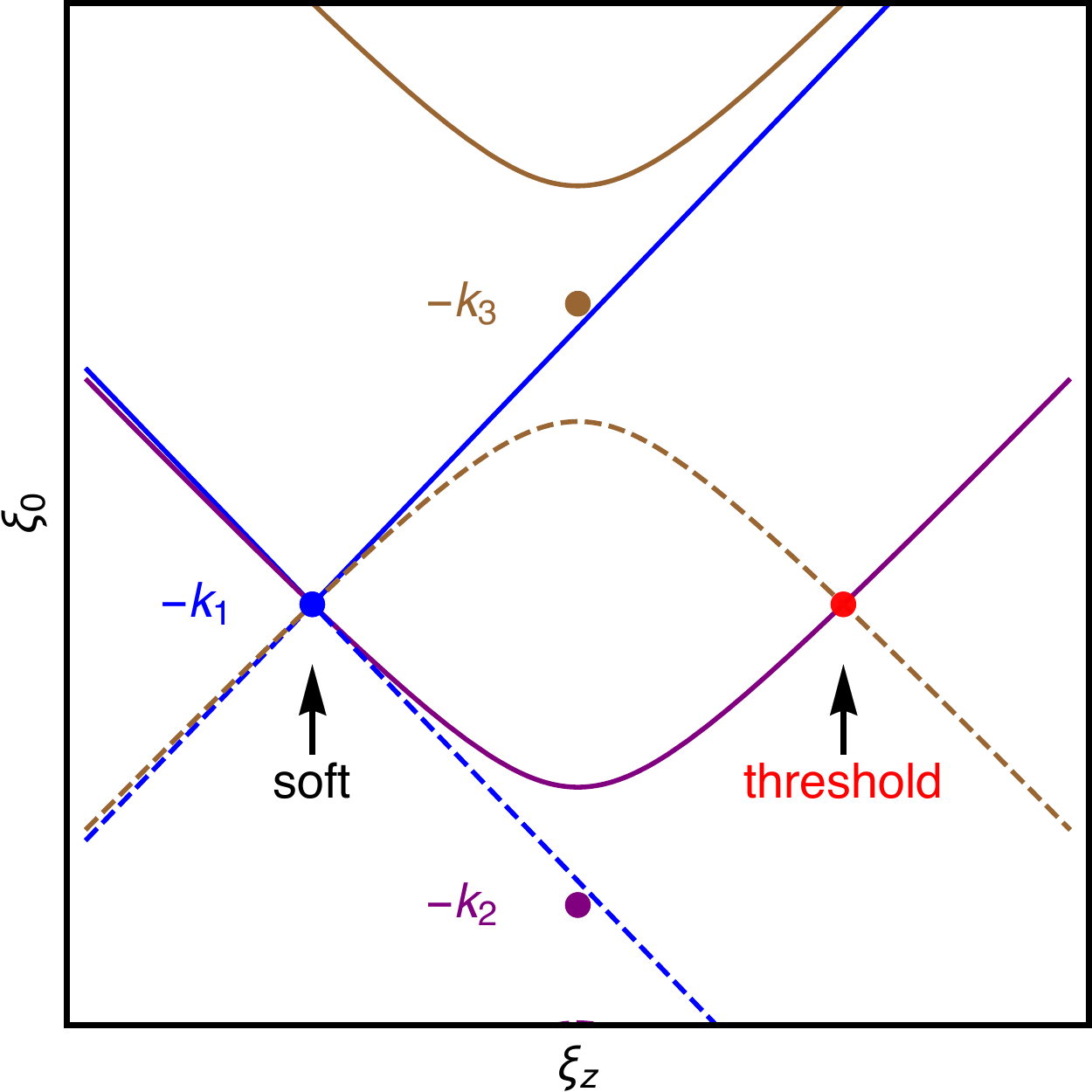} $\quad$
		\includegraphics[width=0.45\textwidth]{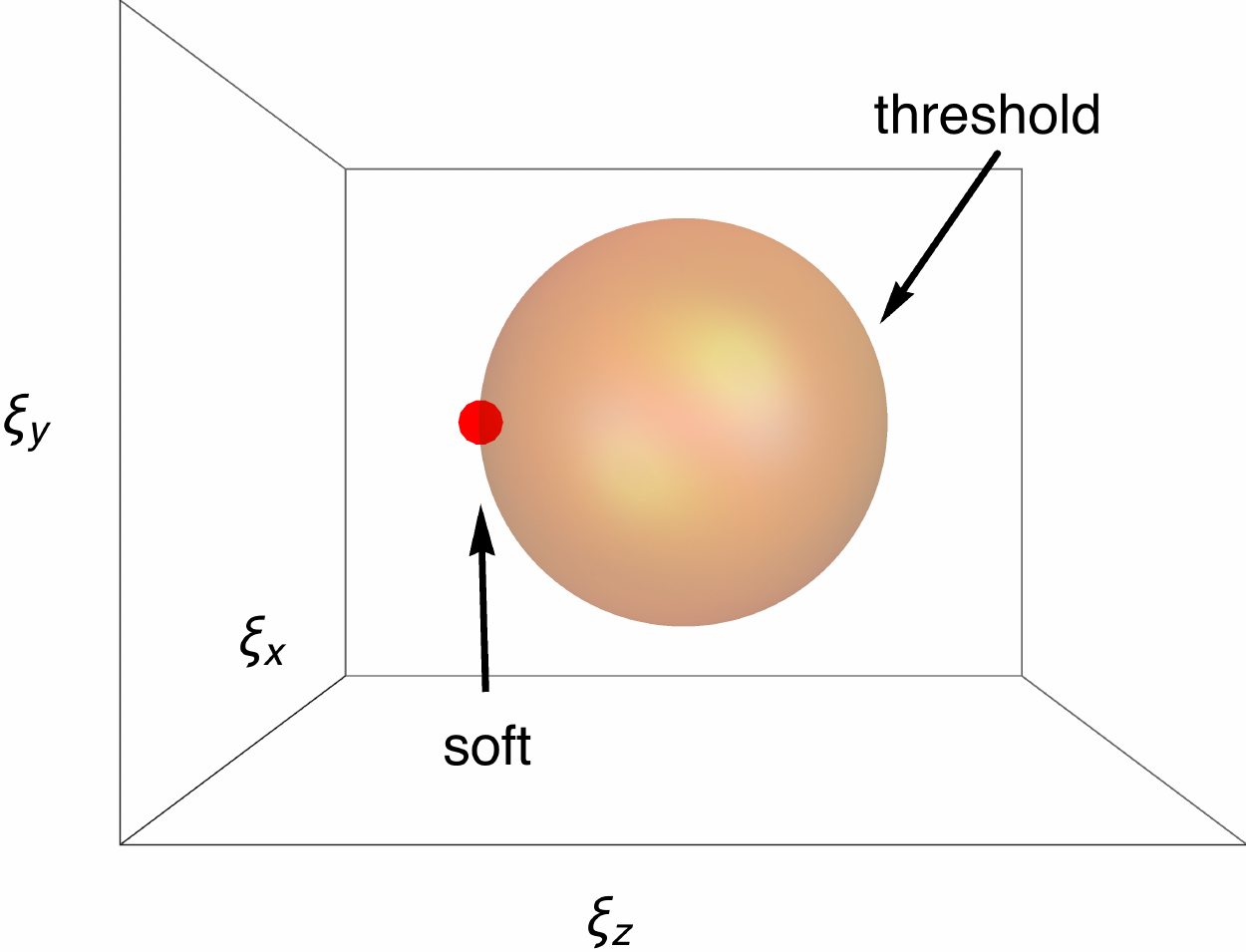}
		\caption{On-shell hyperboloids of the massive three-point function in the loop coordinates $\ell^{\mu}=\sqrt{\s}/2\,(\xi_0,\xi_x,\xi_y,\xi_z)$ in two dimensions (left plot); forward and backward on-shell hyperboloids are represented by solid and dashed lines, respectively. The intersections of on-shell hyperboloids lead to soft and threshold singularities in the loop three-momentum space (right plot), collinear singularities are regulated by the mass.}
		\label{Figure:FDUMRegions}
	\end{center}
\end{figure}

Before attempting to compute an actual physical process involving massive particles, let's start by considering the massive scalar three-point function, so we can familiarise ourselves with the additional notations and tools we need to introduce when applying FDU to massive particles. As for the massless case, the internal lines are written $q_1=\ell+p_1$, $q_2=\ell+p_{12}$ and $q_3=\ell$, where $\ell$ is the loop momentum. We take one internal state, corresponding to $q_1$, to be massless, and the remaining internal and outgoing particles to be massive, of mass $M$. The final-state on-shell momenta are labelled as $p_1$ and $p_2$, with $p_1^2=p_2^2=M^2$, and the incoming one is labelled as $p_3$ -- where $p_3=p_1+p_2=p_{12}$ by momentum conservation -- with virtuality $p_3^2=\s>0$. We chose this configuration because it will later correspond to the QCD corrections to the $\gamma^*\to q\qbar$ physical process, that will be studied in \Section{\ref{Section:FDUMNLO}}. We also define
\begin{equation}\label{Equation:FDUMReducedMasses}
m=\frac{2M}{\s}\;,\qquad\beta=\sqrt{1-m^2}\;,
\end{equation}
as the reduced mass and velocity, respectively. The well-known result of the massive scalar three-point function is given by~\cite{Ellis:2007qk,Rodrigo:1999qg} and reads
\begin{align}\label{Equation:FDUMS3PFLiterature}
L^{(1)}_{m>0}(p_1,p_2,-p_3)=&~\int_\ell\,\,\prod\limits_{i=1}^3\,G_F(q_i)\nn\\
=&~-\frac{\cGamma}{\s\,\beta}\left[\log(X_S)\left(-\frac{1}{\epsilon}-\frac{\log(X_S)}{2}+2\log(1-X_S^2)\right.\right.\nn\\
&~+\left.\left.\log\left(\frac{m^2}{4}\right)\right)+\Li_2(X_S^2)+2\Li_2(1-X_S)-\frac{\pi^2}{6}\right]+\Oep{1}\;,
\end{align}
where
\begin{equation}\label{Equation:FDUMXSDef}
X_S=-x_S-i0\,\textrm{Sgn}(\s)\;,\qquad x_S=\frac{1-\beta}{1+\beta}
\end{equation}
and where we recall
\begin{equation}\label{Equation:FDUMcGamma}
\cGamma=\frac{\Gamma(1+\epsilon)\Gamma^2(1-\epsilon)}{(4\pi)^{2-\epsilon}\Gamma(1-2\epsilon)}\;.
\end{equation}
Applying LTD, we can write
\begin{equation}\label{Equation:FDUMS3PFMassiveLTD}
L^{(1)}_{m>0}(p_1,p_2,-p_3)=\,\sum\limits_{i=1}^3\,I_i\;,
\end{equation}
with the dual contributions
\begin{align}
I_1=&~-\int_\ell\,\frac{\deltatilde{q_1}}{(2q_1\cdot p_2-i0)(-2q_1\cdot p_1+i0)}\;,\label{Equation:FDUMS3PFMassiveDual1}\\
I_2=&~-\int_\ell\,\frac{\deltatilde{q_2}}{(2M^2-2q_2\cdot p_2+i0)(\s-2q_2\cdot p_{12}+i0)}\;,\label{Equation:FDUMS3PFMassiveDual2}\\
I_3=&~-\int_\ell\,\frac{\deltatilde{q_3}}{(2M^2+2q_3\cdot p_1-i0)(\s+2q_3\cdot p_{12}-i0)}\;.\label{Equation:FDUMS3PFMassiveDual3}
\end{align}
The corresponding on-shell hyperboloids are shown in \Fig{\ref{Figure:FDUMRegions}} (left). Due to the rotational symmetry of the problem, only showing the $(\ell_0,\ell_z)$ plane is enough. As discussed previously in \Chapter{\ref{Chapter:LTD}}, the intersection of on-shell hyperboloids is associated with multiple internal propagators vanishing simultaneously. In this case, the forward on-shell hyperboloid of $G_F(q_1)$ and $G_F(q_2)$, and the backward one of $G_F(q_3)$ intersect at a single point, where a soft singularity develops. Another intersection takes place between the forward on-shell hyperboloids of $G_F(q_2)$ and the backward one of $G_F(q_3)$, which in this case corresponds to a threshold singularity that will appear inside $I_2$. Notice that, as explained above, there are no collinear singularities, because of the mass preventing the on-shell hyperboloids to degenerate into light-cones as it was the case for the process studied in \Chapter{\ref{Chapter:FDU}}.\\
\\
As we did for the massless case, and in order to simplify the analytic calculation of the dual integrals in \Eqs{\ref{Equation:FDUMS3PFMassiveDual1}}{\ref{Equation:FDUMS3PFMassiveDual3}}, we work in the centre-of-mass frame of $p_1$ and $p_2$. With this assumption, the external momenta can be written as
\begin{equation}\label{Equation:FDUMMomentaParamMassive}
p_1^\mu=\frac{\sqrt{\s}}{2}(1,\mathbf{0},\beta)\;,\qquad p_2^\mu=\frac{\sqrt{\s}}{2}(1,\mathbf{0},-\beta)\;.
\end{equation}
When parametrising the internal momenta, we must take into account that, while $q_1$ corresponds to a massless state, $q_2$ and $q_3$ are associated with massive particles of mass $M$. This means we need to slightly modify the parametrisations for the latter and write
\begin{align}
q_1^{\mu}=&~\frac{\sqrt{\s}}{2}\,\xi_{1,0}\,\left(1,2\sqrt{v_1(1-v_1)}\,\eiPerp{1},1-2v_1\right)\;,\nn\\
q_i^{\mu}=&~\frac{\sqrt{\s}}{2}\,\left(\xi_{i,0},2\xi_i\,\sqrt{v_i(1-v_i)}\,\eiPerp{i},\xi_{i}(1-2v_i)\right)\;, \quad\xi_{i,0}=\sqrt{\xi_i^2+m^2}\;,\quad i\in\{2,3\}\;,
\end{align}
where $\xi_{1,0},\xi_2,\xi_3\in[0,\infty)$ and $v_i\in[0,1]$. Keep in mind that in the previous chapter, we were using $(\xi_{i,0},v_i)$ as the integration variables for all internal lines, because considering the energy or the modulus three-momentum of an on-shell massless particle is equivalent. However, within the LTD formalism, the integration is performed over the three-momentum $\xi_i$, which is different than integrating over the energy $\xi_{i,0}$ in the case of the second and the third contribution\footnote{We \emph{could} integrate over $\xi_{i,0}$ even for massive particles and rewrite the three-momentum and the measure accordingly, but this would introduce an unnecessary additional layer of complexity.}.\\
\\
With these variables, the loop integration measure is transformed into
\begin{equation}\label{IntMeasureMassive}
\int_\ell\,\deltatilde{q_i}=\s\int_0^\infty\frac{\xi_i^2}{\xi_{i,0}}d[\xi_i]\int_0^1\,\dv{i}\;,
\end{equation}
and the scalar products of internal momenta with external ones are given by
\begin{align}
4\frac{q_i\cdot p_1}{\s}=\xi_{i,0}-\beta_i\,\xi_i(1-2v_i)\;,\nn\\
4\frac{q_i\cdot p_2}{\s}=\xi_{i,0}+\beta_i\,\xi_i(1-2v_i)\;,\nn\\
\end{align}
where $\beta_1=1$ and $\beta_2=\beta_3=\beta$. With this we can rewrite the dual contributions in \Eqs{\ref{Equation:FDUMS3PFMassiveDual1}}{\ref{Equation:FDUMS3PFMassiveDual3}} as
\begin{align}\label{Equation:FDUMDualOtherParam}
I_1=&~\frac{4}{\s}\int\,\frac{\dxi{1}\,\dv{1}}{\xi_{1,0}(1-(1-2 v_1)^2)\beta^2}\;,\nn\\ 
I_2=&~\frac{2}{\s}\int\,\frac{\xi_2^2\,d[\xi_2]\,\dv{2}}{\xi_{2,0}(1-\xi_{2,0}+i0)(\xi_{2,0}+\beta\,\xi_2(1-2v_2)-m^2)}\;,\nn\\ 
I_3=&~-\frac{2}{\s}\int\,\frac{\xi_3^2\,d[\xi_3]\,\dv{3}}{\xi_{3,0}(1+\xi_{3,0})(\xi_{3,0}-\beta\,\xi_3(1-2 v_3)+m^2)}\;.
\end{align}
The threshold singularity in $I_2$ is still present, and occurs this time for $\xi_{2,0}=1$, i.e. $\xi_2=\beta\leq1$. Notice that by taking the massless limit --~defined by $m\to0$ and $\beta\to1$~-- in \Eq{\ref{Equation:FDUMDualOtherParam}}, we recover the dual integrands in \Eq{\ref{Equation:FDUS3PFDualsBis}}.\\
\\
Nevertheless, and as it could be expected, the massive case is a bit more cumbersome because of the presence of an additional scale. Again, $I_1$ vanishes since the energy integral factorises and lacks any characteristic scale. To be more precise, $I_1$ is singular in both the IR and the UV limit; however, the sum of the three dual integrals as well as the equivalent original Feynman integral contain only soft divergences. The other two dual integrals can be integrated in the angular variable analytically, which allows us to keep the exact $\epsilon$ dependence, leading to
\begin{align}\label{Equation:FDUMS3PFI2I3Bis}
I_2=&~\frac{2\Gamma^2(1-\epsilon)}{\s\,\Gamma(2-2\epsilon)}\int\,\dxi{2}\,\frac{\xi_2^2}{\xi_{2,0}}\frac{\,_2F_1\left(1,1-\epsilon;2-2\epsilon;\frac{2\beta\,\xi_2}{\xi_{2,0}+\beta\,\xi_2-m^2}\right)}{(1-\xi_{2,0}+i0)(\xi_{2,0}+\beta\,\xi_2-m^2)}\;,\nn\\
I_3=&~-\frac{2\Gamma^2(1-\epsilon)}{\s\,\Gamma(2-2\epsilon)}\int\,\dxi{3}\,\frac{\xi_3^2}{\xi_{3,0}}\frac{\,_2F_1\left(1,1-\epsilon;2-2\epsilon;-\frac{2\beta\,\xi_3}{\xi_{3,0}+\beta\,\xi_3-m^2}\right)}{(1+\xi_{3,0})(\xi_{3,0}-\beta\,\xi_3+m^2)}\;.
\end{align}
Because of the presence of the hypergeometric functions $\,_2F_1$, it is necessary to perform an $\epsilon$ expansion before attempting to integrate over the loop three-momentum. This will lead to a final result that includes corrections up to $\Oep{0}$. Besides that, the two integrals of \Eq{\ref{Equation:FDUMS3PFI2I3Bis}} are singular in the UV. For this reason, we introduce the following expansion
\begin{equation}\label{Equation:FDUMS3PFUVExpansion}
I_i=\int\,\frac{\dxi{i}}{\xi_{i,0}}g_i(\xi_i)=\int\,\frac{\dxi{i}}{\xi_{i,0}}\,g_\uv+\left[\int\,\frac{\dxi{i}}{\xi_{i,0}}\big(g_i(\xi_i)-g_\uv\big)\right]_{\epsilon=0}\;,
\end{equation}
where we define $g_\uv=\lim_{\xi_i\to\infty}\,g_i(\xi_1)$. The first term in the right-hand side of \Eq{\ref{Equation:FDUMS3PFUVExpansion}} gives the same result for both dual integrals, namely
\begin{equation}\label{S3PFUVTerm}
\int\,\frac{\dxi{i}}{\xi_{i,0}}\,g_\uv=-\frac{\cGamma}{\s}\frac{x_s^{-\epsilon}(1+x_s)^{1+2\epsilon}\,\Gamma(1-2\epsilon)}{2\epsilon(1-2\epsilon)\Gamma^2(1-\epsilon)}\,_2F_1(1,1-\epsilon;2-2\epsilon,1-x_s)\;,\quad i\in\{2,3\}\;.
\end{equation}
For the second term, which is regular in the UV\footnote{Indeed, by subtracting the UV limit $g_\uv$ from the integrand itself, we obtain an integrable quantity, thus validating the limit $\epsilon\to0$.}, we perform the change of variable
\begin{equation}\label{Equation:FDUMS3PFCOV}
\xi_i=\frac{m}{2}\left(z-\frac{1}{z}\right)\;,
\end{equation}
with $z\in[1,\infty)$. The total result for the real part of the dual integrals, up to $\Oep{1}$, reads
\begin{align}\label{Equation:FDUMS3PFDual23Re}
\ReText(I_2)=&~\frac{\cGamma}{2\beta\,\s}\left(\frac{\s}{\mu^2}\right)^{-\epsilon}\left[\log(x_s)\left(\frac{1}{\epsilon}-\frac{1}{2}\log(x_s)-2\log(\beta)\right)-2\Li_2(x_s)-\frac{5\pi^2}{3}\right]+\Oep{1}\;,\nn\\
I_3=&~\frac{\cGamma}{2\beta\,\s}\left(\frac{\s}{\mu^2}\right)^{-\epsilon}\left[\log(x_s)\left(\frac{1}{\epsilon}-\frac{1}{2}\log(x_s)-2\log(\beta)\right)-2\Li_2(x_s)+\frac{\pi^2}{3}\right]+\Oep{1}\;.
\end{align}
The dual contribution $I_3$ is purely real, while $I_2$ generates an imaginary component due the intersection of the forward on-shell hyperboloid of $G_F(q_2)$ with the backward on-shell hyperboloid of $G_F(q_3)$. Its imaginary part can be calculated to all orders in $\epsilon$ from
\begin{align}\label{Equation:FDUMS3PFDual2Im}
i\ImText(I_2)=&~\frac{1}{2}\int_\ell\,\deltatilde{q_2}\,\deltatilde{q_3}\,G_D(q_2;q_1)\nn\\
=&~-\frac{2i\pi}{\s}\int\,\frac{\xi_2^2\,\delta(1-\xi_{2,0})d[\xi_2]\,\dv{2}}{\xi_{2,0}(\xi_{2,0}+\beta\,\xi_2(1-2v_2)-m^2)}=i\frac{\cGammaTilde}{\beta\,\s}\left(\frac{\beta^2\,\s}{\mu^2}\right)^{-\epsilon}\frac{\sin(2\pi\epsilon)}{2\epsilon^2}\;,
\end{align}
which is the expected result that would be obtained through the application of Cutkosky's rules. One can verify that the sum of the contributions in \Eq{\ref{Equation:FDUMS3PFDual23Re}} and \Eq{\ref{Equation:FDUMS3PFDual2Im}} agrees with the literature result in \Eq{\ref{Equation:FDUMS3PFLiterature}}.

\section{Massive scalar decay rate in DREG}\label{Section:FDUMMassiveScalarDecayRate}
\fancyhead[LO]{\ref*{Section:FDUMMassiveScalarDecayRate}~~\nameref*{Section:FDUMMassiveScalarDecayRate}}

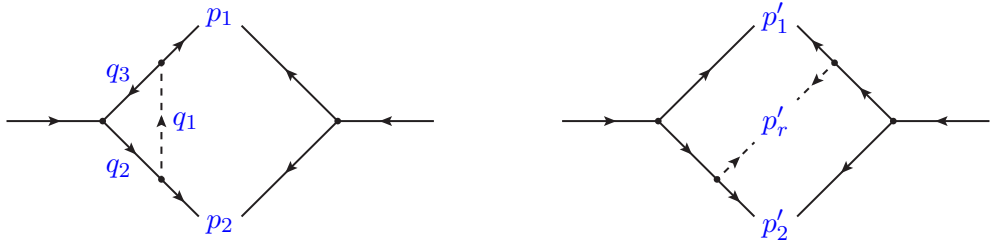
\begin{figure}
	\centering
	\begin{picture}(400,100)(-200,-50)
	\SetWidth{0.75}
	\ArrowLine[arrowscale=0.75](-184,0)(-148,0)
	\ArrowLine[arrowscale=0.75](-126,22)(-148,0)
	\ArrowLine[arrowscale=0.75](-126,22)(-112,36)
	\ArrowLine[arrowscale=0.75](-148,0)(-126,-22)
	\ArrowLine[arrowscale=0.75](-126,-22)(-112,-36)
	\DashLine[arrow,arrowscale=0.75](-126,-22)(-126,22){3}
	\Vertex(-148,0){1.25}
	\Vertex(-126,22){1.25}
	\Vertex(-126,-22){1.25}
	\ArrowLine[arrowscale=0.75](-60,0)(-96,36)
	\ArrowLine[arrowscale=0.75](-60,0)(-96,-36)
	\ArrowLine[arrowscale=0.75](-24,0)(-60,0)
	\Vertex(-60,0){1.25}
	\ArrowLine[arrowscale=0.75](24,0)(60,0)
	\ArrowLine[arrowscale=0.75](60,0)(96,36)
	\ArrowLine[arrowscale=0.75](60,0)(82,-22)
	\DashLine[arrow,arrowscale=0.75](82,-22)(96,-8){3}
	\ArrowLine[arrowscale=0.75](82,-22)(96,-36)
	\Vertex(60,0){1.25}
	\Vertex(82,-22){1.25}
	\ArrowLine[arrowscale=0.75](184,0)(148,0)
	\ArrowLine[arrowscale=0.75](148,0)(126,22)
	\ArrowLine[arrowscale=0.75](126,22)(112,36)
	\DashLine[arrow,arrowscale=0.75](126,22)(112,8){3}
	\ArrowLine[arrowscale=0.75](148,0)(112,-36)
	\Vertex(148,0){1.25}
	\Vertex(126,22){1.25}
	\color{blue}
	\Text(-104,39){$p_1$}
	\Text(-104,-39){$p_2$}
	\Text(-117,0){$q_1$}
	\Text(-142,18){$q_3$}
	\Text(-142,-18){$q_2$}
	\Text(104,39){$p'_1$}
	\Text(104,-39){$p'_2$}
	\Text(104,0){$p'_r$}
	\end{picture}
	\caption{Momentum configuration of the virtual and real contributions to the process $\phi\to\phi\phi$ at NLO. The one-loop contribution is proportional to the scalar three-point function, with a virtual massless particle $\psi$ (represented by a dashed line) inside the loop (left). The real contribution comes from the interference terms originated by the emission of an on-shell massless particle $\psi$ (right). In this case, the momentum configuration is given by $p_3\to p'_1+p'_2+p'_r$.}
	\label{Figure:FDUMNLOConfiguration}
\end{figure}

In order to establish a physical parallelism and understand the subtraction of IR singularities in the massive case, we first work with a simplified toy scalar model with a massive scalar particle $\phi$ that couples to a massless one, $\psi$. Concretely, we consider the decay process $\phi(p_3)\to\phi(p_1)+\phi(p_2)$, with $p_1^2=p_2^2=M^2$ and $p_3^2=p_{12}^2=\s$. The Born-level decay rate is given by
\begin{equation}\label{TMBornLevel}
\Gamma^{(0)}=\frac{g^2}{2\sqrt{\s}}\int\,d\Phi_{1\to2}\;,
\end{equation}
where $g$ is the $\phi\phi\psi$ coupling, and where $\s>4M^2$ to guarantee the physical feasibility of the process. As always, to compute the NLO correction, we need to consider both the virtual contribution and the real contribution. For the former, we will assume the presence of only one massless particle inside the loop; for the latter, we will assume the radiated particle to be massless as well. The relevant Feynman diagrams\footnote{This decay rate does not correspond to any actual physical theory, since the full set of Feynman diagrams has not been taken into account. However, for illustrative purposes, it is enough to restrict the following discussion to virtual and real contributions that have a similar topology~\cite{Hernandez-Pinto:2015ysa,Sborlini:2016gbr}.} are exhibited in \Fig{\ref{Figure:FDUMNLOConfiguration}}.\\
\\
Let's start with the virtual part, which we assume to be proportional to the previously computed massive scalar three-point function, i.e.
\begin{equation}\label{Equation:FDUMTMGammaV}
\Gamma_\V^{(1)}=\frac{1}{2\sqrt{\s}}\int\,d\Phi_{1\to2}\,\ReText\langle\mathcal{M}^{(0)}|\mathcal{M}^{(1)}\rangle=-\Gamma^{(0)}2g^2\,\s\,\ReText\,L_{m>0}^{(1)}(p_1,p_2,-p_3)\;.
\end{equation}
On one hand, since $\s>0$, the virtual decay rate is given by \Eq{\ref{Equation:FDUMS3PFLiterature}} as a function of $x_s$ and reads
\begin{equation}\label{Equation:FDUMTMVirtualResult}
\Gamma_\V^{(1)}=\Gamma^{(0)}\frac{4g^2\,\cGamma}{\beta}\left(\frac{\s}{\mu^2}\right)^{-\epsilon}\left[\log(x_S)\left(-\frac{1}{2\epsilon}+\frac{1}{4}\log(x_S)+\log(\beta)\right)+2\Li_2(x_S)+\frac{\pi^2}{3}\right]+\Oep{1}\;.
\end{equation}
On the other hand, the cancellation of IR singularities is achieved by including the interference terms originated in the process $\phi(p_3)\to\phi(p_1')+\phi(p_2')+\psi(p_r')$. Explicitly,
\begin{align}\label{Equation:FDUMTMGammaR}
\Gamma_\r^{(1)}=&~\frac{1}{2\sqrt{\s}}\int\,d\Phi_{1\to3}\,2\ReText\langle\mathcal{M}_{2r}^{(0)}|\mathcal{M}_{2r}^{(2)}\rangle=\frac{g^4}{2\sqrt{\s}}\int\,d\Phi_{1\to3}\frac{2\s}{(2p_1'\cdot p_r')(2p_2'\cdot p_r')}\nn\\
=&~\Gamma^{(0)}2g^2\frac{(4\pi)^{\epsilon-2}}{\Gamma(1-\epsilon)}\left(\frac{\s}{\mu^2}\right)^{-\epsilon}\beta^{-1+2\epsilon}\int\,\theta(h_p)\,h_p^{-\epsilon}\,\frac{d\yp{1r}\,d\yp{2r}}{\yp{1r}\,\yp{2r}}\;,
\end{align}
where we used the definition of the massive three-body phase space given in \Eq{\ref{Equation:APPPhaseSpace1to3}}. We also recall that the dimensionless scalar products $\yp{ij}$ we already used in \Chapter{\ref{Chapter:FDU}} are defined as $\yp{ir}=s_{ir}'/\s$. To compute the integral in \Eq{\ref{Equation:FDUMTMGammaR}}, we apply the change of variables suggested in \Appendix{\ref{Section:APPTechniquesforPSintegration}} which allows us to factorise the energy and the angular dependence of the integrand. By using \Eq{\ref{Equation:APPPhaseSpaceCOV}}, we obtain
\begin{align}\label{Equation:FDUMTMGammaRzw}
\Gamma_\r^{(1)}=&~\Gamma^{(0)}2g^2\frac{(4\pi)^{\epsilon-2}}{\Gamma(1-\epsilon)}\left(\frac{\s}{\mu^2}\right)^{–\epsilon}\beta^{-1+2\epsilon}(1+x_S)^{6\epsilon}\nn\\
&~\times\int_{x_S}^{x_S^{-1}}\,dz\frac{z^{-1+2\epsilon}(1+z)^{2\epsilon}}{(z-x_S)^{3\epsilon}(1-x_S\,z)^{3\epsilon}}\int_0\,dw\,w^{-1-2\epsilon}(1-w)^{-\epsilon}\;.
\end{align}
The integration in $w$ can be trivially performed, leading to the appearance of an $\epsilon$-pole. The integral in $z$, though, is finite if $x_S>0$ (which is true if $m>0$). We therefore perform an expansion in $\epsilon$ of \Eq{\ref{Equation:FDUMTMGammaRzw}}, resulting after integration in
\begin{align}\label{Equation:FDUMTMRealResult}
\Gamma_\r^{(1)}=&~\Gamma^{(0)}\frac{4g^2\,\cGamma}{\Gamma(1-\epsilon)}\left(\frac{\s}{\mu^2}\right)^{-\epsilon}\left[\log(x_S)\left(\frac{1}{2\epsilon}-\frac{1}{4}\log(x_S)+\log(1+x_S)+\log(1-x_S^2)\right)\right.\nn\\
&~+\left.\Li_2(x_S)+\Li_2(x_S^2)-\frac{\pi^2}{3}\right]+\Oep{1}\;.
\end{align}
Putting together the virtual (\Eq{\ref{Equation:FDUMTMVirtualResult}}) and real (\Eq{\ref{Equation:FDUMTMRealResult}}) contributions, we get
\begin{equation}\label{Equation:FDUMTMFullResult}
\Gamma^{(1)}=\Gamma^{(0)}\frac{4a}{\beta}\Big[\log(x_S)\big(\log(1-x_S)+\log(1-x_S^2)\big)+2\Li_2(x_S)+\Li_2(x_S^2)\Big]+\Oep{1}\;,
\end{equation}
with $a=g^2/(4\pi^2)$.\\
\\
The purpose of the following discussion will be the derivation of a purely four-dimensional representation of this result through the local cancellation of all IR divergences present in the real and virtual contributions. As for the massless case studied in \Chapter{\ref{Chapter:FDU}}, we will show that the cancellation of IR singularities at the integrand level can be achieved by using suitable mapping of momenta.

\section{Phase-space partition and real-virtual mapping with massive particles}\label{Section:FDUMRealVirtual}
\fancyhead[LO]{\ref*{Section:FDUMRealVirtual}~~\nameref*{Section:FDUMRealVirtual}}

\begin{figure}[t]
	\begin{center}
		\includegraphics[width=7cm]{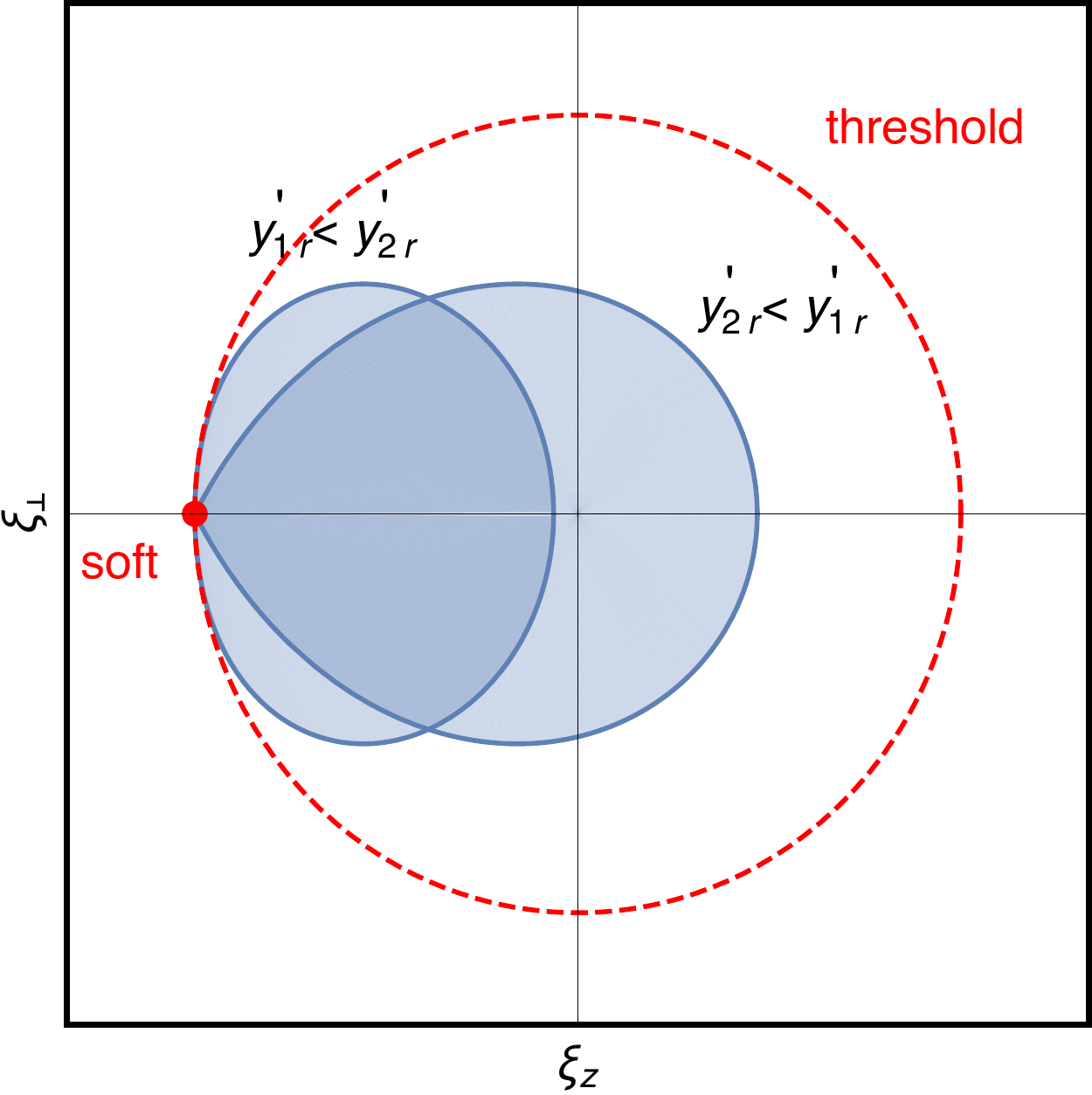}
		\caption{The dual integration regions in the loop three-momentum space, with $\xi_\perp = \sqrt{\xi_x^2+\xi_y^2}$.}
		\label{Figure:FDUMDualRegions}
	\end{center}
\end{figure}

The first step will be to introduce a complete partition of the phase space in such a way that each individual region of that partition contains a single soft, collinear or quasi-collinear configuration. The quasi-collinear configurations are those for which a massless particle becomes collinear with a massive one~\cite{Catani:2002hc}. In this case, the mass acts as a regulator, preventing the emergence of $\epsilon$-poles that would otherwise manifest themselves when performing the computation with DREG. Instead, finite logarithmic terms in the mass will appear at intermediate steps, and will only cancel when considering the total cross section.Consequently, the massless and $\epsilon\to0$ limits do not commute in the traditional approach, when considering virtual and real corrections separately.\\
\\
The kinematics being the same, we use the same separation as for the massless case, i.e. we take advantage of the identity
\begin{equation}\label{Equation:FDUMThetaIdentityMassive}
1=\theta(\yp{2r}-\yp{1r})+\theta(\yp{1r}-\yp{2r})\;.
\end{equation}
We will make sure throughout the following that this partition, together with the well-motivated mapping of momenta, ensures a smooth massless limit -- achieved by an integrand level cancellation of the logarithmic dependences in the mass arising from the quasi-collinear configurations of the real and virtual corrections -- and thus a more stable numerical implementation of the method.\\
\\
The next step is to define the proper momentum mapping in each region to match the singular behaviour of the real and the dual integrands. One the main difficulties when dealing with massive particles is the fact that the on-shell conditions lead to a system of quadratic equations in the mapping parameters. One possible and effective workaround is to rewrite the massive four-vectors as a linear combination of well-chosen massless momenta, simplifying the mapping equations. When the particles have the same mass, the corresponding momenta can be written as
\begin{equation}\label{Equation:FDUMMassivetoHatMomenta}
p_1^\mu=\beta_+\,\hat{p}_1^\mu+\beta_-\,\hat{p}_2^\mu\;,\qquad p_2^\mu=\beta_-\,\hat{p}_1^\mu+\beta_+\,\hat{p}_2^\mu\;,
\end{equation}
with $\hat{p}_1^2=\hat{p}_2^2=0$ and $\beta_\pm=(1\pm\beta)/2$. In addition, the massless momenta fulfil the useful identities
\begin{equation}\label{Equation:FDUMHatMomentaIdentities}
2\hat{p}_1\cdot\hat{p}_2=\s\;,\qquad \hat{p}_1^\mu+\hat{p}_2^\mu=p_1^\mu+p_2^\mu\;.
\end{equation}
In their centre-of-mass frame, they are simply given by
\begin{equation}\label{Equation:FDUMHatMomentaParam}
\hat{p}_1=\frac{\sqrt{\s}}{2}(1,\mathbf{0}_\perp,1)\;,\qquad\hat{p}_2=\frac{\sqrt{\s}}{2}(1,\mathbf{0}_\perp,-1)\;.
\end{equation}
\\
We now go back to the toy example used in \Section{\ref{Section:FDUMS3PF}}, and start with the first region\linebreak $\mathcal{R}_1=\{\yp{1r}<\yp{2r}\}$. Motivated by factorisation properties of QCD in the collinear limit~\cite{Collins:1989gx,Catani:2011st}, and using the momentum decomposition in \Eq{\ref{Equation:FDUMMassivetoHatMomenta}}, we propose the mapping
\begin{align}\label{Equation:FDUMMappingR1}
p_r'^\mu=&~q_1^\mu\;,\nn\\
p_1'^\mu=&~(1-\alpha_1)\hat{p}_1^\mu+(1-\gamma_1)\hat{p}_2^\mu-q_1^\mu\;,\nn\\
p_2'^\mu=&~\alpha_1\,\hat{p}_1^\mu+\gamma_1\,\hat{p}_2\;,
\end{align}
which fulfil momentum conservation by construction. As in the massless case, $p_2'^\mu$ acts as the spectator of the splitting process and is used to balance momentum conservation. The emitters have momenta $p_1$ and $p_1'$, and have the same mass. Although restricted to the three final-state particles, the momentum mapping in \Eq{\ref{Equation:FDUMMappingR1}} can easily be generalised to a multipartonic case, with $p_k'=p_k$ for $k\neq1,2,r$. The parameters $\alpha_1$ and $\gamma_1$ are determined from the two on-shell conditions\footnote{In this context, because $\mathcal{R}_1$ is associated with $I_1$, we have $q_1^2=0$, meaning that the condition\linebreak$(p_r')^2=q_1^2=0$ is automatically fulfilled.}
\begin{align}
(p_1')^2=&~(1-\alpha_1)(1-\gamma_1)\s-2q_1\cdot\big((1-\alpha_1)\hat{p}_1+(1-\gamma_1)\hat{p}_2\big)=M^2\;,\\
(p_2')^2=&~\alpha_1\,\gamma_1\,\s=M^2\;,
\end{align}
whose explicit solutions in terms of the integration variables $(\xi_{1,0},v_1)$ are
\begin{align}\label{Equation:FDUMMappingR1Solutions}
\alpha_1=&~\frac{1-\xi_{1,0}-\sqrt{(1-\xi_{1,0})^2-m^2(1-\xi_{1,0}+v_1(1-v_1)\xi_{1,0}^2)}}{2(1-v_1\,\xi_{1,0})}\nn\;,\\
\gamma_1=&~\frac{1-\xi_{1,0}+\sqrt{(1-\xi_{1,0})^2-m^2(1-\xi_{1,0}+v_1(1-v_1)\xi_{1,0}^2)}}{2(1-(1-v_1)\xi_{1,0})}\;.
\end{align}
As a matter of fact, since we are dealing with quadratic equations, there are two sets of solutions for $(\alpha_1,\gamma_1)$. Only the solution in \Eq{\ref{Equation:FDUMMappingR1Solutions}} is physical and compatible with the soft limit, as it recovers the Born-level kinematics when $\xi_{1,0}\to0$. Indeed, in that limit, $(\alpha_1,\gamma_1)\to(\beta_-,\beta_+)$ and therefore $(p_1',p_2')\to(p_1,p_2)$. Moreover, it also properly reduces to the massless parametrisation defined in \Eq{\ref{Equation:FDUMappingR1}}, as in the limit $m\to0$, we have
\begin{equation}\label{Equation:FDUMMappingR1SolutionsMasslessLimit}
\alpha_1\to0\;,\qquad\gamma_1\to\frac{1-\xi_{1,0}}{1-(1-v_1)\xi_{1,0}}\;.
\end{equation}
Using these definitions, the kinematical invariants $\yp{ij}$ become
\begin{align}\label{Equation:FDUMyijpR1}
\yp{1r}=&~\frac{\xi_{1,0}}{1-(1-v_1)\xi_{1,0}}\big(v_1+\alpha_1(1-2v_1)\big)\;,\nn\\
\yp{2r}=&~\xi_{1,0}-\yp{1r}\;,\nn\\
\yp{12}=&~1-\xi_{1,0}-\frac{m^2}{2}\;,
\end{align}
which fulfil $\yp{1r}+\yp{2r}+\yp{12}=1-m^2/2$. Once again, we easily recover the massless expressions in \Eq{\ref{Equation:FDUyijpR1}} with $\alpha_1=0$. In order to simplify the result, it is convenient to write the mass in terms of $\alpha_1$. According to \Eq{\ref{Equation:FDUMMappingR1Solutions}},
\begin{equation}\label{Equation:FDUMmOfAlpha1}
m^2=\frac{4\alpha_1\big(1-\xi_{1,0}-\alpha_1(1-v_1\,\xi_{1,0})\big)}{1-(1-v_1)\xi_{1,0}}\;.
\end{equation}
The Jacobian associated to the change of variables of \Eq{\ref{Equation:FDUMyijpR1}} is given by
\begin{equation}\label{Equation:FDUMR1Jacobian}
\mathcal{J}_1(\xi_{1,0},v_1)=\frac{\xi_{1,0}\big(1-\xi_{1,0}-\alpha_1(2-\xi_{1,0})\big)^2}{\big(1-(1-v_1)\xi_{1,0}\big)^2\big(1-\xi_{1,0}-2\alpha_1(1-v_1\,\xi_{1,0})\big)}\;,
\end{equation}
with $d\yp{1r}\,d\yp{2r}=\mathcal{J}_1(\xi_{1,0},v_1)\,d\xi_{1,0}\,dv_1$. Notice that this expression is free of square roots, thanks to the fact we rewrote the mass in terms of $\alpha_1$, as suggested in \Eq{\ref{Equation:FDUMmOfAlpha1}}. In the same way, we need to express $\mathcal{R}_1$ in terms of the dual variables. Using the mapping given in \Eq{\ref{Equation:FDUMyijpR1}}, we obtain
\begin{equation}\label{Equation:FDUMR1Dual}
\mathcal{R}_1(\xi_{1,0},v_1)=\theta(1-2v_1)\,\theta\left(\frac{1-2v_1}{1-v_1}\left(1-\frac{1-\sqrt{1-4m^2\,v_1(1-v_1)}}{2v_1}\right)-\xi_{1,0}\right)\;,
\end{equation}
which is the characteristic function associated with the integration domain delimited by the region $\mathcal{R}_1$. Observe that the massless limit agrees with \Eq{\ref{Equation:FDUR1Dual}}. Moreover, the three-body phase-space limits defined by the condition $h_p=0$ are simply determined by $v_1=0$. In \Fig{\ref{Figure:FDUMDualRegions}}, this domain in drawn in the loop three-momentum space.\\
\\
In the complementary region, $\mathcal{R}_2=\{\yp{2r}<\yp{1r}\}$, the mapping is defined by
\begin{align}\label{Equation:FDUMMappingR2}
p_r'^\mu=&~(1-\gamma_2)\hat{p}_1^\mu+(1-\alpha_2)\hat{p}_2^\mu-q_2^\mu\;,\nn\\
p_1'^\mu=&~\gamma_2\,\hat{p}_1^\mu+\alpha_2\,\hat{p}_2\;,\nn\\
p_2'^\mu=&~q_2^\mu\;,
\end{align}
where we have exchanged the role of the radiated particle and the emitter, in order to keep $p_2'$ massive. The associated on-shell conditions\footnote{The condition $(p_2')^2=q_2^2=M^2$ is already fulfilled since $q_2$ is on shell.} are
\begin{align}
(p_1')^2=&~\alpha_2\,\gamma_2\,\s=M^2\;,\nn\\
(p_r')^2=&~M^2+(1-\alpha_2)(1-\gamma_2)\s-2q_2\cdot\big((1-\gamma_2)\hat{p}_1+(1-\alpha_2)\hat{p}_2\big)=0\;,
\end{align}
whose solutions in terms of $(\xi_2,v_2)$ are
\begin{align}\label{Equation:FDUMMappingR2Solutions}
\alpha_2=&~\frac{1-\xi_{2,0}+m^2/2-\sqrt{(1-\xi_{2,0})^2-m^2\,v_2(1-v_2)\xi_2^2}}{2-(1-v_2)\xi_2-\xi_{2,0}}\nn\;,\\
\gamma_2=&~\frac{1-\xi_{2,0}+m^2/2-\sqrt{(1-\xi_{2,0})^2-m^2\,v_2(1-v_2)\xi_2^2}}{2+(1-v_2)\xi_2-\xi_{2,0}}\nn\;.
\end{align}
The consistency of the solution is again verified by the massless limit, where\footnote{Note that in this limit $\xi_{i,0}=\xi_i$.}
\begin{equation}\label{Equation:FDUMMappingR2SolutionsMasslessLimit}
\alpha_2\to0\;,\qquad\gamma_2\to\frac{1-\xi_2}{1-v_2\,\xi_2}\;.
\end{equation}
In this region, the kinematical invariants are therefore rewritten
\begin{align}\label{Equation:FDUMyijpR2}
\yp{1r}=&~1-\xi_{2,0}\;,\nn\\
\yp{2r}=&~\frac{\xi_{2,0}+(1-2v_2)(1-2\alpha_2)\xi_2-m^2}{2+(1-2v_2)\xi_2-\xi_{2,0}}\;,\nn\\
\yp{12}=&~\xi_{2,0}-\frac{m^2}{2}-\yp{2r}\;,
\end{align}
which also fulfils $\yp{1r}+\yp{2r}+\yp{12}=1-m^2/2$. We can verify that we get back \Eq{\ref{Equation:FDUyijpR2}} when $\alpha_2=0$. The Jacobian in this region reads
\begin{equation}\label{Equation:FDUMR2Jacobian}
\mathcal{J}(\xi_2,v_2)=\frac{4\xi_2\big(1-\xi_{2,0}+m^2/2-\alpha_2(2-\xi_{2,0})\big)^2}{\xi_{2,0}\big(2+(1-2v_2)\xi_2-\xi_{2,0}\big)^2\big(1-\xi_{2,0}+m^2/2-\alpha_2(2-(1-2v_2)\xi_2-\xi_{2,0})\big)}\;,
\end{equation}
with $d\yp{1r}\,d\yp{2r}=\mathcal{J}_2(\xi_2,v_2)\,d\xi_2\,dv_2$, and where we made use of the relation
\begin{equation}\label{Equation:FDUMmOfAlpha2}
m^2=\frac{4\alpha_2\big(2(1-\xi_{2,0})-\alpha_2(1-v_2)\xi_2-\xi_{2,0}\big)}{2+(1-2v_2)\xi_2-\xi_{2,0}-4\alpha_2}
\end{equation}
to simplify the expression. Finally, the characteristic function of the integration domain in this region is given by
\begin{equation}\label{Equation:FDUMR2Dual}
\mathcal{R}_2(\xi_2,v_2)=\theta\left(\beta\left(\left(1+\sqrt{(1-v_2)(1-m^2\,v^2)}\right)^2-m^2\,v_2^2\right)^{-1/2}-\xi_2\right)\;,
\end{equation}
which reduces as expected to \Eq{\ref{Equation:FDUR2Dual}} in the massless limit. The corresponding domain in the loop three-momentum space is shown in \Fig{\ref{Figure:FDUMDualRegions}}.

\subsection{General momentum mapping}

The momentum mapping previously presented can easily be extended to the most general multipartonic case in which the emitter and the spectator have different masses, namely $p_i^2=m_i^2$ and $p_j^2=m_j^2$. The decomposition of their momenta in terms of two massless momenta ($\hat{p}_i^2=\hat{p}_j^2=0$) is given by
\begin{align}\label{Equation:FDUMGeneralMapping}
p_i^\mu=&~\beta_+\,\hat{p}_i^\mu+\beta_-\,\hat{p}_j^\mu\nn\;,\\
p_j^\mu=&~(1-\beta_+)\hat{p}_i^\mu+(1-\beta_-)\hat{p}_j^\mu\;,
\end{align}
with
\begin{equation}\label{Equation:FDUMGeneralbetapm}
\beta_\pm=\frac{s_{ij}+m_i^2-m_j^2\pm\lambda(s_{ij},m_i^2,m_j^2)}{2s_{ij}}\;,
\end{equation}
where $\lambda(s_{ij},m_i^2,m_j^2)=\sqrt{(s_{ij}-(m_i-m_j)^2)(s_{ij}-(m_i+m_j)^2)}$, with $s_{ij}=(p_i+p_j)^2$, is the usual Kall\'en function. As before, the massless momenta fulfil the useful condition $\hat{p}_i+\hat{p}_j=p_i+p_j$ by construction. The mapping with the momenta of the real process is essentially identical to the mapping already considered in \Eq{\ref{Equation:FDUMMappingR1}}, and is the massive generalisation of the one in \Eq{\ref{Equation:FDUMappingGeneralisation}}. Explicitly,
\begin{align}\label{Equation:FDUMMappingGeneralisation}
p_r'^\mu=&~q_i^\mu\;,\quad&p_i'^\mu=&~(1-\alpha_i)\hat{p}_i^\mu+(1-\gamma_i)\hat{p}_j^\mu-q_i^\mu\;,\nn\\
p_j'^\mu=&~\alpha_i\,\hat{p}_i^\mu\,+\gamma_i\,\hat{p}_j^\mu\;,\quad&&\nn\\
p_k'^\mu=&~p_k^\mu\;,\quad&k\neq&~i,j,r\;,
\end{align}
which leads to the on-shell conditions
\begin{align}\label{Equation:FDUMGeneralOSCond}
(p_i')^2=&~(1-\alpha_i)(1-\gamma_i)s_{ij}-2q_i\cdot\big((1-\alpha_i)\hat{p}_i+(1-\gamma_i)\hat{p}_j\big)+m_r^2=(m_i')^2\;,\nn\\
(p_j')^2=&~\alpha_i\,\gamma_i\,s_{ij}=m_j^2\;,
\end{align}
where we have imposed that the spectator and the radiated particle have the same flavour -- and thus the same mass -- in the virtual and real processes, i.e. $p_j^2=(p_j')^2=m_j^2$ and $q_i^2=(p_r')^2=m_r^2$, respectively. The emitters, however, might have different flavours\footnote{This case of figure occurs, for instance, when a gluon splits into a massive quark-antiquark pair.}, meaning that we can have $(p_i')^2=(m_i')^2\neq m_i^2$. The solution to \Eq{\ref{Equation:FDUMGeneralOSCond}} for the parameters of the mapping reads
\begin{align}
\alpha_i=&~\frac{(p_i+p_j-q_i)^2+m_j^2-(m_i')^2-\Lambda_{ij}}{2(s_{ij}-2q_i\cdot\hat{p}_i)}\;,\nn\\
\gamma_i=&~\frac{(p_i+p_j-q_i)^2+m_j^2-(m_i')^2+\Lambda_{ij}}{2(s_{ij}-2q_i\cdot\hat{p}_i)}\;,
\end{align}
with
\begin{equation}\label{Equation:FDUMGeneralLambda}
\Lambda_{ij}=\sqrt{\big((p_i+p_j-q_i)^2+m_j^2-(m_i')^2\big)^2-\frac{4m_j^2}{s_{ij}}(s_{ij}-2q_i\cdot\hat{p}_i)(s_{ij}-2q_i\cdot\hat{p}_j)}\;.
\end{equation}
The momentum mapping in \Eq{\ref{Equation:FDUMMappingGeneralisation}} has a smooth limit whenever any of the involved particles become massless; in particular, if the spectator is as massless particle, then $\alpha_i=0$.

\section{Massive scalar decay rate from the four-dimensional unsubtraction}\label{Section:FDUMDecayRateFDU}
\fancyhead[LO]{\ref*{Section:FDUMDecayRateFDU}~~\nameref*{Section:FDUMDecayRateFDU}}

We now have all the necessary ingredients to apply the FDU method to a massive process; we will start with the toy example presented in \Section{\ref{Section:FDUMMassiveScalarDecayRate}}. We have to combine at the integrand level the dual loop contributions (\Section{\ref{Section:FDUMS3PF}}) with the real-radiation terms (\Section{\ref{Section:FDUMMassiveScalarDecayRate}}), with the help of the momentum mapping defined in \Section{\ref{Section:FDUMRealVirtual}}. Since the sum of all contributions is UV and IR finite, the final result is free of $\epsilon$-poles. We would like to emphasise that, in a general situation, this last assertion is not enough to guarantee the integrability of the expressions in four dimensions\footnote{We will indeed encounter a counterexample in \Chapter{\ref{Chapter:HOL}}, when considering the finite one-loop amplitude of the $gg\to H$ and $H\to\gamma\gamma$ processes.}. However, by virtue of the momentum mappings and the unification of the dual coordinates, LTD naturally leads to a local cancellation of divergences, and the limit $\epsilon\to0$ can be considered at the \emph{integrand level}.\\
\\
The LTD representation of the virtual decay rate in this toy example is given by
\begin{equation}\label{Equation:FDUMGammaVLTD}
\Gamma_\V^{(1)}=\frac{1}{2\sqrt{\s}}\,\sum\limits_{i=1}^3\,\int\,d\Phi_{1\to2}2\ReText\langle\mathcal{M}|\mathcal{M}^{(1)}(\deltatilde{q_i})\rangle\;,
\end{equation}
with
\begin{equation}
\langle\mathcal{M}^{(0)}|\mathcal{M}^{(1)}\big(\deltatilde{q_i}\big)\rangle=-g^4\,\s\,I_i\;,
\end{equation}
where the dual integrands $I_i$ are defined in \Eqs{\ref{Equation:FDUMS3PFMassiveDual1}}{\ref{Equation:FDUMS3PFMassiveDual3}}, in which we set $\epsilon$ to 0 inside the integration measure. In order to ensure the cross-cancellation of spurious singularities and get a direct $\epsilon=0$ limit, we must rewrite all the on-shell momenta in terms of the same coordinate system, as we did for the massless case in \Section{\ref{Section:FDUUCS}}. Because of the presence of the mass, the change of variables is slightly modified. We need to solve instead
\begin{align}
\mathbf{q}_1=&~\frac{\s}{2}\,\xi_{1,0}\left(2\sqrt{v_1(1-v_1)}\,\eiPerp{1},1-2v_1\right)\nn\\
=&~\mathbf{q}_3+\mathbf{p}_1=\frac{\s}{2}\left(2\xi_3\sqrt{v_3(1-v_3)}\,\eiPerp{3},\xi_3(1-2v_3)+\beta\right)\;.
\end{align}
We find
\begin{align}\label{Equation:FDUMCOVUnification}
\xi_{1,0}&=~\sqrt{(\beta+\xi)^2-4\beta\,v\,\xi}\;,\nn\\
v_1&=~\dfrac{1}{2}\left(1-\dfrac{\beta+(1-2v)\xi}{\sqrt{(\beta+\xi)^2-4\beta\,v\,\xi}}\right)\;,
\end{align}
where we directly replaced $(\xi_3,v_3)$ by $(\xi,v)$, as we have $q_3=\ell$. It is worth mentioning that this change of reference frame is well defined because the argument of the square root is always positive. Indeed,
\begin{equation}\label{Equation:FDUMUCSIsOk}
(\xi+\beta)^2-4\beta\,\xi\,v\geq(\xi+\beta)^2-4\xi\,\beta=(\xi-\beta)^2\geq0\;,
\end{equation}
since $v\leq1$. The associated Jacobian is given by
\begin{equation}\label{Equation:FDUMUCSJacobian}
\mathcal{J}(\xi,v)=\frac{\xi^2}{(\xi+\beta)^2-4\beta\,v\,\xi}\;,
\end{equation}
whose massless limit -- reached for $\beta\to1$ -- agrees with \Eq{\ref{Equation:FDUUCSJacobian}}. Similarly, we can apply the trivial replacement $(\xi_2,v_2)\to(\xi,v)$, given that $\mathbf{q}_2=\mathbf{q}_3$. With that change of variables, the virtual decay rate in \Eq{\ref{Equation:FDUMGammaVLTD}} becomes a single unconstrained integral in the loop three-momentum.\\
\\
The real counterpart of this process has been previously computed and is given by \Eq{\ref{Equation:FDUMTMGammaR}}. From this, we split the real three-body phase space according to \Eq{\ref{Equation:FDUMThetaIdentityMassive}}, and define
\begin{equation}\label{Equation:FDUMThetaIdentityMassiveGammaRi}
\widetilde{\Gamma}_{\r,i}^{(1)}=\frac{1}{2\s}\int\,d\Phi_{1\to3}\,2\ReText\langle\mathcal{M}_{2r}^{(0)}|\mathcal{M}_{1r}^{(0)}\rangle\mathcal{R}_i\;,\qquad i\in\{1,2\}\;,
\end{equation}
that obviously fulfil
\begin{equation}
\Gamma_{\r}^{(1)}=\widetilde{\Gamma}_{\r,1}^{(1)}+\widetilde{\Gamma}_{\r,2}^{(1)}\;.
\end{equation}
Then, we apply the momentum mappings defined in \Eq{\ref{Equation:FDUMyijpR1}} and \Eq{\ref{Equation:FDUMyijpR2}} in their respective region. The main advantage of these mappings is that they are optimised to smoothly deal with the massless limit in each of the two regions. Rewriting the real contributions in terms of the loop variables gives
\begin{align}
\label{Equation:FDUMGammaR1xiv}\widetilde{\Gamma}_{\r,1}^{(1)}=&~\Gamma^{(0)}\frac{2a}{\beta}\int\,d\xi_{1,0}\,dv_1\frac{\mathcal{R}_1(\xi_{1,0},v_1)\mathcal{J}_1(\xi_{1,0},v_1)\big(1-\xi_{1,0}(1-v_1)\big)^2}{\xi_{1,0}\big(v_1+\alpha_1(1-v_1)\big)\big((1-v_1)(1-\xi_{1,0})-\alpha_1(1-2v_1)\big)}\;,\\
\label{Equation:FDUMGammaR2xiv}\widetilde{\Gamma}_{\r,2}^{(1)}=&~\Gamma^{(0)}\frac{2a}{\beta}\int\,d\xi_2\,dv_2\frac{\mathcal{R}_2(\xi_2,v_2)\mathcal{J}_2(\xi_2,v_2)\big(2+(1-v_2)\xi_2-\xi_{2,0}\big)}{(1-\xi_{2,0})\big(\xi_{2,0}+(1-v_2)(1-2\alpha_2)\xi_2-m^2\big)}\;,
\end{align}
where the Jacobians of the respective transformations are given by \Eq{\ref{Equation:FDUMR1Jacobian}} and \Eq{\ref{Equation:FDUMR2Jacobian}}. The integration domains are restricted by the characteristic functions $\mathcal{R}_i$, given in \Eqs{\ref{Equation:FDUMR1Dual}}{\ref{Equation:FDUMR2Dual}}. After applying the change of variables given previously, we can take $\epsilon=0$ in these expressions\footnote{To be completely rigorous, it is important to recall that the real and virtual contributions are not entirely well-defined by themselves in the limit $\epsilon\to0$, regardless of how we write the integration variables. It is only when considering their sum that we obtain a quantity that is integrable in this limit. To simplify the discussion and the expressions, however, we took the liberty to set $\epsilon=0$ at intermediate steps.}.\\
\\
The sum of the virtual with the real contributions, that is to say \Eq{\ref{Equation:FDUMGammaVLTD}} with \Eq{\ref{Equation:FDUMGammaR1xiv}} and \Eq{\ref{Equation:FDUMGammaR2xiv}}, is a finite and well-defined integral in the $\epsilon\to0$ limit, thanks to the local cancellation of all IR singularities in the loop three-momentum space at the integrand level.\\
\\
Now the last thing we need to deal with is the threshold singularity appearing in the virtual contribution, for $\xi_2=\beta$. This singularity is integrable and can be treated numerically by contour deformation~\cite{Buchta:2015xda,Buchta:2015wna}, but for the toy scalar model that we are considering -- and later in this chapter, the physical examples -- there is a simpler solution. The idea is to compactify the regions where $\xi<\beta$ and $\xi>\beta$ into the sphere unit by using a very simple change of variables. Explicitly, we can write
\begin{equation}\label{Equation:FDUMCOVThreshold}
\int_0^\infty\,d\xi\,g(\xi,v)=\beta\int_0^1\,dx\big[g(\beta\,x,v)+x^{-2}\,g(\beta\,x^{-1},v)\big]\;,
\end{equation}
where the threshold singularity has been mapped onto the upper end-point $x=1$. This approach is very efficient for the numerical implementation.\\
\begin{figure}[t]
	\centering
	\includegraphics[width=0.6\textwidth]{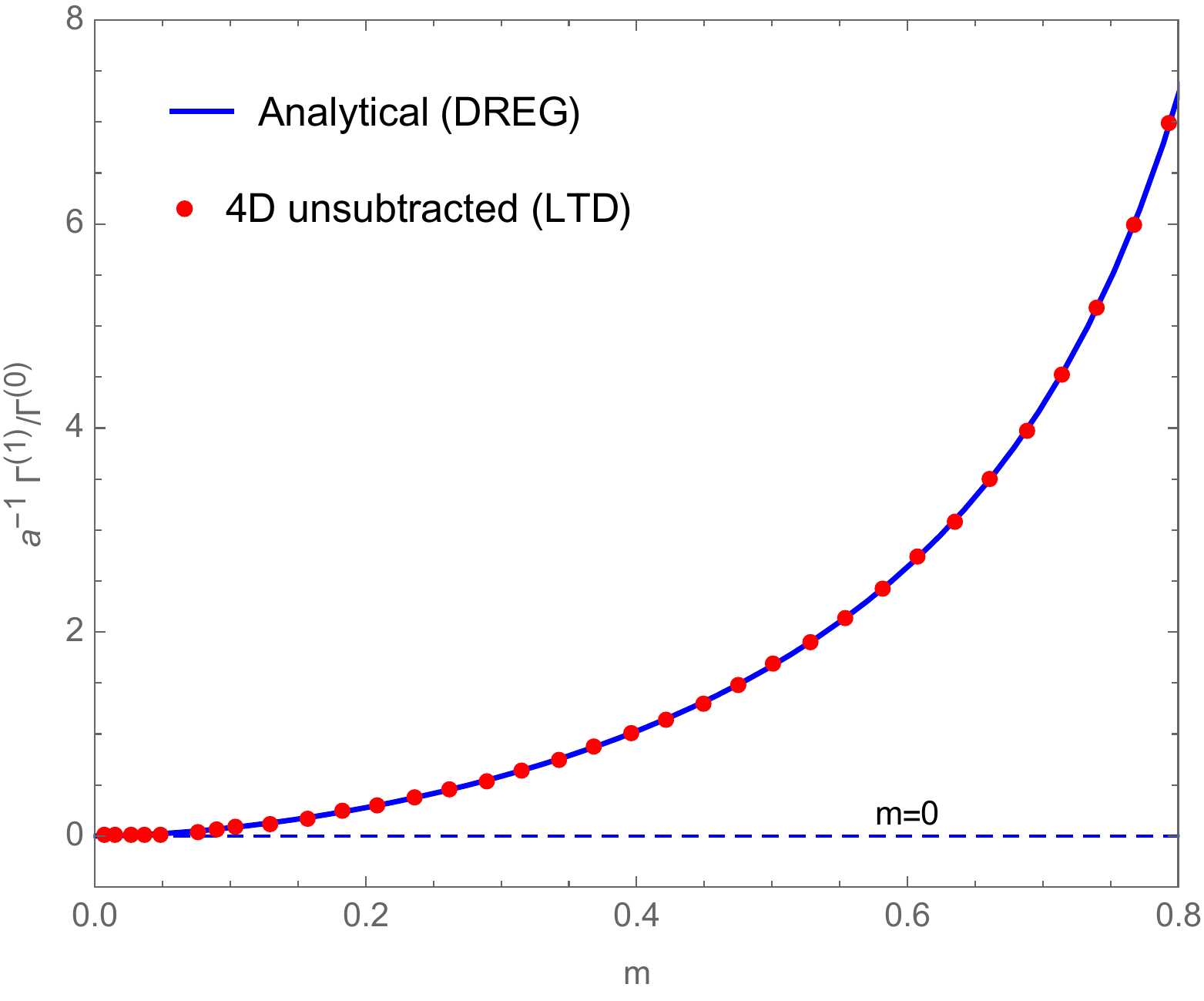}
	\caption{Total decay rate at NLO normalised to the leading order for the toy scalar example,  $a^{-1}\Gamma^{(1)}/\Gamma^{(0)}$, as a function of the dimensionless mass parameter $m$. The horizontal dashed line represents the massless limit, the solid line corresponds to the analytic result obtained through DREG, and the dots are obtained numerically using FDU.}
	\label{Figure:FDUMPlotGammaTildesTM}
\end{figure}
\\
Finally, we numerically integrate simultaneously the virtual and real corrections from \Eq{\ref{Equation:FDUMGammaVLTD}}, and \Eqs{\ref{Equation:FDUMGammaR1xiv}}{\ref{Equation:FDUMGammaR2xiv}} respectively, with the help of \Eq{\ref{Equation:FDUMCOVThreshold}} to obtain the total decay rate at NLO, written $\Gamma^{(1)}$, as a function of the dimensionless parameter $m$. The result is shown in \Fig{\ref{Figure:FDUMPlotGammaTildesTM}}, and is compared with the DREG analytic expression given by \Eq{\ref{Equation:FDUMTMFullResult}}. The agreement is excellent and the computation is very stable numerically. Computing all the points of the plot of \Fig{\ref{Figure:FDUMPlotGammaTildesTM}} took a few minutes, using a standard laptop (2.3GHz quad-core processor). Moreover, the massless transition is very smooth because the momentum mappings are optimised to deal with quasi-collinear configurations. This means, in other words, that the massless limit can directly be taken at the integrand level, which is another very interesting advantage of the LTD approach.

\section{Unintegrated wave function and mass renormalisation for heavy quarks}\label{Section:FDUMRenormalisation}
\fancyhead[LO]{\ref*{Section:FDUMRenormalisation}~~\nameref*{Section:FDUMRenormalisation}}

In order to consider physical processes with heavy quarks, we must also take into account self-energy corrections. In the Feynman gauge with on-shell renormalisation conditions, the well-known expressions of the wave function and mass renormalisation constants\footnote{Note that the notation differs from \Section{\ref{Section:SARUV}}, where we used the symbol $\delta$ instead of $\Delta$.},
\begin{align}\label{Equation:FDUMDeltaZ2DeltaZM}
\Delta Z_2=&~\frac{\alpha_S}{4\pi}C_F\left(-\frac{1}{\epsilon_\uv}-\frac{2}{\epsilon_\ir}+3\log\left(\frac{M^2}{\mu^2}\right)-4\right)\;,\nn\\
\text{and}\quad\Delta Z_M^{\rm OS}=&~\frac{\alpha_S}{4\pi}C_F\left(-\frac{3}{\epsilon_\uv}+3\log\left(\frac{M^2}{\mu^2}\right)-4\right)\;,
\end{align}
are not suitable for the implementation of a local subtraction of the IR singularities. Indeed, what we need are \emph{unintegrated} expressions. The case of the massless quark has been studied in detail in \Section{\ref{Section:FDUNLO}}. In \Eq{\ref{Equation:FDUMDeltaZ2DeltaZM}}, we explicitly identify the origin of the $\epsilon$-poles, and we expect the IR singularities of the wave function to cancel the IR singularities arising from the squared amplitudes of the real process.\\
\\
We consider the process in which there are two on-shell massive fermions with momenta $p_1$ (quark) and $p_2$ (antiquark). The explicit one-loop self-energies are given by
\begin{align}\label{Equation:FDUMSelfEnergies}
-i\Sigma(p_1)=&~i\,g_S^2\,C_F\int_\ell\left(\,\prod\limits_{i=1,3}\,G_F(q_i)\right)\gamma^{\sigma_1}(-\slashed{q}_3+M)\gamma^{\sigma_2}\,d_{\sigma_1\sigma_2}(q_1)\;,\nn\\
-i\Sigma(-p_2)=&~i\,g_S^2\,C_F\int_\ell\left(\,\prod\limits_{i=1,2}\,G_F(q_i)\right)\gamma^{\sigma_1}(-\slashed{q}_2+M)\gamma^{\sigma_2}\,d_{\sigma_1\sigma_2}(q_1)\;.
\end{align}
In these expressions, we used the same momenta $q_i$ used previously in this chapter (their configuration can be found \Fig{\ref{Figure:FDUVirtualRealDiagrams}}). This will allow us to reuse the same momentum mappings we already defined to treat the vertex corrections. Because of the symmetry $\slashed{p}_1\leftrightarrow-\slashed{p}_2$, it is enough to compute the unintegrated expression for the quark self-energy, as the one of the antiquark can be directly deduced from it. Thus, we will only consider $\Sigma(\slashed{p}_1)$ in the following. In the Feynman gauge,
\begin{equation}\label{Equation:FDUMSelfEnergyTraced}
\Sigma(p_1)=g_S^2\,C_F\int_\ell\,\left(\,\prod\limits_{i=1,3}\,G_F(q_i)\right)\big((d-2)\slashed{q}_3+d\,M\big)\;.
\end{equation}
Working in the on-shell (OS) renormalisation scheme\footnote{The OS scheme is better suited for external particles, compared to the $\MSbar$ scheme, since the correction vanishes if the particles are massless.}, the renormalised self-energy must fulfil
\begin{equation}\label{Equation:FDUMSelfEnergyOSConditions}
\Sigma_R(\slashed{p}_1=M)=0\qquad\text{and}\qquad\left.\frac{d\Sigma_R(\slashed{p}_1)}{d\slashed{p}_1}\right|_{\slashed{p}_1=M}=0\;,
\end{equation}
from which the wave function and mass renormalisation corrections are given by
\begin{equation}\label{Equation:FDUMDeltaSelfEnergy}
\Delta Z_2=\left.\frac{\partial}{\partial\slashed{p}_1}\Sigma(\slashed{p}_1)\right|_{\slashed{p}_1=M}\qquad\text{and}\qquad\Delta Z_M^{\rm OS}=-\frac{1}{M}\Sigma(\slashed{p}_1=M)\;.
\end{equation}
Explicitly, from \Eq{\ref{Equation:FDUMSelfEnergyTraced}} we obtain
\begin{align}
\label{Equation:FDUMDeltaZ2Explicit}\Delta Z_2(p_1)=&~-g_S^2\,C_F\int_\ell\,G_F(q_1)\,G_D(q_3)\left((d-2)\frac{q_1\cdot p_2}{p_1\cdot p_2}+4M^2\left(1-\frac{q_1\cdot p_2}{p_1\cdot p_2}\right)G_F(q_3)\right)\;,\\
\label{Equation:FDUMDeltaZMExplicit}\Delta Z_M^{\rm OS}(p_1)=&~-g_S^2\,C_F\int_\ell\,G_F(q_1)\,G_D(q_3)\left((d-2)\frac{q_1\cdot p_2}{p_1\cdot p_2}+2\right)\;.
\end{align}
It is worth stressing that the expression of the wave function renormalisation constant in \Eq{\ref{Equation:FDUMDeltaZ2Explicit}} tends smoothly in the massless limit to the corresponding expression in \Eq{\ref{Equation:FDUSelfEnergiesMasslessTraced}}. It is also relevant to notice that the term proportional to $M^2\big(G_F(q_3)\big)^2$ leads to soft divergences when $q_1$ becomes on shell, although they are expected to cancel the soft divergences of the squared amplitudes of the real corrections. While applying LTD to the mass renormalisation constant is straightforward, the term proportional to $M^2\big(G_F(q_3)\big)^2$ in \Eq{\ref{Equation:FDUMDeltaZ2Explicit}} introduces a double pole that needs to be treated adequately. The dual representations for both renormalisation factors are
\begin{align}\label{Equation:FDUMDeltaZ2DeltaZMCut}
\Delta Z_2(p_1)=&~g_S^2\,C_F\int_\ell\,\Bigg[-\frac{\deltatilde{q_1}}{2q_1\cdot p_1}\left((d-2)\frac{q_1\cdot p_2}{p_1\cdot p_2}-\frac{4M^2}{2q_1\cdot p_1}\left(1-\frac{q_1\cdot p_2}{p_1\cdot p_2}\right)\right)+\frac{\deltatilde{q_3}}{2M^2+2q_3\cdot p_1}\nn\\
&~\times\Bigg((d-2)\left(1+\frac{q_3\cdot p_2}{p_1\cdot p_2}\right)+\frac{4M^2}{p_1\cdot p_2}\Bigg(-\frac{\mathbf{q}_3\cdot \mathbf{p}_2}{2\left(\Sup{q_{3,0}}\right)^2}+\frac{\left(\Sup{q_{3,0}}+p_{1,0}\right)q_3\cdot p_2}{\Sup{q_{3,0}}(2M^2+2q_3\cdot p_1)}\Bigg)\Bigg)\Bigg]\;,\nn\\
\Delta Z_M^{\rm OS}(p_1)=&~g_S^2\,C_F\int\,\left[-\frac{\deltatilde{q_1}}{2q_1\cdot p_1}\left((d-2)\frac{q_1\cdot p_2}{p_1\cdot p_2}+2\right)+\frac{\deltatilde{q_3}}{2M^2+2q_3\cdot p_1}\left((d-2)\frac{q_2\cdot p_2}{p_1\cdot p_2}+d\right)\right]\;,
\end{align}
which, in terms of the dual variables defined in \Section{\ref{Section:FDUMS3PF}}, read
\begin{align}\label{Equation:FDUMDeltaZ2DeltaZMDual}
\Delta Z_2(p_1)=&~g_S^2\,C_F\Bigg[\int\,\frac{2\dxi{1}\,\dv{1}}{1-\beta(1-2v_1)}\left(-(d-2)\frac{\xi_{1,0}\big(1+\beta(1-2v_1)\big)}{1+\beta^2}\right.\nn\\
&~+\left.\frac{2m^2}{1-\beta(1-2v_1)}\left(\frac{1}{\xi_{1,0}}-\frac{1+\beta(1-2v_1)}{1+\beta^2}\right)\right) \nn \\
&~+\int\,\frac{2\xi_3^2\,d[\xi_3]\,\dv{3}}{\xi_{3,0}(\xi_{3,0}-\beta\,\xi_3(1-2v_3)+m^2)}
\left((d-2)\left(1+\frac{\xi_3 +\beta\,\xi_3(1-2v_3)}{1+\beta^2} \right)\right.\nn\\ 
&~+\left.\frac{2m^2}{(1+\beta^2)\xi_{3,0}}\left(\frac{\beta\,\xi_3(1-2v_3)}{\xi_{3,0}}+\frac{(1+\xi_{3,0})\big(\xi_{3,0}+\beta\,\xi_{3,0}(1-2v_3)\big)}{\xi_{3,0}-\beta\,\xi_3(1-2v_3)+m^2}\right)\right)\Bigg]\nn\\
\Delta Z^{\rm OS}_M(p_1)=&~g_S^2\,C_F\Bigg[-\int  
\frac{2\, d[\xi_3]\, d[v_1]}{1-\beta(1-2v_1)} \, \left((d-2) \, \frac{\xi_{1,0}\,(1+\beta(1-2v_1))}{1+\beta^2}+2 \right)\nn \\ 
&~+\int\,\frac{2\xi_3^2\,d[\xi_3]\,\dv{3}}{\xi_{3,0}(\xi_{3,0}-\beta\,\xi_3(1-2v_3)+m^2)}
\left((d-2)\,\frac{\xi_{3,0}+\beta\,\xi_3(1-2v_3)}{1+\beta^2}+ d\right)\Bigg]\;.
\end{align}

\section{Ultraviolet renormalisation}\label{Section:FDUMUV}
\fancyhead[LO]{\ref*{Section:FDUMUV}~~\nameref*{Section:FDUMUV}}

In this section we build the integrand-level counterterms that are necessary to cancel the singularities appearing when considering the physical case. First, we need to remove the UV divergences of the renormalisation constants by defining suitable integrand level counterterms. They are built by expanding \Eqs{\ref{Equation:FDUMDeltaZ2Explicit}}{\ref{Equation:FDUMDeltaZMExplicit}} around the UV propagator $G_F(q_\uv)=1/(q_\uv-\mu_\uv^2+i0)$, as discussed in \Section{\ref{Section:FDUUV}}, where we recall $q_\uv=\ell+k_\uv$ with $k_\uv$ arbitrary. Since it is the simplest choice, we take $k_\uv=0$. We obtain
\begin{align}\label{Equation:FDUMDeltaZ2DeltaZMUV}
Z_2^\uv(p_1)=&~-(d-2)g_S^2\,C_F\int_\ell\,\big(G_D(q_\uv)\big)^2\left(1+\frac{q_\uv\cdot p_2}{p_1\cdot p_2}\right)\nn\\
&~\times\big(1-G_F(q_\uv)(2q_\uv\cdot p_1+\mu_\uv^2)\big)\;,\nn\\
\Delta Z_M^{{\rm OS},\uv}(p_1)=&~-g_S^2\,C_F\int\,\big(G_D(q_\uv)\big)^2\left(d+(d-2)\frac{q_\uv\cdot p_2}{p_1\cdot p_2}\right)\nn\\
&~\times\big(1-G_F(q_\uv)(2q_\uv\cdot p_1+2d^{-1}\mu_\uv^2)\big)\;,
\end{align}
whose integrated forms are
\begin{align}
\Delta Z_2^\uv=&~-g_S^2\,C_F\,\frac{\Se}{16\pi^2}\left(\frac{\mu_\uv^2}{\mu^2}\right)^{-\epsilon}\frac{1-\epsilon}{\epsilon}\;,\nn\\
\Delta Z_M^{{\rm OS},\uv}=&~-g_S^2\,C_F\,\frac{\Se}{16\pi^2}\left(\frac{\mu_\uv^2}{\mu^2}\right)^{-\epsilon}\frac{3}{\epsilon}\;.
\end{align}
The subleading terms in \Eq{\ref{Equation:FDUMDeltaZ2DeltaZMUV}}, which are proportional to $\mu_\uv^2$, have been adjusted in such a way that only the UV poles in \Eq{\ref{Equation:FDUMDeltaZ2DeltaZM}} are subtracted at $\Oep{0}$. Therefore, the quantities
\begin{equation}\label{Equation:FDUMDeltaZ2IRDeltaZMIR}
\Delta Z_2^\ir=\Delta Z_2-\Delta Z_2^\uv\;,\qquad\Delta Z_M^{{\rm OS},\ir}=\Delta Z_M^{\rm OS}-\Delta Z_M^{{\rm OS},\uv}\;,
\end{equation}
only contain IR singularities, including the finite terms which are scheme dependent.\\
\\
Obtaining the dual representation of \Eq{\ref{Equation:FDUMDeltaZ2DeltaZMUV}} requires to deal with poles of second and third order (see \Section{\ref{Section:LTDMultiPole}}), located at $\Sup{q_{\uv,0}}=\sqrt{\mathbf{q}_\uv^2+\mu_\uv^2-i0}$. This leads to
\begin{align}\label{Equation:FDUMDeltaZ2DeltaZMUVDual}
\Delta Z_2^\uv=&~-(d-2)g_S^2\,C_F\int_\ell\,\frac{\deltatilde{q_\uv}}{2(\Sup{q_{\uv,0}})^2}\Bigg[\left(1-\frac{\mathbf{q}_\uv\cdot\mathbf{p}_2}{p_1\cdot p_2}\right)\nn\\
&~\times\left(1-\frac{3(2\mathbf{q}_\uv\cdot\mathbf{p}_1-\mu_\uv^2)}{4(\Sup{q_{\uv,0}})^2}\right)-\frac{p_{1,0}\,p_{2,0}}{2p_1\cdot p_2}\Bigg]\nn\\
\Delta Z_M^{{\rm OS},\uv}=&~-(d-2)g_S^2\,C_F\int_\ell\,\frac{\deltatilde{q_\uv}}{2(\Sup{q_{\uv,0}})^2}\Bigg[\left(d-(d-2)\frac{\mathbf{q}_\uv\cdot\mathbf{p}_2}{p_1\cdot p_2}\right)\nn\\
&~\times\left(1-\frac{3(2\mathbf{q}_\uv\cdot\mathbf{p}_1-2d^{-1}\mu_\uv^2)}{4(\Sup{q_{\uv,0}})^2}\right)-(d-2)\frac{p_{1,0}\,p_{2,0}}{2p_1\cdot p_2}\Bigg]\;.
\end{align}
By using the parametrisation
\begin{align}\label{Equation:FDUMUVParam}
q_\uv^\mu=&~\frac{\sqrt{\s}}{2}\left(\xi_{\uv,0},2\xi_\uv\sqrt{v_\uv(1-v_\uv)}\,\mathbf{e}_{\uv,\perp},\xi_\uv(1-2v_\uv)\right)\;,\nn\\
\xi_{\uv,0}=&~\sqrt{\xi_\uv^2+m_\uv^2}\;,
\end{align}
with $m_\uv=2\mu_\uv/\sqrt{\s}$, the UV counterterms get the form
\begin{align}\label{Equation:FDUMDeltaZ2DeltaZMUVDualParam}
\Delta Z_2^\uv=&~-(d-2)g_S^2\,C_F\int\,d[\xi_\uv]\,\dv{\uv}\frac{2\xi_\uv^2}{\xi_{\uv,0}^3}\left[\left(1+\frac{\beta\,\xi_\uv(1-2v_\uv)}{2(1+\beta^2)}\right)\right.\nn\\
&~\times\left.\left(1-\frac{3(2\beta\,\xi_\uv(1-v_\uv)-m_\uv^2)}{4\xi_{\uv,0}^3}-\frac{1}{2(1+\beta^2)}\right)\right]\;,\nn\\
\Delta Z_M^{{\rm OS},\uv}=&~-g_S^2\,C_F\int\,d[\xi_\uv]\,\dv{\uv}\frac{2\xi_\uv^2}{\xi_{\uv,0}^3}\left[\left(d+(d-2)\frac{\beta\,\xi_\uv(1-2v_\uv)}{2(1+\beta^2)}\right)\right.\nn\\
&~\times\left.\left(1-\frac{3(2\beta\,\xi_\uv(1-v_\uv)-2d^{-1}\,m_\uv^2)}{4\xi_{\uv,0}^3}-\frac{d-2}{2(1+\beta^2)}\right)\right]\;.
\end{align}
\\
Similarly, we need to take into account the UV singularities generated by the $Aq\qbar$ vertex, with $A=\phi,\gamma,Z$, to obtain the corresponding counterterm for the explicit physical examples we will consider later in this chapter. As for the self-energy contributions, the UV counterterm is obtained by expanding the vertex corrections around the UV propagator $G_F(q_\uv)$. In the Feynman gauge, we have
\begin{equation}\label{Equation:FDUMGammaAUV}
\mathbf{\Gamma}_{A,\uv}^{(1)}=g_S^2\,C_F\int_\ell\,\big(G_F(q_\uv)\big)^3\left(\gamma^\sigma\,\slashed{q}_\uv\,\mathbf{\Gamma}_A^{(0)}\,\slashed{q}_\uv\,\gamma_\sigma-d_{A,\uv}\,\mu_\uv^2\,\mathbf{\Gamma}_A^{(0)}\right)\;,
\end{equation}
where the tree-level vertices $\mathbf{\Gamma}_A^{(0)}$ are given in \Eq{\ref{Equation:APPAqqbargVertices}}. Note that we have introduced a subleading term proportional to $\mu_\uv^2$, where $d_{A,\uv}$ is a scheme-fixing parameter depending only on the particle $A$. In the $\MSbar$ scheme, we want to adjust $d_{A,\uv}$ so we subtract only the $\epsilon$-pole (that is to say\footnote{The two conditions are actually not completely equivalent as we may still have $\Oep{n}$ with $n\geq1$ non-vanishing terms. In this particular case, and with the values of $d_{A,\uv}$ provided, the cancellation is exact at all orders in $\epsilon$. But for more complex counterterms, having an exact cancellation at all orders would involve unnecessarily complicated expressions for the scheme-fixing parameters; they are indeed much simpler if we focus on removing solely the $\Oep{0}$ part, which in our framework is enough.} the $\Oep{0}$ part of the integrated counterterm vanishes). Performing the explicit calculation in $d$ dimensions, we find\footnote{In \Chapter{\ref{Chapter:HTL}} where we derive similar expressions, we construct the counterterms in a slightly different way, that is more practical for the process under consideration. Consequently, the definition of the scheme-fixing parameters $d_\uv$ is different.}
\begin{equation}\label{Equation:FDUMdAUVValues}
d_{\phi,\uv}=d+4\;,\qquad d_{\gamma,\uv}=d_{Z,\uv}=d\;.
\end{equation}
Integrating the vertex UV counterterm leads to
\begin{equation}\label{Equation:FDUMVertexAUVResult}
\mathbf{\Gamma}_{A,\uv}^{(1)}=g_S^2\,C_F\,\frac{\Se}{16\pi^2}\left(\frac{\mu_\uv^2}{\mu^2}\right)^{-\epsilon}\frac{c_{A,\uv}}{\epsilon}\,\mathbf{\Gamma}_A^{(0)}\;,
\end{equation}
with
\begin{equation}\label{Equation:FDUMcAUVValues}
c_{\phi,\uv}=4\;,\qquad c_{\gamma,\uv}=c_{Z,\uv}=1\;,
\end{equation}
and the corresponding matrix element reads
\begin{equation}\label{Equation:FDUMMEUVVertex}
\langle\mathcal{M}_A^{(0)}|\mathcal{M}_{A,\uv}^{(1)}\rangle=g_S^2\,C_F\,\frac{\Se}{16\pi^2}|\mathcal{M}_A^{(0)}|^2\left(\frac{\mu_\uv^2}{\mu^2}\right)^{-\epsilon}\frac{c_{A,\uv}}{\epsilon}\;.
\end{equation}
The dual representation of \Eq{\ref{Equation:FDUMGammaAUV}} reads
\begin{align}
\mathbf{\Gamma}_{A,\uv}^{(1)}=&~g_S^2\,C_F\int_\ell\,\frac{\deltatilde{q_\uv}}{8(\Sup{q_{\uv,0}})^2}\Bigg[\gamma^\sigma\,\gamma^0\,\mathbf{\Gamma}_A^{(0)}\,\gamma^0\,\gamma_\sigma\nn\\
&~-\frac{3}{(\Sup{q_{\uv,0}})^2}\left(\gamma^\sigma(\gamma\cdot \mathbf{q}_\uv)\gamma_A^{(0)}(\gamma\cdot \mathbf{q}_\uv)\gamma_\sigma-d_{A,\uv}\,\mu_\uv^2\,\Gamma_A^{(0)}\right)\Bigg]\;.
\end{align}
After an explicit calculation, and by using \Eq{\ref{Equation:FDUMUVParam}}, we obtain for each particle
\begin{align}\label{Equation:FDUMUVVertexDual}
\langle\mathcal{M}_\phi^{(0)}|\mathcal{M}_{\phi,\uv}^{(1)}\rangle=&~g_S^2\,C_F\,|\mathcal{M}_\phi^{(0)}|^2\int\,d[\xi_\uv]\,\dv{\uv}\frac{2\xi_\uv^2}{\xi_{\uv,0}^3}\left(7-2\epsilon-\frac{3\xi_\uv^2}{\xi_{\uv,0}^2}\right)\;,\nn\\
\langle\mathcal{M}_\gamma^{(0)}|\mathcal{M}_{\gamma,\uv}^{(1)}\rangle=&~g_S^2\,C_F\,\int\,d[\xi_\uv]\,\dv{\uv}\left[\frac{2(e\,e_q)^2C_A}{1-\epsilon}f(\xi_\uv,v_\uv)\right.\nn\\
&~+\left.|\mathcal{M}_\gamma^{(0)}|^2\frac{\xi_\uv^2}{\xi_{\uv,0}^3}\left(7-4\epsilon-\frac{3\xi_\uv^2}{\xi_{\uv,0}^2}\big(1+4v_\uv(1-v_\uv)\big)\right)\right]\;,\nn\\
\langle\mathcal{M}_Z^{(0)}|\mathcal{M}_{Z,\uv}^{(1)}\rangle=&~g_S^2\,C_F\,\int\,d[\xi_\uv]\,\dv{\uv}\left[\frac{2g_{V,q}^2\,C_A}{1-\epsilon}f(\xi_\uv,v_\uv)\right.\nn\\
&~+\left.|\mathcal{M}_Z^{(0)}|^2\frac{\xi_\uv^2}{\xi_{\uv,0}^3}\left(7-4\epsilon-\frac{3\xi_\uv^2}{\xi_{\uv,0}^2}\big(1+4v_\uv(1-v_\uv)\big)\right)\right]\;,
\end{align}
where the function
\begin{equation}\label{fUV}
f(\xi_\uv,v_\uv)=24M^2\frac{\xi_\uv^4}{\xi_{\uv,0}^5}\big(\epsilon(1-2v_\uv)^2+6v_\uv(1-v_\uv)-1\big)
\end{equation}
integrates to 0 in $d$ dimensions, but is still necessary to achieve a local cancellation of the UV behaviour in the LTD framework.\\
\\
One can notice that the UV divergences of the wave function cancel exactly the UV divergences of the vertex corrections for photons and $Z$ bosons, because conserved currents or partially conserved currents, as the vector and axial ones, do not get renormalised (this was also the case for the process studied in \Chapter{\ref{Chapter:FDU}}). The corresponding dual representations, however, do not cancel each other at the integrand level. In particular, the wave function renormalisation constant contains linear UV singularities that cancel upon integration, and the vertex UV counterterm exhibit terms that are proportional to the mass and cancel after integrating. The contribution of all these terms though, is crucial to achieve a local cancellation of all the UV singularities.\\
\\
On the other hand, the Yukawa coupling needs to be renormalised, and we have
\begin{equation}
Y_q^0\,\mu_0^\epsilon=Y_q\,\mu^\epsilon\left(1-\alpha_S\,\frac{C_F}{4\pi}\,\frac{3}{\epsilon}\right)+\mathcal{O}(\alpha_S^2)\;.
\end{equation}

\section{Application of the FDU algorithm to physical processes}\label{Section:FDUMNLO}
\fancyhead[LO]{\ref*{Section:FDUMNLO}~~\nameref*{Section:FDUMNLO}}

\begin{figure}[t]
	\begin{center}
		\includegraphics[width=0.45\textwidth]{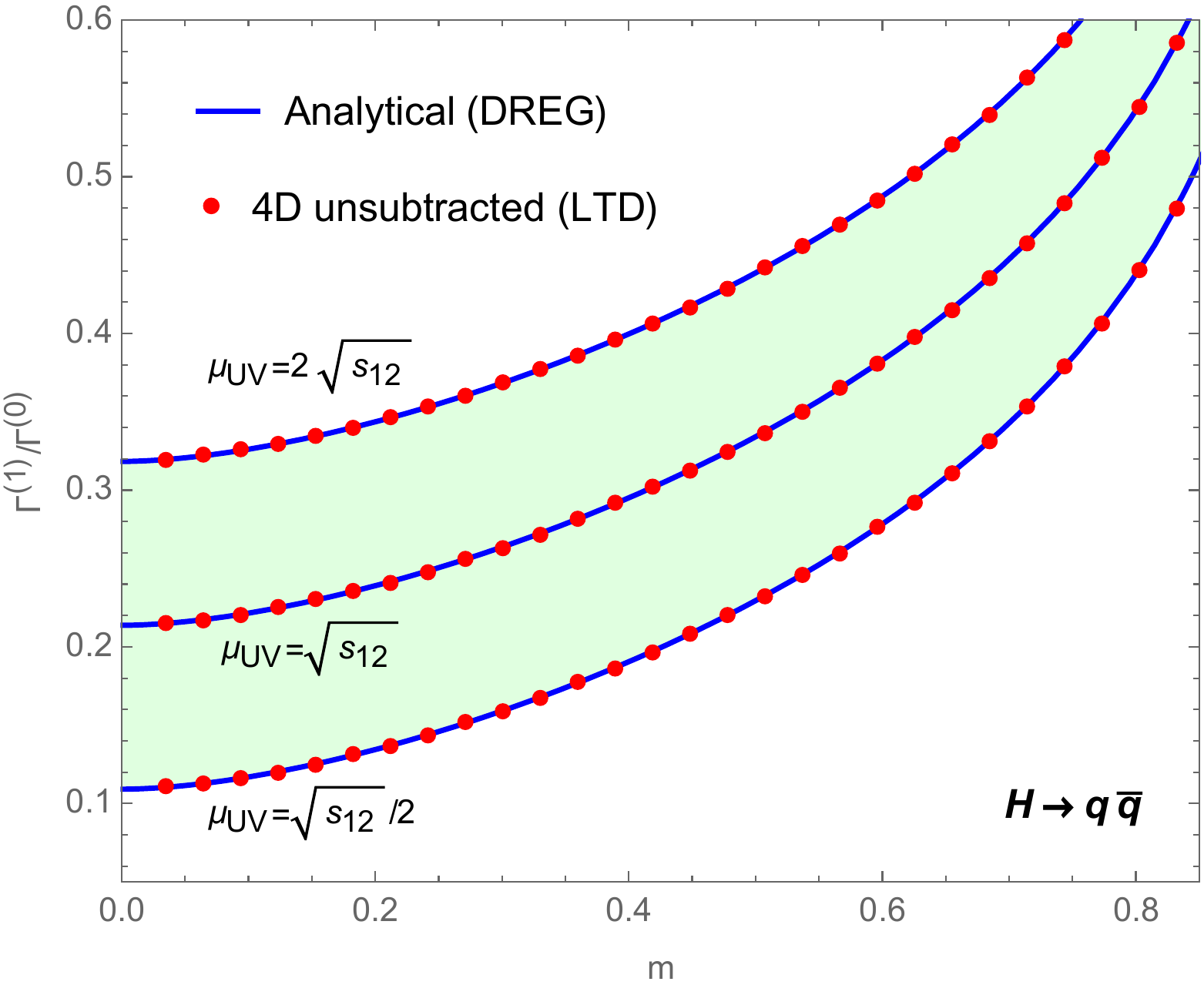} \qquad
		\includegraphics[width=0.45\textwidth]{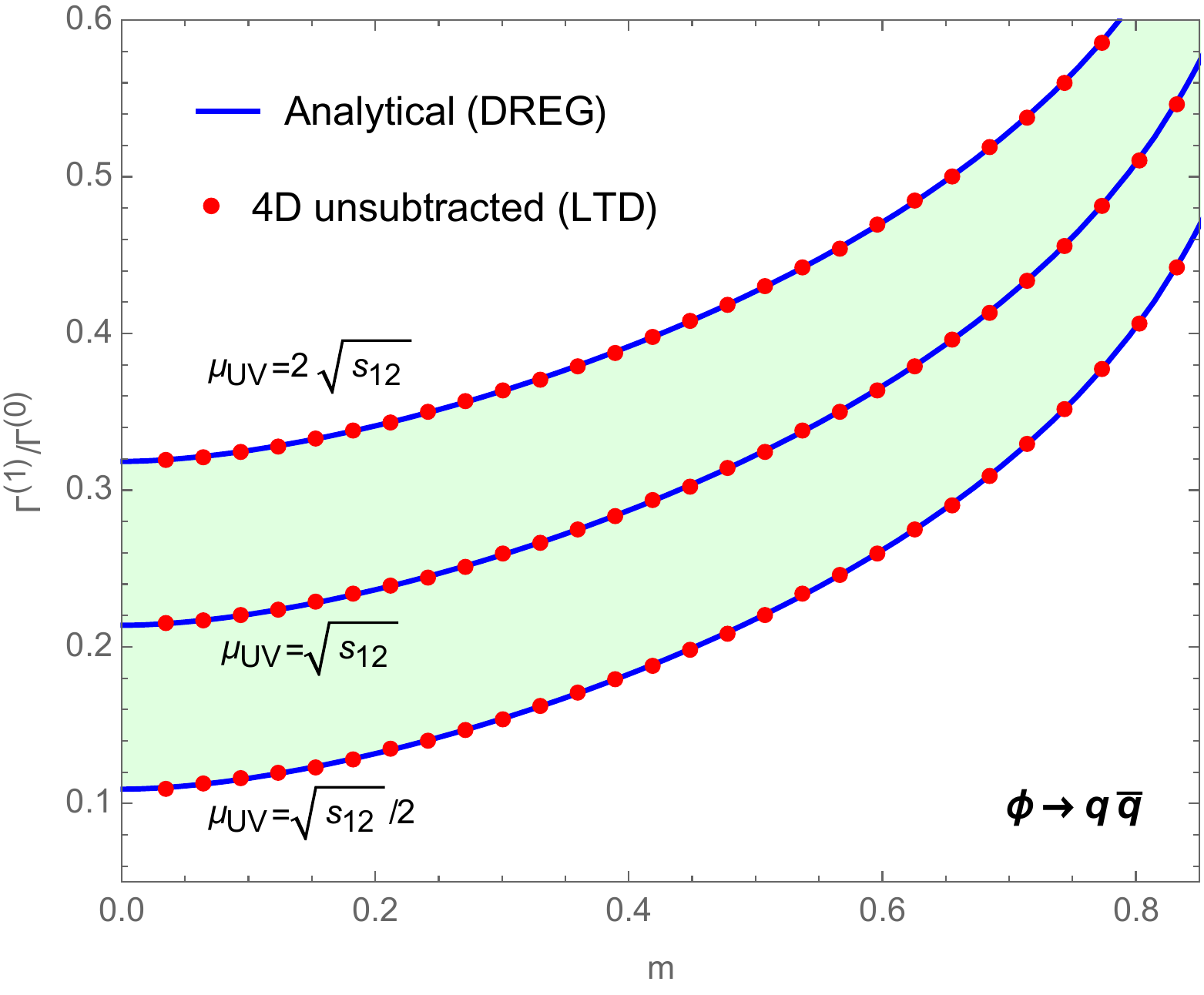}
		\caption{Total decay rate at NLO for scalar and pseudoscalar particles into a pair of heavy quarks as a function of the mass, normalised to the LO. In the left panel, we consider a standard Higgs boson, while in the right panel we plot the decay rate for a pseudoscalar particle ($c_q=1/2$). The solid blue lines correspond to the usual DREG analytic result, while the red dots were computed numerically using FDU. We also consider a renormalisation scale variation, in the range $1/2<\mu_\uv/\sqrt{\s}<2$.}
		\label{Figure:FDUMDecayRatesHiggses}
	\end{center}
\end{figure}

\begin{figure}[h]
	\centering
		\includegraphics[width=0.6\textwidth]{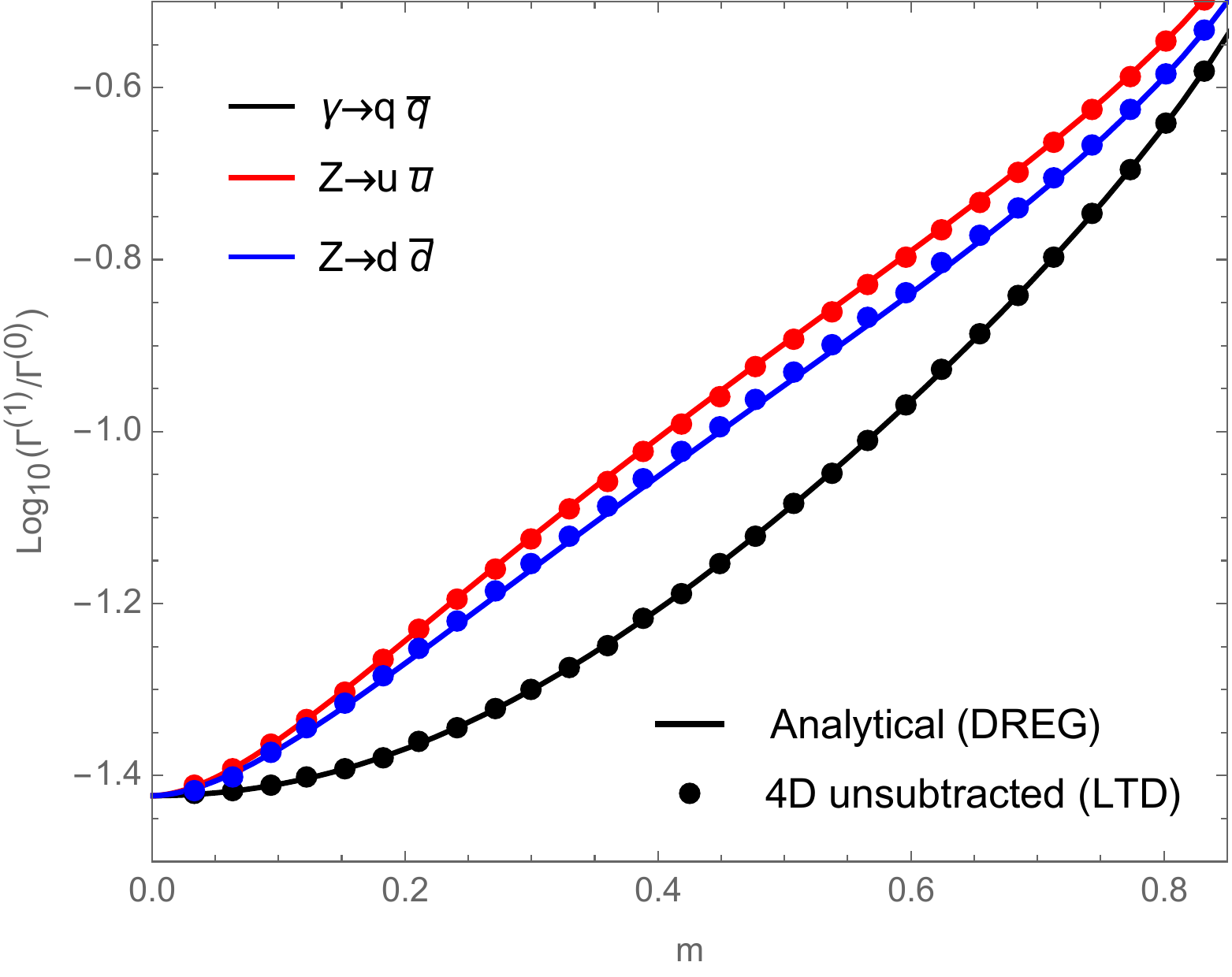}
		\caption{Total decay rate at NLO for off-shell vector particles into a pair of heavy quarks as a function of the mass, normalised to the LO. We consider three physical cases: $\gamma^*\to q\bar q$ (black), $Z^*\to u\bar{u}$ (red, up-type quarks) and $Z^*\to d\bar{d}$ (blue, down-type quarks). Solid lines corresponds to the analytic results obtained with DREG, while the dots were obtained with the FDU method.}
		\label{Figure:DecayRateVectors}
\end{figure}

We now have all the tools necessary to finally test the four-dimensional implementation of NLO corrections to physical process in the LTD framework. In this section, we will compute the NLO QCD corrections to the decay rate $A^*\to q\qbar(g)$, where $A=\phi,\gamma,Z$. As we will see, the general procedure is completely independent of the decaying particle.\\
\\
For a given particle $A$, the renormalised one-loop amplitude is given by
\begin{equation}\label{Equation:FDUMRenormalisedOneLoopA}
|\mathcal{M}_A^{(1,\r)}\rangle=|\mathcal{M}_A^{(1)}\rangle-|\mathcal{M}_A^{(1,\uv)}\rangle+\frac{1}{2}\left(\Delta Z_2^\ir(p_1)+\Delta Z_2^\ir(p_2)\right)|\mathcal{M}_A^{(0)}\rangle\;,
\end{equation}
where $|\mathcal{M}_A^{(1,\uv)}\rangle$ is the unintegrated UV counterterm of the one-loop vertex correction $|\mathcal{M}_A^{(1)}\rangle$, and $\Delta Z_2^\ir(p_1)$ and $\Delta Z_2^\ir(p_2)$ are the IR components of the quark and antiquark self-energy corrections, respectively. From the renormalised one-loop amplitude $|\mathcal{M}_A^{(1,\r)}\rangle$, which contains only IR singularities by definition, we construct the LTD representation of the renormalised decay rate
\begin{equation}\label{Equation:GammaVAR}
\Gamma_{\V,A}^{(1,\r)}=\frac{1}{2\sqrt{\s}}\,\sum\limits_{i=1}^3\,\int\,d\Phi_{1\to2}\,2\ReText\langle\mathcal{M}|\mathcal{M}_A^{(1,\r)}\big(\deltatilde{q_i}\big)\rangle\;.
\end{equation}
The corresponding dual amplitudes for the vertex corrections are given in \Eqss{\ref{Equation:APPDualphiqq}}{\ref{Equation:APPDualgammaqq}}{\ref{Equation:APPDualZqq}}. In a similar way as for the toy scalar example presented in \Section{\ref{Section:FDUMDecayRateFDU}}, the real contributions are implemented by splitting the real phase space into two domains
\begin{equation}\label{Equation:GammaRAi}
\widetilde{\Gamma}_{\r,A,i}^{(1)}=\frac{1}{2\sqrt{\s}}\int\,d\Phi_{1\to3}|\mathcal{M}_{A\to q\qbar g}^{(0)}|^2\,\mathcal{R}_i(\yp{ir}<\yp{jr})\;,\qquad i,j\in\{1,2\}\;,
\end{equation}
with $\widetilde{\Gamma}_{\r,A,1}^{(1)}+\widetilde{\Gamma}_{\r,A,2}^{(1)}=\widetilde{\Gamma}_{\r,A}^{(1)}$ being the total real decay rate. The real emission squared amplitudes are given in \Eq{\ref{Equation:APPAqqbargReal}}. In each of the real phase-space domains, we introduce the corresponding momentum mapping defined in \Section{\ref{Section:FDUMRealVirtual}}. The sum of the virtual and real corrections in \Eqs{\ref{Equation:GammaVAR}}{\ref{Equation:GammaRAi}} is a single integral in the loop-three momentum. It is locally and completely UV and IR finite, and thus can be numerically computed with $\epsilon=0$. For the numerical implementation, we follow the same procedure --~i.e. we compactify the three-momentum integration domain~-- as for the toy example. Our results, normalised to the LO decay rate $\Gamma_A^{(0)}$, are presented in \Fig{\ref{Figure:FDUMDecayRatesHiggses}} for an incoming scalar or pseudo-scalar, and \Fig{\ref{Figure:DecayRateVectors}} for an incoming vector boson, and are compared with the analytic total decay rate
\begin{equation}\label{Equation:GammaALiterature}
\Gamma_A^{(1)}=\frac{\alpha_S}{4\pi}\,C_F\left(\Gamma_A^{(0)}\left(F(x_S)+2(c_{A,\uv}-1)\log\left(\frac{\mu_\uv^2}{\s}\right)\right)+G_A(x_S)\right)+\Oep{1}\;.
\end{equation}
\\
The computing time is of the order a few minutes, similar to that of the toy scalar example. The agreement with the analytic prediction is excellent for all three processes. Moreover, the massless limit  -- reached for $x_S\to0$ -- is also very smooth and well-defined. We want to emphasise the importance of this last point, as in the traditional approach, individual contributions are in general not well-defined in this limit.

\section{Conclusion}\label{Section:FDUMConclusion}
\fancyhead[LO]{\ref*{Section:FDUMConclusion}~~\nameref*{Section:FDUMConclusion}}

In this chapter, we have generalised the FDU method~\cite{Hernandez-Pinto:2015ysa,Sborlini:2016gbr,Sborlini:2016hat,Rodrigo:2016hqc,Hernandez-Pinto:2016uwx,Sborlini:2016bod,Driencourt-Mangin:2016dpf,Sborlini:2017nee} to deal with massive particles. We have showed that in the massive case, singularities are still restricted to a compact region of the loop three-momentum integration domain. In particular, the momentum mapping used to locally cancel IR singularities have been generalised, and properly deals with the quasi-collinear configurations.\\
\\
We have started by studying the scalar three-point function with massive particles within the LTD approach. After recovering previously known results, we have analysed the integration domain of the dual contributions, which allowed us to understand the origin of its singular structure. Then, we illustrated the local cancellation of IR and quasi-collinear configurations with a toy scalar example.\\
\\
The full cancellation of IR singularities requires the explicit contribution of the self-energy corrections. For this reason, we have had to define unintegrated versions of the quark wave function and mass renormalisation factors in the on-shell renormalisation scheme. Compared to the massless case, this procedure is much less trivial since the on-shell scheme has to be built at the integrand level. Nonetheless, the unintegrated renormalisation constants are completely general and lead to a fully local cancellation of the remaining IR singularities. The treatment of UV divergences has also been carefully discussed, and suitable unintegrated UV counterterms have been built for both self-energy and vertex corrections, successfully reproducing the $\MSbar$ conditions and leading to a fully local cancellation of UV singularities.\\
\\
Finally, we have put the FDU algorithm into practice to compute the NLO QCD corrections to the decay of scalar and vector particles into a pair of massive quarks. Our results have been compared to the standard DREG expressions, and we have found an impressive agreement. In particular, the transition to the massless limit is completely smooth because the quasi-collinear configurations of the real and virtual corrections are matched at the integrand level. With the results presented in this chapter, FDU can be applied to any multipartonic process involving heavy quarks in particular, or any massive particle in general.

\newpage
\hspace{0pt}
\thispagestyle{empty}
\clearpage

\chapter{Universal dual amplitudes and asymptotic expansions in four dimensions}\label{Chapter:HOL}
\thispagestyle{fancychapter}
\fancyhead[RE]{\nameref*{Chapter:HOL}}

Although the one-loop amplitude of the Higgs boson to massless gauge bosons is finite because of the lack of direct interaction at tree-level in the SM, a well-defined regularisation scheme is still required for their correct evaluation. In this chapter, we reanalyse these amplitudes in the framework of FDU and LTD, and show how a local renormalisation solves potential regularisation ambiguities \cite{Driencourt-Mangin:2017gop}. The Higgs boson interactions are also used to illustrate new additional advantages of this formalism. We show that LTD naturally leads to very compact integrand expressions in four space-time dimensions of the one-loop amplitude with virtual electroweak gauge bosons. They exhibit the same functional form as the amplitudes with top quarks and charged scalars, thus opening further possibilities for simplifications in higher-order computations. Another application presented here is the implementation of asymptotic expansions of dual amplitudes, by taking advantage of the Euclidean nature of the integration domain of dual integrals.

\section{Introduction}\label{Section:HOLIntro}
\fancyhead[LO]{\ref*{Section:HOLIntro}~~\nameref*{Section:HOLIntro}}

The $gg\to H$ and $H\to\gamma\gamma$ channels are among the most important channels for the experimental study of the Higgs particle. It is therefore essential, from the theoretical point of view, to achieve the highest possible level of precision when evaluating their respective cross section and decay rate. The one-loop contribution to the $Hgg$ vertex have been known since a long time ago~\cite{Wilczek:1977zn,Georgi:1977gs,Rizzo:1980a}, as well as the Higgs boson decay into a photon pair~\cite{Ellis:1975ap,Ioffe:1976sd,Shifman:1979eb}. It is also well-known that these amplitudes are finite due to the absence of a direct interaction at tree-level in the SM. However, contrary to what is naively expected, DREG -- or any other regularisation technique~\cite{Gnendiger:2017pys} -- and a well-defined renormalisation scheme are still required for their correct evaluation. Indeed, a naive calculation in four space-time dimensions not only leads to the incorrect result~\cite{Gastmans:2011wh}, but also spoils gauge invariance and produces inconsistent physical effects, such as the absence of decoupling in the limit where the mass of the particle inside the loop becomes much greater that the mass of the Higgs boson.\\
\\
In this chapter, we use the LTD/FDU formalism to recalculate the $gg\to H$ and $H\to\gamma\gamma$ scattering amplitudes at one-loop level, and their asymptotic expansion in the loop particle mass. For the Higgs boson production through gluon fusion, we consider a top quark inside the loop, while for the decay to two photons, we consider a charged scalar, a top quark, or a $W$ gauge boson. The four-dimensional nature of the FDU approach allows for an alternative insight into the structure of these amplitudes, unveiling for instance the origin of local UV singularities that vanish in the integrated amplitude but still lead to finite contributions.\\
\\
We start by recalling how to extract scalar quantities from tensor amplitudes, in \Section{\ref{Section:HOLTensorProjection}}. Then, in \Section{\ref{Section:HOLUniversalDualAmplitudes}}, we apply the LTD theorem to obtain compact expressions for the dual integrands, and show that they exhibit the same functional form regardless of the internal particle. This is a highly non-trivial result as intermediate expressions with gauge bosons diverge faster in the UV than those with scalars and fermions.  After that, in \Section{\ref{Section:HOLRenormalisation}}, we discuss the local renormalisation of the one-loop amplitude by introducing a suitable counterterm that locally cancels the UV behaviour of the one-loop integrand and allows for a direct integration of the amplitude in $d=4$ dimensions. In \Section{\ref{Section:HOLAsymptoticExpansions}}, we show that the simplicity and well-behaved convergence of the large-mass and small-mass asymptotic expansions of the Higgs boson amplitudes in the LTD formalism avoids considering complementary expansions in different regions of the loop-momentum, simplifying the procedure by a significant margin. Finally, we draw our conclusions in \Section{\ref{Section:HOLConclusion}}.

\section{Tensor projection}\label{Section:HOLTensorProjection}
\fancyhead[LO]{\ref*{Section:HOLTensorProjection}~~\nameref*{Section:HOLTensorProjection}}

The one-loop scattering amplitudes of the Higgs boson production through gluon fusion has the form
\begin{equation}\label{Equation:HOLAmpggH}
|\mathcal{M}_{gg\to H}^{(1)}\rangle=i\,g_S^2\,\Tr(\mathbf{T}^a\,\mathbf{T}^b)\,\mathcal{A}_{\mu\nu}^{(1,t)}\,\varepsilon_a^\mu(p_1)\,\varepsilon_b^\nu(p_2)\;,
\end{equation}
while the one of the Higgs boson decay to two photons has the form
\begin{equation}\label{Equation:HOLAmpHgammagamma}
|\mathcal{M}_{H\to\gamma\gamma}^{(1)}\rangle=i\,e^2\left(\,\sum\limits_{f=\phi,t,W}\,e_f^2\,N_C^f\,\mathcal{A}_{\mu\nu}^{(1,f)}\right)(\varepsilon_a^\mu(p_1))^*\,(\varepsilon_b^\nu(p_2))^*\;,
\end{equation}
where $e$ (respectively $g_S$) is the electromagnetic (respectively strong) coupling, $\varepsilon$ is the polarisation vector of the external gluons and photons, $\Tr(\mathbf{T}^a\,\mathbf{T}^b)=T_R\,\delta^{ab}$ is the colour factor and $e_f$ (respectively $N_C^f$) is the electric charge (respectively the number of colours) of the particle $f$. To be completely rigorous, in \Eq{\ref{Equation:HOLAmpggH}} we should not only consider the top quark, but the other quarks as well; however, their mass being much smaller than the one of the top quark, their corresponding contributions are heavily suppressed. Similarly, the sum in \Eq{\ref{Equation:HOLAmpHgammagamma}} should include all massive leptons. In any case, this has no impact on the discussion carried out in this chapter.\\
\\
By Lorentz invariance, the colour and electric charge stripped tensor amplitude is given by
\begin{equation}\label{Equation:HOLTensorDecomposition}
\mathcal{A}_{\mu\nu}^{(1,f)}=\,\sum\limits_{i=1}^5\,\mathcal{A}_i^{(1,f)}\,T_{\mu\nu}^i\;,
\end{equation}
where the tensor basis is
\begin{equation}\label{Equation:HOLTensorBasis}
T^{\mu\nu,i}=\left\{g^{\mu\nu}-\frac{2p_1^\nu\,p_2^\mu}{\s},g^{\mu\nu},\frac{2p_1^\mu\,p_2^\nu}{\s},\frac{2p_1^\mu\,p_1^\nu}{\s},\frac{2p_2^\mu\,p_2^\nu}{\s}\right\}\;,
\end{equation}
with $\s=(p_1+p_2)^2$. If the Higgs boson is considered to be on shell, we have $\s=M_H^2$.
Now in order to work with scalar quantities, we need to be able to extract the coefficients $\mathcal{A}_i^{(1,f)}$ introduced in \Eq{\ref{Equation:HOLTensorDecomposition}}. There is no need to consider $\mathcal{A}_i^{(1,f)}$ for $i\in\{3,4,5\}$ though, as they vanish after contraction with the polarisation vectors, and therefore do not contribute to the scattering amplitude\footnote{Note that this statement is valid only if the external photons and gluons are on shell, which is the case for the process we are considering here.}. In addition, because of gauge invariance, $\mathcal{A}_2^{(1,f)}$ must vanish after integration, which leaves $\mathcal{A}_1^{(1,f)}$ as the only relevant physical term. It is still interesting, however, to calculate an integrand expression for $\mathcal{A}_2^{(1,f)}$, as it can be used to simplify expressions at intermediate steps, as we will see. To extract these two coefficients, we use the projectors
\begin{equation}\label{Equation:HOLProjectors}
P_1^{\mu\nu}=\frac{1}{d-2}\left(g^{\mu\nu}-\frac{2p_2^\mu\,p_1^\nu}{\s}-(d-1)\frac{2p_1^\mu\,p_2^\nu}{\s}\right)\qquad\text{and} \qquad P_2^{\mu\nu}=\frac{2p_1^\mu\,p_2^\nu}{\s}\,
\end{equation}
that satisfy $P_i^{\mu\nu}\,\mathcal{A}_{\mu\nu}^{(1,f)}=\mathcal{A}_i^{(1,f)}$ for $i\in\{1,2\}$.\\
\\
For $f=\phi,t,W$, the analytic expressions for $\mathcal{A}_1^{(1,f)}$ are available in e.g.~\cite{Spira:1995rr,Harlander:2005rq,Aglietti:2006tp} and read
\begin{align}\label{Equation:HOLLiteratureResults}
\mathcal{A}_1^{(1,\phi)}=&~\frac{g_\phi\,\s}{16\pi^2}\left(1+\frac{2}{r_\phi}\log^2\left(\frac{\beta_\phi-1}{\beta_\phi+1}\right)\right)\;,\nn\\
\mathcal{A}_1^{(1,t)}=&~\frac{g_t\,\s}{16\pi^2}\left(-2+\left(1-\frac{4}{r_t}\right)\log^2\left(\frac{\beta_t-1}{\beta_t+1}\right)\right)\;,\nn\\
\mathcal{A}_1^{(1,W)}=&~\frac{g_W\,\s}{16\pi^2}\left(3+\frac{r_W}{2}+	\left(-3+\frac{6}{r_W}\right)\log^2\left(\frac{\beta_W-1}{\beta_W+1}\right)\right)\;,
\end{align}
where we define
\begin{eqnarray}
g_f=\frac{2M_f^2}{\vev\,\s}\;,\qquad r_f=\frac{\s}{M_f^2}\qquad\text{and}\qquad\beta_f=\sqrt{1-\frac{4M_f^2}{\s+i0}}\;,
\end{eqnarray}
with $M_f$ the mass of the particle $f$, and $\vev$ the vacuum expectation value of the Higgs field.

\section{Universal dual amplitudes for $gg\to H$ and $H\to\gamma\gamma$}\label{Section:HOLUniversalDualAmplitudes}
\fancyhead[LO]{\ref*{Section:HOLUniversalDualAmplitudes}~~\nameref*{Section:HOLUniversalDualAmplitudes}}

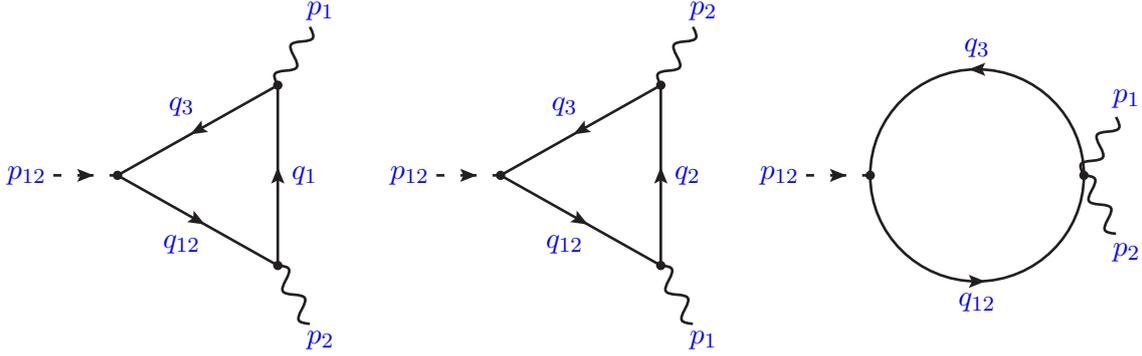
\begin{figure}
		\centering
		\begin{picture}(140,150)(-20,-75)
		\SetWidth{1}
		\DashLine[arrow](0,0)(24,0){6}
		\Line[arrow](84,34)(24,0)
		\Line[arrow](24,0)(84,-34)
		\Line[arrow](84,-34)(84,34)
		\Photon(84,34)(96,56){3}{2}
		\Photon(84,-34)(96,-56){3}{2}
		\Vertex(84,34){1.75}
		\Vertex(24,0){1.75}
		\Vertex(84,-34){1.75}
		\color{blue}
		\Text(-10,0){$p_{12}$}
		\Text(100,62){$p_1$}
		\Text(100,-62){$p_2$}
		\Text(48,26){$q_3$}
		\Text(94,0){$q_1$}
		\Text(48,-26){$q_{12}$}
		\end{picture}
		\hfill
		\begin{picture}(140,150)(-20,-75)
		\SetWidth{1}
		\DashLine[arrow](0,0)(24,0){6}
		\Line[arrow](84,34)(24,0)
		\Line[arrow](24,0)(84,-34)
		\Line[arrow](84,-34)(84,34)
		\Photon(84,34)(96,56){3}{2}
		\Photon(84,-34)(96,-56){3}{2}
		\Vertex(84,34){1.75}
		\Vertex(24,0){1.75}
		\Vertex(84,-34){1.75}
		\color{blue}
		\Text(-10,0){$p_{12}$}
		\Text(100,62){$p_2$}
		\Text(100,-62){$p_1$}
		\Text(48,26){$q_3$}
		\Text(94,0){$q_2$}
		\Text(48,-26){$q_{12}$}
		\end{picture}
		\hfill
		\begin{picture}(140,150)(-15,-75)
		\SetWidth{1}
		\Arc[arrow](64,0)(40,0,180)
		\Arc[arrow](64,0)(40,180,360)
		\DashLine[arrow](0,0)(24,0){6}
		\Photon(104,0)(116,22){3}{2}
		\Photon(104,0)(116,-22){3}{2}
		\Vertex(104,0){1.75}
		\Vertex(24,0){1.75}
		\color{blue}
		\Text(-10,0){$p_{12}$}
		\Text(120,29){$p_1$}
		\Text(120,-29){$p_2$}
		\Text(64,48){$q_3$}
		\Text(64,-48){$q_{12}$}
		\end{picture}
	\caption{One-loop Feynman diagrams contributing to the $H\to\gamma\gamma$ amplitude. The particle running inside the loop can be a top quark (left and middle diagram only), a charged scalar (all three diagrams) or a $W$ gauge boson (idem).}
	\label{Figure:HOLDiagrams}
\end{figure}

For the evaluation of the relevant one-loop quantities, we write the internal momenta as $q_1=\ell+p_1$, $q_2=\ell+p_2$, $q_3=\ell$ and $q_{12}=\ell+p_{12}$, with $\ell$ the loop momentum. Note that these notations differ from the previous chapters\footnote{They also differ from~\cite{Driencourt-Mangin:2017gop}. We modified them so the labellings match the ones we will use in \Chapter{\ref{Chapter:HTL}}.}, as this time we need to work with an additional internal propagator whose momenta is $\ell+p_2$, to account for the diagrams where the two photons or gluons are exchanged (see \Fig{\ref{Figure:HOLDiagrams}}).\\
\\
In that respect, the one-loop amplitude with virtual top quarks as internal particles can be written
\begin{align}\label{Equation:HOLAmptop}
\mathcal{A}_{\mu\nu}^{(1,t)}=&~\frac{M_t}{\vev}\int_\ell\left(\,\prod\limits_{i=3,12}\,G_F(q_i)\right)\Tr\left[(\slashed{q}_3+M_t)\,\gamma^\mu\,(\slashed{q}_1+M_t)\,\gamma^\nu\,(\slashed{q}_{12}+M_t)\right]G_F(q_1)\nn\\
&~+\{(1,\mu)\leftrightarrow(2,\nu)\}
\end{align}
before projection. Similarly, we can write for charged scalars
\begin{align}\label{Equation:HOLAmpphi}
\mathcal{A}_{\mu\nu}^{(1,\phi)}=&~\frac{2M_\phi}{\vev}\int_\ell\left(\,\prod\limits_{i=3,12}\,G_F(q_i)\right)\Big[\left(q_1+q_3\right)^\mu\left(q_1+q_{12}\right)^\nu\,G_F(q_1)-g^{\mu\nu}\Big]\nn\\
&~+\{(1,\mu)\leftrightarrow(2,\nu)\}\;,
\end{align}
where the contribution of the third diagram of \Fig{\ref{Figure:HOLDiagrams}} has been split into two equal, symmetric parts\footnote{One can notice that for the charged scalar, the third diagram does not contribute whatsoever to the amplitude, as we have $g_{\mu\nu}\,P_1^{\mu\nu}=0$. Nevertheless, it still contributes to $\mathcal{A}_2^{(1,\phi)}$.}. We do not provide here the explicit expression of $\mathcal{A}_{\mu\nu}^{(1,W)}$, as it is quite heavy. It can easily be obtained using the Feynman rules of \Appendix{\ref{Section:APPFeynmanRules}} though. It is worth mentioning that for the amplitude involving $W$ bosons, we work in the unitary gauge\footnote{In this gauge, the propagating degrees of freedom are physical, i.e. there is no need to consider ghost particles.} for which the internal propagators
\begin{equation}\label{Equation:HOLPropW}
-i\left(g^{\mu\nu}-\frac{q_i^\mu\,q_i^\nu}{M_W^2}\right)\frac{1}{q_i^2-M_W^2+i0}
\end{equation}
do not introduce additional poles in the complex loop momentum space, allowing for an easier application of the LTD theorem. This choice of gauge, however, makes partial results for the $W$ more divergent in the UV compared to the top quark and the charged scalar. Indeed, additional powers are introduced by the $WW\gamma$ vertex, which is linear in the loop momentum, and the $W$ propagator, as shown in \Eq{\ref{Equation:HOLPropW}}. Besides, the $W$ amplitude receives a contribution from the bubble diagrams that involves a $WW\gamma\gamma$ interaction vertex, which does not exist for the top quark. However, it is a very remarkable feature of LTD that by setting the internal propagators on shell, the degree of divergence of intermediate expressions is reduced automatically. Moreover, the cross-cancellations between the three diagrams --~which are expected from gauge invariance~-- are explicit within the LTD formalism, without extra manipulations needed. In the end, each dual amplitude for all three different internal particles have the same degree of divergence in the UV.\\
\\
Even more remarkably, and as a partial consequence of what has just been discussed, the expressions of $\mathcal{A}_1^{(1,f)}$ and $\mathcal{A}_2^{(1,f)}$ exhibit the exact same functional form, regardless of the internal particle $f$. Indeed, the expressions of the dual contributions read
\begin{align}\label{Equation:HOLA2Dual}
\mathcal{A}_2^{(1,f)}\big(\deltatilde{q_i}\big)=&~g_f\frac{c_3^{(f)}}{2}\int_\ell\,\deltatilde{q_i}\;,\quad &&~i\in\{1,2\}\;,\nn\\
\mathcal{A}_2^{(1,f)}\big(\deltatilde{q_j}\big)=&~-g_f\frac{c_3^{(f)}}{2}\int_\ell\,\deltatilde{q_j}\;,\quad &&~j\in\{3,12\}\;,
\end{align}
for the second, simpler coefficient, and
\begin{align}\label{Equation:HOLA1Dual}
\mathcal{A}_1^{(1,f)}\big(\deltatilde{q_i}\big)=&~g_f\int_\ell\,\deltatilde{q_i}\left(\frac{\s\,M_f^2}{(2q_i\cdot p_1)(2q_i\cdot p_2)}\,c_1^{(f)}+c_2^{(f)}\right)\;,\quad i\in\{1,2\}\;,\nn\\
\mathcal{A}_1^{(1,f)}\big(\deltatilde{q_3}\big)=&~g_f\int_\ell\,\deltatilde{q_3}\frac{\s}{\s+2q_3\cdot p_{12}}\nn\\
&~\times\left(\left(-\frac{\s\,M_f^2}{(2q_3\cdot p_1)(2q_3\cdot p_2)}\,c_1^{(f)}-c_2^{(f)}\right)\frac{2q_3\cdot p_{12}}{\s}+c_3^{(f)}\right)\;,\nn\\
\mathcal{A}_1^{(1,f)}\big(\deltatilde{q_{12}}\big)=&~g_f\int_\ell\,\deltatilde{q_{12}}\frac{\s}{\s-2q_{12}\cdot p_{12}+i0}\nn\\
&~\times\left(\left(\frac{\s\,M_f^2}{(2q_{12}\cdot p_1)(2q_{12}\cdot p_2)}\,c_1^{(f)}+c_2^{(f)}\right)\frac{2q_{12}\cdot p_{12}}{\s}+c_3^{(f)}\right)\;,
\end{align}
for the first one. The parameters $c_i^{(f)}$ are scalar coefficients, and they are written $c_i^{(f)}=c_{i,0}^{(f)}+c_{i,1}^{(f)}\,r_f$ where we recall $r_f=\s/M_f^2$ is the dimensionless ratio of the two scales associated with this process. For the three flavours we consider, namely $f=\phi,t,W$, these coefficients are given by
\begin{align}\label{Equation:HOLcif}
c_{1,0}^{(f)}=&~\left(\frac{4}{d-2},-\frac{8}{d-2},\frac{4(d-1)}{d-2}\right)\;,\quad&c_{1,1}^{(f)}=&~\left(0,1,\frac{2(5-2d)}{d-2}\right)\;,\nn\\
c_{2,0}^{(f)}=&~\left(-\frac{2d}{d-2},\frac{4d}{d-2},-\frac{2d(d-1)}{d-2}\right)\;,\quad&c_{2,1}^{(f)}=&~\left(0,0,\frac{d-4}{d-2}\right)\;,\nn\\
c_{3,0}^{(f)}=&~(d-2)c_{1,0}^{(f)}\;,\quad&c_{3,1}^{(f)}=&~0\;.
\end{align}
We also define the coefficient $c_{23}^{(f)}=c_2^{(f)}+c_3^{(f)}$ that fulfils
\begin{equation}\label{Equation:HOLc23f}
c_{23,0}^{(f)}=\frac{d-4}{2}\,c_{1,0}^{(f)}\;,\qquad c_{23,1}^{(f)}=\left(0,0,\frac{d-4}{d-2}\right)\;.
\end{equation}
It will be used in future expressions.\\
\\
The fact that the unintegrated amplitudes for different flavours all share the same structure suggests that the amplitude for the different virtual states could be reduced to the determination of the scalar coefficients $c_i^{(f)}$. The universality of the expressions in \Eqs{\ref{Equation:HOLA2Dual}}{\ref{Equation:HOLA1Dual}} could be supported by supersymmetric Ward identities at tree level similar to those relating amplitudes with heavy quarks and heavy scalars~\cite{Schwinn:2006ca,Ferrario:2006np}, as the dual representation is indeed a tree-level like object. It is also interesting to notice that the two-loop amplitudes for scalar and pseudoscalar Higgs bosons decay to two photons have been calculated in~\cite{Harlander:2005rq} based on the assumption that if two physical processes correspond to a similar set of Feynman diagrams, then their cross sections must be described by a common set of analytic functions. Their calculation is thus simply reduced to the determination of the coefficients of a linear combination of those functions, which is done by solving a large set of linear equations arising from comparing the asymptotic expansions of a given ansatz and a one-dimensional integral representation of the amplitude. Such argument could also be used for the case we are currently discussing, since similar physical processes should be described by similar integrand representations, albeit with different coefficients. In \Chapter{\ref{Chapter:HTL}}, in which we compute the two-loop corrections of the $H\to\gamma\gamma$ process, we will see that, for the most part, a similar universality holds.\\
\\
In what follows, we show that it is possible to simplify by a significant margin the expressions in \Eq{\ref{Equation:HOLA1Dual}} -- and therefore their sum. First, by working in the centre-of-mass frame of the Higgs boson, in which
$p_{1,0}=p_{2,0}$ and $\mathbf{p}_1=-\mathbf{p}_2$, we can see that
\begin{align}\label{Equation:HOLSimpDen}
\frac{1}{(2q_1\cdot p_1)(2q_1\cdot p_2)}=&~\frac{1}{4\big((\Sup{q_{1,0}}\,p_{1,0})^2-(\mathbf{q}_1\cdot\mathbf{p}_1)\big)^2}=\frac{1}{4\big(((\boldsymbol{\ell}+\mathbf{p}_1)^2+M_f^2)p_{1,0}^2-((\boldsymbol{\ell}+\mathbf{p}_1)\cdot\mathbf{p}_1)^2\big)}\nn\\
=&~\frac{1}{4\big((\boldsymbol{\ell}^2+M_f^2)p_{1,0}^2-(\boldsymbol{\ell}\cdot\mathbf{p}_1)^2\big)}=\frac{1}{(2\ell\cdot p_1)(2\ell\cdot p_2)}\;,
\end{align}
by recalling that $\mathbf{p}_1^2=p_{1,0}^2$. The same exercise can be carried out for $q_2$, by simply exchanging the role of $p_1$ and $p_2$. We now need to unify the coordinate system, but in this case, as we did not parametrise and integrate each dual integrand individually, the unification is straightforward. Indeed, it is enough to simply rewrite all momenta in terms of the internal momentum $\ell$, as long as the respective on-shell $\deltatilde{q_i}$ dependence is kept. And finally, by noting that in the centre-of-mass frame
\begin{equation}\label{Equation:HOLDeltatildesEquality}
\deltatilde{q_3}F(q_3)=\deltatilde{q_{12}}F(q_{12})=\deltatilde{\ell}F(\ell)
\end{equation}
for any Lorentz-invariant function $F$, as well as
\begin{equation}
\deltatilde{q_i}=\deltatilde{\ell}\,\frac{\Sup{\ell_0}}{\Sup{q_{i,0}}}\qquad\text{for }i\in\{1,2\}\;,
\end{equation}
we obtain
\begin{align}\label{HOLA2}
\mathcal{A}_2^{(1,f)}=&~\mathcal{A}_2^{(1,f)}\big(\deltatilde{q_1}\big)+\mathcal{A}_2^{(1,f)}\big(\deltatilde{q_2}\big)+\mathcal{A}_2^{(1,f)}\big(\deltatilde{q_3}\big)+\mathcal{A}_2^{(1,f)}\big(\deltatilde{q_{12}}\big)\nn\\
=&~g_f\frac{c_3^{(f)}}{2}\int_\ell\,\deltatilde{\ell}\left(\frac{\Sup{\ell_0}}{\Sup{q_{1,0}}}+\frac{\Sup{\ell_0}}{\Sup{q_{2,0}}}-2\right)\;,
\end{align}
and
\begin{align}\label{Equation:HOLA1}
\mathcal{A}_1^{(1,f)}=&~\mathcal{A}_1^{(1,f)}\big(\deltatilde{q_1}\big)+\mathcal{A}_1^{(1,f)}\big(\deltatilde{q_2}\big)+\mathcal{A}_1^{(1,f)}\big(\deltatilde{q_3}\big)+\mathcal{A}_1^{(1,f)}\big(\deltatilde{q_{12}}\big)\nn\\
=&~g_f\int_\ell\,\deltatilde{\ell}\Bigg[\left(\frac{\Sup{\ell_0}}{\Sup{q_{1,0}}}+\frac{\Sup{\ell_0}}{\Sup{q_{2,0}}}+\frac{2(2\ell\cdot p_{12})^2}{\s^2-(2\ell\cdot p_{12}-i0)^2}\right)\left(\frac{\s\,M_f^2}{(2\ell\cdot p_1)(2\ell\cdot p_2)}\,c_1^{(f)}+c_2^{(f)}\right)\nn\\
&~+\frac{2\s^2}{\s^2-(2\ell\cdot p_{12}-i0)^2}\,c_3^{(f)}\Bigg]\;,
\end{align}
where the on-shell loop energies are given by
\begin{equation}\label{Equation:HOLOnShellEnergies}
\Sup{q_{1,0}}=\sqrt{(\boldsymbol{\ell}+\mathbf{p}_1)^2+M_f^2}\;,\quad\Sup{q_{2,0}}=\sqrt{(\boldsymbol{\ell}+\mathbf{p}_2)^2+M_f^2}\quad\text{and}\quad\Sup{\ell_0}=\sqrt{\boldsymbol{\ell}^2+M_f^2}\;.
\end{equation}
Furthermore, we know that the coefficient $\mathcal{A}_2^{(1,f)}$ vanishes upon integration in $d$ dimensions\footnote{A naive calculation with $d=4$, though, would lead to $\mathcal{A}_2^{(1,f)}\neq0$, violating gauge invariance. As will we see in \Section{\ref{Section:HOLRenormalisation}}, it is only after properly renormalising $\mathcal{A}_2^{(1,f)}$ that it vanishes if one takes the limit $d=4$ before integrating.}. We can exploit this property to simplify the integrand-level expressions of $\mathcal{A}_1^{(1,f)}$ by subtracting a well-chosen multiple of $\mathcal{A}_2^{(1,f)}$ before integrating. The transformation
\begin{equation}\label{Equation:HOLTransfA1}
\mathcal{A}^{(1,f)}\to\mathcal{A}_1^{(1,f)}-2\frac{c_2^{(f)}}{c_3^{(f)}}\mathcal{A}_2^{(1,f)}
\end{equation}
reduces the number of independent scalar coefficients that are necessary to describe $\mathcal{A}_1^{(1,f)}$ from three to two. Indeed, applying this transformation leads to
\begin{align}\label{Equation:HOLA}
\mathcal{A}^{(1,f)}=&~g_f\,\s\int_\ell\,\deltatilde{\ell}\Bigg[\left(\frac{\Sup{\ell_0}}{\Sup{q_{1,0}}}+\frac{\Sup{\ell_0}}{\Sup{q_{2,0}}}+\frac{2(2\ell\cdot p_{12})^2}{\s^2-(2\ell\cdot p_{12}-i0)^2}\right)\frac{M_f^2}{(2\ell\cdot p_1)(2\ell\cdot p_2)}\,c_1^{(f)}\nn\\
&~+\frac{2\s}{\s^2-(2\ell\cdot p_{12}-i0)^2}\,c_{23}^{(f)}\Bigg]\;,
\end{align}
which depends only on $c_1^{(f)}$ and $c_{23}^{(f)}$, that have been defined in \Eq{\ref{Equation:HOLcif}} and \Eq{\ref{Equation:HOLc23f}}, respectively. Notice that while they are labelled differently because they do not have the same integrand-level expressions, we have $\mathcal{A}^{(1,f)}=\mathcal{A}_1^{(1,f)}$ after integration in $d$ dimensions.\\
\\
In the next section, we will discuss how to implement a completely local renormalisation to achieve integrability in four dimensions, and explain why it is necessary in order to recover the result obtained in the traditional approach.

\section{Local renormalisation and four-dimensional representation}\label{Section:HOLRenormalisation}
\fancyhead[LO]{\ref*{Section:HOLRenormalisation}~~\nameref*{Section:HOLRenormalisation}}

Since there is no tree-level equivalent for the $gg\to H$ and $H\to\gamma\gamma$ processes, the one-loop amplitude is expected to be finite. However, a naive four-dimensional approach --~that is, simply setting $d=4$ at the \emph{integrand level} --~leads to a different result than the one obtained using the traditional approach. In order to understand the discrepancy, let's have a closer look at \Eq{\ref{Equation:HOLA}}. The part proportional to $c_1^{(f)}$ is finite in the UV thanks to cancellations occurring inside the parenthesis. This means that evaluating this integral in $d$ or 4 dimensions would lead to the same result, up to $\Oep{1}$. On the other hand, the second part, proportional to $c_{23}^{(f)}$ exhibits a logarithmically divergent term, that would manifest itself as an $\epsilon$-pole. This pole, however, is multiplied by $c_{23}^{(f)}\propto (d-4)$, leading to a non-zero finite part that has to be taken into account. Therefore, since the coefficient $c_{23}^{(f)}$ depends on the nature of the particle circulating inside the loop through the associated Feynman rules, a consistent treatment of the dimensional extension of Dirac and Lorentz algebra is necessary to avoid a potential mismatch in the final result. In fact, if we take $d=4$ from the very first steps of the calculation, any term proportional to $c_{23}^{(f)}$ does not appear, which would eventually be the source of the discrepancy. From the mathematical point of view, this behaviour is due to the \emph{non-integrability} of the unregularised corresponding integral. Without a proper counterterm, the limit $d\to4$ is not continuous, and does not commute with the integral sign. The integrated result being finite is therefore not enough to guarantee a straightforward four-dimensional treatment of a given integral\footnote{This is the counterexample we talked about at the beginning of \Section{\ref{Section:FDUMDecayRateFDU}}.}.\\
\\
In what follows, we will build the appropriate counterterm needed to cancel the local UV behaviour of the one-loop amplitude. We will also ensure that this counterterm integrates to 0 in DREG, as in the traditional method no actual renormalisation is needed. To do so, we use a slightly different approach that the one used in the previous chapters. Instead of expanding around the UV propagator $G_F(q_\uv)$ at Feynman integral level, we do it after applying LTD, i.e. directly from \Eq{\ref{Equation:HOLA}}. As it was discussed before, the leading contribution is the one proportional to $c_{23}^{(f)}$. We therefore directly obtain the dual representation of the UV counterterm, which reads
\begin{equation}\label{Equation:HOLAUV}
\mathcal{A}_{\uv}^{(1,f)}=-g_f\,\s\,c_{23}^{(f)}\int_\ell\,\deltatilde{q_\uv}\frac{1}{2(\Sup{q_\uv})^{2}}\left(1+\frac{1}{(\Sup{q_\uv})^{2}}\frac{3\mu_\uv^2}{d-4}\right)\;,
\end{equation}
with $\Sup{q_\uv}=\sqrt{\boldsymbol{\ell}^2+\mu_\uv^2}$. The term proportional to $\mu_\uv^2$ in \Eq{\ref{Equation:HOLAUV}} is subleading\footnote{Strictly speaking, it is only \emph{locally} subleading, as both terms will generate the same, opposite finite parts after integration.} in the UV and has been adjusted to fix the finite part so the counterterm indeed integrates to 0 in $d$ dimensions. We want to emphasise that this last step was the only one of the entire calculation for which DREG was necessary. Moreover, as we explained in previous chapters, once a given unintegrated local counterterm has been derived, it can be used to regularise any other similar process. Note that in this particular case, the renormalisation scale $\mu_\uv^2$ is arbitrary, as its dependence vanishes after integration.\\
\\
Finally, the renormalised amplitude is obtained by subtracting \Eq{\ref{Equation:HOLAUV}} from \Eq{\ref{Equation:HOLA}}, and setting $d$ to 4. Explicitly,
\begin{equation}\label{Equation:HOLAR}
\left.\mathcal{A}_{\r}^{(1,f)}\right|_{d=4}\left.=\left(\mathcal{A}^{(1,f)}-\mathcal{A}_\uv^{(1,f)}\right)\right|_{d=4}\;.
\end{equation}
Notice that, because it is proportional to $c_{23}^{(f)}$, the first term of the integrand in \Eq{\ref{Equation:HOLAUV}} vanishes in four dimensions. The second term, however, is proportional to $\hat{c}_{23}^{(f)}=c_{23}^{(f)}/(d-4)$ and leads to a finite contribution.\\
\\
We could also apply the same procedure to locally regularise $\mathcal{A}_2^{(1,f)}$, which would make it integrate to 0 in four dimensions. It is not needed for the rest of the calculation, though, and would actually be redundant, since this term has already been included inside the definition of $\mathcal{A}^{(1,f)}$ given in \Eq{\ref{Equation:HOLTransfA1}}. Moreover, the structure of the local UV behaviours of $\mathcal{A}_1^{(1,f)}$ and $\mathcal{A}_2^{(1,f)}$ taken separately is more complex than the one of $\mathcal{A}^{(1,f)}$, as many cancellations occur when applying \Eq{\ref{Equation:HOLTransfA1}}.\\
\\
Before moving forward, it is interesting to note that Dyson's prescription --~which consists in subtracting the amplitude evaluated with vanishing external momenta~-- has a similar regularisation effect for internal top quarks and charged scalars, for which the correct result is recovered. It fails, however, for internal $W$ bosons. Indeed, although the non-decoupling term in the limit $M_W/\s\to\infty$ is properly subtracted, there is a mismatch inside the $\mathcal{O}\big((M_W/\s)^0\big)$ part. In our formalism, the discrepancy is very easily explained by the fact that $c_{23,1}^{(W)}\neq0$ (while $c_{23,1}^{(t)}=c_{23,1}^{(\phi)}=0$). Setting $p_1^2=p_2^2=0$ is more or less equivalent to taking $r\ll1$, and in this limit, $c_{23,1}^{(f)}$ is subleading and thus not accounted for. One can check that the difference between the two approaches is indeed generated by the locally divergent term proportional to $c_{23,1}^{(W)}$.\\
\\
Now that we have everything we need, we can write the explicit expression of the unintegrated renormalised amplitude. Explicitly,
\begin{align}\label{Equation:HOLARFinal}
\left.\mathcal{A}_\r^{(1,f)}\right|_{d=4}=&~g_f\,\s\int_{\boldsymbol{\ell}}\,\Bigg[\frac{1}{2\Sup{\ell_0}}\left(\frac{\Sup{\ell_0}}{\Sup{q_{1,0}}}+\frac{\Sup{\ell_0}}{\Sup{q_{2,0}}}+\frac{2(2\ell\cdot p_{12})^2}{\s-(2\ell\cdot p_{12}-i0)^2}\right)\frac{M_f^2}{(2\ell\cdot p_1)(2\ell\cdot p_2)}\,c_1^{(f)}\nn\\
&+\frac{3\mu_\uv^2}{4(\Sup{q_{\uv,0}})^5}\,\hat{c}_{23}^{(f)}\Bigg]\;,
\end{align}
where the two coefficients $c_1^{(f)}$ and $\hat{c}_{23}^{(f)}$ have been evaluated at $d=4$, i.e.
\begin{equation}\label{Equation:HOLCoeffsd}
c_1^{(f)}=(2,-4+r_t,6-r_W)\;,\qquad\text{and}\qquad\hat{c}_{23}^{(f)}=(1,-2,3+r_W/2)\;.
\end{equation}
The measure of the integral in \Eq{\ref{Equation:HOLARFinal}} is defined in the spatial component of the loop momentum, $\boldsymbol{\ell}$, for which we have $\int_\ell=\int\,d^{d-1}\boldsymbol{\ell}/(2\pi)^{d-1}$. For the explicit integration, we parametrise the loop three-momentum as
\begin{equation}\label{Equation:HOLParam}
\boldsymbol{\ell}=M_f\,\xi\,(2\sqrt{v(1-v)}\,\mathbf{e}_\perp,1-2v)\;,
\end{equation}
with $\xi$ its normalised modulus\footnote{Notice that the normalisation is different from the one we have used in previous chapters. In this case, normalising by the mass $M_f$ instead of the centre-of-mass frame energy $\sqrt{\s}/2$ slightly simplifies the integration. Of course, it does not modify the final result.}, and $\mathbf{e}_\perp$, the unit vector in the transverse plane. The dual integration measure thus becomes
\begin{equation}\label{Equation:HOLDualMeasure}
\int_\ell\,\deltatilde{\ell}=\int_{\boldsymbol{\ell}}\,\frac{1}{2\Sup{\ell_0}}=\frac{M_f^2}{4\pi^2}\int_0^\infty\,\xi_0^{-1}\,\xi^2\,d\xi\int_0^1\,dv\;,
\end{equation}
with $\xi_0=\sqrt{\xi^2+1}$. The square roots that appear in \Eq{\ref{Equation:HOLARFinal}} can be transformed into rational functions of a new variable $x$ by implementing the change of variables
\begin{equation}\label{Equation:HOLCOV}
\xi=\frac{1}{2}\left(\sqrt{x}-\frac{1}{\sqrt{x}}\right)\;,
\end{equation}
with $x\in[1,\infty)$. The integrated amplitude reads
\begin{equation}\label{Equation:HOLARIntegrated}
\left.\mathcal{A}_\r^{(1,f)}\right|_{d=4}=\frac{g_f\,\s}{16\pi^2}\left(\frac{M_f^2}{\s}\log\left(\frac{\beta_f-1}{\beta_f+1}\right)c_1^{(f)}+2\hat{c}_{23}^{(f)}\right)\;,
\end{equation}
which agrees with the results given in \Eq{\ref{Equation:HOLLiteratureResults}}.

\section{Asymptotic expansions in the Euclidean space of the loop three-momentum}\label{Section:HOLAsymptoticExpansions}
\fancyhead[LO]{\ref*{Section:HOLAsymptoticExpansions}~~\nameref*{Section:HOLAsymptoticExpansions}}

As already said several times in this thesis, LTD at one loop reduces the original $d$-dimensional integration domain with Minkowski metric into a $(d-1)$-dimensional domain with Euclidean metric. In the particular case where $d=4$, this domain corresponds to the loop three-momentum space. This is a very interesting feature that allows one to circumvent potential difficulties that arise when performing asymptotic expansions of the integrand in a Minkowski space~\cite{Beneke:1997zp,Smirnov:2002pj}. The $gg\to H$ and $H\to\gamma\gamma$ processes can be used as benchmark examples to illustrate the ease of performing asymptotic expansions within the LTD formalism. The method that will be discussed in the following is also applicable to other, more complicated processes.\\
\\
We will first start by considering the large-mass limit, i.e. $M_f^2\gg\s$, or equivalently $r_f\ll1$. As an example and starting point, let's consider the dual contribution where $q_3$ in on shell. More specifically, we consider the quantity
\begin{equation}\label{Equation:HOLDualq3BeforeExpansion}
\deltatilde{q_3}\,G_D(q_3;q_2)=\frac{\deltatilde{q_3}}{\s+2q_3\cdot p_{12}}\;,
\end{equation}
where $q_3\cdot p_{12}=\Sup{q_{3,0}}\,\sqrt{\s}$ in the centre-of-mass frame, with the on-shell energy $\Sup{q_{3,0}}$ found in \Eq{\ref{Equation:HOLOnShellEnergies}}. By remarking that we always have $\Sup{q_{3,0}}\geq M_f$ regardless of the value of $\boldsymbol{\ell}$, asymptotically expanding \Eq{\ref{Equation:HOLDualq3BeforeExpansion}} is straightforward and leads to
\begin{equation}\label{Equation:HOLDualq3Expanded}
\deltatilde{q_3}\,G_D(q_3;q_2)=\frac{\deltatilde{q_3}}{2q_3\cdot p_{12}}\,\sum\limits_{n=0}^{\infty}\,\left(-\frac{\s}{2q_3\cdot p_{12}}\right)^n\;.
\end{equation}
Likewise, we have
\begin{equation}\label{Equation:HOLEnergyFracExpanded}
\frac{\Sup{\ell_0}}{\Sup{q_{1,0}}}=\,\sum\limits_{n=0}^\infty\,\frac{\Gamma(2n+1)}{\Gamma(n+1)}\left(-\frac{2\boldsymbol{\ell}\cdot \mathbf{p}_1+\mathbf{p}_1^2}{2(\Sup{\ell_0})^2}\right)^n
\end{equation}
which is also a valid expansion since $\Sup{\ell_0}\geq M_f$. Notice that each consecutive term of the expansions in \Eqs{\ref{Equation:HOLDualq3Expanded}}{\ref{Equation:HOLEnergyFracExpanded}} is less singular in the UV, and completely regular in the IR. Taking into account the aforementioned, we can expand the unintegrated renormalised amplitude in \Eq{\ref{Equation:HOLARFinal}}, which reads, for the high-mass limit,
\begin{equation}\label{Equation:HOLARHighMassExpansion}
\left.\mathcal{A}_\r^{(1,f)}\right|_{d=4}^{r_f\ll1}=g_f\,\s\int_\ell\,\left[\frac{M_f^2}{4(\Sup{\ell_0})^5}\left(\,\sum\limits_{n=1}^{\infty}\,\mathcal{Q}_n(z)\left(\frac{\s}{(2\Sup{\ell_0})^2}\right)^{n-1}\right)c_1^{(f)}+\frac{3\mu_\uv^2}{4(\Sup{q_{\uv,0}})^5}\,\hat{c}_{23}^{(f)}\right]\;,
\end{equation}
with
\begin{equation}\label{Equation:HOLQn}
\mathcal{Q}_n(z)=\frac{P_{2n}(z)-1}{1-z^2}\;,
\end{equation}
where $P_{2n}$ are the Legendre polynomials, and
\begin{equation}\label{Equation:HOLz}
z=\frac{2\boldsymbol{\ell}\cdot \mathbf{p}_1}{\Sup{\ell_0}\,\sqrt{\s}}=\frac{\xi(1-2v)}{\xi_0}\;,
\end{equation}
where we used the parametrisation in \Eq{\ref{Equation:HOLParam}} to write the last equality. Integrating the asymptotic expansion in \Eq{\ref{Equation:HOLARHighMassExpansion}} can be very easily done analytically, and we find for the lowest orders
\begin{align}\label{Equation:HOLHighMassExpansionIntegrated}
\left.\mathcal{A}_\r^{(1,f)}\right|_{d=4}^{r_f\ll1}=&~\frac{\s}{8\pi^2\vev}\Bigg[\frac{2\hat{c}_{23,0}^{(f)}-c_{1,0}^{(f)}}{r_f}+\hat{c}_{23,1}^{(f)}-\frac{c_{1,0}^{(f)}}{12}-c_{1,1}^{(f)}-\left(\frac{c_{1,0}^{(f)}}{90}+\frac{c_{1,1}^{(f)}}{12}\right)r_f\nn\\
&~-\left(\frac{c_{1,0}^{(f)}}{560}+\frac{c_{1,1}^{(f)}}{90}\right)r_f^2-\left(\frac{c_{1,0}^{(f)}}{3150}+\frac{c_{1,1}^{(f)}}{560}\right)r_f^3-\left(\frac{c_{1,0}^{(f)}}{16632}+\frac{c_{1,1}^{(f)}}{3159}\right)r_f^4+\mathcal{O}(r_f^5)\Bigg]\;.
\end{align}
As we can see, the $1/r_f$, non-decoupling term vanishes thanks to the fact $\hat{c}_{23,0}^{(f)}=c_{1,0}^{(f)}/2$ (see \Eq{\ref{Equation:HOLc23f}}). For all the three different internal particles, we explicitly have
\begin{align}
\left.\mathcal{A}_\r^{(1,\phi)}\right|_{d=4}^{r\ll1}=&~\frac{\s}{16\pi^2\vev}\left(-\frac{1}{3}-\frac{2}{45}r_\phi-\frac{1}{140}r_\phi^2-\frac{2}{1575}r_\phi^3-\frac{1}{4158}r_\phi^4+\mathcal{O}(r_\phi^5)\right)\;,\nn\\
\left.\mathcal{A}_\r^{(1,t)}\right|_{d=4}^{r\ll1}=&~\frac{\s}{16\pi^2\vev}\left(-\frac{4}{3}-\frac{7}{90}r_t-\frac{1}{126}r_t^2-\frac{13}{12600}r_t^3-\frac{8}{51975}r_t^4+\mathcal{O}(r_t^5)\right)\;,\nn\\
\left.\mathcal{A}_\r^{(1,W)}\right|_{d=4}^{r\ll1}=&~\frac{\s}{16\pi^2\vev}\left(7+\frac{11}{30}r_W+\frac{19}{420}r_W^2+\frac{29}{4200}r_W^3+\frac{41}{34650}r_W^4+\mathcal{O}(r_W^5)\right)\;,
\end{align}
which agrees with~\cite{Aglietti:2006tp}.\\
\\
A similar work can be carried out for the small mass limit, i.e. $M_f^2\ll\s$, or $r_f\gg1$ for which we have
\begin{align}\label{Equation:HOLARSmallMassExpansion}
\left.\mathcal{A}_\r^{(1,f)}\right|_{d=4}^{r_f\gg1}=&~g_f\,\s\int_\ell\Bigg[-\frac{m_f^2}{(\Sup{\ell_0})^3(1-z^2)}\left(1-\,\sum\limits_{n=1}^\infty\frac{4(\Sup{\ell_0})^2\big(\s\,m_f^2(2-m_f^2)\big)^{n-1}}{\big(4\boldsymbol{\ell}^2-\s(1+m_f^2)^2\big)^n}\right)c_1^{(f)}\nn\\
&~+\frac{3\mu_\uv^2}{4(\Sup{q_{\uv,0}})^5}\,\hat{c}_{23}^{(f)}\Bigg]\;,
\end{align}
with $m_f^2=-M_f^2/(\s+i0)$, and $z$ given in \Eq{\ref{Equation:HOLz}}. Once again, each consecutive term of the expansion in \Eq{\ref{Equation:HOLARSmallMassExpansion}} is less singular in the UV, allowing for a full calculation with $d=4$. However, it is important to note that the parametrisation given in \Eq{\ref{Equation:HOLParam}} is not any more suitable in the small mass limit, as $\boldsymbol{\ell}\sim M_f\,\xi$ would be neglected before $\s$, although $|\boldsymbol{\ell}|\in[0,\infty)$ is unconstrained. Instead, we use
\begin{equation}\label{Equation:HOLParamBis}
\boldsymbol{\ell}=\frac{\sqrt{\s}}{2}\,\xi\,(2\sqrt{v(1-v)}\,\mathbf{e}_\perp,1-2v)\;,
\end{equation}
for which $\xi_0=\sqrt{\xi^2-m_f^2}$. Integrating \Eq{\ref{Equation:HOLARSmallMassExpansion}} leads to 
\begin{align}\label{Equation:HOLSmallMassExpansionIntegrated}
\left.\mathcal{A}_\r^{(1,f)}\right|_{d=4}^{r_f\gg1}=&~\frac{\s}{8\pi\vev}\Bigg[2\hat{c}_{23,1}^{(f)}-\left(2\hat{c}_{23,0}^{(f)}+c_{1,1}^{(f)}\,L_f^2\right)m_f^2\nn\\
&~+\left(c_{1,0}^{(f)}\,L_f^2+4c_{1,1}^{(f)}\,L_f\right)m_f^4-\left(4c_{1,0}^{(f)}+2c_{1,1}^{(f)}(2+3L_f)\right)m_f^6\nn\\
&~+\left(2c_{1,0}^{(f)}(2+3L_f)+4c_{1,1}^{(f)}\left(3+\frac{10}{3}L_f\right)\right)m_f^8+\mathcal{O}(m_f^{10})\Bigg]\;,
\end{align}
where logarithmic contributions, given by $L_f=\log(m_f^2)$, appear. As expected, the leading term in \Eq{\ref{Equation:HOLSmallMassExpansionIntegrated}} vanishes for the charged scalar and the top quark, because indeed for these particles, $\hat{c}_{23,1}^{(f)}=0$; it leads to a constant for the $W$ for which $\hat{c}_{23,1}^(W)=1/2$. Replacing the coefficients gives, for all three different internal particles,
\begin{align}
\left.\mathcal{A}_\r^{(1,\phi)}\right|_{d=4}^{r_f\gg1}=&~\frac{M_\phi^2}{8\pi^2\vev}\big(2-2L_\phi^2\,m_\phi^2+8L_\phi\,m_\phi^4-4(2+3L_\phi)m_\phi^6\big)+\mathcal{O}(m_\phi^{10})\;,\nn\\
\left.\mathcal{A}_\r^{(1,t)}\right|_{d=4}^{r_f\gg1}=&~\frac{M_t^2}{8\pi^2\vev}\big(-4+L_t^2-4(1-L_t)L_t\,m_t^2+2(2-5L_t)m_t^4\nn\\
&~+4(1+8L_t/3)m_t^6\big)+\mathcal{O}(m_t^{10})\;,\nn\\
\left.\mathcal{A}_\r^{(1,W)}\right|_{d=4}^{r_f\gg1}=&~\frac{M_W^2}{8\pi^2\vev}\big(-m_W^{-2}+3(2-L_W^2)+6(2-L_W)L_W\,m_W^2\nn\\
&~-6(2-L_W)m_W^4+4(3+L_W)m_W^6\big)+\mathcal{O}(m_W^{10})\;,
\end{align}
which are once again in agreement with~\cite{Aglietti:2006tp}.\\
\\
In both cases -- the small and large mass limits -- all the integrand-level asymptotic expansions have been integrated directly in four dimensions. This is achievable thanks to the fact that in the Euclidean space of the loop three-momentum, we only had to consider a single kinematical region to achieve the correct asymptotic expansion in either of the two limits. In fact, this is a direct consequence of dealing with integrable and locally regularised representations of the scattering amplitudes, as there is a strict commutativity between the integral sign and the series expansions.

\section{Conclusion}\label{Section:HOLConclusion}
\fancyhead[LO]{\ref*{Section:HOLConclusion}~~\nameref*{Section:HOLConclusion}}

In this chapter, we have presented a very compact and universal integrand-level representation of the one-loop amplitude for the Higgs boson to two massless gauge bosons. The amplitude exhibits the same functional form for internal scalars, fermions and vector bosons. Presumably, if there was a way to obtain a priori the coefficients $c_i^{(f)}$ for a given process and a given particle, this universality could be exploited to bypass heavy calculations involved when dealing with more complicated gauge structures. Admittedly, at one-loop level the applications would be rather inconsequential. But if the universality holds at higher orders (and for two loops, it seems it is mostly the case, see \Chapter{\ref{Chapter:HTL}}), this could potentially open very interesting possibilities.\\
\\
The amplitude has also been locally renormalised such that a pure four-dimensional expression free from potential scheme subtleties is obtained. To do so, we have had to introduce a counterterm that integrates to 0 in $d$ dimensions. Yet, there is some arbitrariness here. Even though in the traditional approach, renormalisation is not needed for this process, the local divergence is still present. In DREG, it does not matter at all in the end since the limit $d\to4$ is taken \emph{after} integration. But in order to be completely rigorous, a counterterm with no pole and a scheme-dependent finite part should still be introduced. In the particular case of the $\MSbar$ scheme, this finite part is zero, justifying the absence of actual renormalisation in the traditional approach. Other schemes, however, may give birth to a non-trivial finite part.\\
\\
Since the integration of the FDU/LTD amplitude effectively occurs in an Euclidean space -- the loop three-momentum space --, asymptotic expansions are easily implemented.  In fact, the local regularisation in an Euclidean space implies that the series expansion of the integrand commutes with the integral symbol. Thus, expanding the integrand in any parameter (for instance, the mass of the particle circulating inside the loop) and integrating the series expansion order by order will lead to the correct result. This procedure, although focused here on the Higgs boson interactions, can be generalised to other processes. In particular, the methods presented in this chapter open new horizons for more efficient implementations and further simplifications of higher-order computations and asymptotic expansions.

\chapter{Universal four-dimensional representation at two loops}\label{Chapter:HTL}
\thispagestyle{fancychapter}
\fancyhead[RE]{\nameref*{Chapter:HTL}}

In this chapter, we extend useful properties of the $H\to\gamma\gamma$ unintegrated dual amplitudes from one- to two-loop level, using the Loop-Tree Duality formalism. In particular, we show that the universality of the functional form -- regardless of the nature of the internal particle -- still holds at this order. We also present an algorithmic way to renormalise two-loop amplitudes, by locally cancelling the ultraviolet singularities at the integrand level, thus allowing for a full four-dimensional numerical implementation of the method. Our results are compared with analytic expressions already available in the literature, and a perfect numerical agreement is found. The success of this computation plays a crucial role for the development of a fully local four-dimensional framework to compute physical observables at NNLO and beyond.

\section{Introduction}\label{Section:HTLIntro}
\fancyhead[LO]{\ref*{Section:HTLIntro}~~\nameref*{Section:HTLIntro}}

The two-loop QCD corrections to the decay process $H \to \gamma \gamma$ have been first evaluated in the heavy-top limit~\cite{Harlander:2005rq,Zheng:1990qa,Djouadi:1990aj,Dawson:1992cy} and with the full top-mass dependence~\cite{Fleischer:2004vb,Aglietti:2006tp}. The two-loop electroweak corrections have been investigated in~\cite{Aglietti:2004nj,Actis:2008ts,Passarino:2007fp,Degrassi:2005mc,Fugel:2004ug}. Combining the two-loop QCD and electroweak corrections, it is possible to observe a nearly complete cancellation between these two contributions for $M_{H}=126$ GeV~\cite{Maierhofer:2012vv}. At NNLO the non-singlet~\cite{Steinhauser:1996wy} and singlet
QCD contributions~\cite{Maierhofer:2012vv} have been calculated in the heavy top quark limit.\\
\\
In view of the enormous success of the SM with the detection of the Higgs boson, new directions have been taken to discuss in more details the consequences of this discovery. In particular, from the phenomenological point of view, the background of the experiment has to be removed. Thus, QCD predictions up to the Next-to-Next-to-Next-to-Leading order (N$^{3}$LO) have been provided in an effective theory~\cite{Anastasiou:2016cez}. Also, it has been shown that the mixed effects of QCD-electroweak contribution to the amplitude are relevant~\cite{Bonetti:2018ukf}.\\ 
\\
In \Chapter{\ref{Chapter:HOL}}, we extensively studied hidden mathematical properties of the amplitudes $gg\to H$ and $H\to\gamma\gamma$ at LO, and we showed that these amplitudes exhibited remarkable properties when computed using the LTD theorem. The dual contributions we obtained for different internal particles -- charged electroweak gauge bosons, massive fermions and charged scalars -- featured the very same functional forms, and could be written in a universal way using scalar parameters depending only on the space-time dimension $d$, and the mass of the particles involved in the process. We also obtained a pure four-dimensional ($d=4$) representation of the renormalised amplitude and recovered the well-known results found in the literature.\\
\\
In this chapter, we push the computation further by considering the $H\to\gamma\gamma$ process at two-loop level, and show that the above-mentioned properties are still present. In order to obtain the renormalised amplitude, we perform a local UV renormalisation
that leads to a finite integrand in four space-time dimensions. This algorithm is based on a refinement of the \emph{expansion around the UV propagator}~\cite{Becker:2010ng,Sborlini:2016gbr,Sborlini:2016hat} to account for the different singular behaviours of the internal loop momenta in the UV region. Furthermore, since this amplitude is IR safe, we can directly treat the virtual integrand in four dimensions. We point out that the calculation of this amplitude, done below the mass threshold limit and in the $\overline{\text{MS}}$
renormalisation scheme, is the first two-loop application to a physical process done through
LTD. We note that for individual diagrams, unphysical threshold singularities appear but they cancel among themselves when the full amplitude is considered.\\
\\
In the same spirit of the universality that these amplitudes exhibit at LO, we consider as internal particles charged scalars and top quarks. While we only consider QED corrections, they can be straightforwardly promoted to QCD ones by replacing the couplings accordingly. We compare our results with known analytic expressions~\cite{Fleischer:2004vb,Aglietti:2006tp}, finding full agreement.\\
\\
We verify that the LTD approach holds at multi-loop level and, therefore, that N$^{k}$LO predictions involving virtual amplitudes can be achieved using its formalism. Additionally, we remark that the traditional approach based on the use of integration-by-parts identities~\cite{Chetyrkin:1981qh,Laporta:2001dd} is not needed to evaluate the actual amplitude. In fact, we overcome the calculation of the latter making our procedure much lighter as we shall describe here.\\
\\
This chapter is organised as follows. In \Section{\ref{Section:HTLReduction}}, we sketch the algorithm to algebraically reduce integrand-level expressions of two-loop dual amplitudes, and rewrite every scalar product involved in terms of denominators. Then, we provide the tensor structure of the $H\to\gamma\gamma$ amplitude in \Section{\ref{Section:HTLTensorProjection}}. In \Section{\ref{Section:HTLDualAmplitudes}}, we collect and write the universal coefficients involved in the universal structure of the two-loop dual expressions. We discuss in \Section{\ref{Section:HTLThresholds}} the cancellation of unphysical threshold singularities that appear among the dual contributions, and we explicitly show how they occur. In \Section{\ref{Section:HTLUV}} , we discuss an algorithmic approach to locally renormalise two-loop amplitudes within the LTD formalism. In particular, we focus on the determination of the scheme-fixing parameters in the $\overline{\text{MS}}$ scheme. Finally in \Section{\ref{Section:HTLNumericalIntegration}}, we present our numerical results and show a complete agreement with the analytic expressions. We draw our conclusions and discuss future directions of this work in \Section{\ref{Section:HTLConclusion}}.

\section{Algebraic reduction of two-loop dual amplitudes}\label{Section:HTLReduction}
\fancyhead[LO]{\ref*{Section:HTLReduction}~~\nameref*{Section:HTLReduction}}

\begin{figure}
	\begin{minipage}{0.5\textwidth}
		\centering
		\begin{picture}(250,200)(-125,-100)
		\SetWidth{1}
		\Arc[arrow](0,0)(50,-45,45)
		\Arc[arrow](0,0)(50,45,83)
		\Arc[arrow](0,0)(50,-83,-45)
		\Arc[arrow](0,0)(50,83,120)
		\Arc[arrow](0,0)(50,-120,-83)
		\Arc[arrow](0,0)(50,120,180)
		\Arc[arrow](0,0)(50,180,240)
		\Line[arrow](35,35)(57,57)
		\Line[arrow](35,-35)(57,-57)
		\Line[arrow](-25,43)(-40,69)
		\Line[arrow](-50,0)(-80,0)
		\Line[arrow](-25,-43)(-40,-69)
		\Arc[arrow](180,0)(180,164,196)
		\Vertex(60,25){1.5}
		\Vertex(65,0){1.5}
		\Vertex(60,-25){1.5}
		\Vertex(-46,46){1.5}
		\Vertex(-57,33){1.5}
		\Vertex(-63,17){1.5}
		\Vertex(-63,-17){1.5}
		\Vertex(-57,-33){1.5}
		\Vertex(-46,-46){1.5}
		\color{blue}
		\Text(64,64){$p_i$}
		\Text(-44,76){$p_{i-1}$}
		\Text(-88,0){$p_N$}
		\Text(-44,-76){$p_{j+1}$}
		\Text(64,-64){$p_j$}
		\Text(27,56){$q_{\overline{1,i-1}}$}
		\Text(-12,59){$q_{1,i-1}$}
		\Text(-12,-59){$q_{1,j}$}
		\Text(27,-56){$q_{\overline{1,j}}$}
		\Text(-16,0){$q_{N+1}$}
		\end{picture}
	\end{minipage}
	\begin{minipage}{0.5\textwidth}
		\centering
		\begin{picture}(250,200)(-125,-100)
		\SetWidth{1}
		\Arc[arrow](0,0)(50,-45,12)
		\Arc[arrow](0,0)(50,12,66)
		\Arc[arrow](0,0)(50,-83,-45)
		\Arc[arrow](0,0)(50,66,120)
		\Arc[arrow](0,0)(50,-120,-83)
		\Arc[arrow](0,0)(50,120,180)
		\Arc[arrow](0,0)(50,180,240)
		\Line[arrow](49,10)(78,17)
		\Line[arrow](20.4,45)(34,73)
		\Line[arrow](35,-35)(57,-57)
		\Line[arrow](-25,43)(-40,69)
		\Line[arrow](-50,0)(-80,0)
		\Line[arrow](-25,-43)(-40,-69)
		\Arc[arrow](158,-24)(154,153,189.5)
		\Vertex(65,-2){1.5}
		\Vertex(62,-18){1.5}
		\Vertex(56,-33){1.5}
		\Vertex(-46,46){1.5}
		\Vertex(-57,33){1.5}
		\Vertex(-63,17){1.5}
		\Vertex(-63,-17){1.5}
		\Vertex(-57,-33){1.5}
		\Vertex(-46,-46){1.5}
		\color{blue}
		\Text(91,18){$p_{i+1}$}
		\Text(36,80){$p_i$}
		\Text(-44,76){$p_{i-1}$}
		\Text(-88,0){$p_N$}
		\Text(-44,-76){$p_{j+1}$}
		\Text(64,-64){$p_j$}
		\Text(49,38){$q_{\overline{1,i}}$}
		\Text(-3,59){$q_{1,i-1}$}
		\Text(-12,-59){$q_{1,j}$}
		\Text(27,-56){$q_{\overline{1,j}}$}
		\Text(-11,0){$q_{N+1}$}
		\end{picture}
	\end{minipage}
	\begin{minipage}{0.5\textwidth}
		\centering
		\begin{picture}(250,200)(-125,-100)
		\SetWidth{1}
		\Arc[arrow](0,0)(50,-35,35)
		\Arc[arrow](0,0)(50,35,75)
		\Arc[arrow](0,0)(50,-75,-35)
		\Arc[arrow](0,0)(50,74,180)
		\Arc[arrow](0,0)(50,-180,-75)
		\Line[arrow](13,48)(21,77)
		\Line[arrow](13,-48)(21,-77)
		\Line[arrow](-50,0)(-80,0)
		\Vertex(-12,64){1.5}
		\Vertex(-38,52){1.5}
		\Vertex(-58,29){1.5}
		\Vertex(-58,-29){1.5}
		\Vertex(-38,-52){1.5}
		\Vertex(-12,-64){1.5}
		\Arc[arrow](50,0)(30,108,252)
		\color{blue}
		\Text(23,85){$p_i$}
		\Text(23,-85){$p_{i+1}$}
		\Text(-88,0){$p_N$}
		\Text(35,-50){$q_{1,i}$}
		\Text(62,0){$q_{\overline{1,i}}$}
		\Text(34,49){$q_{1,i}$}
		\Text(4,0){$q_{N+1}$}
		\end{picture}
	\end{minipage}
	\begin{minipage}{0.5\textwidth}
		\centering
		\begin{picture}(250,200)(-125,-100)
		\SetWidth{1}
		\Arc[arrow](0,0)(50,0,75)
		\Arc[arrow](0,0)(50,-75,0)
		\Arc[arrow](0,0)(50,74,180)
		\Arc[arrow](0,0)(50,-180,-75)
		\Line[arrow](13,48)(21,77)
		\Line[arrow](13,-48)(21,-77)
		\Line[arrow](-50,0)(-80,0)
		\Vertex(-12,64){1.5}
		\Vertex(-38,52){1.5}
		\Vertex(-58,29){1.5}
		\Vertex(-58,-29){1.5}
		\Vertex(-38,-52){1.5}
		\Vertex(-12,-64){1.5}
		\Arc[arrow](32,0)(18,0,360)
		\color{blue}
		\Text(23,85){$p_i$}
		\Text(23,-85){$p_{i+1}$}
		\Text(-88,0){$p_N$}
		\Text(50,-36){$q_{1,i}$}
		\Text(50,36){$q_{1,i}$}
		\Text(-14,0){$q_{N+1}/q_{\overline{1,i}}$}
		\end{picture}
	\end{minipage}
	\begin{minipage}{0.5\textwidth}
		\centering
		\begin{picture}(250,200)(-125,-100)
		\SetWidth{1}
		\Arc[arrow](-23,0)(27,-60,15)
		\Arc[arrow](-23,0)(27,15,90)
		\Arc[arrow](-23,0)(27,90,180)
		\Arc[arrow](-23,0)(27,180,-60)
		\Line[arrow](-9,-23)(6,-49)
		\Line[arrow](-50,0)(-80,0)
		\Line[arrow](3,7)(17,11)
		\Line[arrow](-23,27)(-23,57)
		\Line[arrow](52,20)(67,24)
		\Vertex(-39,38){1.5}
		\Vertex(-53,30){1.5}
		\Vertex(-63,16){1.5}
		\Vertex(-59,-21){1.5}
		\Vertex(-44,-36){1.5}
		\Vertex(-23,-42){1.5}
		\Arc[arrow](35,16)(18,15,195)
		\Arc[arrow](35,16)(18,195,15)
		\color{blue}
		\Text(-23,64){$p_{i-1}$}
		\Text(9,-56){$p_{i+1}$}
		\Text(75,25){$p_i$}
		\Text(-88,0){$p_N$}
		\Text(-2,30){$q_{i-1}$}
		\Text(10,-13){$q_i$}
		\Text(31,43){$q_{\overline{i-1}}$}
		\Text(40,-11){$q_{\overline{i}}$}
		\end{picture}
	\end{minipage}
	\begin{minipage}{0.5\textwidth}
		\centering
		\begin{picture}(250,200)(-125,-100)
		\SetWidth{1}
		\Arc[arrow](-23,0)(27,-60,15)
		\Arc[arrow](-23,0)(27,15,90)
		\Arc[arrow](-23,0)(27,90,180)
		\Arc[arrow](-23,0)(27,180,-60)
		\Line[arrow](-9,-23)(6,-49)
		\Line[arrow](-50,0)(-80,0)
		\Line[arrow](3,7)(67,24)
		\Line[arrow](-23,27)(-23,57)
		\Vertex(-39,38){1.5}
		\Vertex(-53,30){1.5}
		\Vertex(-63,16){1.5}
		\Vertex(-59,-21){1.5}
		\Vertex(-44,-36){1.5}
		\Vertex(-23,-42){1.5}
		\Arc[arrow,arrowpos=0.25](30,33)(18,15,375)
		\color{blue}
		\Text(-23,64){$p_{i-1}$}
		\Text(9,-56){$p_{i+1}$}
		\Text(75,25){$p_i$}
		\Text(-88,0){$p_N$}
		\Text(-2,30){$q_{i-1}$}
		\Text(10,-13){$q_i$}
		\Text(26,60){$q_{\overline{i-1}}/q_{\overline{i}}$}
		\end{picture}
	\end{minipage}
	\caption{Assignment of momenta in two-loop planar diagrams.}
	\label{Figure:HTLAssignmentOfMomenta}
\end{figure}
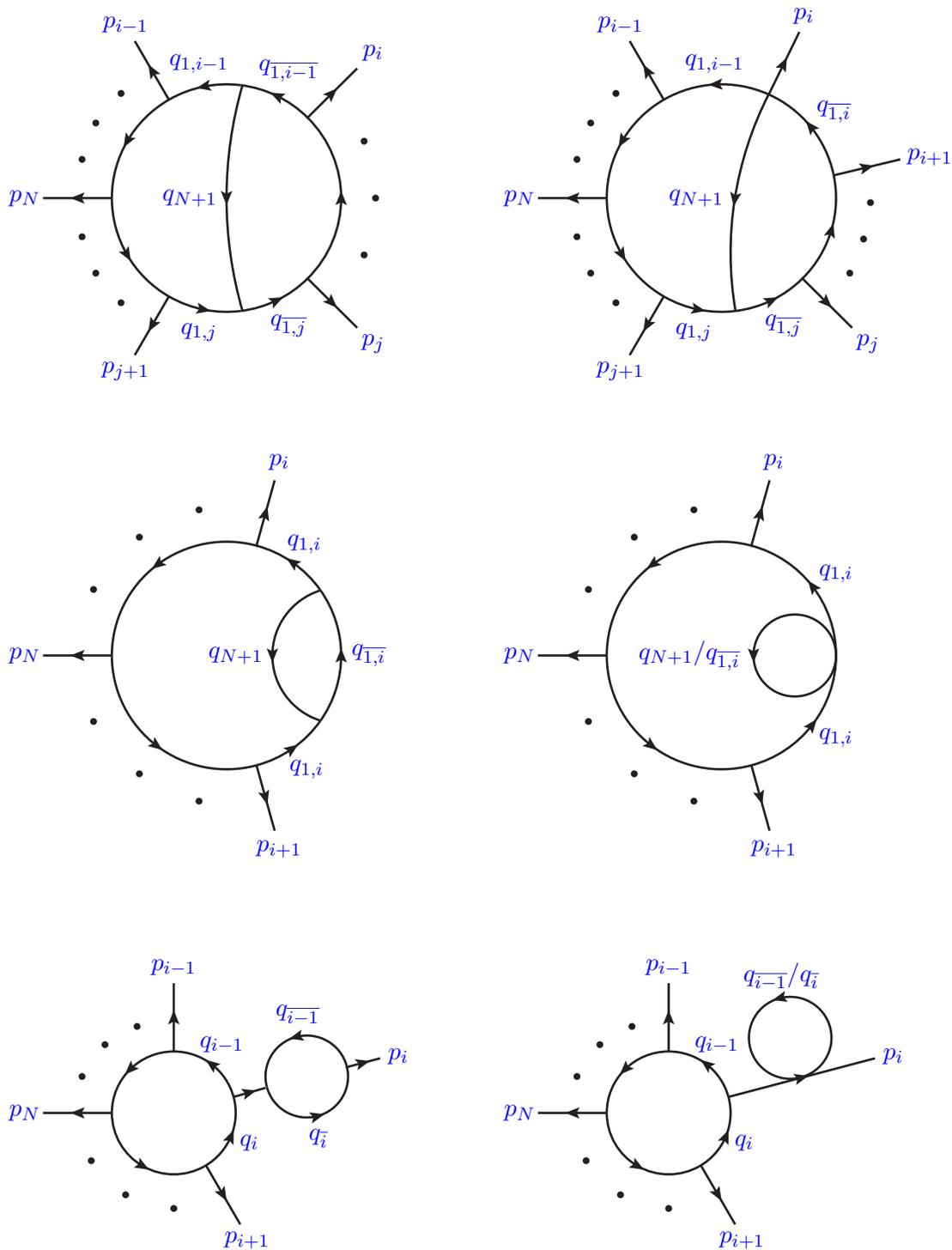

In order to make the two-loop expressions more compact, we will perform an algebraic reduction of the dual amplitudes to integrals that involve both positive and negative powers of dual propagators. In this section, we analyse only the case of planar diagrams\footnote{In the more general case, it might not always be possible to completely express a two-loop amplitude in terms of propagators. Irreducible scalar products may indeed appear in the numerator.}, as they are those that appear in the practical example that we present in this chapter.\\
\\
Let's first consider a scattering amplitude with $N$ external legs with ordered external momenta $\{p_1,p_2,\dots,p_N\}$. At one-loop, we have $N$ different propagators and $N-1$ independent scalar products $\ell_1\cdot p_i$ (indeed, because of momentum conservation, $\ell_1\cdot p_N=-\,\sum_{i=1}^{N-1}\,\ell_1\cdot p_i$, with $\ell_1$ the loop four-momentum). In the Feynman representation, the propagators are quadratic in $\ell_1$, whereas after applying LTD the dual propagators are linear in $\ell_1$. In both formalisms, however, it is always possible to write numerators in terms of propagators diagram by diagram. Now, we consider the set of all the two-loop planar Feynman diagrams that can be constructed from the ordered one-loop seed diagram. These diagrams are generated by attaching a line with a single propagator with momenta $q_{N+1}=\ell_2$ in all possible ways while keeping the same ordering of the external momenta in the loop formed by the other two loop lines $\alpha_1$ and $\alpha_3$ (see \Fig{\ref{Figure:HTLAssignmentOfMomenta}} for the assignment of loop momenta). This means that these planar two-loop Feynman diagrams can be constructed from the sets of propagators
\begin{equation}
\alpha_1=\{q_1,q_{12},\dots,q_{1,N}\}\;,\qquad\alpha_2=\{q_{N+1}\}\;,\qquad\alpha_3=\{q_{\overline{1}}, q_{\overline{12}},\dots,q_{\overline{1,N}}\}\;,
\end{equation}
with $q_{1,i}=\ell_1+\,\sum_{j=1}^i\,p_j$, $q_{N+1}=\ell_2$ and $q_{\overline{1,i}}=\ell_{12}+\,\sum_{j=1}^i\,p_j$, with $\ell_{12}=\ell_1+\ell_2$. This sums to $N(\alpha_1+\alpha_2+\alpha_3)=2N+1$ possible propagators. If there are only three-point interactions, each individual Feynman diagram will contain $N+3$ propagators from these sets. If there are $V_4$ four-point interaction vertices, each individual Feynman diagram will contain $N+3-V_4$ propagators. In any case, this means that the maximum number of propagators a given diagram can exhibit is $N+3$.\\
\\
Now, because of momentum conservation, we have $2N-1$ independent scalar products that are
\begin{equation}
\{\ell_1\cdot p_i,\ell_2\cdot p_j,\ell_1\cdot\ell_2~~|~~i,j\in\{1,\dots, N-1\}\}\;.
\end{equation}
In LTD at two-loops, two internal particles are set on shell, which means there only remain $2N-1$ dual propagators for a given double-cut. Moreover, dual propagators are linear in each of the loop momenta, and the dual numerators do not involve squared loop-momenta. Therefore, for each double-cut, and considering all the Feynman diagrams with the same ordering of the external particles, it is possible to rewrite all the scalar products involved (and thus the numerators) in terms of dual propagators, and this in a unique way. There is no need to introduce irreducible scalar products (ISP) because 
the set of Feynman diagrams contains all the necessary propagators to perform the algebraic reduction.\\
\\
The algebraic reduction of a planar two-loop dual amplitude with $N$ external legs,
and at most squared propagators in one single loop line, leads to
\begin{align}
\mathcal{A}^{(2)}_N=&~\int_{\ell_1}\,\int_{\ell_2}\,\mathcal{N}(\ell_1,\ell_2,\{p_i\}_N)\,G_F(\alpha_1\cup\alpha_2\cup\alpha_3)+{\rm perm.}\nn\\
=&~\int_{\ell_1}\,\int_{\ell_2}\,\,\sum_{j,k}\,\left[\frac{c_{a_0;a_1,\dots,a_{2N-1}}(\{p_i\}_N)}{(q_{j,0}^{(+)})^{a_0}(d_{i_1})^{a_1}(d_{i_2})^{a_2}\cdots(d_{i_{2N-1}})^{a_{2N-1}}}\right]\deltatilde{q_j,q_k}+{\rm perm.}\;,
\end{align}
where $d_{i_l}=\big(G_D(q_j,q_k;q_{i_l})\big)^{-1}$ is the dual propagator of the internal line with momentum $q_{i_l}$, evaluated with both $q_j$ and $q_k$ on shell\footnote{In the general case, we should differentiate $G_D(q_j;q_{i_l})$ from $G_D(q_k;q_{i_l})$, as they may have distinct prescriptions. In our case however, since we are dealing with real scattering amplitudes, it is not necessary to make the distinction.}, $\deltatilde{q_j,q_k}=\deltatilde{q_j}\,\deltatilde{q_k}$, and
\begin{equation}
\,\sum_{i=1}^{2N-1}\,a_i\leq N+1\;,\qquad a_0\in\{2,1,0\}\;.
\end{equation}
The scalar coefficients $c_{a_0; a_1, \ldots, a_{2N-1}}$ depend only on the external momenta, and are not necessarily independent. Our purpose is to rearrange the expressions for the dual amplitudes  in order to obtain the minimal set of independent coefficients $c_{a_0; a_1, \ldots, a_{2N-1}}$. Another relevant issue to obtain the most compact integrand expressions is to label the internal momenta in the most symmetric way. In \Fig{\ref{Figure:HTLAssignmentOfMomenta}}, we show the assignments that we will use in the most general case for planar two-loop diagrams. Computer algebra programs for the automatic generation of two-loop amplitudes, like \textsc{FeynArts}~\cite{Hahn:2000kx}, might use a different criteria which will require a relabelling of the internal propagators to achieve the most suitable assignment. We will with illustrate in the next sections the full procedure with the benchmark amplitude $H\to\gamma\gamma$.

\section{Tensor projection}\label{Section:HTLTensorProjection}
\fancyhead[LO]{\ref*{Section:HTLTensorProjection}~~\nameref*{Section:HTLTensorProjection}}

The scattering amplitude describing the Higgs boson decay to two photons is given by 
\begin{equation}\label{Equation:HTLVirtualAmplitude}
|\mathcal{M}_{H\to\gamma\gamma}\rangle=i\,e^2\,\left(\,\sum_{f=\phi,t,W}\,e^2_f\,N_C^f\,\mathcal{A}_{\mu\nu}^{(f)}\right)(\varepsilon^\mu(p_1))^*(\varepsilon^\nu(p_2))^*\;,
\end{equation}
with $e$ the electromagnetic coupling, $e_f$ (respectively $N_C^f$) the electric charge (respectively the number of colours) of the virtual particle $f$, and $\varepsilon(p_i)$ the polarisation vectors of the external on-shell photons. Here, we assume that both photons are coupled to the same flavour. In the following, we restrict ourselves to the corrections at two loops with $f=\phi,t$. The tensor amplitude $\mathcal{A}^{(f)}_{\mu\nu}$
fulfils the perturbative expansion
\begin{equation}
\mathcal{A}_{\mu\nu}^{(f)}=\mathcal{A}_{\mu\nu}^{(1,f)}+(e\,e_f)^2\,\mathcal{A}_{\mu\nu}^{(2,f)}+\mathcal{O}(e^4)\;,
\end{equation}
where $\mathcal{A}^{(1,f)}_{\mu\nu}$ is the one-loop amplitude, and $\mathcal{A}^{(2,f)}_{\mu\nu}$ is the two-loop QED correction. The tensor amplitudes $\mathcal{A}_{\mu\nu}^{(L,f)}$ can be decomposed through Lorentz and gauge invariance as
\begin{equation}
\mathcal{A}_{\mu\nu}^{(L,f)}=\,\sum\limits_{i=1}^6\,\mathcal{A}_i^{(L ,f)}\,T_{i,\mu\nu}\;,
\end{equation}
in terms of the tensor basis
\begin{equation}
T_i^{\mu\nu}=\left\{g^{\mu\nu}-\frac{2p_2^\mu\,p_1^\nu}{\s},g^{\mu\nu},\frac{2p_1^{\mu}\,p_2^{\nu}}{\s},\frac{2p_1^{\mu}\,p_1^{\nu}}{\s},\frac{2p_2^{\mu}\,p_2^{\nu}}{\s},\epsilon^{\mu\nu\sigma\rho} \frac{2p_{1,\sigma}\,p_{2,\rho}}{\s}\right\}\;,
\end{equation}
with $\s=(p_1+p_2)^2$. The tensor structure $T_6^{\mu\nu}$ may appear for the first time at two-loop order because of the potential presence of a $\gamma^5$, but its interference with the one-loop amplitude vanishes. As in \Chapter{\ref{Chapter:HOL}}, we use the projectors
\begin{equation}\label{Equation:HTLProjectors}
P_1^{\mu\nu}=\frac{1}{d-2}\left(g^{\mu\nu}-\frac{2p_2^\mu\,p_1^\nu}{\s}-(d-1)\frac{2p_1^\mu\,p_2^\nu}{\s}\right)\qquad\text{and} \qquad P_2^{\mu\nu}=\frac{2p_1^\mu\,p_2^\nu}{\s}\,
\end{equation}
to extract the scalar amplitudes $\mathcal{A}_1^{(L,f)}$ and $\mathcal{A}_2^{(L,f)}$, as $\mathcal{A}_i^{(1,f)}=P_i^{\mu\nu}\,\mathcal{A}_{\mu\nu}^{(L,f)}$. There is no need to compute $\mathcal{A}_i^{(L,f)}$, for $i\in\{3,4,5\}$, as they vanish after contracting with the polarisation vectors, and therefore do not contribute to the scattering amplitude for on-shell photons. Moreover, because of gauge invariance, $\mathcal{A}_2^{(L,f)}$ is expected to vanish after integration, which leaves $\mathcal{A}_1^{(L,f)}$ as the only relevant physical term. It is still interesting to consider and get integrand expressions for $\mathcal{A}_2^{(L,f)}$ though, as it can be used to simplify expressions. Indeed, we already showed at one-loop level (see \Section{\ref{Section:HOLUniversalDualAmplitudes}}) that the transformation
\begin{equation}\label{Equation:HTLTransfA1}
\mathcal{A}^{(L,f)}\to\mathcal{A}_1^{(L,f)}-2\frac{c_2^{(f)}}{c_3^{(f)}}\mathcal{A}_2^{(L,f)}
\end{equation}
with $c_2^{(f)}=c_{23}^{(f)}-c_3^{(f)}$, and
\begin{equation}
\frac{2c_2^{(t)}}{c_3^{(t)}}=\frac{2c_2^{(\phi)}}{c_3^{(\phi)}}=-\frac{d}{d-2}
\end{equation}
reduces the number of independent scalar coefficients $c_i^{(f)}$ to describe $\mathcal{A}_1^{(1,f)}$ from three to two. Remarkably enough, the same transformation can be used at two-loop level to drastically reduce the number of coefficients needed to write $\mathcal{A}_1^{(2,f)}$.

\section{Dual amplitude for $H\to \gamma\gamma$ at two loops}\label{Section:HTLDualAmplitudes}
\fancyhead[LO]{\ref*{Section:HTLDualAmplitudes}~~\nameref*{Section:HTLDualAmplitudes}}

\begin{figure}
	\begin{picture}(140,150)(-20,-75)
		\SetWidth{1}
		\Arc[arrow](64,0)(40,-80,40)
		\Arc[arrow](64,0)(40,40,160)
		\Arc[arrow](64,0)(40,160,280)
		\DashLine[arrow](0,0)(24,0){6}
		\Photon(84,34.5)(96,56){3}{2}
		\Photon(84,-34.5)(96,-56){3}{2}
		\SetColor{Blue}
		\Arc(104,0)(24,108,252)
		\Arc(44,34)(24,228,373)
		\Arc(44,-34)(24,347,492)
		\Arc(104,70)(70,210,270)
		\Arc(104,-70)(70,90,150)
		\Arc(-17,0)(70,330,390)
		\color{blue}
		\Text(-10,0){$p_{12}$}
		\Text(100,62){$p_1$}
		\Text(100,-62){$p_2$}
		\SetColor{Black}
		\Vertex(104,0){1.75}
		\Vertex(44,34.5){1.75}
		\Vertex(44,-34.5){1.75}
		\Vertex(84,34.5){1.75}
		\Vertex(84,-34.5){1.75}
		\Vertex(24,0){1.75}
		\Vertex(97,23){1.75}
		\Vertex(97,-23){1.75}
		\Vertex(27.5,16){1.75}
		\Vertex(27.5,-16){1.75}
		\Vertex(67,-40){1.75}
		\Vertex(67,40){1.75}
	\end{picture}
	\hfill
	\begin{picture}(140,150)(-18,-75)
		\SetWidth{1}
		\Arc[arrow](64,0)(40,-80,40)
		\Arc[arrow](64,0)(40,40,160)
		\Arc[arrow](64,0)(40,160,280)
		\DashLine[arrow](0,0)(24,0){6}
		\Photon(84,34.5)(96,56){3}{2}
		\Photon(84,-34.5)(96,-56){3}{2}
		\SetColor{Blue}
		\Arc(104,0)(24,108,252)
		\Arc(44,34)(24,228,373)
		\Arc(44,-34)(24,347,492)
		\Arc(104,70)(70,210,270)
		\Arc(104,-70)(70,90,150)
		\Arc(-17,0)(70,330,390)
		\Arc(145,0)(70,150,210)
		\Arc(64,48)(24,213,327)
		\Arc(64,-48)(24,33,147)
		\Arc(106,24)(24,153,267)
		\Arc(106,-24)(24,93,207)
		\Arc(96,0)(8,0,360)
		\Arc(48,28)(8,0,360)
		\Arc(48,-28)(8,0,360)
		\Line(44,34.5)(84,-34.5)
		\Line(44,-34.5)(84,34.5)
		\color{blue}
		\Text(-10,0){$p_{12}$}
		\Text(100,62){$p_1$}
		\Text(100,-62){$p_2$}
		\SetColor{Black}
		\Vertex(104,0){1.75}
		\Vertex(44,34.5){1.75}
		\Vertex(44,-34.5){1.75}
		\Vertex(84,34.5){1.75}
		\Vertex(84,-34.5){1.75}
		\Vertex(24,0){1.75}
		\Vertex(97,23){1.75}
		\Vertex(97,-23){1.75}
		\Vertex(27.5,16){1.75}
		\Vertex(27.5,-16){1.75}
		\Vertex(67,-40){1.75}
		\Vertex(67,40){1.75}
	\end{picture}
	\hfill
	\begin{picture}(140,150)(-16,-75)
		\SetWidth{1}
		\Arc[arrow](64,0)(40,55,235)
		\Arc[arrow](64,0)(40,235,415)
		\DashLine[arrow](0,0)(24,0){6}
		\Photon(104,0)(116,22){3}{2}
		\Photon(104,0)(116,-22){3}{2}
		\SetColor{Blue}
		\Line(64,-40)(64,40)
		\Arc(64,40)(24,198,342)
		\Arc(64,-40)(24,18,162)
		\Arc(64,32)(8,0,360)
		\Arc(64,-32)(8,0,360)
		\color{blue}
		\Text(-10,0){$p_{12}$}
		\Text(120,29){$p_1$}
		\Text(120,-29){$p_2$}
		\SetColor{Black}
		\Vertex(64,40){1.75}
		\Vertex(64,-40){1.75}
		\Vertex(104,0){1.75}
		\Vertex(24,0){1.75}
		\Vertex(87,33){1.75}
		\Vertex(41,33){1.75}
		\Vertex(87,-33){1.75}
		\Vertex(41,-33){1.75}
	\end{picture}
	\caption{Two-loop mandala Feynman diagrams for $H\to\gamma\gamma$. The black solid lines are quarks (left diagram) or scalars (middle and right diagrams). The blue solid lines are virtual photons.}
	\label{Figure:HTLMandala}
\end{figure}
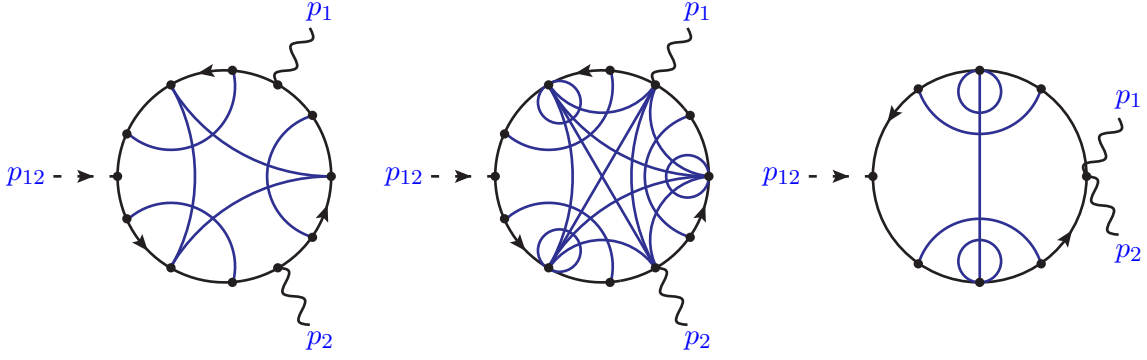

At two-loop level, there are 12 Feynman diagrams contributing to the $H\to \gamma\gamma$ scattering amplitude with internal top quarks. For internal charged scalars, there are 37 Feynman diagrams. They are drawn in \Fig{\ref{Figure:HTLMandala}} where all diagrams sharing the same global topology are superimposed in so-called mandala diagrams. In this chapter, we only consider QED corrections, and therefore photons as the extra internal particle, and do not take into account ``mixed'' diagrams where different massive particles may appear. In the traditional approach, massless snail diagrams are usually ignored, since they integrate to zero. However, within our approach, we need them to preserve the universal structure of the integrands.\\
\\
All the diagrams are planar and can be constructed from the following internal momenta: 
\begin{align}
\alpha_1:q_i=&~\ell_1+p_i\;,\qquad q_{12}=\ell_1+p_{12}\;,\qquad q_3=\ell_1\;,\nn\\ 
\alpha_2:q_4=&~\ell_2\;,\nn\\ 
-\alpha_2:q_{\overline{4}}=&~-\ell_2\;,\nn\\ 
\alpha_3:q_{\overline{i}}=&~\ell_{12}+p_i\;,\qquad q_{\overline{12}}=\ell_{12}+p_{12}\;,\qquad q_{\overline{3}}=\ell_{12}\;.
\end{align}
Only $q_4$ (and $q_{\overline{4}}$), which labels the photon, is massless, while all the other internal momenta have mass $M_f$.\\
\\
If the Higgs boson is on shell, the loop amplitude is below threshold and is therefore purely real. In that kinematical regime the dual prescriptions become irrelevant, and the dual functions fulfil the identity
\begin{equation}
G_D(\alpha_i \cup \alpha_j) = G_D(\alpha_i) \, G_F(\alpha_j) + G_F(\alpha_i) \, G_D(\alpha_j)~.
\end{equation}
Hence, the LTD representation in \Eq{\ref{Equation:LTDL2SecondStep}} adopts the simpler form
\begin{align}\label{Equation:HTLDualityRelationNoThreshold}
\mathcal{A}_N=&\int_{\ell_1}\,\int_{\ell_2}\,\mathcal{N}(\ell_1,\ell_2,\{p_i\}_N)\,\otimes\,\bigg[G_D(\alpha_1)\,G_D(\alpha_2)\,G_F(\alpha_3)\nn\\
&~+G_F(\alpha_1)\,G_D(-\alpha_2)\,G_D(\alpha_3)+G_D(\alpha_1)\,G_F(\alpha_2)\,G_D(\alpha_3)\bigg]\;. 
\end{align}
Following the algebraic reduction defined in \Section{\ref{Section:HTLReduction}},
\begin{align}\label{Equation:HTLA2Reduced}
\mathcal{A}_1^{(2,f)}\propto\int_{\ell_1}\,\int_{\ell_2}\,\,\sum_{j,k}\,\left[\frac{c^{(f)}_{a_0; a_1,\dots,a_5}(p_1,p_2)}{(\kappa_j)^{a_0}(D_{i_1})^{a_1}(D_{i_2})^{a_2}\cdots(D_{i_5})^{a_{5}}} \right]\,\deltatilde{q_j,q_k}+{\rm perm.} 
\end{align}
with $\kappa_j=q_{j,0}^{(+)}/M_f$,
\begin{equation}
\sum_{i=1}^{5}a_i\le 4\;,\quad\text{and}\quad a_0\in\{2,1,0\}\;.
\end{equation}
For a given dual or Feynman propagator, we have defined the dimensionless denominator $D_{i_l}=\big(M_f^2\,G_{F/D}(q_{i_l})\big)^{-1}$. For example, in terms of these dimensionless denominators, the one-loop amplitude found in \Eq{\ref{Equation:HOLA}} takes the form 
\begin{align}
\mathcal{A}^{(1,f)}=&~g_f^{(1)}\,\int_{\ell_1}\,\Bigg[-\left(\frac{\deltatilde{q_1}}{D_{12}\,D_3}+\frac{\deltatilde{q_{12}}}{D_3\,D_1}+\frac{\deltatilde{q_3}}{D_{12}\,D_1}\right)\,c_1^{(f)}\nn\\
&~+\left(\frac{\deltatilde{q_{12}}}{D_3}+\frac{\deltatilde{q_3}}{D_{12}}\right)\,\frac{c_{23}^{(f)}}{2}\Bigg]+\{1\leftrightarrow 2\}\;,
\label{Equation:compactdi}
\end{align}
with
\begin{equation}
g_f^{(1)}=\frac{2}{\vev}\;.
\end{equation}
From now on, we use a different global factor depending on whether the expressions it multiplies has been algebraically reduced or not. The usual $g_f$ will be used for unreduced expressions, and $g_f^{(L)}$ for reduced ones.\\
\\
Before moving forward, it is important to note that for several double-cuts (for instance $\tilde{\delta}(q_3,q_4)$), diagrams with different external ordering will contribute, which means that we have, in our case, two extra propagators while having the same number of scalar products. Their expressions will therefore not be unique, as we will have the relation
\begin{equation}
D_3+D_{12}-D_1-D_2=D_{\overline{3}}+D_{\overline{12}}-D_{\overline{1}}-D_{\overline{2}}=r_f\;.
\end{equation}
One workaround is to compute such a double-cut for the top quark (because it does not involve diagrams with four-point interaction vertices involving both external photons) by separating the diagrams with ordering $\{p_1,p_2\}$ and the ones with ordering $\{p_2,p_1\}$, applying the reduction on both sets, and using their sum as an ansatz for the charged scalar.

\subsection{The two-loop amplitude $\mathcal{A}_2^{(2,f)}$}\label{Section:HTLA2}

The two-loop amplitude $A_2^{(2,f)}$ is obtained by projecting $\mathcal{A}_{\mu \nu}^{(2,f)}$
using the projector $P_2^{\mu\nu}$ defined in \Eq{\ref{Equation:HTLProjectors}}, namely $A_2^{(2,f)}=P_2^{\mu\nu}\,\mathcal{A}_{\mu \nu}^{(2,f)}$. Due to gauge invariance, it has to vanish after integration. Still, it is interesting to obtain an explicit expression because it establishes a useful integrand relation that can be used afterwards. Remarkably, it can be written in a very compact form for $f=\phi,t$, namely
\begin{align}\label{Equation:HTLA2Universal}
\mathcal{A}_2^{(2,f)}=&~g_f^{(2)}\,\int_{\ell_1}\,\int_{\ell_2}\,\frac{c_3^{(f)}}{2}\,\sum_{i=1, 2, 12, 3}\,\sigma(i)\Bigg[\left(G(D_{\overline{i}},\kappa_i,c_{4,u}^{(f)})+F(D_{\overline{i}},\kappa_4/\kappa_i)\right)\,\deltatilde{q_i,q_4}\nn\\ 
&~-F(-D_i,0)\,\deltatilde{q_{\overline{4}},q_{\overline{i}}}+\left(G(D_4, \kappa_i, -c_{4,nu}^{(f)})+F(D_4,-\kappa_{\overline{i}}/\kappa_i)\right)\deltatilde{q_i,q_{\overline{i}}}\Bigg]\;,
\end{align}
where
\begin{align}\label{Equation:HTLFG}
g_f^{(2)} =2\s\,g_f^{(1)}=\frac{4\s}{\vev}\;, &\qquad\sigma(i)=
\begin{cases}
+1\,,~~~i\in\{1,2\}\;,\\
-1\,,~~~i\in\{3,12\}\;,
\end{cases}\nn\\
G(D_j,\kappa_i,c)=\frac{1}{\kappa_i^2}\left(\frac{1}{D_j}+c\right)\;,&\qquad F(D_i,r_\kappa)=\frac{1}{D_i}\left(\frac{2}{D_i}\left(1+r_\kappa\right)-1\right)\;,\nn\\
\end{align}
and where the coefficients $c_{4,u}^{(f)}$ and $c_{4,nu}^{(f)}$ can be found in \Eqs{\ref{Equation:HTLctop}}{\ref{Equation:HTLcphi}}. Notice there is a difference in mass dimensions between $g_f^{(1)}$ and $g_f^{(2)}$. This is due to the presence of the second loop measure and the additional $\tilde{\delta}$, whose product has dimension of mass squared. The contribution
\begin{align}\label{Equation:HTLS2Snails}
\mathcal{S}_2^{(2,f)}&=g_f^{(2)}\,\int_{\ell_1}\,\int_{\ell_2}\,\frac{c_3^{(f)}}{2}\,\sum_{i=1, 2, 12, 3}\,\sigma(i)\,\frac{1}{\kappa_i^2}\,\left(c_{4,u}^{(f)}\,\deltatilde{q_i,q_4}-c_{4,nu}^{(f)}\,\deltatilde{q_i,q_{\overline{i}}}\right)=0\;,
\end{align}
that originally appears inside $\mathcal{A}_2^{(2,f)}$ vanishes because the subintegrals,
\begin{equation}
\int_{\ell_1}\,\frac{\deltatilde{q_i}}{\kappa_i^n}=\int_{\ell_1}\,\frac{\deltatilde{q_j}}{\kappa_j^n}=\int_{\ell_2}\,\frac{\deltatilde{q_{\overline{i}}}}{\kappa_{\overline{i}}^n}=\int_{\ell_2}\,\frac{\deltatilde{q_{\overline{j}}}}{\kappa_{\overline{j}}^n}\;,\qquad i\neq j\;.
\end{equation}
 with $n\in\{0,1,2\}$, are equivalent. Consequently, the sum over integrals in \Eq{\ref{Equation:HTLS2Snails}} vanishes.

\subsection{The two-loop amplitude $\mathcal{A}^{(2,f)}$}\label{Section:HTLA1}

In this section, we apply the same transformation given in \Eq{\ref{Equation:HTLTransfA1}} at two loops ($L=2$) to simplify the expressions for the amplitude $\mathcal{A}_1^{(2,f)}$. Then, and as explained in \Section{\ref{Section:HTLReduction}}, we perform an algebraic reduction to express the dual representation in the form \Eq{\ref{Equation:HTLA2Reduced}} and extract the scalar coefficients $c^{(f)}_{a_0;a_1,\ldots, a_5}$. As very few of them are indeed independent, they are simply relabelled as $c_i^{(f)}$. We obtain very compact expressions for all the double-cuts of the LTD representation. The full expressions for the unrenormalised amplitude $\mathcal{A}^{(2,f)}$ are collected in \Appendix{\ref{Section:APPUnrenormalisedTwoLoop}}. As for the one-loop case, the same expressions are valid regardless of the virtual particle circulating in the loop as a function of the flavour-dependent coefficients $c_i^{(f)}$.\\
\\
For the top quark and the charged scalar, the two independent coefficients that appeared at one loop are (see \Eqs{\ref{Equation:HOLcif}}{\ref{Equation:HOLc23f}})
\begin{align}
c_1^{(t)}=&~-\frac{8}{d-2}+r_t\;,\quad& c_{23}^{(t)}=&~-\frac{4(d-4)}{d-2}\;,\\
c_1^{(\phi)}=&~\frac{4}{d-2}\;,\quad& c_{23}^{(\phi)}=&~\frac{2(d-4)}{d-2}\;.
\end{align}
At two loops, they are still present, along with the following extra coefficients:
\begin{align}\label{Equation:HTLctop}
&c_{4,u}^{(t)}=-\frac{d-2}{4}\,,&&c_{4,nu}^{(t)}=-\frac{d-2}{4}\,, &&c_7^{(t)}=-\frac{1}{4}(c_1^{(t)}-r_t)\;, \nn \\
&c_8^{(t)}=c_1^{(t)}+\frac{(d-6)d+10}{2(d-2)}\, r_t\,,&&c_9^{(t)}=c_1^{(t)}-\frac{(d-8)d+10}{2(d-2)}\, r_t\,,&&c_{10}^{(t)}=c_1^{(t)}-\frac{(d-8)d+14}{2(d-2)}\,r_t\;,\nn\\
&c_{11}^{(t)}=c_1^{(t)}+\frac{(d-8)d+18}{2(d-2)}\, r_t\,,&&c_{12}^{(t)}=-\frac{(d-4)(d-5)}{d-2}\, r_t\,,&&c_{13}^{(t)}=-\frac{(d-6)d+12}{2(d-2)}\, r_t\;,\nn\\
&c_{14}^{(t)}=\frac{3}{4}\left(c_1^{(f)}-\frac{d}{3(d-2)}\,r_t\right)\,,&&c_{15}^{(t)}=-\frac{1}{2}\left(c_1^{(f)}+\frac{r_t}{2}\right)\,,&&c_{16}^{(t)}=\frac{d-8}{4}\;,\nn\\
&c_{17}^{(t)}=\frac{d-4}{4}\,,&&c_{18}^{(t)}=-\frac{(d-4)^2}{4(d-2)}\,,&&c_{19}^{(t)}=\frac{1}{2}\left(c_1^{(t)}+\frac{1}{d-2}\,r_t\right)\;,\nn\\
&c_{20}^{(t)}=\frac{1}{4}(c_1^{(t)}+r_t)\,,&&c_{21}^{(t)}=-\frac{2(d-4)}{d-2}+\frac{(d-10)d+18}{4(d-2)}\,r_t\,,&&c_{22}^{(t)}=-2+\frac{(d-4)d}{4(d-2)}\,r_t\;,\\
\label{Equation:HTLcphi}
&c_{4,u}^{(\phi)}=-\frac{d-2}{4}\,,&&c_{4,nu}^{(\phi)}=\frac{1}{4}\,,&&c_7^{(\phi)}=-\frac{1}{4}\, c_1^{(\phi)}\;,\nn\\
&c_8^{(\phi)}=c_1^{(\phi)}\,,&&c_9^{(\phi)}=c_1^{(\phi)}-\frac{1}{(d-2)}\,r_\phi,&&c_{10}^{(\phi)}=c_1^{(\phi)}\;,\nn\\
&c_{11}^{(\phi)}=c_1^{(\phi)}+\frac{d-4}{d-2}\, r_\phi\,,&&c_{12}^{(\phi)}=-\frac{3(d-4)}{2(d-2)}\, r_\phi\,,&&c_{13}^{(\phi)}=\frac{1}{d-2}\,r_\phi\;,\nn\\
&c_{14}^{(\phi)}=\frac{3}{4}\,c_1^{(\phi)}\;,&&c_{15}^{(\phi)}=-\frac{1}{2}\,c_1^{(\phi)}\,,&&c_{16}^{(\phi)}=\frac{1}{2}\;,\nn\\
&c_{17}^{(\phi)}=0\,,&&c_{18}^{(\phi)}=0\,,&&c_{19}^{(\phi)}=\frac{1}{2}\, c_1^{(\phi)}\;,\nn\\
&c_{20}^{(\phi)}=\frac{1}{4}\,c_1^{(\phi)}\,,&&c_{21}^{(\phi)}=-\frac{3}{d-2}\,,&&c_{22}^{(\phi)}=-\frac{1}{d-2}\;.
\end{align}
Notice that these coefficients can be highly simplified in the particular case $d=4$, motivating further the study of a full four-dimensional representation of this process at two-loop level.

\section{Cancellation of integrand singularities at two loops}\label{Section:HTLThresholds}
\fancyhead[LO]{\ref*{Section:HTLThresholds}~~\nameref*{Section:HTLThresholds}}

In addition to be infrared safe, the scattering amplitude for $H\to \gamma\gamma$ is purely real if the Higgs boson is on shell. It is therefore completely free of soft and physical threshold singularities. Still, some of the dual propagators might go on shell inside 
the integration domain, leading to singularities of the integrand. As it has been demonstrated in \Section{\ref{Section:LTDCancellationOfThreshold}} for the one-loop level, one of the advantages of LTD is the partial cancellation of potential singularities of the integrand. In this section, we extend the analysis of the integrand singularities of $H\to \gamma\gamma$ at two-loop level, using the example and notations previously introduced in this chapter.\\
\\
If a dual propagator $G_D(q_i;q_j)$ becomes on shell where both internal momenta $q_i$ and $q_j$ belong to the same loop line $\alpha_k$, then the corresponding singular behaviour is equivalent to the one-loop case. For example, consider the dual propagator $G_D(q_{\overline{12}};q_{\overline{3}})$.  A physical threshold (forward-backward singularity\footnote{We recall that a forward-backward singularity in the loop momentum space arises in the intersection of the forward on-shell hyperboloid or positive energy mode of one propagator with the backward on-shell hyperboloid or negative energy mode of another propagator.}) occurs if (see \Eq{\ref{Equation:LTDlambdappCondition}})
\begin{equation}\label{Equation:HTLFB}
k_{ji}^2-(m_j+m_i)^2\geq0\;,\qquad k_{ji,0}<0\;,
\end{equation}
with $k_{ji}=q_{\overline{3}}-q_{\overline{12}}=-p_{12}$ and $m_j=m_i=M_f$. Therefore, $- p_{12,0}<0$ but\linebreak $\s-4M_f^2<0$ for $\s=M_H^2$. An integrand singularity (forward-forward singularity) occurs if (see \Eq{\ref{Equation:LTDlambdapmCondition}})
\begin{equation}
k_{ji}^2-(m_j-m_i)^2\leq0\;.
\end{equation}
This would be the case of, e.g., $G_D(q_{\overline{1}};q_{\overline{12}})$ since in this case $k_{ji}=p_2$, but this integrand singularity cancels in the sum of the two dual contributions involving $\deltatilde{q_{\overline{1}}}$ and $\deltatilde{q_{\overline{12}}}$. The analysis can be extended easily to the other dual cuts, and similar conclusions are found.\\
\\
Now, let's consider the genuine two-loop case where a dual propagator becomes on shell in the double-cut of two propagators that do not belong to the same loop line. For the rest of this section, we consider
\begin{equation}
q_i\in\alpha_1\;,\quad q_j\in\alpha_2\quad\text{and}\quad q_k\in\alpha_3\;.
\end{equation}
We study the quantity
\begin{equation}
S_{ijk}=\frac{1}{(2\pi i)^2}\,\tilde{\delta}(q_i,q_j)\, G_D(q_i,q_j;q_k)+\{(i,j,k)\to(\overline{j},k,i)\}+\{(i,j,k)\to(k,i,j)\}\;,
\end{equation}
where $\overline{j}$ indicates that we reverse the momentum flow of $q_j$, as explained in \Section{\ref{Section:LTDMultiLoop}}, namely
\begin{equation}\label{Equation:HTLReverseFlowExample}
q_{\overline{j}}=-q_j\;,\qquad\Sup{q_{\overline{j},0}}=\Sup{q_{j,0}}\;.
\end{equation}
We therefore have
\begin{align}\label{Equation:HTLSijkValue}
S_{ijk}=&~\frac{\delta(q_{i,0}-\Sup{q_{i,0}})}{2\Sup{q_{i,0}}}\frac{\delta(q_{j,0}-\Sup{q_{j,0}})}{2\Sup{q_{j,0}}}\frac{1}{(\Sup{q_{i,0}}+\Sup{q_{j,0}}+k_{k(ij),0})^2-(\Sup{q_{k,0}})^2}\nn\\
&~+\frac{\delta(q_{j,0}+\Sup{q_{j,0}})}{2\Sup{q_{j,0}}}\frac{\delta(q_{k,0}-\Sup{q_{k,0}})}{2\Sup{q_{k,0}}}\frac{1}{(\Sup{q_{j,0}}+\Sup{q_{k,0}}-k_{k(ij),0})^2-(\Sup{q_{i,0}})^2}\nn\\
&~+\frac{\delta(q_{i,0}-\Sup{q_{i,0}})}{2\Sup{q_{i,0}}}\frac{\delta(q_{k,0}-\Sup{q_{k,0}})}{2\Sup{q_{k,0}}}\frac{1}{(\Sup{q_{k,0}}-\Sup{q_{i,0}}-k_{k(ij),0})^2-(\Sup{q_{j,0}})^2}\;,
\end{align}
where
\begin{equation}\label{Equation:HTLkkij}
k_{k(ij)}=q_k-q_i-q_j
\end{equation}
only depends on the external momenta. The quantity $S_{ijk}$ becomes singular in the limits summarised by the condition $\lambda_{ijk}^{\pm\pm\pm}\to0$, with 
\begin{equation}
\lambda_{ijk}^{\pm\pm\pm}=\pm\Sup{q_{i,0}}\pm\Sup{q_{j,0}}\pm\Sup{q_{k,0}}+k_{k(ij),0}\;.
\end{equation}
According to \Eq{\ref{Equation:HTLSijkValue}}, only four independent solutions have to be considered, namely $\lambda_{ijk}^{+++}$, $\lambda_{ijk}^{++-}$, $\lambda_{ijk}^{+--}$ and $\lambda_{ijk}^{---}$.
For two of these limits, there is a perfect cancellation of the integrand singularities, as indeed
\begin{equation}
\lim_{\lambda_{ijk}^{++-}\to0}S_{ijk}=\mathcal{O}\left((\lambda_{ijk}^{++-})^0\right)\;,\qquad\lim_{\lambda_{ijk}^{+--}\to0}S_{ijk}=\mathcal{O}\left((\lambda_{ijk}^{+--})^0\right)\;,
\end{equation}
while for the remaining two limits,
\begin{align}
\lim_{\lambda_{ijk}^{+++}\to0}S_{ijk}=&~-\theta(-k_{k(ij),0})\,(\lambda_{ijk}^{+++})^{-1}_{+++}\,\frac{\delta(q_{i,0}-\Sup{q_{i,0}})\,\delta(q_{j,0}-\Sup{q_{j,0}})}{(2\Sup{q_{i,0}})(2\Sup{q_{j,0}})(2\Sup{q_{k,0}})}+\mathcal{O}\left((\lambda_{ijk}^{+++})^0\right)\;,\nn\\
\lim_{\lambda_{ijk}^{---}\to 0}S_{ijk}=&~\theta(k_{k(ij),0})\,(\lambda_{ijk}^{---})^{-1}\,\frac{\delta(q_{j,0}+\Sup{q_{j,0}})\,\delta(q_{k,0}-\Sup{q_{k,0}})}{(2\Sup{q_{i,0}})(2\Sup{q_{j,0}})(2\Sup{q_{k,0}})}+\mathcal{O}\left((\lambda_{ijk}^{---})^0\right)\;.
\end{align}
Although these singularities remain in the general case, it is possible to show that $\lambda_{ijk}^{+++}$ (resp. $\lambda_{ijk}^{---}$) can only cancel if, in the particular case where $m_i=m_k=M_f$ and $m_j=0$,
\begin{equation}\label{Equation:LambdaCond}
\begin{cases}
k_{k(ij),0}<0\\
k_{k(ij)}^2-4M_f^2\geq0
\end{cases}
\qquad\left(\text{resp.}
\quad
\begin{cases}
k_{k(ij),0}>0\\
k_{k(ij)}^2-4M_f^2\geq0
\end{cases}
\right)\;.
\end{equation}
Because of the fact $k$ does not depend on the loop momenta, the highest possible value of $k_{k(ij)}^2$ is $(p_1+p_2)^2=\s$, reached for instance when considering $G_D(q_1,q_4;q_{\overline{12}})$. This means that the second condition of \Eq{\ref{Equation:LambdaCond}} reduces to
\begin{equation}
\frac{4M_f^2}{\s}<1\;,
\end{equation}
which is not fulfilled if, as stated above, the Higgs boson is assumed to be on shell, i.e. $\s=M_H^2$.\\
\\
Now, we consider the possibility to encounter soft singularities as one of the internal particles is massless. The integrand becomes soft in the loop momentum $\ell_2$ if $\Sup{q_{j,0}}=0$. In that case, the analysis of the singular behaviour is very similar to the one-loop case. We must solve the condition
\begin{equation}\label{Equation:HTLSoftCondition}
\lambda_{ik}^{\pm\pm}=\pm\Sup{q_{i,0}}\pm\Sup{q_{k,0}}+k_{ki,0}
\end{equation}
in $\ell_1$. If the three propagators are attached to the same vertex, then $k_{k(ij),0}=0$
and $\Sup{q_{i,0}}=\Sup{q_{k,0}}$. In that case $\lambda_{ik}^{+-}$ and $\lambda_{ik}^{-+}$ can vanish, then enhancing the integrand singularity in $\ell_2$, but the overall singularity cancels between dual contributions. The other solution, with $\lambda_{ik}^{++}=0$, is not possible because $\Sup{q_{i,0}}=\Sup{q_{k,0}}\geq M_f$. If the three propagators do not interact in the same vertex, then $k_{ki,0}\neq0$. This configuration includes the cases where $q_i$ and $q_k$ belong to the same or to different loop lines. There are solutions to \Eq{\ref{Equation:HTLSoftCondition}}, but again either they cancel among dual contributions
or \Eq{\ref{Equation:HTLFB}} is not fulfilled. In all the cases, the soft singularities of the integrand in $\ell_2$ do not translate into soft singularities of the amplitude because of the integration measure.

\section{Algorithmic approach to two-loop local UV renormalisation}\label{Section:HTLUV}
\fancyhead[LO]{\ref*{Section:HTLUV}~~\nameref*{Section:HTLUV}}

In this section we introduce a novel method to build local integrand-level UV counterterms, applicable to any two-loop process. We also explain how to recover well-known renormalisation schemes by introducing scheme-fixing parameters. In particular, we apply this algorithm to the $H\to\gamma\gamma$ process studied in this chapter.\\
\\
Let's consider a generic unrenormalised two-loop amplitude $\mathcal{A}^{(2)}$, written as
\begin{equation}
\mathcal{A}^{(2)}=\int_{\ell_ 1}\,\int_{\ell_2}\,\mathcal{I}(\ell_1,\ell_2)\;,
\end{equation}
which we will assume to be completely free of any infrared singularity. Thus, only UV singularities may appear when either or both of $|\boldsymbol{\ell}_1|$ and $|\boldsymbol{\ell}_2|$ go to infinity. The local two-loop UV counterterms are built recursively by first fixing one of the two loop momenta, say $\ell_j$, and expanding the integrand $\mathcal{I}(\ell_1,\ell_2)$ up to logarithmic order around the UV propagator~\cite{Becker:2010ng}
\begin{equation}
G_F(q_{i,\uv})=\frac{1}{q_{i,\uv}^2-\mu_\uv^2+i0}\;,
\end{equation}
where the arbitrary scale $\mu_\uv$ represents the renormalisation scale, and 
$q_{i,\uv} = \ell_i+k_{i,\uv}$. For simplicity, we take $k_{i,\uv}=0$. The quantity
\begin{equation}\label{Equation:HTLACT1CT2}
\mathcal{A}^{(2)}-\mathcal{A}_{1,\uv}^{(2)}-\mathcal{A}_{2,\uv}^{(2)}\;,
\end{equation}
where $\mathcal{A}_{i,\uv}^{(2)}$ denotes the two-loop amplitude $\mathcal{A}^{(2)}$ in the limit $|\boldsymbol{\ell}_i|\to\infty$, is not necessarily UV finite when both loop momenta are simultaneously large. It is necessary to subtract also the double UV behaviour of \Eq{\ref{Equation:HTLACT1CT2}}. With this contribution, which is represented by $\mathcal{A}^{(2)}_{\uv^2}$, the final renormalised amplitude reads 
\begin{equation}\label{Equation:HTLACT1CT2CT112}
\mathcal{A}^{(2)}_{\r}= \mathcal{A}^{(2)}-\mathcal{A}_{1,\uv}^{(2)}-\mathcal{A}_{2,\uv}^{(2)}-\mathcal{A}_{\uv^2}^{(2)}\;,
\end{equation}
and is UV safe in all the limits.  As an example, we consider
\begin{equation}
\mathcal{I}(\ell_1,\ell_2)=\frac{1}{(\ell_1^2-M^2+i0)(\ell_2^2-M^2+i0)^2((\ell_1+\ell_2)^2-M^2+i0)}\;.
\end{equation}
This integrand produces a UV singularity when $|\boldsymbol{\ell_1}|\to\infty$, but it is superficially regular in $\ell_2$, meaning $\mathcal{A}_{2,\uv}^{(2)} =0$. Computing the remaining counterterm gives
\begin{equation}
\mathcal{A}_{1,\uv}^{(2)}=\int_{\ell_ 1}\,\frac{1}{(\ell_{1}^2-\mu_\uv^2+i0)^2}\int_{\ell_2}\,\frac{1}{(\ell_2^2-M^2+i0)^2}\;,
\end{equation}
which effectively removes the UV behaviour in $\ell_1$, but at the same time also introduces a singularity in $\ell_2$. 
It is therefore necessary to introduce the additional counterterm $\mathcal{A}_{\uv^2}^{(2)}$ to fix the UV behaviour when both loop momenta go to infinity. This is done by expanding \Eq{\ref{Equation:HTLACT1CT2}} for very high values of $\ell_1$ and $\ell_2$, while never neglecting one compared to the other. In this example, we get
\begin{equation}
\mathcal{A}_{\uv^2}^{(2)}=\int_{\ell_ 1}\,\int_{\ell_2}\,\frac{1}{(\ell_1^2-\mu_\uv^2+i0)(\ell_2^2-\mu_\uv^2+i0)^2}\left(\frac{1}{(\ell_1+\ell_2)^2-\mu_\uv^2+i0}-\frac{1}{\ell_1^2-\mu_\uv^2+i0}\right)~.
\end{equation}
Then, the renormalised amplitude, as defined in \Eq{\ref{Equation:HTLACT1CT2CT112}}, is finite in the UV. It is still necessary, though, to introduce subleading contributions to fix the renormalisation scheme. This is better explained in the following for the $H\to\gamma\gamma$ two-loop amplitude.\\
\\
With the labelling of the internal momenta that we have adopted for the $H\to\gamma\gamma$ amplitude, it is more convenient to express the UV behaviour at two loops in terms of $q_{1,\uv}=\ell_1$ and $q_{12,\uv}=\ell_{12}$, with $\ell_2=\ell_{12}-\ell_1$. Explicitly, the single and double UV behaviours are implemented by making use of the following transformations
\begin{align}\label{Equation:HTLTransfUV}
\mathcal{S}_{j,\uv}:&~\{\ell_j^2~|~\ell_j\cdot k_i\}\to\{\lambda^2\,q_{j,\uv}^2+(1-\lambda^2)\,\mu_\uv^2~|~\lambda\,q_{j,\uv}\cdot k_i\}\;,\qquad j,k\in\{1,12\}\;,\nn\\ 
\mathcal{S}_{\uv^2}:&~\{\ell_j^2~|~\ell_j\cdot\ell_k~|~\ell_j\cdot k_i\}\to\nn\\ 
&~\{\lambda^2\,q_{j,\uv}^2+(1-\lambda^2)\,\mu_\uv^2~|~\lambda^2\,q_{j,\uv}\cdot q_{k,\uv}+(1-\lambda^2)/2\,\mu_\uv^2~|~\lambda\,q_{j,\uv}\cdot k_i\}\;,  
\end{align}
then expanding for $\lambda\to\infty$ and truncating the corresponding series in $\lambda$ up to logarithmic degree. This last operation is represented by the function $L_\lambda$. In particular, the UV counterterms are defined as
\begin{align}
\mathcal{A}^{(2,f)}_{j,\uv}=&~L_\lambda\left(\left.\mathcal{A}^{(2,f)}\right|_{\mathcal{S}_{j,\uv}}\right)-(e\,e_f)^2\left(d_{j,\uv}^{(f)}\,\mu_\uv^2\,\int_{\ell_j}\,\left(G_F(q_{j,\uv})\right)^3\right)\mathcal{A}^{(1,f)}\;,\quad j\in \{1, 12\}\;,\\
\mathcal{A}^{(2,f)}_{\uv^2}=&~L_\lambda\left(\left.\left(\mathcal{A}^{(2,f)}-\,\sum_{j=1,12}\,\mathcal{A}^{(2,f)}_{j,\uv}\right)\right|_{\mathcal{S}_{\uv^2}}\right)\nn\\
&~-4g_f\,\s\,(e\, e_f)^2\left(d_{\uv^2}^{(f)}\,\mu_\uv^4\int_{\ell_1}\,\int_{\ell_2}\,\left(G_F(q_{1,\uv})\right)^3\left(G_F(q_{12,\uv})\right)^3\right)\;,\label{Equation:HTLAUV2}
\end{align}
where $\mathcal{A}^{(1,f)}$ is the unintegrated one-loop amplitude written in terms of $\ell_i$ ($\ell_1$ or $\ell_{12}$) with $i\neq j$, and where $d_{j,\uv}^{(f)}$ and $d_{\uv^2}^{(f)}$ are scalar coefficients used to fix the renormalisation scheme. Note that the integrals they multiply integrate to finite quantities. The factor 4 appearing in the second line of \Eq{\ref{Equation:HTLAUV2}} is arbitrary and has been introduced to conveniently rescale $d_{\uv^2}^{(f)}$. 

\subsection{Higgs boson vertex renormalisation}\label{Section:HTLHiggsVertex}

\begin{table}
	\begin{center}
		\begin{tabular}{|c|ccc|ccc|}
			\hline
			& $c_{H,\uv}^{(f)}$ & $d_{H, \uv}^{(f)}$ & $C_{H, \uv}^{(f)}$ 
			& $c_{\gamma, \uv}^{(f)}$& $d_{\gamma, \uv}^{(f)}$ & $C_{\gamma, \uv}^{(f)}$ \\ \hline \hline
			$t \bar{t}$ & $d$ & $4$ & $4$ & $(d-2)/2$ & $2$ & $1$  \\
			$\phi\phi^\dagger$ & 1 & $0$ & $1$ & $-2~~~	$ & $0$ & $-2~~~$ \\
			\hline
		\end{tabular}
		\caption{Values of the scheme fixing parameters in the $\MSbar$ for the single UV counterterms.}
		\label{Table:HTLCoefSingleUV}
	\end{center}
\end{table}

In the Feynman gauge, the one-loop QED correction to the Higgs boson vertex exhibits the UV behaviour 
\begin{align}\label{Equation:HTLGenericHiggsVertex}
\mathbf{\Gamma}^{(1,f)}_{H,\uv}=&~(e\,e_f)^2\,\int_{\ell_1}\,\big(G_F(q_{1,\uv})\big)^2\left(c_{H,\uv}^{(f)}-G_F(q_{1,\uv})\,d_{H,\uv}^{(f)}\,\mu_\uv^2\right)\,\mathbf{\Gamma}^{(0,f)}_{H}\nn\\
=&~(e\,e_f)^2\frac{\Se}{16\pi^2}\left(\frac{\mu_\uv^2}{\mu^2}\right)^{-\epsilon}\frac{C_{H,\uv}^{(f)}}{\epsilon}\,\mathbf{\Gamma}^{(0,f)}_{H}\;,
\end{align}
where $\mathbf{\Gamma}^{(0,t)}_H=-i\,M_t/\vev$ and $\mathbf{\Gamma}^{(0,\phi)}_H=-2i\,M_\phi^2/\vev$ are the tree-level vertex interactions, and $\Se=(4\pi)^\epsilon\Gamma(1+\epsilon)$. The coefficients $d_{H,\uv}^{(f)}$ are subleading and are necessary to fix the renormalisation scheme. The values of these coefficients in the $\MSbar$ renormalisation scheme\footnote{We distinguish $\Se= (4\pi)^\epsilon\, \Gamma(1+\epsilon)$ from the usual $\MSbar$ scheme factor $S_\epsilon^{\MSbar} = (4\pi)^\epsilon e^{-\epsilon\gamma_E}$ or $S_\epsilon= (4\pi)^\epsilon/\Gamma(1-\epsilon)$ as used in~\cite{Bolzoni:2010bt}. At NLO all these definitions lead to the same expressions. At NNLO, they lead to slightly different bookkeeping of the IR and UV poles at intermediate steps, but physical cross sections of infrared-safe observables are the same.} are summarised in \Table{\ref{Table:HTLCoefSingleUV}}, and they are related
to the coefficient of the integrated vertex counterterm through
\begin{equation}\label{Equation:HTLIntegratedHiggsVertex}
C_{H,\uv}^{(f)} = c_{H,\uv}^{(f)}+\frac{\epsilon}{2} \, d_{H, \uv}^{(f)}\;.
\end{equation}
From the expression of the vertex counterterm in \Eq{\ref{Equation:HTLGenericHiggsVertex}} we can construct the UV counterterm of the two-loop scattering amplitude in the limit $|\boldsymbol{\ell}_1|\to\infty$ with $|\boldsymbol{\ell}_{12}|$ fixed. It reads
\begin{equation}\label{Equation:HTLA1UV}
\mathcal{A}_{H,\uv}^{(2,f)}=\int_{\ell_1}\,\big(G_F(q_{1,\uv})\big)^2\left(c_{H,\uv}^{(f)}-G_F(q_{1,\uv})\,d_{H,\uv}^{(f)}\,\mu_\uv^2\right)\mathcal{A}^{(1,f)}(\ell_{12})\;,
\end{equation}
where $\mathcal{A}^{(1,f)}$ is the unrenormalised one-loop $H\to\gamma\gamma$ amplitude that can be found in \Eq{\ref{Equation:HOLA}}. Note that $\mathcal{A}^{(1,f)}$ is locally divergent and should be renormalised as well in \Eq{\ref{Equation:HTLA1UV}}. However, by definition, we want $\mathcal{A}_{H,\uv}^{(2,f)}$ to exactly cancel the singularities arising when  $|\boldsymbol{\ell_1}|\to\infty$. Putting $\mathcal{A}_{\r}^{(1,f)}$ instead of $\mathcal{A}^{(1,f)}$ in \Eq{\ref{Equation:HTLA1UV}} would therefore alter the UV behaviour of the single counterterm and not properly remove the corresponding infinities. It is only when considering the double UV counterterm (\Section{\ref{Section:HTLDoubleUV}}) that the one-loop amplitude implicitly gets renormalised.\\
\\
The corresponding dual representation is
\begin{equation}
\mathcal{A}_{H,\uv}^{(2,f)}(q_{1,\uv},q_{\overline{i}})=\int_{\ell_1}\,\frac{\deltatilde{q_{1,\uv}}}{2\,(\Sup{q_{1,\uv}})^2}\,\left(c_{H,\uv}^{(f)}+d_{H,\uv}^{(f)}\,\frac{3\mu_\uv^2}{4\,(\Sup{q_{1,\uv}})^2}\right)\mathcal{A}^{(1,f)}(q_{\overline{i}})\;,
\end{equation}
where $\Sup{q_{1,\uv}}=\sqrt{\boldsymbol{\ell}_1^2+\mu_\uv^2}$. Since the diagrams (2 for the top quark, 3 for the charged scalar) that contribute to the Higgs boson vertex correction are the only ones that are divergent when $|\boldsymbol{\ell_1}|\to\infty$, we directly have $\mathcal{A}_{1,\uv}^{(2,f)}=\mathcal{A}_{H,\uv}^{(2,f)}$.

\subsection{Photon vertex renormalisation}\label{Section:HTLPhotonVertex}

The one-loop correction to the photon interaction vertex to top quarks in the UV is given in the Feynman gauge by
\begin{align}\label{Equation:HTLttPhoton}
\mathbf{\Gamma}_{\gamma,\uv}^{(1,t)}=&~(e\,e_t)^2\,\int_{\ell_2}\,\big(G_F(q_{12,\uv})\big)^2\left(\left(c_{\gamma,\uv}^{(t)}-G_F(q_{12,\uv})\,d_{\gamma, \uv}^{(t)}\,\mu_\uv^2\right)\mathbf{\Gamma}_{\gamma}^{(0,t)}+c_{\gamma,\uv}^{(t)}\,\mathbf{\Delta}_{\gamma,\uv}^{(1,t)}\right)\nn\\
=&~(e\,e_t)^2\frac{\Se}{16\pi^2}\left(\frac{\mu_\uv^2}{\mu^2}\right)^{-\epsilon}\frac{C_{\gamma,\uv}^{(t)}}{\epsilon}\,\mathbf{\Gamma}_\gamma^{(0,t)}\;,
\end{align}
with
\begin{equation}\label{Equation:HTLttPhotonDelta}
\mathbf{\Delta}_{\gamma,\uv}^{(1,t)}=\mathbf{\Gamma}_\gamma^{(0,t)}-2\,G_F(q_{12,\uv})\,\slashed{q}_{12,\uv}\{\slashed{q}_{12,\uv},\mathbf{\Gamma}_\gamma^{(0,t)}\}\;.
\end{equation}
In \Eq{\ref{Equation:HTLttPhotonDelta}}, the term proportional to $\mathbf{\Delta}_{\gamma,\uv}^{(1,t)}$ integrates to zero in $d$ space-time dimensions. Similarly to \Eq{\ref{Equation:HTLIntegratedHiggsVertex}}, the coefficient of the integrated vertex counterterm is given by 
\begin{equation}
C_{\gamma,\uv}^{(f)}=c_{\gamma,\uv}^{(f)}+\frac{\epsilon}{2}\,d_{\gamma,\uv}^{(f)}\;.
\end{equation}
Although the integrated UV vertex correction is proportional to the tree-level vertex $\mathbf{\Gamma}_{\gamma}^{(0,t)}=i\,e\,e_t\,\gamma^{\mu_i}$, thanks to the replacement $q_{12,\uv}^{\mu_1}\,q_{12,\uv}^{\mu_2}\to q_{12,\uv}^2 \, g^{\mu_1\mu_2}/d$, we cannot use this replacement in the unintegrated form because it would alter the local UV behaviour. We must keep the full expression in \Eq{\ref{Equation:HTLttPhoton}}, including especially the term proportional to $\mathbf{\Delta}^{(1,t)}_{\gamma,\uv}$, to construct the local UV counterterm of the two-loop scattering amplitude.\\
\\
For charged scalars as internal particles, we need to consider both the three-point and the four-point interaction vertices. For the three-point vertex there are three contributing diagrams\footnote{There are actually 12 contributing diagrams, but they can be divided into 4 sets of 3 diagrams, as we have two vertices and two possible ordering for the external particles.}, and the corresponding counterterm reads
\begin{align}\label{Equation:HTLScalarPhoton}
\mathbf{\Gamma}_{\gamma,\uv}^{(1,\phi)}=&~(e\,e_\phi)^2\,\int_{\ell_2}\,\left( G_F(q_{12,\uv})\right)^2c_{\gamma, \uv}^{(\phi)}\left(\mathbf{\Gamma}_\gamma^{(0,\phi)}+ \mathbf{\Delta}_{\gamma,\uv}^{(1,\phi)}\right)\nn\\
=&(e\,e_\phi)^2\frac{\Se}{16\pi^2}\left(\frac{\mu_\uv^2}{\mu^2}\right)^{-\epsilon}\frac{C_{\gamma,\uv}^{(\phi)}}{\epsilon}\,\mathbf{\Gamma}_\gamma^{(0,\phi)}\;,
\end{align}
where
\begin{equation}\label{Equation:HTLScalarPhotonDelta}
\mathbf{\Delta}_{\gamma,\uv}^{(1,\phi)}=\frac{1}{2}\left(\mathbf{\Gamma}_{\gamma}^{(0,\phi)}(q_{12,\uv},p_i)+4i(e\,e_\phi)\,G_F(q_{12,\uv})\,(q_{12,\uv}\cdot(k_{i-1}+k_i))\,q_{12,\uv}^{\mu_i}\right)\;,
\end{equation}
with $\mathbf{\Gamma}_{\gamma}^{(0,\phi)}=-i\,(e\,e_\phi)\left(q_{i-1}+q_i\right)^{\mu_i}$ and $\mathbf{\Gamma}^{(0,\phi)}_{\gamma}(q_{12,\uv},p_i)=i\,(e\,e_\phi)\left(q_{\overline{i-1}}+q_{\overline{i}}\right)^{\mu_i}$, where $q_{i-1}$ and $q_i$ are the outgoing and incoming internal momenta, respectively. For example, $q_{i-1}=q_3+p_1$ and $q_i=q_3+p_{12}$ for the vertex correction with emission of a photon with momentum $p_2$ in the lower corner of the two-loop Feynman diagram.\\
\\
For the four-point interaction vertex, there are nine contributing diagrams\footnote{Note that these diagrams include the ones contributing to the Higgs boson vertex correction. However, the singular regime considered here is different since we study the limit $|\boldsymbol{\ell_{12}}|\to\infty$.}, and we have
\begin{align}\label{Equation:HTLScalarPhotonPhoton}
\mathbf{\Gamma}_{\gamma \gamma,\uv}^{(1,\phi)}=&~(e\,e_\phi)^2\,\int_{\ell_2}\,\big(G_F(q_{12,\uv})\big)^2c_{\gamma,\uv}^{(\phi)}\left(\mathbf{\Gamma}_{\gamma\gamma}^{(0,\phi)}+\mathbf{\Delta}_{\gamma\gamma,\uv}\right)^{(1,\phi)}\nn\\
=&~(e\,e_\phi)^2\frac{\Se}{16\pi^2}\left(\frac{\mu_\uv^2}{\mu^2}\right)^{-\epsilon}\frac{C_{\gamma,\uv}^{(\phi)}}{\epsilon}\,\mathbf{\Gamma}^{(0,\phi)}_{\gamma\gamma}\;,
\end{align}
where
\begin{equation}\label{Equation:HTLScalarPhotonPhotonDelta}
\mathbf{\Delta}_{\gamma \gamma,\uv}^{(1,\phi)}=\frac{1}{2}\mathbf{\Gamma}_{\gamma\gamma}^{(0,\phi)}-4i\,(e\,e_\phi)^2\,G_F(q_{12, \uv})\,q_{12,\uv}^{\mu_1}\,q_{12, \uv}^{\mu_2}\;.
\end{equation}
with  $\mathbf{\Gamma}^{(0,\phi)}_{\gamma\gamma}=2i\,(e\, e_\phi)^2\,g^{\mu_1\mu_2}$. Remarkably, the coefficient $c_{\gamma, \uv}^{(\phi)}$ is the same as for the three-point interaction vertex. Again, in \Eq{\ref{Equation:HTLScalarPhoton}} and \Eq{\ref{Equation:HTLScalarPhotonPhoton}}, we cannot apply the replacement $q_{12,\uv}^{\mu_1}\,q_{12,\uv}^{\mu_2}\to q_{12,\uv}^2\,g^{\mu_1\mu_2}/d$ at the integrand level even though the terms $\mathbf{\Delta}^{(1,\phi)}_{\gamma}$ in \Eq{\ref{Equation:HTLScalarPhoton}}  and  $\mathbf{\Delta}^{(1,\phi)}_{\gamma\gamma}$ in \Eq{\ref{Equation:HTLScalarPhotonPhoton}} integrate to zero. Also notice that it was not necessary to introduce subleading terms for the scalar vertices because the finite part of the corresponding integrated counterterm is already 0.\\
\\
The integrated counterterm reads
\begin{align}
\mathcal{A}_{\gamma,\uv}^{(2,f)} (q_i, q_{12,\uv})=&~\left( \frac{\Se}{16\pi^2} \left( \frac{\mu_\uv^2}{\mu^2}\right)^{-\epsilon} \frac{2C_{\gamma,\uv}^{(f)}}{\epsilon} \right)\mathcal{A}_\gamma^{(1,f)}(q_i)\;,\label{Equation:HTLAGammaUV}\\
\mathcal{A}_{\gamma\gamma,\uv}^{(2,\phi)}(q_i,q_{12,\uv})=&~\left(\frac{\Se}{16\pi^2}\left(\frac{\mu_\uv^2}{\mu^2}\right)^{-\epsilon}\frac{C_{\gamma,\uv}^{(\phi)}}{\epsilon}\right)\mathcal{A}_{\gamma\gamma}^{(1,\phi)}(q_i)\;,\label{Equation:HTLAGammaGammaUV}
\end{align}
where $\mathcal{A}_{\gamma}^{(1,f)}$ is the sum of the two one-loop amplitudes involving triangle diagrams and $\mathcal{A}_{\gamma\gamma}^{(1,\phi)}$ is the one-loop bubble amplitude (which appears only for the charged scalar). The relative factors 2 in \Eq{\ref{Equation:HTLAGammaUV}} comes from the fact there are two three-point vertices to renormalise for each contributing diagram.\\
\\
The corresponding dual representations are
\begin{align}
\mathcal{A}_{\gamma,\uv}^{(2,f)} (q_i,q_{12,\uv})=&\int_{\ell_1}\,\frac{\deltatilde{q_{12,\uv}}}{(\Sup{q_{12,\uv}})^2}\,\Bigg[\left(c_{\gamma,\uv}^{(f)}+d_{\gamma,\uv}^{(f)}\,\frac{3\mu_\uv^2}{4(\Sup{q_{12,\uv}})^2}\right)\mathcal{A}_\gamma^{(1,f)}(q_i)\nn\\
&+c_{\gamma,\uv}^{(f)}\,\mathbf{\Delta}^{(1,f)}_{\gamma,\uv}(q_i,q_{12,\uv})\Bigg]\;,\label{Equation:HTLAgammaUVDual}\\
\mathcal{A}_{\gamma\gamma,\uv}^{(2,\phi)} (q_i,q_{12,\uv})=&\int_{\ell_1}\,\frac{\deltatilde{q_{12,\uv}}}{2(\Sup{q_{12,\uv}})^2}\,c_{\gamma,\uv}^{(\phi)}\left[\mathcal{A}_{\gamma\gamma}^{(1,\phi)}(q_i)+\mathbf{\Delta}^{(1,\phi)}_{\gamma\gamma,\uv}(q_i,q_{12,\uv}) \right]\;,\label{Equation:HTLAgammagammaUVDual}
\end{align}
where $\Sup{q_{12,\uv}}=\sqrt{\boldsymbol{\ell}_{12}^2+\mu_\uv^2}$.

\subsection{Self-energy renormalisation}

With our labelling of the momenta the self-energy insertions are defined in terms of the internal momenta $q_4=\ell_2$ and $q_{\overline{i}}=\ell_{12}+k_i$, with $i\in\{1,2,3,12\}$. Explicitly, in the Feynman gauge we have (notice the relative sign in $q_{\overline{i}}$ with respect to~\cite{Sborlini:2016hat} because of the fact the momentum flows in the opposite direction)
\begin{equation}\label{Equation:HTLSelfqi}
\Sigma^{(1,t)}(q_i)=(e\,e_t)^2\,\int_{\ell_2}\,G_F(q_4, q_{\overline{i}})\left(-2c_{\gamma,\uv}^{(t)}\,\slashed{q}_{\overline{i}}+c_{H,\uv}^{(t)}\,M_t\right)\;.
\end{equation}
The UV expansion of \Eq{\ref{Equation:HTLSelfqi}} reads
\begin{align}\label{Equation:HTLNewSelfUV}
\Sigma_\uv^{(1,t)}(q_i)=&(e\,e_t)^2\int_{\ell_2}\,\big(G_F(q_{12,\uv})\big)^2\,\bigg[-\left(c_{\gamma,\uv}^{(t)}-G_F(q_{12,\uv})\,d_{\gamma,\uv}^{(t)}\,\mu_\uv^2\right)\,\slashed{q}_{i}\nn\\
&~+\left(c_{H,\uv}^{(t)}-G_F(q_{12,\uv})\,d_{H,\uv}^{(t)}\,\mu_\uv^2\right)M_t+c_{\gamma,\uv}^{(t)}\,\mathbf{\Delta}^{(1,t)}_{\Sigma,\uv}\bigg]\nn\\
=&(e\,e_t)^2\frac{\Se}{16\pi^2}\left(\frac{\mu_\uv^2}{\mu^2}\right)^{-\epsilon}\frac{1}{\epsilon}\left(-C_{\gamma,\uv}^{(t)}\,\slashed{q}_i+C_{H,\uv}^{(t)}\,M_t\right)\;,
\end{align}
with 
\begin{equation}\label{Equation:HTLNewSelfUVDelta}
\mathbf{\Delta}_{\Sigma,\uv}^{(1,t)}=2\left(\frac{\slashed{q}_i}{2}-\slashed{k}_i-\slashed{q}_{12,\uv}-4G_F(q_{12,\uv})\left(q_{12,\uv}\cdot\left(\frac{q_i}{2}-k_i\right)\right)\,\slashed{q}_{12,\uv}\right)\;,
\end{equation}
where the coefficients $d_{k,\uv}^{(t)}$ are subleading contributions to be fixed through the renormalisation scheme. It is remarkable that it has been possible to write the quark self-energy in terms of the same coefficients that appear in the Higgs boson and photon vertices. Notice that the expression in \Eq{\ref{Equation:HTLNewSelfUV}} is simpler than the corresponding expression calculated in \Chapter{\ref{Chapter:FDUM}}, and only differs at $\Oep{1}$, which does not have any consequence at the considered order.\\
\\
The scalar self-energy corrections, which also include the snail diagrams, is written
\begin{equation}\label{Equation:HTLSelfqiScalar}
\Sigma^{(1,\phi)}(q_i)=(e\,e_\phi)^2\int_{\ell_2}\,G_F(q_4)\left(-c_{T,\uv}^{(\phi)}+G_F(q_{\overline{i}})\,\left(-c_{\gamma,\uv}^{(\phi)}\,\left(q_{\overline{i}}+\frac{q_i}{2}\right)\cdot q_i+c_{H,\uv}^{(\phi)}\,M_\phi^2\right)\right)\;,
\end{equation}
where $c_{T,\uv}^{(\phi)}=d-1$, or, equivalently, $c_{T,\uv}^{(\phi)}=4(c_{4,nu}^{(\phi)}-c_{4,u}^{(\phi)})$. One possibility would be to subtract the contribution which is proportional to $c_{T,\uv}^{(\phi)}$ before expanding in the UV, which would be equivalent to subtract a zero (actually this would not only work for the term generated by the snail diagrams, but also for any term that exclusively contains propagators depending only on $\ell_1$). However, it would modify only the double-cuts $\deltatilde{q_i,q_4}$. The UV expansion of \Eq{\ref{Equation:HTLSelfqiScalar}} reads 
\begin{align}\label{Equation:HTLNewSelfUVScalar}
\Sigma_\uv^{(1,\phi)}(q_i)=&~(e\,e_\phi)^2\,\int_{\ell_2}\,G_F(q_{12,\uv})\bigg[G_F(q_{12,\uv})\left(-c_{\gamma,\uv}^{(\phi)}\,q_i^2+c_{H,\uv}^{(\phi)}\,M_\phi^2+c_{\gamma,\uv}^{(\phi)}\,\mathbf{\Delta}^{(1,\phi)}_{\Sigma,\uv}\right)\nn\\
&+c_{T,\uv}^{(\phi)}\,\mathbf{\Delta}^{(1,\phi)}_{T,\uv}\bigg]\nn\\
=&~(e\,e_\phi)^2\frac{\Se}{16\pi^2}\left(\frac{\mu_\uv^2}{\mu^2}\right)^{-\epsilon}\frac{1}{\epsilon}\left(-C_{\gamma,\uv}^{(\phi)}\,q_i^2+C_{H,\uv}^{(\phi)}\,M_\phi^2\right)\;,
\end{align}
with
\begin{equation}\label{Equation:HTLNewSelfUVScalarDelta}
\mathbf{\Delta}_{\Sigma,\uv}^{(1,\phi)}=\left(\frac{q_i}{2}-k_i-q_{12,\uv}\right)\cdot q_i-4G_F(q_{12,\uv})\left(q_{12,\uv}\cdot\left(\frac{q_i}{2}-k_i\right)\right)q_{12,\uv}\cdot q_i\;, 
\end{equation}
\begin{align}\label{Equation:HTLNewSelfUVScalarSnail}
\mathbf{\Delta}^{(1,\phi)}_{T, \uv}=&-2+G_F(q_{12,\uv})\left((q_{12,\uv}+k_i)^2+2\left(\frac{q_i}{2}-k_i-q_{12,\uv}\right)\cdot q_i\right)\nn\\
&~-4\big(G_F(q_{12,\uv})\big)^2\left(\left(q_{12,\uv}\cdot\left(q_i-k_i\right)\right)^2+\frac{\mu_\uv^4}{d-2}\right)\;.
\end{align}
The terms $\mathbf{\Delta}^{(1,\phi)}_{\Sigma,\uv}$ and $\mathbf{\Delta}^{(1,\phi)}_{T, \uv}$ integrate to zero independently in $d$ dimensions. In the latter, i.e. \Eq{\ref{Equation:HTLNewSelfUVScalarSnail}}, it was necessary to include a subleading contribution, proportional to $\mu^4_\uv$. We will not provide the integrated and dual expressions for $\mathcal{A}_{\Sigma,\uv}^{(2,f)}(q_i,q_{12,\uv})$ because they are quite heavy.\\
\\
Finally, the counterterm $\mathcal{A}_{12,\uv}^{(2,f)}$ is simply obtained by considering the photon vertex and self-energy contributions together, namely
\begin{align}
\mathcal{A}_{12,\uv}^{(2,t)}(q_i,q_{12,\uv})=&\mathcal{A}_{\gamma,\uv}^{(2,t)} (q_i,q_{12,\uv})+\mathcal{A}_{\Sigma,\uv}^{(2,t)} (q_i,q_{12,\uv})\;,\nn\\
\mathcal{A}_{12,\uv}^{(2,\phi)}(q_i,q_{12,\uv})=&\mathcal{A}_{\gamma,\uv}^{(2,\phi)} (q_i,q_{12,\uv})+\mathcal{A}_{\gamma\gamma,\uv}^{(2,\phi)} (q_i,q_{12,\uv})+\mathcal{A}_{\Sigma,\uv}^{(2,\phi)} (q_i,q_{12,\uv})\;.
\end{align}

\subsection{The double UV counterterm}\label{Section:HTLDoubleUV}

While it is entirely possible to compute $\mathcal{A}_{\uv^2}^{(2,f)}$ by directly taking the sum of all contributions, it is more interesting to consider well-chosen subsets of diagrams -- it also lightens intermediate expressions. For the top as the internal particle, it is logical to consider all the contributions to the Higgs boson vertex corrections, all the contributions to the photon vertex corrections and all the contributions to the self-energy corrections, as there is no ambiguity or cross-contributions for the $H\to\gamma\gamma$ process at two-loop. They will be written $\mathcal{A}_{H,\uv^2}^{(2,t)}$, $\mathcal{A}_{\gamma,\uv^2}^{(2,t)}$, $\mathcal{A}_{\Sigma,\uv^2}^{(2,t)}$, and account for 2, 4 and 6 diagrams, respectively. For the charged scalar as an internal particle, there is a subtlety. The three diagrams that contribute to the Higgs boson vertex correction also contribute to the $\gamma\gamma\phi\phi^\dagger$ vertex correction. For this reason, if we want to split $\mathcal{A}^{(2,\phi)}_{\uv^2}$, we have to consider both corrections together. For the photon and self-energy corrections, though, there is no ambiguity whatsoever. Thus, we define $\mathcal{A}_{H+\gamma\gamma,\uv^2}^{(2,\phi)}=\mathcal{A}_{H,\uv^2}^{(2,\phi)}$, $\mathcal{A}_{\gamma,\uv^2}^{(2,\phi)}$, $\mathcal{A}_{\Sigma,\uv^2}^{(2,\phi)}$, that account for 9, 12 and 16 diagrams, respectively.\\
\\
According to \Eq{\ref{Equation:HTLAUV2}}, the unintegrated double UV counterterms have the form
\begin{align}\label{Equation:HTLDoubleUVForm}
\mathcal{A}_{\{H,\gamma,\Sigma\},\uv^2}^{(2,f)}=&~g_f\,\s(e\,e_f)^2\,\int_{\ell_1}\,\int_{\ell_{2}}\,\bigg[\big(G_F(q_{1,\uv})\big)^{n_1}\,\big(G_F(q_{2,\uv})\big)^{n_2}\,\big(G_F(q_{12,\uv})\big)^{n_{12}}\,\mathcal{N}_{\{H,\gamma,\Sigma\}}^{(f)}\nn\\
&~-4\left(G_F(q_{1,\uv})\right)^3\left(G_F(q_{12,\uv})\right)^3d_{\{H,\gamma,\Sigma\},\uv^2}^{(f)}\,\mu_\uv^4\bigg]\;,
\end{align}
where
\begin{equation}
G_F(q_{2,\uv})=\frac{1}{(q_{12,\uv}-q_{1,\uv})^2-\mu_\uv^2 +i0}\;,
\end{equation}
and with $n_i$ being positive integers. Note that even though $\mathcal{N}_{\{H,\gamma,\Sigma\}}^{(f)}(q_{1,\uv},q_{12,\uv})$ should be expected to also depend on the external momenta $p_1$ and $p_2$, it is a remarkable feature that, thanks to welcome cancellations, it does not when considering the sum of all contributing diagrams. The expression in \Eq{\ref{Equation:HTLDoubleUVForm}} is therefore free of irreducible scalar products and can very easily be reduced through integrations by parts to the form
\begin{equation}\label{Equation:HTLAHgammaSigmaUV2}
\mathcal{A}^{(2,f)}_{\{H,\gamma,\Sigma\},\uv^2}= g_f\,\s(e\,e_f)^2\left(c_{\{H,\gamma,\Sigma\},\ominus}^{(f)}\,I_\ominus+c_{\{H,\gamma,\Sigma\},\odot}^{(f)}\,I_\odot^2 \right)\;,
\end{equation}
where $I_\ominus$ is the sunrise scalar integral with missing external momenta and all the internal masses equal, and $I_\odot$ is the massive tadpole. Their expressions are available in e.g.~\cite{Caffo:1998du,Laporta:2004rb} and read
\begin{align}\label{Equation:HTLMISunrise}
I_\ominus=&~\frac{1}{\mu_\uv^2}\int_{\ell_1}\,\int_{\ell_2}\,G_F(q_{1,\uv},q_{12, \uv},q_{2,\uv})\nn\\
=&~\left(\frac{\Se}{16\pi^2}\right)^2\left(\frac{\mu_\uv^2}{\mu^2}\right)^{-2\epsilon}\left(-\frac{3}{2\epsilon^2}-\frac{9}{2\epsilon}+K_\ominus+\Oep{1}\right)\;,
\end{align}
with
\begin{equation}\label{Equation:HTLKominus}
K_\ominus=-\frac{21}{2}+2\sqrt{3}\text{Cl}_2\left(\frac{\pi}{3}\right)\;,
\end{equation}
where $\text{Cl}_2$ is the Clausen function of order 2, and
\begin{equation}\label{Equation:HTLMITadpole}
I_\odot=\frac{1}{\mu_\uv^2}\int_{\ell_i}\,G_F(q_{i,\uv})=\frac{\Se}{16\pi^2}\left(\frac{\mu_\uv^2}{\mu^2}\right)^{-\epsilon}\frac{1}{\epsilon(1-\epsilon)}\;.
\end{equation}
Because of the presence of the double pole inside \Eq{\ref{Equation:HTLMISunrise}}, it is, in the general case, necessary to keep track of the different normalisations as choosing one over another could lead to a shift in the finite part. As we are working in the $\MSbar$ scheme, we should rather factorise the usual $S_\epsilon^{\MSbar}=(4\pi)^\epsilon e^{-\epsilon\,\gamma_E}$, but doing so, a global factor equal to $\Se/S_\epsilon^{\MSbar}=1+\pi^2\,\epsilon^2/12+\mathcal{O}(\epsilon^3)$ would appear. We will see that in the end considering one normalisation over the other does not introduce any mismatch, so we chose to keep $\Se$ for simplicity.\\
\\
These two master integrals (MI) are the only ones needed to evaluate $\mathcal{A}^{(2,f)}_{\uv^2}$ and fix the subleading terms $d_{\uv^2}^{(f)}$. Note that in \Eq{\ref{Equation:HTLAHgammaSigmaUV2}}, the dependence in $d_{\{H,\gamma,\Sigma\},\uv^2}^{(f)}$ is implicitly included in $c_{\{H,\gamma,\Sigma\},\odot}^{(f)}$.\\
\\
The unintegrated double UV counterterm for the diagrams with loop corrections in the Higgs boson vertex and internal top quarks reads
\begin{align}\label{Equation:HTLAHUV2}
\mathcal{A}_{H,\uv^2}^{(2,t)}=&~g_f\,\s\,(e\,e_t)^2\,\int_{\ell_1}\,\int_{\ell_{2}}\,\big(G_F(q_{1,\uv})\big)^2\,\big(G_F(q_{12,\uv})\big)^2\nn\\
&~\bigg[-4\frac{d-4}{d-2}\left(c_{H,\uv}^{(t)}-G_F(q_{1,\uv})\,d_{H,\uv}^{(t)}\,\mu_\uv^2\right)\nn\\
&~+\frac{4G_F(q_{2,\uv})}{d-2}\left(d(d-4)q_{1,\uv}^2-2(d-2)^2q_{1,\uv}\cdot q_{12,\uv}+4\mu_\uv^2\right)\nn\\
&~-4G_F(q_{1,\uv})\,G_F(q_{12,\uv})\,d_{H,\uv^2}^{(t)}\,\mu_\uv^4\bigg]\;,
\end{align}
and gives
\begin{equation}
c_{H,\ominus}^{(t)}=-\frac{2(d-3)((d-2)^2+4)}{3(d-2)}\;,\qquad c_{H,\odot}^{(t)}=(d-2)\left(4+\frac{(d-4)^2}{4}\left(d_{H,\uv}^{(t)}-\frac{d-2}{4}d_{H,\uv^2}^{(t)}\right)\right)\;.
\end{equation}
where the parameter $c_{H,\uv}^{(t)}$ has been replaced by its value, given in \Table{\ref{Table:HTLCoefSingleUV}}. After integration, we have
\begin{equation}
\mathcal{A}_{H,\uv^2}^{(2,t)}=g_f\,\s(e\,e_t)^2\,\left(\frac{\Se}{16\pi^2}\right)^2\left(\frac{\mu_\uv^2}{\mu^2}\right)^{-2\epsilon}\left(52+\frac{16K_\ominus}{3}+2d_{H,\uv}^{(t)}-d_{H,\uv^2}^{(t)}+\Oep{1}\right)\;.
\end{equation}
The unintegrated expressions for the double UV counterterm for the photon and self-energy corrections are a bit heavy, so we will only provide the MI coefficients. For the photon vertex corrections, they read
\begin{align}
c_{\gamma,\ominus}^{(t)}=&~\frac{8d(d-4)(d-2)}{3(d-2)}\;,\nn\\
c_{\gamma,\odot}^{(t)}=&~-\frac{(d-4)(d-2)}{2}\left(2d-(d-4)d_{\gamma,\uv}^{(t)}+\frac{(d-4)(d-2)}{8}d_{\gamma,\uv^2}^{(t)}\right)\;,
\end{align}
while for the self-energy corrections, they read
\begin{align}
c_{\Sigma,\ominus}^{(t)}=&~-\frac{4(d-4)^2(d-3)}{3(d-2)}\;,\nn\\
c_{\Sigma,\odot}^{(t)}=&~\frac{(d-4)^2(d-2)}{4}\left(\frac{(d-4)(d-2)}{2}+d_{H,\uv}^{(t)}-4d_{\gamma,\uv}^{(t)}+4-\frac{d-2}{4}d_{\Sigma,\uv^2}^{(t)}\right)\;,
\end{align}
where once again we replaced $c_{H,\uv}^{(t)}$ and $c_{\gamma,\uv}^{(t)}$ by their value. Integrating the counterterms gives
\begin{equation}
\mathcal{A}_{\gamma,\uv^2}^{(2,t)}=g_f\,\s\,(e\,e_t)^2\left(\frac{\Se}{16\pi^2}\right)^2\left(\frac{\mu_\uv^2}{\mu^2}\right)^{-2\epsilon}\left(-16+4d_{\gamma,\uv}^{(t)}-d_{\gamma,\uv^2}^{(t)}+\Oep{1}\right)\;,
\end{equation}
and
\begin{equation}
\mathcal{A}_{\Sigma,\uv^2}^{(2,t)}=g_f\,\s\,(e\,e_t)^2\left(\frac{\Se}{16\pi^2}\right)^2\left(\frac{\mu_\uv^2}{\mu^2}\right)^{-2\epsilon}\left(4+2d_{H,\uv}^{(t)}-8d_{\gamma,\uv}^{(t)}-d_{\Sigma,\uv^2}^{(t)}+\Oep{1}\right)\;,
\end{equation}
leading to
\begin{equation}
\mathcal{A}_{\uv^2}^{(2,t)}=g_f\,\s\,(e\,e_t)^2\left(\frac{\Se}{16\pi^2}\right)^2\left(\frac{\mu_\uv^2}{\mu^2}\right)^{-2\epsilon}\left(40+\frac{16K_\ominus}{3}+4(d_{H,\uv}^{(t)}-d_{\gamma,\uv}^{(t)})-d_{\uv^2}^{(t)}+\Oep{1}\right)\;,
\end{equation}
where we used $d_{H,\uv^2}^{(t)}+d_{\gamma,\uv^2}^{(t)}+d_{\Sigma,\uv^2}^{(t)}=d_{\uv^2}^{(t)}$.\\
\\
For the scalar, the unintegrated double UV counterterm for the Higgs boson and $\gamma\gamma\phi\phi^\dagger$ vertex corrections read
\begin{align}
\mathcal{A}_{H+\gamma\gamma,\uv^2}^{(\phi)}=&~g_f\,\s\,(e\,e_\phi)^2\int_{\ell_1}\,\int_{\ell_2}\,\big(G_F(q_{1,\uv})\big)^2\,\big(G_F(q_{12,\uv})\big)^2\nn\\
&~\bigg[2\frac{d-4}{d-2}c_{H,\uv}^{(\phi)}-\frac{c_{\gamma,\uv}^{(\phi)}}{d-2}\left(4-3d+4\mu^2G_F(q_{12,\uv})\right)\nn\\
&~-\frac{2G_F(q_{2,\uv})}{d-2}\left((d-4)q_{1,\uv}^2+(4-3d)q_{12,\uv}^2+2d\,q_{1,\uv}\cdot q_{12,\uv}+d\,\mu_\uv^2\right)\nn\\
&~-4G_F(q_{1,\uv})G_F(q_{12,\uv})d_{H+\gamma\gamma,\uv^2}^{(\phi)}\,\mu_\uv^4\bigg]\;,
\end{align}
which after IBP reduction gives
\begin{equation}
c_{H+\gamma\gamma,\ominus}^{(\phi)}=0\;,\qquad c_{\gamma,\odot}^{(\phi)}=-(d-2)\left(2+\frac{(d-4)^2(d-2)}{16}d_{H+\gamma\gamma,\uv^2}^{(\phi)}\right)\;.
\end{equation}
We obtain
\begin{equation}
\mathcal{A}_{H+\gamma\gamma,\uv^2}^{(2,\phi)}=g_f\,\s\,(e\,e_\phi)^2\left(\frac{\Se}{16\pi^2}\right)^2\left(\frac{\mu_\uv^2}{\mu^2}\right)^{-2\epsilon}\left(-\frac{4}{\epsilon^2}-\frac{4}{\epsilon}-4-d_{H+\gamma\gamma,\uv^2}^{(\phi)}+\Oep{1}\right)\;.
\end{equation}
As for the top, the unintegrated expressions for $\mathcal{A}^{(2,\phi)}_{\gamma,\uv^2}$ and $\mathcal{A}^{(2,\phi)}_{\Sigma,\uv^2}$ are a bit heavy. Their corresponding MI coefficients are
\begin{equation}
c_{\gamma,\ominus}^{(\phi)}=0\;,\qquad c_{\gamma,\odot}^{(\phi)}=-(d-2)\left(4-\frac{(d-4)^2(d-2)}{16}d_{\gamma,\uv^2}^{(\phi)}\right)\;.
\end{equation}
and
\begin{align}
c_{\Sigma,\ominus}^{(\phi)}=&~-\frac{8(d-6)(d-3)}{3(d-2)}\;,\nn\\
c_{\Sigma,\odot}^{(\phi)}=&~-\frac{d-2}{6}\left((d-6)\left((d-4)^2(d-2)-32\right)+(d-4)^2(d-2)d_{\Sigma,\uv^2}^{(\phi)}\right)\;.
\end{align}
After integration, we find
\begin{equation}
\mathcal{A}^{(2,\phi)}_{\gamma,\uv^2}=g_f\,\s\,(e\,e_\phi)^2\left(\frac{\Se}{16\pi^2}\right)^2\left(\frac{\mu_\uv^2}{\mu^2}\right)^{-2\epsilon}\left(\frac{8}{\epsilon^2}+\frac{8}{\epsilon}+8-d_{\gamma,\uv^2}^{(\phi)}+\Oep{1}\right)\;,
\end{equation}
and
\begin{equation}
\mathcal{A}^{(2,\phi)}_{\Sigma,\uv^2}=g_f\,\s\,(e\,e_\phi)^2\left(\frac{\Se}{16\pi^2}\right)^2\left(\frac{\mu_\uv^2}{\mu^2}\right)^{-2\epsilon}\left(-\frac{4}{\epsilon^2}-\frac{4}{\epsilon}-22-\frac{8K_\ominus}{3}-d_{\Sigma,\uv^2}^{(\phi)}+\Oep{1}\right)\;,
\end{equation}
leading to
\begin{equation}
\mathcal{A}_{\uv^2}^{(2,\phi)}=g_f\,\s\,(e\,e_\phi)^2\left(\frac{\Se}{16\pi^2}\right)^2\left(\frac{\mu_\uv^2}{\mu^2}\right)^{-2\epsilon}\left(-18-\frac{8K_\ominus}{3}-d_{\uv^2}^{(\phi)}+\Oep{1}\right)\;,
\end{equation}
The coefficients $d_{\{H,\gamma,\Sigma\},\uv^2}^{(f)}$ can now very easily be adjusted to obtain the desired $\mathcal{O}(\epsilon^0)$ part. Their value in the $\MSbar$ scheme are listed in \Table{\ref{Table:HTLCoefDoubleUV}}.\\
\begin{table}[t]
	\begin{center}
		\begin{tabular}{|c|ccc|c|}
			\hline
			& $d_{H,\uv^2}^{(f)}$ & $d_{\gamma, \uv^2}^{(f)}$ & $d_{\Sigma, \uv^2}^{(f)}$ & $d_{\uv^2}^{(f)}$\\
			\hline \hline
			$t \bar{t}$ & $60+16K_\ominus/3$ & $-8~~~$ & $-4~~~$ & $48+16K_\ominus/3$ \\
			$\phi\phi^\dagger$ & $-4~~~$ & $8$ & $-22-8K_\ominus/3$ & $-18-8K_\ominus/3~~~$ \\
			\hline
		\end{tabular}
		\caption{Values of the scheme fixing parameters in the $\MSbar$ for the double UV counterterms. The parameter $K_\ominus$ is given in \Eq{\ref{Equation:HTLKominus}}.}
		\label{Table:HTLCoefDoubleUV}
	\end{center}
\end{table}
\\
It is remarkable that for both particles, the full double UV counterterm does not exhibit any $\epsilon$-poles, justifying in the meantime that using $\Se$ instead of $S_\epsilon^{\MSbar}$ does not introduce any discrepancy in the final result. While for the top quark each intermediate double UV counterterm is finite, for the charged scalar the cancellation of divergences only occurs when taking into account the sum of all 37 diagrams. This is due to the fact there are more subtle interplays between different contributing topologies, because of the more complex gauge structure. In both cases, the absence of divergences after integration means that in the traditional approach, these double UV counterterms should not be needed. In our formalism however, they are essential in order to cancel local divergent behaviours appearing inside the amplitude and the single UV counterterms. Note also that the unintegrated counterterms exhibit terms proportional to $d-4$ that vanish when taking the four-dimensional limit at the integrand level, even though they still lead to finite parts when keeping the $d$ dependence, because they multiply quantities that generate $\epsilon$-poles. This means that while the counterterms $\mathcal{A}^{(2,f)}_{\uv^2}$ are finite, they will not lead to the same result if computed in $d$ or 4 dimensions. It is only when considering the renormalised amplitude $\mathcal{A}^{(2,f)}_{\r}$ that we have a strict and rigorous commutativity between integrating and taking the limit $\epsilon\to0$. The same happens at one-loop level (see \Chapter{\ref{Chapter:HOL}}), where the counterterm $A_{\uv}^{(1,f)}$ is not needed  in $d$ dimensions (it integrates to 0 in the $\MSbar$). One can wonder if this property holds at three-loop order and beyond, if it is exclusive to the $H\to\gamma\gamma$ process and why, and if there is an underlying reason behind it.

\section{Numerical integration}\label{Section:HTLNumericalIntegration}
\fancyhead[LO]{\ref*{Section:HTLNumericalIntegration}~~\nameref*{Section:HTLNumericalIntegration}}

\begin{figure}[t]
	\begin{center}
		\includegraphics[width=0.45\textwidth]{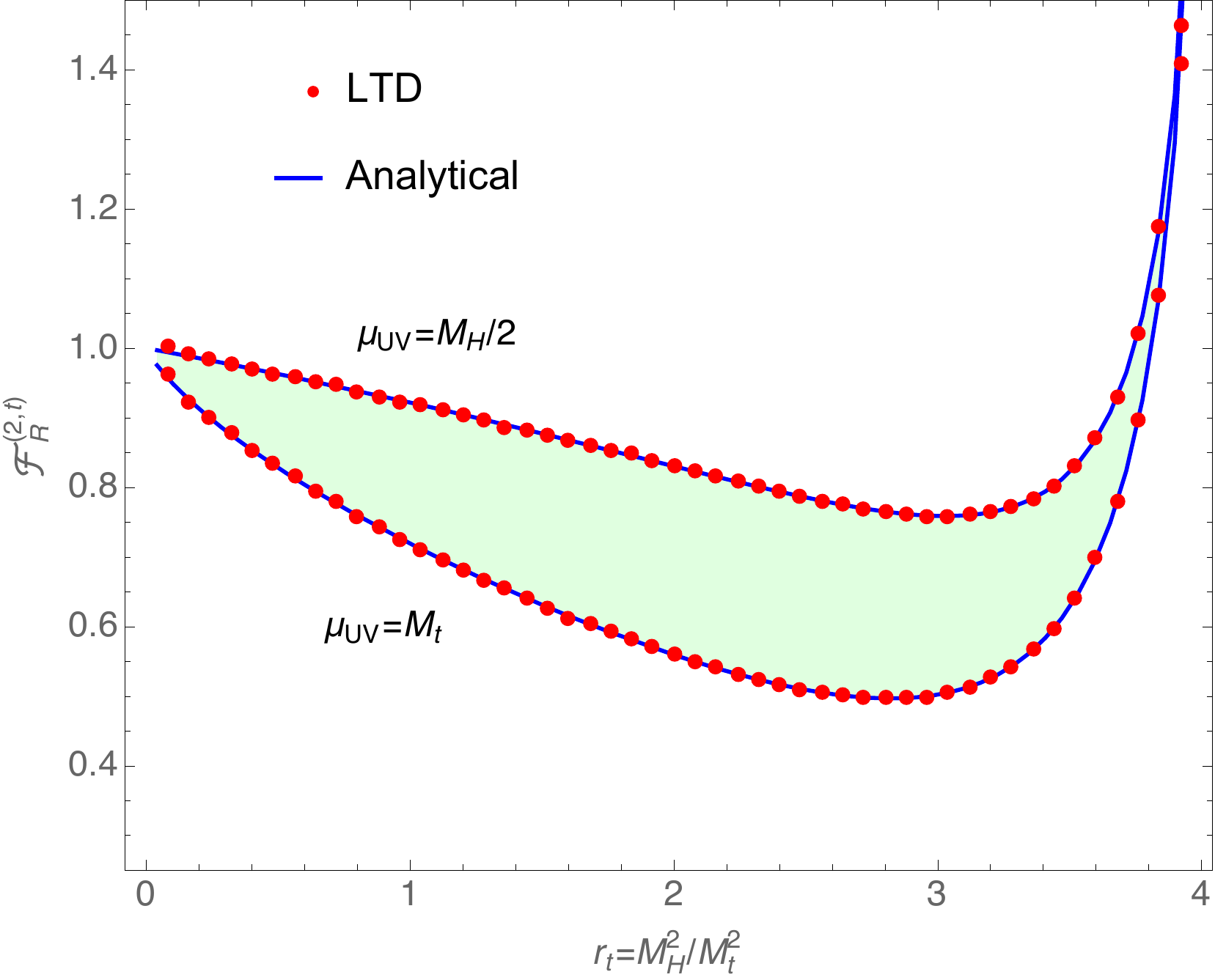} 
		$\qquad$
		\includegraphics[width=0.462\textwidth]{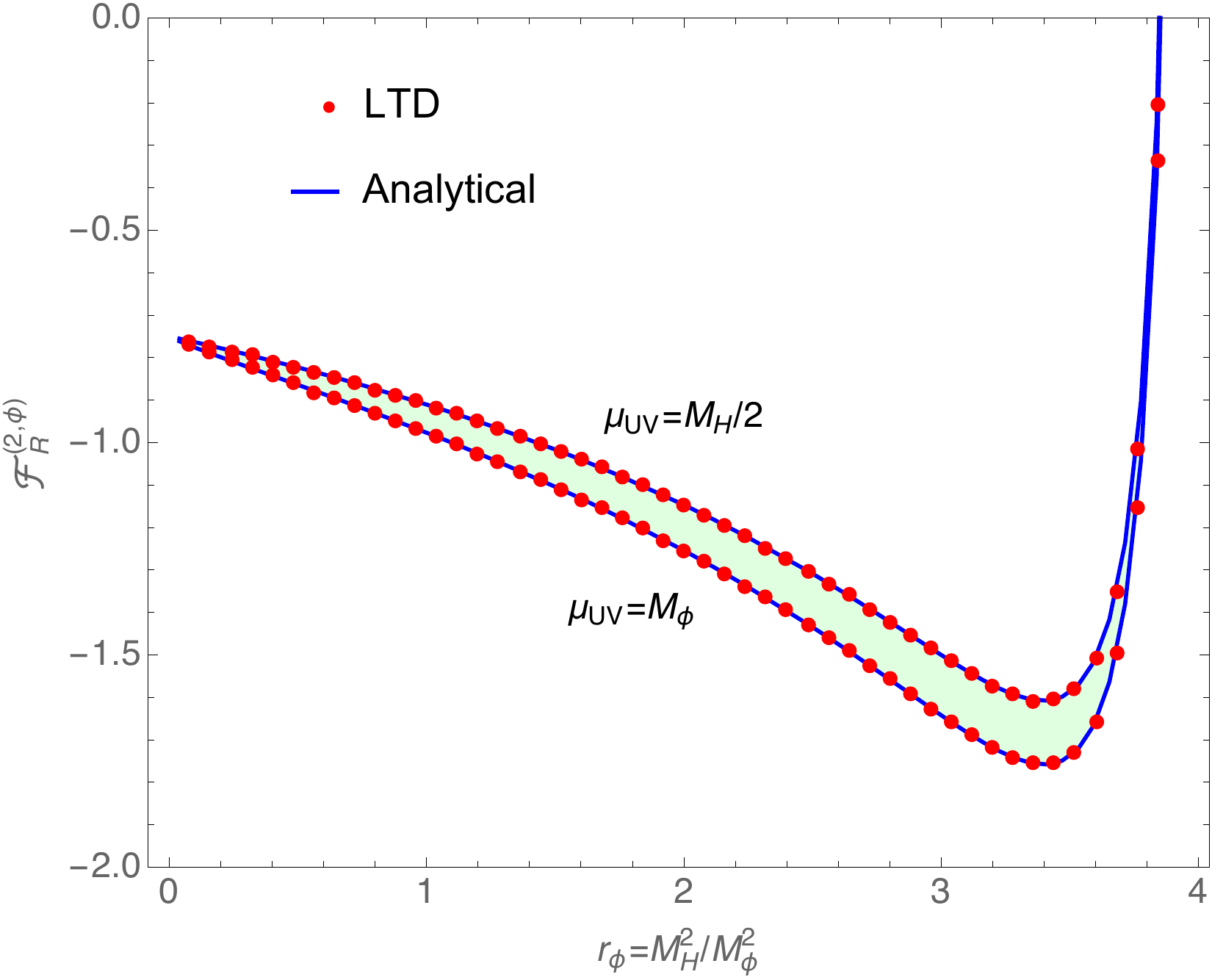}
		\caption{Integrated renormalised amplitude of the two-loop corrections to the $H\to\gamma\gamma$ process, as a function of the inverse mass square of the particle running inside the loop. On the left we show the top quark contribution and on the right, the charged scalar contribution. The solid blue lines represent the analytic result using DREG, while the red dots have been obtained numerically with the LTD formalism. For each particle, two values of the renormalisation scale have been considered, namely $\mu_\uv=M_H/2=\sqrt{\s}/2$, and $\mu_\uv=M_f$. We have rescaled our results as we used different normalisations; explicitly, $\mathcal{F}_\r^{(2,f)}=64\pi^6\,\mathcal{A}^{(2,f)}_{\r}$.}
		\label{Figure:HTLPlots}
	\end{center}
\end{figure}

The renormalised unintegrated amplitude $\mathcal{A}_{\r}^{(2,f)}=\mathcal{A}^{(2,f)}-\mathcal{A}_{1,\uv}^{(2,f)}-\mathcal{A}_{12,\uv}^{(2,f)}-\mathcal{A}_{\uv^2}^{(2,f)}$ is completely free of local UV (and IR) singularities and can safely be evaluated in four dimensions. In the centre-of-mass frame of the decaying Higgs boson, we parametrise the three-momenta as
\begin{align}
\boldsymbol{\ell}_1&=\frac{\sqrt{\s}}{2}\,\xi_1\,(\sin(\theta_1)\,\sin(\varphi_1),\sin(\theta_1)\,\cos(\varphi_1),\cos(\theta_1))\;,\qquad&\mathbf{p}_1&=\frac{\sqrt{\s}}{2}\,(0,0,1)\nn\;,\\
\boldsymbol{\ell}_2&=\frac{\sqrt{\s}}{2}\,\xi_2\,(\sin(\theta_2)\,\sin(\varphi_2),\sin(\theta_2)\,\cos(\varphi_2),\cos(\theta_2))\;,\qquad&
\mathbf{p}_2&=\frac{\sqrt{\s}}{2}\,(0,0,-1)\;.
\end{align}
Note that we can assume $\boldsymbol{\ell}_1$, for instance, to belong to the $(x,z)$ plane, as there is a global rotational symmetry; this means we can trivially perform the integration over one azimuthal angle -- leading to a factor $2\pi$ -- and set $\varphi_1$ to 0. Furthermore, it is better to use $\boldsymbol{\ell}_{12}$ instead of $\boldsymbol{\ell}_2$ as integration variable, because of the way the UV counterterms have been defined.
We therefore integrate over the two three-momenta
\begin{align}
\boldsymbol{\ell}_1=&~\frac{\sqrt{\s}}{2}\,\xi_1\,(\sin(\theta_1),0,\cos(\theta_1))\;,\nn\\
\boldsymbol{\ell}_{12}=&~\frac{\sqrt{\s}}{2}\,\xi_{12}\,(\sin(\theta_{12})\,\cos(\varphi_{12}),\sin(\theta_{12})\,\sin(\varphi_{12}),\cos(\theta_{12}))\;,
\end{align}
which leads to five integration variables, namely $\xi_1,\xi_{12}\in[0,\infty)$, $\theta_1,\theta_{12}\in[0,\pi]$ and $\varphi_{12}\in[0,2\pi]$. In addition, the usual compactification of the integration domain is performed, where the domains of $\xi_1$ and $\xi_{12}$ are mapped from $[0,\infty)$ onto to $[0,1]$, thanks to the change of variables
\begin{equation}\label{Equation:HTLCOVNum}
\xi_i\to\frac{x_i}{1-x_i}\;,
\end{equation}
with $x_i\in[0,1]$, and $i\in\{1,12\}$. This increases the stability of the numerical integration for very high energies, because it restricts the local cancellation of UV singularities to a compact region. The integration measure, after applying this change of variables and in four dimensions, reads
\begin{equation}\label{Equation:HTLIntegrationMeasure}
\frac{d\boldsymbol{\ell}_1\,d\boldsymbol{\ell}_2}{(2\pi)^8}=\frac{\s^3}{64}\frac{x_1^2\,x_{12}^2\,\sin(\theta_1)\sin(\theta_{12})}{(1-x_1)^4(1-x_{12})^4}\frac{dx_1\,dx_{12}\,d\theta_1\,d\theta_{12}\,d\varphi_{12}}{(2\pi)^7}\;.
\end{equation}
Directly writing all the dual cuts explicitly in terms of the integration variables would not lead to a reasonable computational time, as the integration would be too heavy numerically speaking. Moreover, it would be far from being optimal, since many identical terms would have to be evaluated more than once. Instead, we take advantage of the fact it was possible to write all dual cuts in terms of the reduced denominators $D_i$, as explained in \Section{\ref{Section:HTLReduction}} and shown in \Appendix{\ref{Section:APPUnrenormalisedTwoLoop}}. The first step, which only needs to be done once, is to express all the reduced denominators in terms of the 5 integration variables, and this for each dual cut\footnote{Recall that the expression of a given $D_i$ differs from dual cut to dual cut.}. We can then compute their numerical values for a given point in the integration domain, which allows us to quickly evaluate the integrand at this very point by appropriately replacing each $D_i$. Thus, regardless of the complexity of a given integrand, only a limited amount of objects have to be numerically evaluated. Although quite simple to implement, this strategy helped decreasing the integration time by more than one order of magnitude. To perform the actual integration, we used the in-built \mathematica~function \verb"NIntegrate". Our results are shown in \Fig{\ref{Figure:HTLPlots}}, where they are compared with the analytic results given in \Appendix{\ref{Section:APPTwoLoopAnalytical}}. The integration time is of a few minutes for each point. The biggest source of numerical error comes from the cancellation, between the amplitude and the double UV counterterm, of the non-decoupling term going as $\mathcal{O}(M_f^2/\s)$. For $M_t^2=2\s$ for example, the part being removed from the amplitude by the counterterm is two orders of magnitude higher than the actual result, effectively multiplying the relative numerical error by a factor 100, roughly. Note that this error is twice as big for the top quark as for the charged scalar, because of the relative factor -2 between their respective $\mathcal{O}(M_f^2/\s)$ terms. Nevertheless, the agreement with the analytic result is excellent for all values of the internal and renormalisation masses considered. Numerical instabilities may however appear when considering $r$ very close to 4 (or equivalently $M_f$ very close to $\sqrt{\s}/2$), as we approach the mass threshold. Beyond this limit ($r>4$), contour deformation would be needed. Within the LTD formalism, numerical implementations of contour deformation have already been successfully implemented in~\cite{Buchta:2015wna}.

\section{Conclusion}\label{Section:HTLConclusion}
\fancyhead[LO]{\ref*{Section:HTLConclusion}~~\nameref*{Section:HTLConclusion}}

In this chapter, we have built a purely four-dimensional representation of the renormalised Higgs boson decay amplitudes at two-loop order, which can be directly evaluated numerically. To do so, we have applied the LTD theorem to the $H\to\gamma\gamma$ process, with charged scalars and top quarks as internal massive particles. Working at two-loop level within this formalism involves performing double-cuts and several algebraic manipulations of the expressions. This has been done through a fully automatised \textsc{Mathematica} code, that can be adapted to deal with any two-loop scattering amplitude.\\
\\
After generating the full sets of double-cuts, we have taken advantage of the one-loop results obtained in \Chapter{\ref{Chapter:HOL}} in order to infer the universal integrand-level structure of the two-loop expressions. Surprisingly, we have found many similarities between both cases, which has allowed us to write the full two-loop amplitudes at the integrand level using the same functional form independently of the nature of the particles circulating inside the loop. As in the one-loop case, the explicit process dependence is coded into specific scalar coefficients. We have kept the $d$ dependence in all intermediate steps, although a noticeable simplification of these universal coefficients takes place in the limit $d=4$.\\
\\
 We have examined the singular structure of the two-loop amplitudes to understand how to achieve a purely four-dimensional representation of the finite parts. In the first place, we have studied the cancellation of spurious threshold singularities that appear in individual dual contributions after applying the LTD theorem. When considering the $H\to\gamma\gamma$ process below threshold, the amplitude is guaranteed to be infrared safe as well as free of any physical threshold singularity. In fact, we have managed to prove that all the spurious singularities vanish in this configuration after putting together all possible double cuts.\\
 \\
 Then, we have developed a fully local framework to remove UV singularities. We have implemented an algorithmic approach to renormalise, at the integrand level, the two-loop amplitudes. It is based on a refinement of the expansion around the UV propagator strategy already used in the previous chapters, where we included an additional iteration to remove all the possible UV divergences. A careful study of the UV structure of vertices and self-energies has been performed, which has allowed for the imposition of constraints on the finite remainders containing the specific renormalisation scheme dependence. Furthermore, we have also shown that the UV scale we use in the derivation of the local counterterms actually corresponds to the renormalisation scale used in the traditional approach. These last two points allowed us to build the required counterterms at the integrand level to reproduce the $\MSbar$ results.\\
 \\
 Finally, we have proceeded to combine the universal dual representation of the $H\to\gamma\gamma$ amplitudes together with the local UV counterterms, achieving a four-dimensional representation of the fully renormalised integrand. This has allowed for a purely numerical implementation which fully agrees with the available results in the literature. On top of that, intermediate checks with the scalar sunrise diagram have been performed in order to test the reliability of our code, where the results were compared with ones provided by \textsc{SecDec} \cite{Carter:2010hi,Borowka:2015mxa}.\\
 \\
 The developments presented in this chapter also constitute a major advance in the extension of the FDU formalism at NNLO, by providing a reliable and systematic way to renormalise two-loop amplitudes in a local way.

\chapter{Summary and outlook}\label{Chapter:CONCLUSION}
\thispagestyle{fancychapter}
\fancyhead[RE]{\nameref*{Chapter:CONCLUSION}}

The last two decades have seen a tremendous amount of progress in theoretical high-precision physics. Many advances in particular have been made in the evaluation of multi-loop diagrams, but the main challenge lies in the treatment of IR divergences through efficient subtraction schemes. The complexity of the procedure increasing exponentially with the number of scales, it has become necessary to approach the issue from a different angle, hence calling for the development of new techniques.\\
\\
The Loop-Tree Duality (LTD) provides a new framework for the computation of loop amplitudes. By modifying the customary $+i0$ prescription of the Feynman propagators, it establishes duality relations between loop and phase-space integrals. It reduces the loop integration domain to an Euclidean space, where the integration is performed along on-shell hyperboloids or light-cones. After properly regularising IR and UV divergences, it allows for a direct numerical evaluation of the amplitude. The vast majority of the work developed in this thesis is based on this formalism. In particular, the Four-Dimensional Unsubtraction (FDU) method is intimately linked with LTD, and takes best advantage of the very interesting features of this formalism.\\
\\
Our first step has been to show that within the LTD formalism, the IR singularities of both the real and virtual contributions were restricted to a compact region of the momentum space. We have exploited this crucial fact to build a mapping between real and virtual kinematics, in order to make the sum of both contributions locally free from any IR divergences. To illustrate the method, we have started by computing the NLO corrections to a very simple $1\longrightarrow2$ toy model with massless particles. After splitting the real phase space into two regions, we have combined different pieces of the real and virtual contributions together, and shown that none of the quantities obtained exhibited any more IR divergences. In parallel, we have explained how to cancel the UV singularities of the virtual contributions in a local way, using the two-point scalar function as a starting example. By using an expansion of the integrand around a UV propagator, we have built a counterterm that exactly reproduces the singular behaviour of the original integrand in the high-energy region, rendering their difference free of UV divergences. Then, we have applied the aforementioned techniques to compute the NLO QCD decay-rate of a virtual photon into a pair of massless quarks, and have obtained a pure four-dimensional expression of the amplitude at the integrand level. We have also briefly commented on the generalisation to more loops and external legs.\\
\\
The next logical step was to extend the FDU algorithm to massive particles. The real-virtual mapping mentioned above has been extended to the massive case (and to an arbitrary number of external particles), and we have made sure that the quasi-collinear configuration were properly dealt with, so we could achieve a smooth massless limit. We have started by computing the decay rate at NLO of a massive scalar toy model, before applying the technique to the decay of a scalar or vector boson into two massive quarks. In order to properly account for the complete UV singular behaviour, we have had to build an integral-level representation of the wave-function and mass renormalisation factors. The four-dimensional integrands we have obtained have been evaluated numerically, reproducing the standard DREG results with very good accuracy. If successfully extended at NNLO, the FDU formalism could represent a very good alternative to more traditional approaches.\\
\\
Other advantages of the LTD framework have also been presented in this thesis. We have shown that for the amplitudes of the Higgs boson to two massless gauge bosons at one-loop level, the functional form of the integrand could be written in a universal way for a top quark, a charged scalar, and a $W$ gauge boson as internal particles. Moreover, after properly taking care of (hidden) local UV singularities, we have performed a straightforward asymptotic expansion of the unintegrated amplitude, without having to split the integration domain into several regions. If applicable to other, more complicated processes, it could open new opportunities for a more efficient implementation of higher-order computations. This research direction is currently being investigated.\\
\\
Finally, we have presented the very first application of LTD at two-loop level, with the Higgs boson decay to two photons process. We have shown that, remarkably, the above-mentioned universality of the integrand was still holding at this order. It would be interesting to investigate if this property remains true beyond two loops, and --~if it is the case~-- how it can be exploited to facilitate the computations of processes sharing the same topologies and set of contributing diagrams. We have also proposed a technique to compute local UV counterterms at two-loop level, in any renormalisation scheme. This procedure would be the first ingredient for an extension of FDU beyond NLO. In particular, it has been applied to the process under consideration, where a perfect agreement between our numerical implementation and results available in the literature has been found.\\
\\
LTD and FDU --~even though they were developed only very recently~-- have already proven to be reliable alternatives to DREG and the traditional subtraction methods. Yet a great deal of work remains, and hopefully this thesis establishes solid foundations for the future development of these two very promising formalisms.

\appendix

\chapter{Feynman rules}\label{Section:APPFeynmanRules}
\thispagestyle{fancychapter}
\fancyhead[RE]{\nameref*{Section:APPFeynmanRules}}
\fancyhead[LO]{\ref*{Section:APPFeynmanRules}~~\nameref*{Section:APPFeynmanRules}}

In this appendix, we put a list of all the Feynman rules used in this thesis for the evaluation of Feynman diagrams. For the $W$ gauge boson, we work in the unitary gauge, which means that there is no need to take into account additional diagrams involving ghost particles.\\
\\
\begin{minipage}{0.5\textwidth}
	\centering
	\begin{picture}(0,25)(75,27)
	\DashLine(25,35)(125,35){5}
	\Text(75,51){$\phi$}
	\LongArrow(60,25)(90,25)
	\Text(75,10){$k$}
	\end{picture}
\end{minipage}
\begin{minipage}{0.5\textwidth}
	\centering
	\begin{equation*}
	\frac{i}{k^2-M_\phi^2+i0}
	\end{equation*}
\end{minipage}
\vspace{25pt}
\\
\begin{minipage}{0.5\textwidth}
	\centering
	\begin{picture}(0,25)(75,32)
	\ArrowLine(25,35)(125,35)
	\Text(75,51){$t$}
	\LongArrow(60,25)(90,25)
	\Text(75,10){$k$}
	\end{picture}
\end{minipage}
\begin{minipage}{0.5\textwidth}
	\centering
	\begin{equation*}
	\frac{i(\slashed{k}+M_t)}{k^2-M_t^2+i0}
	\end{equation*}
\end{minipage}
\vspace{25pt}
\\
\begin{minipage}{0.5\textwidth}
	\centering
	\begin{picture}(0,25)(75,27)
	\Photon(25,35)(125,35){3}{7}
	\Text(75,51){$\gamma$}
	\Text(17,35){$\mu$}
	\Text(133,35){$\nu$}
	\LongArrow(60,25)(90,25)
	\Text(75,10){$k$}
	\end{picture}
\end{minipage}
\begin{minipage}{0.5\textwidth}
	\centering
	\begin{equation*}
	-\frac{i\,g^{\mu\nu}}{k^2+i0}
	\end{equation*}
\end{minipage}
\vspace{25pt}
\\
\begin{minipage}{0.5\textwidth}
	\begin{center}
		\begin{picture}(0,25)(75,27)
		\ZigZag(25,35)(125,35){3}{7}
		\Text(75,51){$W$}
		\Text(17,35){$\mu$}
		\Text(133,35){$\nu$}
		\LongArrow(60,25)(90,25)
		\Text(75,10){$k$}
		\end{picture}
	\end{center}
\end{minipage}
\begin{minipage}{0.5\textwidth}
	\centering
	\begin{equation*}
	\frac{-i}{k^2-M_W^2+i0}\left(g^{\mu\nu}-\frac{k^\mu\,k^\nu}{M_W^2}\right)
	\end{equation*}
\end{minipage}
\vspace{45pt}
\\
\begin{minipage}{0.5\textwidth}
	\centering
	\begin{picture}(0,72)(75,38)
	\DashLine(25,72)(80,72){5}
	\Text(55.5,88){$h$}
	\DashLine(80,72)(125,117){5}
	\Text(100,108){$\phi^\pm$}
	\DashLine(125,27)(80,72){5}
	\Text(100,36){$\phi^\mp$}
	\end{picture}
\end{minipage}
\begin{minipage}{0.5\textwidth}
	\begin{equation*}
	-i\,\frac{2M_\phi^2}{\vev}
	\end{equation*}
\end{minipage}
\vspace{45pt}
\\
\begin{minipage}{0.5\textwidth}
	\centering
	\begin{picture}(0,72)(75,38)
	\DashLine(25,72)(80,72){5}
	\Text(55.5,88){$h$}
	\ArrowLine(80,72)(125,117)
	\Text(100,108){$q$}
	\ArrowLine(125,27)(80,72)
	\Text(100,36){$\bar{q}$}
	\end{picture}
\end{minipage}
\begin{minipage}{0.5\textwidth}
	\begin{equation*}
	-i\,\frac{M_q}{\vev}
	\end{equation*}
\end{minipage}
\vspace{45pt}
\\
\begin{minipage}{0.5\textwidth}
	\centering
	\begin{picture}(0,72)(75,38)
	\DashLine(25,72)(80,72){5}
	\Text(55.5,88){$h$}
	\ZigZag(80,72)(125,117){3}{6}
	\Text(100,108){$W^\pm$}
	\Text(132,124){$\alpha$}
	\ZigZag(125,27)(80,72){3}{6}
	\Text(100,36){$W^\mp$}
	\Text(132,20){$\beta$}
	\end{picture}
\end{minipage}
\begin{minipage}{0.5\textwidth}
	\begin{equation*}
	i\,\frac{2M_W}{\vev}\,g^{\alpha\beta}
	\end{equation*}
\end{minipage}
\vspace{45pt}
\\
\begin{minipage}{0.5\textwidth}
	\centering
	\begin{picture}(0,72)(75,38)
	\DashLine(25,72)(80,72){5}
	\Text(55.5,88){$\phi$}
	\ArrowLine(80,72)(125,117)
	\Text(100,108){$q$}
	\ArrowLine(125,27)(80,72)
	\Text(100,36){$\bar{q}$}
	\end{picture}
\end{minipage}
\begin{minipage}{0.5\textwidth}
	\begin{equation*}
	i\,Y_q\,(1+c_q\,\gamma^5)
	\end{equation*}
\end{minipage}
\vspace{45pt}
\\
\begin{minipage}{0.5\textwidth}
	\centering
	\begin{picture}(0,72)(75,38)
	\Photon(25,72)(80,72){3}{5}
	\Text(55.5,88)[]{$\gamma$}
	\Text(15,72)[]{$\mu$}
	\DashLine(80,72)(125,117){5}
	\Text(100,108)[]{$\phi^\pm$}
	\LongArrow(122,105)(102,85)
	\Text(120,87)[]{$k_a$}
	\Text(100,29)[]{$\phi^\mp$}
	\DashLine(125,27)(80,72){5}
	\LongArrow(122,39)(102,59)
	\Text(120,57)[]{$k_b$}
	\end{picture}
\end{minipage}
\begin{minipage}{0.5\textwidth}
	\begin{equation*}
	-i\,e\,e_\phi\,(k_a-k_b)^{\mu}
	\end{equation*}
\end{minipage}
\vspace{45pt}
\\
\begin{minipage}{0.5\textwidth}
	\centering
	\begin{picture}(0,72)(75,38)
	\Photon(25,72)(80,72){3}{5}
	\Text(55.5,88)[]{$\gamma$}
	\Text(15,72)[]{$\mu$}
	\ArrowLine(80,72)(125,117)
	\Text(100,108)[]{$q$}
	\ArrowLine(125,27)(80,72)
	\Text(100,36)[]{$\bar{q}$}
	\end{picture}
\end{minipage}
\begin{minipage}{0.5\textwidth}
	\begin{equation*}
	i\,e\,e_q\,\gamma^\mu
	\end{equation*}
\end{minipage}
\vspace{45pt}
\\
\begin{minipage}{0.5\textwidth}
	\centering
	\begin{picture}(0,72)(75,38)
	\Photon(25,72)(80,72){3}{5}
	\Text(55.5,88){$\gamma$}
	\Text(15,72){$\mu$}
	\LongArrow(37.5,62)(67.5,62)
	\Text(52.5,52){$k_c$}
	\ZigZag(80,72)(125,117){3}{6}
	\Text(100,108){$W^\pm$}
	\Text(132,124){$\alpha$}
	\LongArrow(122,105)(102,85)
	\Text(120,87){$k_a$}
	\ZigZag(125,27)(80,72){3}{6}
	\Text(100,29){$W^\mp$}
	\Text(132,20){$\beta$}
	\LongArrow(122,39)(102,59)
	\Text(120,57){$k_b$}
	\end{picture}
\end{minipage}
\begin{minipage}{0.5\textwidth}
	\begin{equation*}
	-i\,e\,(g^{\beta\alpha}(k_b-k_a)^\mu+g^{\alpha\mu}(k_a-k_c)^\beta+g^{\mu\beta}(k_c-k_b)^\alpha)
	\end{equation*}
\end{minipage}
\vspace{45pt}
\\
\begin{minipage}{0.5\textwidth}
	\centering
	\begin{picture}(0,72)(75,38)
	\ZigZag(25,72)(80,72){3}{5}
	\Text(55.5,88){$Z$}
	\Text(15,72){$\mu$}
	\ArrowLine(80,72)(125,117)
	\Text(100,108){$q$}
	\ArrowLine(125,27)(80,72)
	\Text(100,29){$\bar{q}$}
	\end{picture}
\end{minipage}
\begin{minipage}{0.5\textwidth}
	\begin{equation*}
	i\,\gamma^\mu\,(g_{V,q}+g_{A,q}\,\gamma^5)
	\end{equation*}
\end{minipage}
\vspace{45pt}
\\
\begin{minipage}{0.5\textwidth}
	\centering
	\begin{picture}(0,72)(75,38)
	\Photon(25,117)(75,72){3}{6}
	\Text(50,108){$\gamma$}
	\Text(18,124){$\mu$}
	\Photon(25,27)(75,72){3}{6}
	\Text(50,29){$\gamma$}
	\Text(18,20){$\nu$}
	\DashLine(75,72)(125,117){5}
	\Text(100,108){$\phi^\pm$}
	\DashLine(125,27)(75,72){5}
	\Text(100,29){$\phi^\mp$}
	\end{picture}
\end{minipage}
\begin{minipage}{0.5\textwidth}
	\begin{equation*}
	2i\,e^2\,g^{\mu\nu}
	\end{equation*}
\end{minipage}
\vspace{45pt}
\\
\begin{minipage}{0.5\textwidth}
	\centering
	\begin{picture}(0,72)(75,38)
	\Photon(25,117)(75,72){3}{6}
	\Text(50,108){$\gamma$}
	\Text(18,124){$\mu$}
	\Photon(25,27)(75,72){3}{6}
	\Text(50,29){$\gamma$}
	\Text(18,20){$\nu$}
	\ZigZag(75,72)(125,117){3}{6}
	\Text(100,108){$W^\pm$}
	\Text(132,124){$\alpha$}
	\ZigZag(125,27)(75,72){3}{6}
	\Text(100,29){$W^\mp$}
	\Text(132,20){$\beta$}
	\end{picture}
\end{minipage}
\begin{minipage}{0.5\textwidth}
	\begin{equation*}
	-i\,e^2\,(2g^{\mu\nu}\,g^{\alpha\beta}-g^{\mu\alpha}\,g^{\nu\beta}-g^{\mu\beta}\,g^{\nu\alpha})
	\end{equation*}
\end{minipage}

\newpage
\hspace{0pt}
\thispagestyle{empty}
\clearpage

\chapter{Useful formulae for loop and phase-space integration}\label{Section:APPFormulae}
\thispagestyle{fancychapter}
\fancyhead[RE]{\nameref*{Section:APPFormulae}}

\section{The dual integration measure}\label{Section:APPDualMeasure}
\fancyhead[LO]{\ref*{Section:APPDualMeasure}~~\nameref*{Section:APPDualMeasure}}

Using spherical coordinates in $d$-dimensions, the dual integration measure is rewritten as
\begin{equation}\label{Equation:APPIntegrationMeasure}
\int_\ell\,\deltatilde{q_i}=\frac{\mu^{2\epsilon}}{(2\pi)^{d-1}}\int\,d^dq_i\,\theta(q_{i,0})\delta(q_i^2-m_i^2)=\frac{\mu^{2\epsilon}}{(2\pi)^{d-1}}\int\,\frac{(\mathbf{q}_i^2)^{1-\epsilon}}{2\Sup{q_{i,0}}}d|\mathbf{q}_i|d\Omega_i^{d-2}\;,
\end{equation}
where we recall $\Sup{q_{i,0}}=\sqrt{\mathbf{q}_i^2+m_i^2-i0}$. If the azimuthal integration is trivial, the solid angle is given by
\begin{equation}\label{Equation:APPSolidAngleIntegration}
d\Omega_i^{d-2}=\frac{(4\pi)^{1-\epsilon}}{\Gamma(1-\epsilon)}\int_0^1\,\dv{i}\;,\qquad \dv{i}=(v_i(1-v_i))^{-\epsilon}dv_i\;,
\end{equation}
where $\cos(\theta_i)=1-2v_i$ is the cosine of the polar angle. This leads to
\begin{equation}\label{Equation:APPIntegrationMeasureNoAngle}
\int_\ell\,\deltatilde{q_i}=\frac{\mu^{2\epsilon}(4\pi)^{\epsilon-2}}{\Gamma(1-\epsilon)}\int\,\frac{(4\mathbf{q}_i^2)^{1-\epsilon}}{\Sup{q_{i,0}}}d|\mathbf{q}_i|\dv{i}\;.
\end{equation}
For massless internal propagators, we have $\Sup{q_{i,0}}=\sqrt{\mathbf{q}_i^2-i0}$, and the dual integration measure simplifies to
\begin{equation}\label{Equation:APPIntegrationMeasureNoAngleMasless}
\int_\ell\,\deltatilde{q_i}=\frac{\mu^{2\epsilon}(4\pi)^{\epsilon-2}}{\Gamma(1-\epsilon)}\int\,(2q_{i,0})^{1-2\epsilon}d(2q_{i,0})\dv{i}\;.
\end{equation}

\section{Phase space}\label{Section:APPPS}
\fancyhead[LO]{\ref*{Section:APPPS}~~\nameref*{Section:APPPS}}

Following the notations of this thesis, the $d$-dimensional phase space associated with a $N$-leg scattering process, with $m$ final states particles is given by
\begin{equation}\label{Equation:APPPhaseSpaceGeneral}
d\Phi_m=\mu^{d-4}\left(\,\prod_{i=1}^m\,\int_{p_i}\,\deltatilde{p_i}\right)(2\pi)^d\delta\left(\,\sum_{i=1}^m\,p_i-p\right)\;,
\end{equation}
with $p$ being the sum of incoming momenta for either $1\longrightarrow2$ or $2\longrightarrow m$ processes. In particular, the phase space for a $1\longrightarrow2$ decay with final-state particles of equal masses, $p_1^2=p_2^2=M^2$, and $\s$ the virtuality of the decaying particles is given by
\begin{equation}\label{Equation:APPPhaseSpace1to2}
\int\,d\Phi_{1\to2}=\frac{\Gamma(1-\epsilon)\beta^{1-2\epsilon}}{2(4\pi)^{1-\epsilon}\Gamma(2-2\epsilon)}\left(\frac{\s}{\mu^2}\right)^{-\epsilon}\;,
\end{equation}
with $\beta=\sqrt{1-m^2}$ and $m^2=4M^2/\s$. The corresponding $1\longrightarrow3$ phase space of the real radiation correction with an addition massless particle in the final state, $(p'_r)^2=0$, is given by
\begin{equation}\label{Equation:APPPhaseSpace1to3}
\int\,d\Phi_{1\to3}=\int\,\frac{(4\pi)^{\epsilon-2}\,\s}{\Gamma(1-\epsilon)}\left(\frac{\s}{\mu^2}\right)^{-\epsilon}\left(\int\,d\Phi_{1\to2}\right)\beta^{-1+2\epsilon}\theta(h_p)\,h_p^{-\epsilon}\,d\yp{1r}\,d\yp{2r}\;,
\end{equation}
with
\begin{equation}\label{Equation:APPPhaseSpacehp}
h_p=(1-\yp{1r}-\yp{2r})\yp{1r}\,\yp{2r}-\frac{m^2}{4}(\yp{1r}+\yp{2r})^2\;,
\end{equation}
where $\yp{1r}=2p'_i\cdot p'_r/\s$.

\section{Techniques for phase-space integration}\label{Section:APPTechniquesforPSintegration}
\fancyhead[LO]{\ref*{Section:APPTechniquesforPSintegration}~~\nameref*{Section:APPTechniquesforPSintegration}}

We start from~\cite{Frixione:1995ms}
\begin{equation}\label{Equation:APPSIntegrationBasicFormula}
\int_0^1\,dx\,x^{-1+a\,\epsilon}f(x,\epsilon)=\int_0^1\,dx\left(\frac{x_C^{a\,\epsilon}}{a\,\epsilon}+\left(\frac{1}{x}\right)_C+a\,\epsilon\left(\frac{\log(x)}{x}\right)_C\right)f(x,\epsilon)+\Oep{2}\;,
\end{equation}
where $x_C\in(0,1]$ is an arbitrary cut, and $f$ is a generic test function regular at $x=0$ and free of $\epsilon$-poles. The $C$-distributions are defined according to
\begin{equation}\label{Equation:APPCDistribution}
\int_0^1\,dx\left(\frac{\log^n(x)}{x}\right)_Cf(x,\epsilon)=\int_0^1\,dx\,\log^n(x)\frac{f(x,\epsilon)-f(0,\epsilon)\theta(x_C-x)}{x}\;,
\end{equation}
i.e. the cancellation of the integrand is forced in a neighbourhood of the singular point $x=0$. Notice that the test function must be an entire function of $\epsilon$ in order to avoid additional $\epsilon$-poles. Besides that, \Eqs{\ref{Equation:APPSIntegrationBasicFormula}}{\ref{Equation:APPCDistribution}} can be adapted to several domains and measures. In particular, the $\epsilon$-expansion of the phase space at large loop momentum can be expressed as
\begin{align}
\int_\ell\,\deltatilde{q_i}f(q_i,\epsilon)=&~4\cGamma\int\,q_{i,0}\,dq_{i,0}\int\,dv\Bigg[\big(1+2\log(q_{i,0})(v\,\delta(v)+(1-v)\delta(1-v))\big)f(q_i,0)\nn\\
&-\big(v\,\delta(v)+(1-v)\delta(1-v)\big)\left.\frac{\partial f}{\partial\epsilon}\right|_{\epsilon=0}\Bigg]+\Oep{1}
\end{align}
under the assumption that $f(q_i,\epsilon)$ is an entire function of $\epsilon$ and it has a vanishing soft limit (i.e. $f\to0$ as $q_{i,0}\to0$). Appreciate the presence of extra contributions given by the collinear residues at $v=0$ or $v=1$, as well as additional terms introduced by the linear $\epsilon$-dependence of the integrand.\\
\\
To analytically integrate the real radiation contributions, it is convenient to use the change of variables suggested in \cite{Bilenky:1994ad}, i.e.
\begin{equation}\label{Equation:APPPhaseSpaceCOV}
\yp{1r}=g(z)\,w\;,\qquad\yp{2r}=g(z)\,z\,w\;,\quad\text{with}\quad g(z)=\frac{(z-x_S)(1-x_S\,z)}{z(1+z)(1+x_S)^2}\;,
\end{equation}
which allows one to rewrite the function $h_p$ in \Eq{\ref{Equation:APPPhaseSpacehp}} as
\begin{equation}\label{Equation:APPPhaseSpacehpzw}
h_p=g(z)^3\,z(1+z)\,w^2(1-w)\;,
\end{equation}
Consequently, the phase-space limits, which are determined by the quadratic function $h_p$, simplify to $z\in[x_S,x_S^{-1}]$ and $w\in[0,1]$. The first integral in $w$ can easily be obtained by keeping the exact $\epsilon$ dependence. The second integral in $z$, however, requires to expand the expression up to $\Oep{0}$ before integration.

\chapter{Expressions of the NLO corrections to $\gamma^*\to q\qbar(g)$ in the massless case}\label{Appendix:FourDimensionalRepresentationsMassless}
\thispagestyle{fancychapter}
\fancyhead[RE]{\nameref*{Appendix:FourDimensionalRepresentationsMassless}}

In this appendix, we collect the four-dimensional representation of the integrands associated to the integrals in \Eq{\ref{Equation:FDUIntegratedCrossSections}}. Explicitly, we have
\begin{align}
\sigmatilde_1^{(1)}=&~\sigma^{(0)}\,\frac{\alpha_S}{4\pi}\,C_F\int_0^1\,d\xi_{1,0}\int_0^{1/2}\,dv_14\,\mathcal{R}_1(\xi_{1,0},v_1)\left(2\left(\xi_{1,0}-(1-v_1)^{-1}\right)-\frac{\xi_{1,0} (1-\xi_{1,0})}{\left(1-(1-v_1) \, \xi_{1,0} \right)^2} \right)\;,\nn\\
\sigmatilde_2^{(1)}=&~\sigma^{(0)}\,\frac{\alpha_S}{4\pi}\,C_F\int_0^1\,d\xi_{2,0}\,\int_0^1dv_2\,2\,\mathcal{R}_2(\xi_{2,0},v_2)(1-v_2)^{-1}\Bigg(\frac{2v_2\,\xi_{2,0}\left(\xi_{2,0}(1-v_2)-1\right)}{1-\xi_{2,0}}\nn\\
&~-1+v_2\,\xi_{2,0}+\frac{1}{1-v_2\,\xi_{2,0}}\left(\frac{(1-\xi_{2,0})^2}{(1-v_2\,\xi_{2,0})^2}+\xi_{2,0}^2\right) \Bigg)
\end{align}
for the real-virtual combinations, and
\begin{align}\label{Equation:APPSigmaBarV}
\sigmabar^{(1)}_\V=&~\sigma^{(0)}\,\frac{\alpha_S}{4\pi}\,C_F\int_0^\infty\,d\xi\int_0^1\,dv\Bigg[-2\left(1-\mathcal{R}_1(\xi,v)\right)v^{-1}(1-v)^{-1}\frac{\xi^2(1-2v)^2+1}{\sqrt{(1+\xi)^2-4v\,\xi}}\nn\\
&~+2\left(1-\mathcal{R}_2(\xi,v)\right)(1-v)^{-1}\left[2v\,\xi\left(\xi(1-v)-1\right)\left(\frac{1}{1-\xi+i0}+i\,\pi\,\delta(1-\xi)\right)-1+v\,\xi\right]\nn\\
&~+2v^{-1}\left(\frac{\xi(1-v)(\xi(1-2v)-1)}{1+\xi}+1\right)-\frac{(1-2v)\xi^3(12-7m_{\uv}^2-4\xi^2)}{(\xi^2+m_{\uv}^2)^{5/2}}\nn\\ 
&~-\frac{2\xi^2(m_{\uv}^2+4\xi^2(1-6v(1-v)))}{(\xi^2 + m_{\uv}^2)^{5/2}}\Bigg]
\end{align}
for the dual virtual remnant. In \Eq{\ref{Equation:APPSigmaBarV}}, we have identified all the integration variables, $\xi_{2,0}=\xi_{2,0}=\xi_\uv=\xi$ and $v_2=v_3=v_\uv=v$, while $(\xi_{1,0},v_1)$ are expressed in terms of $(\xi_3,v_3)=(\xi,v)$ by using the change of variables given in \Eq{\ref{Equation:FDUCOVUnification}}. The integration regions are defined in \Eqs{\ref{Equation:FDUR1Dual}}{\ref{Equation:FDUR2Dual}}, and we used \Eq{\ref{Equation:FDUR1Expansion}} to simplify the analytic integration. Notice that the integrand of the dual virtual remnant behave as
\begin{equation}
\frac{d\sigmabar_v^{(1)}}{d\xi\,dv}\propto\frac{1-2v}{\xi^2}+\mathcal{O}(\xi^{-3})
\end{equation}
in the high-energy limit, and as $\mathcal{O}(\xi^{-3})$ after angular integration, thanks to the presence of the local UV counterterm.

\newpage
\hspace{0pt}
\thispagestyle{empty}
\clearpage

\chapter{Dual amplitudes for $A^*\to q\qbar(g)$}\label{Appendix:DualAmplitudes}
\thispagestyle{fancychapter}
\fancyhead[RE]{\nameref*{Appendix:DualAmplitudes}}

In this appendix, we write the dual virtual amplitudes and real squared amplitudes contributing to the NLO QCD corrections to the process $A*\to q\qbar(g)$, with $A=\phi,\gamma,Z$, calculated in \Chapter{\ref{Chapter:FDUM}}. The tree-level vertices are given by (see \Appendix{\ref{Section:APPFeynmanRules}})
\begin{align}\label{Equation:APPAqqbargVertices}
\mathbf{\Gamma}_\phi^{(0)}=&~i\,Y_q(1+c_q\,\gamma^5)\;,\nn\\
\mathbf{\Gamma}_\gamma^{(0)}=&~i\,e\,e_q\,\gamma^\mu\;,\nn\\
\mathbf{\Gamma}_Z^{(0)}=&~i\,\gamma^\mu(g_{V,q}+g_{A,q}\,\gamma^5)\;.
\end{align}
The corresponding Born squared amplitudes, averaged of the initial-state polarisations, read
\begin{align}\label{Equation:APPAqqbargBorn}
|\mathcal{M}_{\phi\to q\qbar}^{(0)}|^2=&~2\s\,Y_q^2\,C_A\,(\beta^2+c_q^2)\;,\nn\\
|\mathcal{M}_{\gamma\to q\qbar}^{(0)}|^2=&~2\s(e\, e_q)^2\,C_A\,\left(1+\frac{m^2}{2(1-\epsilon)}\right)\;,\nn\\
|\mathcal{M}_{Z\to q\qbar}^{(0)}|^2=&~2\s\,C_A\,\frac{2(1-\epsilon)}{3-2\epsilon}\left(g_{V,q}^2\left(1+\frac{m^2}{2(1-\epsilon)}\right)+g_{A,q}^2\,\beta^2\right)\;.
\end{align}
The squared amplitudes for the real process $A^*\to q(p'_1)+\qbar(p'_2)+g(p'_r)$, are given by
\begin{align}\label{Equation:APPAqqbargReal}
|\mathcal{M}_{\phi\to q\qbar g}^{(0)}|^2=&~4g_S^2\ C_F\,\left[|\mathcal{M}_{\phi\to q\qbar}^{(0)}|^2\frac{h_p}{2\s(\yp{1r}\,\yp{2r})^2}\right.\nn\\
&~+\left.Y_q^2(1+c_q^2)C_A(1-\epsilon)\left(1+\frac{\yp{2r}}{\yp{1r}}\right)\right]+\{1\leftrightarrow2\}\;,\nn\\
|\mathcal{M}_{\gamma\to q\qbar g}^{(0)}|^2=&~4g_S^2\,C_F\left[|\mathcal{M}_{\gamma\to q\qbar}^{(0)}|^2\frac{h_p}{2\s(\yp{1r}\,\yp{2r})^2}+(e\,e_q)^2\,C_A\left((1-\epsilon)\frac{\yp{2r}}{\yp{1r}}-\epsilon\right)\right]+\{1 \leftrightarrow2\}\;,\nn\\
|\mathcal{M}_{Z\to q\qbar g}^{(0)}|^2=&~4\,g_S^2\,C_F\left[|\mathcal{M}_{Z\to q\qbar}^{(0)}|^2\frac{h_p}{2\s(\yp{1r}\,\yp{2r})^2}+C_A\frac{2(1-\epsilon)}{3-2\epsilon}\right.\nn\\
&~\times\left.\left((g_{V,q}^2+g_{A,q}^2)\left((1-\epsilon)\frac{\yp{2r}}{\yp{1r}}-\epsilon\right)+g_{A,q}^2\frac{m^2}{2}\left(1+\frac{\yp{2r}}{\yp{1r}}\right)\right)\right]+\{1\leftrightarrow2\}\;,
\end{align}
The dual amplitudes of the vertex corrections to the process $A^*\to q(p'_1)+\qbar(p'_2)$ are given by
\begin{align}\label{Equation:APPAqqbargVirtual}
\langle\mathcal{M}_A^{(0)}|\mathcal{M}_A^{(1)}\big(\deltatilde{q_1}\big)\rangle=&~g_S^2\,C_F\int_\ell\,\deltatilde{q_1}\left(-\frac{\s|\mathcal{M}_A^{(0)}|^2(1+\beta^2)}{(2q_1\cdot p_1)(2q_1\cdot p_2)}+\mathcal{G}_A\big(\deltatilde{q_1}\big)\right)\nn\;,\\
\langle\mathcal{M}_A^{(0)}|\mathcal{M}_A^{(1)}\big(\deltatilde{q_2}\big)\rangle=&~g_S^2\,C_F\int_\ell\,\deltatilde{q_2}\left(\frac{4q_2\cdot p_1|\mathcal{M}_A^{(0)}|^2}{(2M^2-2q_2\cdot p_2)(\s-2q_2\cdot p_{12}+i0)}+\mathcal{G}_A\big(\deltatilde{q_2}\big)\right)\nn\;,\\
\langle\mathcal{M}_A^{(0)}|\mathcal{M}_A^{(1)}\big(\deltatilde{q_3}\big)\rangle=&~g_S^2\,C_F\int_\ell\,\deltatilde{q_3}\left(-\frac{4q_3\cdot p_2|\mathcal{M}_A^{(0)}|^2}{(2M^2+2q_3\cdot p_1)(\s+2q_3\cdot p_{12})}+\mathcal{G}_A\big(\deltatilde{q_3}\big)\right)\nn\;,\\
\end{align}
where $|\mathcal{M}_A^{(0)}|^2$ are the Born squared amplitudes given in \Eq{\ref{Equation:APPAqqbargBorn}}, and where the functions $\mathcal{G}_A\big(\deltatilde{q_i}\big)$ are process dependant. For the process $\phi^*\to q\qbar$, they read
\begin{align}\label{Equation:APPDualphiqq}
\mathcal{G}_\phi\big(\deltatilde{q_1}\big)=&~2Y_q\,C_A\left(1+\beta^2+2c_q^2\left(\frac{\s}{2q_1\cdot p_1}-\frac{\s}{2q_1\cdot p_2}\right)\right)\;,\nn\\
\mathcal{G}_\phi\big(\deltatilde{q_2}\big)=&~2Y_q\,C_A\left(-\frac{m^2\,\s}{2M^2-2q_2\cdot p_2}+\frac{2\s\big(1-(2-\epsilon)\beta^2-c_q^2(1-\epsilon)\big)}{\s-2q_2\cdot p_{12}+i0}\right)\;,\nn\\
\mathcal{G}_\phi\big(\deltatilde{q_3}\big)=&~2Y_q\,C_A\left(-\frac{m^2\,\s}{2M^2+2q_3\cdot p_1}+\frac{2\s\big(1-(2-\epsilon)\beta^2-c_q^2(1-\epsilon)\big)}{\s+2q_3\cdot p_{12}}\right)\;.
\end{align}
For $\gamma^*\to q\qbar$, they read
\begin{align}\label{Equation:APPDualgammaqq}
\mathcal{G}_\gamma\big(\deltatilde{q_1}\big)=&~2(e\,e_q)^2\,C_A\left(\left(2+\frac{m^2}{2(1-\epsilon)}\right)\left(\frac{\s}{2q_1\cdot p_1}-\frac{\s}{2q_1\cdot p_2}\right)+2\right)\;,\nn\\
\mathcal{G}_\gamma\big(\deltatilde{q_2}\big)=&~2(e\,e_q)^2\,C_A\left(\frac{m^2\,\s}{2(1-\epsilon)(2M^2-2q_2\cdot p_2)}+\frac{2\epsilon\,\s+4q_2\cdot p_1}{\s-2q_2\cdot p_{12}+i0}\right)\;,\nn\\
\mathcal{G}_\gamma\big(\deltatilde{q_2}\big)=&~2(e\,e_q)^2\,C_A\left(\frac{m^2\,\s}{2(1-\epsilon)(2M^2+2q_3\cdot p_1)}+\frac{2\epsilon\,\s-4q_3\cdot p_2}{\s+2q_3\cdot p_{12}}\right)\;.
\end{align}
Finally, for $Z^*\to q\qbar$, they read
\begin{align}\label{Equation:APPDualZqq}
\mathcal{G}_Z\big(\deltatilde{q_1}\big)=&~2C_A\frac{2(1-\epsilon)}{3-2\epsilon}\left[g_{v,q}^2\left(\left(2+\frac{m^2}{2(1-\epsilon)}\right)\left(\frac{\s}{2q_1\cdot p_1}-\frac{\s}{2q_1\cdot p_2}\right)+2\right)\right.\nn\\
&~+\left.g_{A,q}^2\left((1+\beta^2)\left(\frac{\s}{2q_1\cdot p_1}-\frac{\s}{2q_1\cdot p_2}+1\right)-\frac{m^2}{2}\left(\frac{q_1\cdot p_2}{q_1\cdot p_1}-\frac{q_1\cdot p_1}{q_1\cdot p_2}\right)\right)\right]\;,\nn\\
\mathcal{G}_Z\big(\deltatilde{q_2}\big)=&~2C_A\frac{2(1-\epsilon)}{3-2\epsilon}\left[g_{v,q}^2\left(\frac{m^2\,\s}{2(1-\epsilon)(2M^2-2q_2\cdot p_2)}+\frac{2\epsilon\,\s+4q_2\cdot p_1}{\s-2q_2\cdot p_{12}+i0}\right)\right.\nn\\
&~+\left.g_{A,q}^2\left(-\frac{m^2(2q_2\cdot p_{12}+\s)}{2(2M^2-2q_2\cdot p_2)}+\frac{(m^2+2\epsilon\,\beta^2)\s+4q_2\cdot p_1}{\s-2q_2\cdot p_{12}+i0}\right)\right]\;,\nn\\
\mathcal{G}_Z\big(\deltatilde{q_3}\big)=&~2C_A\frac{2(1-\epsilon)}{3-2\epsilon}\left[g_{v,q}^2\left(\frac{m^2\,\s}{2(1-\epsilon)(2M^2+2q_3\cdot p_1)}+\frac{2\epsilon\,\s-4q_3\cdot p_2}{\s+2q_3\cdot p_{12}}\right)\right.\nn\\
&~+\left.g_{A,q}^2\left(-\frac{m^2(2q_3\cdot p_{12}+\s)}{2(2M^2+2q_3\cdot p_1)}+\frac{(m^2+2\epsilon\,\beta^2)\s-4q_3\cdot p_2}{\s+2q_3\cdot p_{12}+i0}\right)\right]\;.
\end{align}

\chapter{Two-loop explicit expressions for the $H\to \gamma\gamma$ amplitude}\label{Appendix:TwoLoopExpressions}
\thispagestyle{fancychapter}
\fancyhead[RE]{\nameref*{Appendix:TwoLoopExpressions}}

\section{Unrenormalised two-loop dual amplitudes for $H\to \gamma\gamma$}\label{Section:APPUnrenormalisedTwoLoop}
\fancyhead[LO]{\ref*{Section:APPUnrenormalisedTwoLoop}~~\nameref*{Section:APPUnrenormalisedTwoLoop}}

In this appendix, we collect the explicit expressions for the unrenormalised two-loop dual amplitudes for $H\to \gamma\gamma$. The subindices take the values $i\in\{1,2\}$ and $j,k\in\{3,12\}$, with $j\neq k$. The functions $G$ and $F$ have been defined in \Eq{\ref{Equation:HTLFG}}. Here, we introduce the auxiliary function 
\begin{equation}\label{Equation:APPH}
H(X,\kappa)=-\frac{\kappa}{X}\frac{\partial X}{\partial\kappa}\;.
\end{equation}
For example
\begin{align}
H(D_3\,D_{12},\kappa_i)=&~\kappa_i\,b_{1,0}\left(\frac{1}{D_3}-\frac{1}{D_{12}}\right)\;,\nn\\
H(D_1\,D_k,\kappa_j)=&~\kappa_j\,b_{1,0}\left(\frac{1}{D_1}+\frac{2}{D_k}\right)\;,\nn\\
H(D_k^2,\kappa_j)=&~-\kappa_j\,b_{1,0}\,\frac{4}{D_k}\;,
\end{align}
with $b_{1,0}=2\,p_{1,0}/M_f$ (recall that $p_{1,0}=p_{2,0}$ in the centre-of-mass frame of the decaying Higgs boson). It is very important to note that the permutation inside the following expressions have to be applied on the \emph{arguments} of the function $H$, instead of the \emph{expression} of $H$ itself after derivation, as the symmetry is not any more explicit.

\subsection{Double cuts from $G_D(\alpha_1)\,G_D(\alpha_2)\,G_F(\alpha_3)$}

This is the only set with direct snail contributions for scalars, the only terms that do not 
depend on $\ell_2$ are those proportional to $c_{4,u}$ and $c_{16}$. This subset generates 4 different double cuts that are obtained from the following expressions:
\begin{align}
\mathcal{A}^{(2,f)}(q_i,q_4)=&~g_f^{(2)}\,\int_{\ell_1}\,\int_{\ell_2}\,\deltatilde{q_i,q_4}\BIGG[-\frac{r_f\,c_1^{(f)}}{D_3\,D_{12}}\Bigg(G(D_{\overline{i}},\kappa_i,c_{4,u}^{(f)})\big(1+H(D_3\,D_{12},\kappa_i)\big)+F(D_{\overline{i}},\kappa_4/\kappa_i)\Bigg)\nn\\
&~+\Bigg(c_7^{(f)}\left(\frac{1}{D_{\overline{i}}}-\frac{1}{D_{\overline{3}}}\left(1-\frac{D_3}{D_{12}}\left(1-\frac{D_{\overline{12}}}{D_{\overline{i}}}\right)\right)\right)\nn\\
&~+\frac{1}{D_3}\left(c_8^{(f)}\left(\frac{1}{D_{\overline{3}}}-\frac{1}{D_{\overline{i}}}\right)-\frac{1}{D_{\overline{12}}}\left(c_9^{(f)}-c_{10}^{(f)}\frac{D_{\overline{3}}}{D_{\overline{i}}}\right)\right)\nn\\
&~+2r_f\Bigg[\frac{1}{D_3\,D_{12}}\left(c_1^{(f)}\left(\frac{1}{D_3\,D_{\overline{3}}}+\frac{1}{D_{\overline{i}}}\left(\frac{1}{D_{\overline{3}}}-\frac{1}{D_3} \right)\right)+\frac{c_{14}^{(f)}}{D_{\overline{3}}}+\frac{c_{20}^{(f)}}{D_{\overline{i}}}-c_{16}^{(f)}\right.\nn\\
&~+\left.c_{17}^{(f)}\left(\frac{D_{\overline{i}}-D_{\overline{12}}}{D_{\overline{3}}}+\frac{D_{\overline{3}}}{D_{\overline{i}}}\right)\right)-\frac{1}{D_{\overline{i}}\,D_{\overline{3}}}\left(\frac{c_7^{(f)}}{D_{12}}+c_{18}^{(f)}\right)\Bigg]+\{3\leftrightarrow12\}\Bigg)\BIGG]\;,	
\end{align}
and
\begin{align}
\mathcal{A}^{(2,f)}(q_j,q_4)=&~g_f^{(2)}\,\int_{\ell_1}\,\int_{\ell_2}\tilde{\delta}(q_j,q_4)\BIGG[-\frac{r_f}{D_k}\Bigg[\frac{c_1^{(f)}}{D_1}\Bigg(G(D_{\overline{j}},\kappa_j,c_{4,u}^{(f)})\big(1+H(D_1\,D_k,\kappa_j)\big)+F(D_{\overline{j}},\kappa_4/\kappa_j)\Bigg)\nonumber\\
&~-\frac{c_{23}^{(f)}}{2}\Bigg(G(D_{\overline{j}},\kappa_j,c_{4,u}^{(f)})\big(1+\frac{1}{2}H(D_k^2,\kappa_j)\big)+F(D_{\overline{j}},\kappa_4/\kappa_j)\Bigg)\Bigg]\nn\\
&~+ c_7^{(f)}\left(\frac{1}{D_{\overline{k}}}\left(\frac{D_{\overline{j}}}{D_{\overline{1}}}+\frac{D_k}{D_1}\left(1-\frac{D_{\overline{j}}}{D_{\overline{1}}}\right)\right)-\frac{1}{D_{\overline{1}}}\right)+
\frac{1}{D_1} \left( c_8^{(f)}\left(\frac{1}{D_{\overline{j}}}-\frac{1}{D_{\overline{1}}}\right)-\frac{c_9^{(f)}}{D_{\overline{k}}} \right) \nn\\
&~+ \frac{1}{D_{\overline{1}}} \left(
c_{10}^{(f)}\left(\frac{D_{\overline{j}}}{D_{\overline{k}}}\left(\frac{1}{D_1}-\frac{1}{D_k}\right)-\frac{D_{\overline{k}}}{D_k\, D_{\overline{j}}}\right)+\frac{2c_{11}^{(f)}}{D_k} 
+\frac{c_{13}^{(f)}}{D_{\overline{k}}}
\right)+\frac{c_{12}^{(f)}}{D_k}\left(\frac{1}{D_{\overline{j}}}+\frac{1}{D_{\overline{k}}}\right) \nn\\
&~+2\, r_f\Bigg[\frac{1}{D_k}\Bigg(c_1^{(f)}\left(\frac{1}{D_1}\left(\frac{1}{D_{\overline{j}}}\left(\frac{1}{D_{\overline{1}}}-\frac{1}{D_k}-\frac{1}{D_1}-\frac{1}{2}\right)+\frac{1}{D_{\overline{1}}\,D_{\overline{k}}}\left(1+\frac{D_{\overline{k}}}{D_1}+\frac{D_{\overline{1}}}{D_k}\right)\right) \right.\nn\\
&~+\left. \frac{2-r_f}{2\, D_{\overline{1}}\,D_{\overline{j}}\,D_{\overline{k}}}\right)
- \frac{2c_{16}^{(f)}}{D_1}+c_{17}^{(f)}\left(\frac{1}{D_1}\left(\frac{D_{\overline{j}}+D_{\overline{k}}}{D_{\overline{1}}}-\frac{D_{\overline{k}}}{D_{\overline{j}}}-\frac{D_{\overline{j}}}{D_{\overline{k}}}\right) \right.\nn \\
&~+\left. \left(\frac{D_1}{D_{\overline{1}}}+\frac{D_{\overline{1}}}{D_1}\right)\left(\frac{1}{D_{\overline{j}}}+\frac{1}{D_{\overline{k}}}\right)+\frac{c_7^{(f)}\, r_f}{D_{\overline{j}}\,D_{\overline{k}}}\left(1-\frac{D_1}{D_{\overline{1}}}\right)\right)\nonumber\\
&~+\left(\frac{c_{14}^{(f)}}{D_1}+\frac{c_{19}^{(f)}}{D_{\overline{1}}}\right)\left(\frac{1}{D_{\overline{j}}}+\frac{1}{D_{\overline{k}}}\right)\Bigg)+\frac{1}{D_{\overline{1}}}\left(\frac{c_{18}^{(f)}}{D_{\overline{j}}}-\frac{2c_{15}^{(f)}}{D_1\,D_k}-\frac{1}{D_{\overline{k}}}\left(\frac{c_7^{(f)}}{D_1}-\frac{c_{20}^{(f)}}{D_{\overline{j}}}\right)\right)\nn\\
&~-\frac{c_{23}^{(f)}}{8}\left(\frac{1}{D_{\overline{j}}\,D_{\overline{k}}}+\frac{4}{D_k}\left(\frac{1}{D_k\,D_{\overline{k}}}-\frac{1}{D_{\overline{j}}}\left(\frac{1}{D_k}+\frac{1}{2}-\frac{2-r_f}{2\, D_{\overline{k}}}\right)\right)\right)\Bigg]+\{1\leftrightarrow2\}\BIGG]\;.
\end{align}

\subsection{Double cuts from $G_F(\alpha_1)\,G_D(-\alpha_2)\,G_D(\alpha_3)$}

There are also 4 double cuts in this subset that are obtained from the expressions:
\begin{align}
\mathcal{A}^{(2,f)}(q_{\overline{4}},q_{\overline{i}})=&~g_f^{(2)}\,\int_{\ell_1}\,\int_{\ell_2}\,\deltatilde{q_{\overline{4}},q_{\overline{i}}}\BIGG[-c_7^{(f)}\left(\frac{1}{D_3}\left(1-\frac{D_{\overline{3}}}{D_{\overline{12}}}\left(1-\frac{D_{12}}{D_i}\right)\right) - \frac{1}{D_i}\right) \nn \\ 
&~+\frac{1}{D_3}\left(-\frac{c_8^{(f)}}{D_i}+c_{10}^{(f)}\frac{D_{\overline{3}}}{D_{\overline{12}}}\left(\frac{1}{D_i}-\frac{1}{D_{12}}\right)+\frac{c_{11}^{(f)}}{D_{12}}+\frac{c_{13}^{(f)}}{D_{\overline{12}}}\right)\nn \\ 
&~+2r_f\Bigg[\frac{1}{D_3\,D_{12}}\left(c_1^{(f)}\left(\frac{1}{D_i}\left(\frac{1}{2D_i}+\frac{1}{D_{\overline{3}}}\right)+\frac{2-r_f}{4D_{\overline{3}}\,D_{\overline{12}}}\right)-\frac{c_{15}^{(f)}}{D_i} \right. \nn \\ 
&~+\left.c_{17}^{(f)}\left(\frac{D_i}{D_{\overline{12}}}+\frac{D_{\overline{12}}}{D_i}-c_7^{(f)}\frac{r_f\,D_i}{2D_{\overline{3}}D_{\overline{12}}}\right)+\frac{c_{19}^{(f)}}{D_{\overline{12}}} \right)\nn\\
&~+\frac{1}{D_{\overline{12}}}\left(\frac{1}{D_3}\left(-\frac{c_7^{(f)}}{D_i}+\frac{c_{20}^{(f)}}{D_{\overline{3}}}\right)+c_{18}^{(f)}\left(\frac{1}{D_{12}}-\frac{1}{D_i}\right)\right)\Bigg]+\{3\leftrightarrow12\}\BIGG]\;,
\end{align}
and
\begin{align}
\mathcal{A}^{(2,f)}(q_{\overline{4}},q_{\overline{j}})=&~g_f^{(2)}\,\int_{\ell_1}\,\int_{\ell_2}\,\deltatilde{q_{\overline{4}},q_{\overline{j}}}\BIGG[c_7^{(f)}\left(\frac{1}{D_k}\left(\frac{D_j}{D_1}+\frac{D_{\overline{k}}}{D_{\overline{1}}}\left(1-\frac{D_j}{D_1}\right)\right)-\frac{1}{D_1}\right)\nn\\
&~+\frac{c_8^{(f)}}{D_1\,D_j}+\frac{1}{D_k}\left(-\frac{c_9^{(f)}}{D_1}+c_{10}^{(f)}\,\frac{D_{\overline{k}}}{D_{\overline{1}}}\left(\frac{1}{D_1}-\frac{1}{D_j}\right)+\frac{c_{12}^{(f)}}{D_j}+\frac{c_{13}^{(f)}}{D_{\overline{1}}}\right) \nn \\ 
&~+2r_f\Bigg[\frac{1}{D_j\,D_k}\left(c_1^{(f)}\left(\frac{1}{D_1}\left(\frac{1}{D_j}+\frac{1}{D_{\overline{1}}}\right)+\frac{2-r_f}{2D_{\overline{1}}\,D_{\overline{k}}}\right)+\frac{c_{14}^{(f)}}{D_1}\right.\nn \\ 
&~+\left.c_{17}^{(f)}\left(\frac{D_1}{D_{\overline{1}}}+\frac{D_{\overline{1}}-D_{\overline{k}}}{D_1}+c_{7}^{(f)}\,\frac{r_f}{D_{\overline{k}}}\left( 1-\frac{D_1}{D_{\overline{1}}}\right)\right)+\frac{c_{19}^{(f)}}{D_{\overline{1}}}\right)\nn\\
&~+\frac{1}{D_{\overline{1}}}\left(-\frac{c_7^{(f)}}{D_1\,D_k}+\frac{c_{20}^{(f)}}{D_{\overline{k}}}\left(\frac{1}{D_j}+\frac{1}{D_k}\right)+c_{18}^{(f)}\left(\frac{1}{D_j}-\frac{1}{D_1} \right) \right)\nn\\
&~-\frac{c_{23}^{(f)}}{8}\left(\frac{1}{D_{\overline{k}}}\left(\frac{1}{D_j}+\frac{1}{D_k}\right)+\frac{4}{D_j\,D_k}\left(\frac{1}{D_j}+\frac{2-r_f}{2D_{\overline{k}}}\right)\right)\Bigg]+\{1\leftrightarrow2\}\BIGG]\;.
\end{align}

\subsection{Double cuts from $G_D(\alpha_1)\,G_F(\alpha_2)\,G_D(\alpha_3)$}

In this subset there are 14 double cuts. The terms that do not depend on $D_4$ or $D_{\overline{k}}$ integrate to a massive snail. The generating expressions are:
\begin{align}
\mathcal{A}^{(2,f)}(q_i,q_{\overline{i}})=&~g_f^{(2)}\,\int_{\ell_1}\,\int_{\ell_2}\,\deltatilde{q_i,q_{\overline{i}}}\nn\\
&~\times\BIGG[-\frac{r_f\,c_1^{(f)}}{D_3\,D_{12}}\Bigg(G(D_4,\kappa_i,-c_{4,nu}^{(f)})\big(1+H(D_3\,D_{12},\kappa_i)\big)+F(D_4,-\kappa_{\overline{i}}/\kappa_i)\Bigg)\nn\\
&~+\Bigg(c_7^{(f)}\left(\frac{1}{D_4}\left(1-\frac{D_{12}\,D_{\overline{3}}}{D_3\,D_{\overline{12}}}\right)-\frac{4\,c_{4,nu}^{(f)}}{D_3}\left(1-\frac{D_{\overline{3}}}{D_{\overline{12}}}\right)\right)-\frac{1}{D_3\,D_4}\left(c_8^{(f)}-c_{10}^{(f)}\frac{D_{\overline{3}}}{D_{\overline{12}}}\right)\nn\\
&~+2r_f\Bigg[\frac{1}{D_3\,D_{12}}\Bigg(c_1^{(f)}\left(\frac{1}{D_4}\left(\frac{1}{D_{\overline{3}}}-\frac{1}{D_3}\right)+\frac{c_{4,nu}^{(f)}}{D_3}\right)+\frac{1}{D_4}\left(c_{20}^{(f)}+c_{17}^{(f)}\,D_{\overline{3}}\right)+\frac{c_{21}^{(f)}}{D_{\overline{3}}}\Bigg)\nn\\
&~+\frac{c_7^{(f)}}{D_{12}}\left(\frac{1}{D_{\overline{3}}}\left(\frac{1}{2}-\frac{1}{D_4}\right)+\frac{2c_{4,nu}^{(f)}}{D_3}\left(1-\frac{D_4}{D_{\overline{3}}}\right)\right)-\frac{c_{18}^{(f)}}{D_4\,D_{\overline{3}}}\Bigg]+\{3\leftrightarrow12\} \Bigg)\BIGG]\;,
\end{align}
\begin{align}
\mathcal{A}^{(2,f)}(q_j,q_{\overline{j}})=&~g_f^{(2)}\,\int_{\ell_1}\,\int_{\ell_2}\deltatilde{q_j,q_{\overline{j}}}\nn\\
&~\times\BIGG[-\frac{r_f}{D_k}\Bigg[\frac{c_1^{(f)}}{D_1}\Bigg(G(D_4,\kappa_j,-c_{4,nu}^{(f)})\big(1+H(D_1\,D_k,\kappa_j)\big)+F(D_4,-\kappa_{\overline{j}}/\kappa_j)\Bigg)\nn\\
&~-\frac{c_{23}^{(f)}}{2}\Bigg(G(D_{\overline{j}},\kappa_j,-c_{4,nu}^{(f)})\big(1+\frac{1}{2}H(D_k^2,\kappa_j)\big)+F(D_4,-\kappa_{\overline{j}}/\kappa_j)\Bigg)\Bigg]\nn\\
&~+4c_7^{(f)}\,c_{4,nu}^{(f)}\left(\frac{1}{D_1}-\frac{D_{\overline{k}}}{D_{\overline{1}}\, D_k}\right)+\frac{1}{D_4}\left(\frac{c_8^{(f)}}{D_1}-\frac{1}{D_k}\left(c_{10}^{(f)}\frac{D_{\overline{k}}}{D_{\overline{1}}}-c_{12}^{(f)}\right)\right)\nn\\
&~+2r_f\Bigg[\frac{1}{D_k}\Bigg(c_1^{(f)}\Bigg(\frac{1}{D_4}\left(\frac{1}{D_1}\left(\frac{1}{D_{\overline{1}}}-\frac{1}{D_1}-\frac{1}{D_k}-\frac{1}{2}\right)+\frac{2-r_f}{2D_{\overline{1}}\,D_{\overline{k}}}\right)\nn\\
&~+\frac{c_{4,nu}^{(f)}}{D_1}\left(\frac{1}{D_1}+\frac{1}{D_k}\right)\Bigg)+\frac{1}{D_4}\left(\frac{c_{14}^{(f)}}{D_1}+c_{17}^{(f)}\left(\frac{D_1}{D_{\overline{1}}}+\frac{D_{\overline{1}}-D_{{\overline{k}}}}{D_1}+\frac{c_7^{(f)}\,r_f}{D_{\overline{k}}}\left(1-\frac{D_{\overline{1}}}{D_1}\right)\right)\right)\nn\\ 
&~+2c_7^{(f)}\,c_{4,nu}^{(f)}\left(\frac{1}{D_1}+\frac{1}{D_{\overline{1}}}\left(1-\frac{D_4}{D_1}\right)\right)+\frac{1}{D_{\overline{1}}}\left(\frac{c_{19}^{(f)}}{D_4}+\frac{c_{21}^{(f)}}{D_1}+\frac{c_{22}^{(f)}}{D_{\overline{k}}}\right)\Bigg)\nn\\
&~+\frac{1}{D_4\,D_{\overline{1}}}\left(c_{18}^{(f)}+\frac{c_{20}^{(f)}}{D_{\overline{k}}}\right)-\frac{c_{23}^{(f)}}{8}\Bigg(\frac{1}{D_4}\left(\frac{1}{D_{\overline{k}}}-\frac{4}{D_k}\left(\frac{1}{D_k}+\frac{1}{2}-\frac{2-r_f}{2D_{\overline{k}}}\right)\right)\nn\\
&~+\frac{4c_{4,nu}^{(f)}}{D_k}\left(\frac{1}{D_k}-\frac{1}{D_{\overline{k}}}\right)\Bigg)\Bigg]+\{1\leftrightarrow2\}\BIGG]\;,
\end{align}
\begin{align}
\mathcal{A}^{(2,f)}(q_i,q_{\overline{j}})=&~ g_f^{(2)}\,\int_{\ell_1}\,\int_{\ell_2} \, \deltatilde{q_i,q_{\overline{j}}}\BIGG[-c_7^{(f)}\left(\frac{1}{D_4}-\frac{4c_{4,nu}^{(f)}}{D_j} \right)\left(1-\frac{D_j}{D_k}\left(1-\frac{D_{\overline{k}}}{D_{\overline{i}}}\right)\right)\nn\\
&~+\frac{1}{D_4}\left(\frac{c_8^{(f)}}{D_j}-\frac{c_9^{(f)}}{D_k}+c_{10}^{(f)}\frac{D_{\overline{k}}}{D_k\,D_{\overline{i}}}\right)\nn\\
&~+2r_f\Bigg[\frac{1}{D_j\,D_k}\left(c_1^{(f)}\left(\frac{1}{D_4}\left(\frac{1}{D_j}+\frac{1}{D_{\overline{i}}}\right)-\frac{c_{4,nu}^{(f)}}{D_j}\right)\right.\nn\\
&~+\left.\frac{1}{D_4}\left(c_{14}^{(f)}+c_{17}^{(f)}(D_{\overline{i}}-D_{\overline{k}})\right)+\frac{c_{21}^{(f)}}{D_{\overline{i}}}\right)\nn \\ 
&~+\frac{c_7^{(f)}}{D_k}\left(\frac{1}{D_{\overline{i}}}\left(\frac{1}{2}-\frac{1}{D_4}\right)+\frac{2c_{4,nu}}{D_j}\left(1-\frac{D_4}{D_{\overline{i}}}\right)\right)-\frac{c_{18}^{(f)}}{D_4\,D_{\overline{i}}}\Bigg]\BIGG]\;,
\end{align}
\begin{align}
\mathcal{A}^{(2,f)}(q_j,q_{\overline{i}})=&~g_f^{(2)}\,\int_{\ell_1}\,\int_{\ell_2}\,\deltatilde{q_j,q_{\overline{i}}}\BIGG[-c_7^{(f)}\left(\left(\frac{1}{D_4}-\frac{4c_{4,nu}^{(f)}}{D_k}\right)\left(1-\frac{D_{\overline{j}}}{D_{\overline{k}}}\left(1-\frac{D_k}{D_i}\right)\right)\right.\nn\\
&~+\left.4c_{4,nu}^{(f)}\left(\frac{1}{D_i}-\frac{1}{D_k}\left(1-\frac{D_{\overline{k}}}{D_{\overline{j}}}\right)\right)\right)\nn\\
&~+\frac{1}{D_4}\left(-\frac{c_8^{(f)}}{D_i}+c_{10}^{(f)}\left(\frac{D_{\overline{j}}}{D_{\overline{k}}}\left(\frac{1}{D_i}-\frac{1}{D_k}\right)-\frac{D_{\overline{k}}}{D_k\,D_{\overline{j}}}\right)+\frac{2c_{11}^{(f)}}{D_k}+\frac{c_{13}^{(f)}}{D_{\overline{k}}}\right)\nn\\
&~+2r_f\Bigg[\frac{1}{D_k}\left(c_1^{(f)}\left(\frac{1}{D_4}\left(\frac{1}{D_i}\left(\frac{1}{D_i}+\frac{1}{D_{\overline{j}}}+\frac{1}{D_{\overline{k}}}\right)+\frac{2-r_f}{2D_{\overline{j}}\,D_{\overline{k}}}\right)-\frac{c_{4,nu}^{(f)}}{D_i^2}\right)\right.\nn\\
&~+\left.\frac{c_{21}^{(f)}}{D_i}\left(\frac{1}{D_{\overline{j}}}+\frac{1}{D_{\overline{k}}}\right)+\frac{c_{22}^{(f)}}{D_{\overline{j}}\,D_{\overline{k}}}\right)+c_7^{(f)}\left(\frac{1}{D_i\,D_{\overline{k}}}\left(\frac{1}{2}-\frac{1}{D_4}\right)\right.\nn\\
&~+\left.\frac{2c_{4,nu}^{(f)}}{D_k}\left(\frac{2}{D_i}+\left(1-\frac{D_4}{D_i}\right)\left(\frac{1}{D_{\overline{j}}}+\frac{1}{D_{\overline{k}}}\right)\right)\right)\nn\\
&~+\frac{1}{D_4}\left(\frac{1}{D_{\overline{j}}}\left(c_{18}^{(f)}+\frac{c_{20}^{(f)}}{D_{\overline{k}}}\right)+\frac{1}{D_k}\left(-\frac{2c_{15}^{(f)}}{D_i}+c_{17}^{(f)}\left(D_i\left(\frac{1}{D_{\overline{j}}}+\frac{1}{D_{\overline{k}}}-c_7^{(f)}\,\frac{r_f}{D_{\overline{j}}\,D_{\overline{k}}}\right)\right.\right.\right.\nn\\
&~+\left.\left.\left.\frac{D_{\overline{j}}+D_{\overline{k}}}{D_i}\right)+c_{19}^{(f)}\left(\frac{1}{D_{\overline{j}}}+\frac{1}{D_{\overline{k}}}\right)\right)\right)\Bigg]\BIGG]\;,
\end{align}
and
\begin{align}
A^{(2,t)}(q_j,q_{\overline{k}})=&~g_f^{(2)}\,\int_{\ell_1}\,\int_{\ell_2}\,\BIGG[c_7^{(f)}\left(\frac{1}{D_4}-\frac{4c_{4,nu}^{(f)}}{D_k}\right)\left(\frac{D_{\overline{j}}}{D_{\overline{1}}}+\frac{D_k}{D_1}\left(1-\frac{D_{\overline{j}}}{D_{\overline{1}}}\right)\right)\nn\\
&~+\frac{1}{D_4}\left(-\frac{c_9^{(f)}}{D_1}+c_{10}^{(f)}\frac{D_{\overline{j}}}{D_{\overline{1}}}\left(\frac{1}{D_1}-\frac{1}{D_k}\right)+\frac{c_{12}^{(f)}}{D_{12}}+\frac{c_{13}^{(f)}}{D_{\overline{1}}}\right)\nn\\
&~+2r_f\Bigg[\frac{1}{D_k}\left(c_1^{(f)}\left(\frac{1}{D_4}\left(\frac{1}{D_1}\left(\frac{1}{D_k}+\frac{1}{D_{\overline{1}}}\right)+\frac{2-r_f}{2D_{\overline{1}}\,D_{\overline{j}}}\right)-\frac{c_{4,nu}^{(f)}}{D_1\,D_k}\right)\right.\nn\\
&~+\left.\frac{1}{D_4}\left(\frac{c_{14}^{(f)}}{D_1}+c_{17}^{(f)}\left(\frac{D_1}{D_{\overline{1}}}+\frac{D_{\overline{1}}-D_{\overline{j}}}{D_1}+c_7^{(f)}\,\frac{r_f}{D_{\overline{j}}}\left(1-\frac{D_1}{D_{\overline{1}}}\right)\right)+\frac{c_{19}^{(f)}}{D_{\overline{1}}}\right)\right)\nn\\
&~+c_7^{(f)}\left(\frac{1}{D_1\,D_{\overline{1}}}\left(\frac{1}{2}-\frac{1}{D_4}\right)+\frac{2c_{4,nu}^{(f)}}{D_k}\left(\frac{1}{D_1}+\frac{1}{D_{\overline{1}}}\left(1-\frac{D_4}{D_1}\right)\right)\right)\nn\\
&~+\frac{1}{D_{\overline{1}}}\left(\frac{c_{20}^{(f)}}{D_{\overline{j}}\,D_4}+\frac{c_{21}^{(f)}}{D_1\,D_k}+\frac{c_{22}^{(f)}}{D_k\,D_{\overline{j}}}\right)-\frac{c_{23}^{(f)}}{8}\Bigg(\frac{1}{D_4}\left(\frac{1}{D_{\overline{j}}}+\frac{4}{D_k}\left(\frac{1}{D_k}+\frac{2-r_f}{2D_{\overline{j}}}\right)\right)\nn\\
&~-\frac{4c_{4,nu}^{(f)}}{D_k}\left(\frac{1}{D_k}+\frac{1}{D_{\overline{j}}}\right)\Bigg)\Bigg]+\{1\leftrightarrow2\}\BIGG]\;.
\end{align}

\section{Known analytic results for $H\to \gamma\gamma$ at two loops}\label{Section:APPTwoLoopAnalytical}
\fancyhead[LO]{\ref*{Section:APPTwoLoopAnalytical}~~\nameref*{Section:APPTwoLoopAnalytical}}

In this appendix, we write the analytic results obtained from~\cite{Aglietti:2006tp}, for the top quark and the charged scalar in the $\overline{\text{MS}}$ scheme. We first define
\begin{equation}
x_f=\frac{\sqrt{1-4/r_f}-1}{\sqrt{1-4/r_f}+1}
\end{equation}
where we recall $r_f=\s/M_f^2$. We also write the auxiliary function
\begin{align}
\mathcal{H}_1(x)=&~\frac{9}{10}\zeta_2^2+2\zeta_3\,H(0,x)+\zeta_2\,H(0,0,x)+\frac{1}{4}H(0,0,0,0,x)+\frac{7}{2}H(0,1,0,0,x)\nn\\
&~-2H(0,-1,0,0,x)+4H(0,0,-1,0,x)-H(0,0,1,0,x)\;,
\end{align}
where the standard harmonic polylogarithm notations~\cite{Remiddi:1999ew} have been used. For the top quark,
\begin{align}
\mathcal{F}_\r^{(2,t)}(x)=&~\frac{36x}{(x-1)^2}-\frac{4x(1-14x+x^2)}{(x-1)^4}\zeta_3-\frac{4x(1+x)}{(x-1)^3}H(0,x)-\frac{8x(1+9x+x^2)}{(x-1)^4}H(0,0,x)\nn\\
&~+\frac{2x(3+25x-7x^2+3x^3)}{(x-1)^5}H(0,0,0,x)\nn\\
&~+\frac{4x(1+2x+x^2)}{(x-1^4)}\big(\zeta_2\,H(0,x)+4H(0,-1,0,x)-H(0,1,0,x)\big)\nn\\
&~+\frac{4x(5-6x+5x^2)}{(x-1)^4}H(1,0,0,x)-\frac{8x(1+x+x^2+x^3)}{(x-1)^5}\mathcal{H}_1(x)\nn\\
&~-\left(\frac{12x}{(x-1)^2}-\frac{6x(1+x)}{(x-1)^3}H(0,x)+\frac{6x(1+6x+x^2)}{(x-1)^4}H(0,0,x)\right)\log\left(\frac{M_t^2}{\mu_\uv^2}\right)\;,
\end{align}
whereas for the charged scalar,
\begin{align}
\mathcal{F}_\r^{(2,\phi)}(x)=&~-\frac{14x}{(x-1)^2}-\frac{24x^2}{(x-1)^4}\zeta_3+\frac{x(3-8x+3x^2)}{(x-1)^3(x+1)}H(0,x)+\frac{34x^2}{(x-1)^4}H(0,0,x)\nn\\
&~-\frac{8x^2}{(x-1)^4}\big(\zeta_2\,H(0,x)+4H(0,-1,0,x)-H(0,1,0,x)+H(1,0,0,x)\big)\nn\\
&~-\frac{2x^2(5-11x)}{(x-1)^5}H(0,0,0,x)+\frac{16x^2(1+x^2)}{(x-1)^5(x+1)}\mathcal{H}_1(x)\nn\\
&~+\left(\frac{6x^2}{(x-1)^3(x+1)}H(0,x)-\frac{6x^2}{(x-1)^4}H(0,0,x)-\frac{3}{4}\mathcal{F}_\r^{(1,\phi)}(x)\right)\log\left(\frac{M_\phi^2}{\mu_\uv^2}\right)\;,
\end{align}
with
\begin{equation}
\mathcal{F}_\r^{(1,\phi)}(x)=\frac{4}{r_\phi}\left(1+\frac{2}{r_\phi}H(0,0,x)\right)\;.
\end{equation}

\newpage
\hspace{0pt}
\thispagestyle{empty}
\clearpage

\fancyhead[LO,RE]{Bibliography}
\bibliographystyle{JHEP}
\bibliography{References}

\providecommand{\href}[2]{#2}\begingroup\raggedright\begin{thebibliography}{100}

\bibitem{Sborlini:2016gbr}
G.~F.~R. Sborlini, F.~Driencourt-Mangin, R.~Hern\'andez-Pinto and G.~Rodrigo,
  \emph{{Four-dimensional unsubtraction from the loop-tree duality}},
  \href{http://dx.doi.org/10.1007/JHEP08(2016)160}{\emph{JHEP} {\bf 08} (2016)
  160}, [\href{http://arxiv.org/abs/1604.06699}{{\tt 1604.06699}}].

\bibitem{Sborlini:2016hat}
G.~F.~R. Sborlini, F.~Driencourt-Mangin and G.~Rodrigo, \emph{{Four-dimensional
  unsubtraction with massive particles}},
  \href{http://dx.doi.org/10.1007/JHEP10(2016)162}{\emph{JHEP} {\bf 10} (2016)
  162}, [\href{http://arxiv.org/abs/1608.01584}{{\tt 1608.01584}}].

\bibitem{Driencourt-Mangin:2017gop}
F.~Driencourt-Mangin, G.~Rodrigo and G.~F.~R. Sborlini, \emph{{Universal dual
  amplitudes and asymptotic expansions for $gg\to H$ and $H\to\gamma\gamma$ in
  four dimensions}},
  \href{http://dx.doi.org/10.1140/epjc/s10052-018-5692-5}{\emph{Eur. Phys. J.}
  {\bf C78} (2018) 231}, [\href{http://arxiv.org/abs/1702.07581}{{\tt
  1702.07581}}].

\bibitem{Gnendiger:2017pys}
C.~Gnendiger et~al., \emph{{To ${d}$, or not to ${d}$: recent developments and
  comparisons of regularization schemes}},
  \href{http://dx.doi.org/10.1140/epjc/s10052-017-5023-2}{\emph{Eur. Phys. J.}
  {\bf C77} (2017) 471}, [\href{http://arxiv.org/abs/1705.01827}{{\tt
  1705.01827}}].

\bibitem{Driencourt-Mangin:2019aix}
F.~Driencourt-Mangin, G.~Rodrigo, G.~F.~R. Sborlini and W.~J. Torres~Bobadilla,
  \emph{{Universal four-dimensional representation of $H\to\gamma\gamma$ at two
  loops through the Loop-Tree Duality}},
  \href{http://dx.doi.org/10.1007/JHEP02(2019)143}{\emph{JHEP} {\bf 02} (2019)
  143}, [\href{http://arxiv.org/abs/1901.09853}{{\tt 1901.09853}}].

\bibitem{Rodrigo:2016hqc}
G.~Rodrigo, F.~Driencourt-Mangin, G.~F.~R. Sborlini and R.~J.
  Hern\'andez-Pinto, \emph{{Applications of the loop-tree duality}},
  \href{http://dx.doi.org/10.22323/1.260.0037}{\emph{PoS} {\bf LL2016} (2016)
  037}, [\href{http://arxiv.org/abs/1608.01800}{{\tt 1608.01800}}].

\bibitem{Hernandez-Pinto:2016uwx}
R.~J. Hernández-Pinto, F.~Driencourt-Mangin, G.~Rodrigo and G.~F.~R. Sborlini,
  \emph{{NLO cross sections in 4 dimensions without DREG}},
  \href{http://dx.doi.org/10.1088/1742-6596/761/1/012021}{\emph{J. Phys. Conf.
  Ser.} {\bf 761} (2016) 012021}, [\href{http://arxiv.org/abs/1609.02454}{{\tt
  1609.02454}}].

\bibitem{Sborlini:2016bod}
G.~F.~R. Sborlini, F.~Driencourt-Mangin, R.~Hernandez-Pinto and G.~Rodrigo,
  \emph{{Towards regularized higher-order computations in QFT without DREG}},
  \href{http://dx.doi.org/10.22323/1.282.0353}{\emph{PoS} {\bf ICHEP2016}
  (2016) 353}, [\href{http://arxiv.org/abs/1611.04824}{{\tt 1611.04824}}].

\bibitem{Driencourt-Mangin:2016dpf}
F.~Driencourt-Mangin, \emph{{Computation of NLO Processes Involving Heavy
  Quarks Using Loop-Tree Duality}},
  \href{http://dx.doi.org/10.1063/1.4977166}{\emph{AIP Conf. Proc.} {\bf 1819}
  (2017) 060010}, [\href{http://arxiv.org/abs/1611.07352}{{\tt 1611.07352}}].

\bibitem{Sborlini:2017mhc}
G.~F.~R. Sborlini, F.~Driencourt-Mangin and G.~Rodrigo, \emph{{On the universal
  structure of Higgs amplitudes mediated by heavy particles}},
  \href{http://dx.doi.org/10.22323/1.314.0767}{\emph{PoS} {\bf EPS-HEP2017}
  (2017) 767}, [\href{http://arxiv.org/abs/1709.09860}{{\tt 1709.09860}}].

\bibitem{Sborlini:2017nee}
G.~F.~R. Sborlini, F.~Driencourt-Mangin, R.~Hernandez-Pinto and G.~Rodrigo,
  \emph{{Four-dimensional regularization of higher-order computations: FDU
  approach}}, \href{http://dx.doi.org/10.22323/1.314.0547}{\emph{PoS} {\bf
  EPS-HEP2017} (2017) 547}, [\href{http://arxiv.org/abs/1710.04516}{{\tt
  1710.04516}}].

\bibitem{Rodrigo:2018jme}
G.~Rodrigo, F.~Driencourt-Mangin, G.~F.~R. Sborlini and R.~J. Hernández-Pinto,
  \emph{{Recent developments from the loop-tree duality}},
  \href{http://dx.doi.org/10.22323/1.290.0013}{\emph{PoS} {\bf RADCOR2017}
  (2018) 013}, [\href{http://arxiv.org/abs/1801.04465}{{\tt 1801.04465}}].

\bibitem{Bendavid:2018nar}
J.~R. Andersen et~al., \emph{{Les Houches 2017: Physics at TeV Colliders
  Standard Model Working Group Report}},
  \href{http://arxiv.org/abs/1803.07977}{{\tt 1803.07977}}.

\bibitem{Bambade:2019fyw}
P.~Bambade et~al., \emph{{The International Linear Collider: A Global
  Project}},  \href{http://arxiv.org/abs/1903.01629}{{\tt 1903.01629}}.

\bibitem{deBlas:2018mhx}
J.~de~Blas et~al., \emph{{The CLIC Potential for New Physics}},
  \href{http://arxiv.org/abs/1812.02093}{{\tt 1812.02093}}.

\bibitem{Charles:2018vfv}
{\scshape CLICdp, CLIC} collaboration, \emph{{The Compact Linear Collider
  (CLIC) - 2018 Summary Report}},  2018.

\bibitem{Mangano:2018mur}
{\scshape FCC} collaboration, \emph{{Future Circular Collider: Vol. 1 Physics
  opportunities}},  2018.

\bibitem{Benedikt:2018qee}
{\scshape FCC} collaboration, \emph{{Future Circular Collider: Vol. 2 The
  Lepton Collider (FCC-ee)}},  2018.

\bibitem{Benedikt:2018csr}
{\scshape FCC} collaboration, \emph{{Future Circular Collider: Vol. 3 The
  Hadron Collider (FCC-hh)}},  2018.

\bibitem{Zimmermann:2018wdi}
{\scshape FCC} collaboration, \emph{{Future Circular Collider: Vol. 4 The
  High-Energy LHC (HE-LHC)}},  2018.

\bibitem{CEPC-SPPCStudyGroup:2015csa}
{\scshape CEPC-SPPC Study Group} collaboration, \emph{{CEPC-SPPC Preliminary
  Conceptual Design Report. 1. Physics and Detector}},  2015.

\bibitem{CEPC-SPPCStudyGroup:2015esa}
{\scshape CEPC-SPPC Study Group} collaboration, \emph{{CEPC-SPPC Preliminary
  Conceptual Design Report. 2. Accelerator}},  2015.

\bibitem{Alwall:2014hca}
J.~Alwall, R.~Frederix, S.~Frixione, V.~Hirschi, F.~Maltoni, O.~Mattelaer
  et~al., \emph{{The automated computation of tree-level and next-to-leading
  order differential cross sections, and their matching to parton shower
  simulations}}, \href{http://dx.doi.org/10.1007/JHEP07(2014)079}{\emph{JHEP}
  {\bf 07} (2014) 079}, [\href{http://arxiv.org/abs/1405.0301}{{\tt
  1405.0301}}].

\bibitem{Campbell:2011bn}
J.~M. Campbell, R.~K. Ellis and C.~Williams, \emph{{Vector boson pair
  production at the LHC}},
  \href{http://dx.doi.org/10.1007/JHEP07(2011)018}{\emph{JHEP} {\bf 07} (2011)
  018}, [\href{http://arxiv.org/abs/1105.0020}{{\tt 1105.0020}}].

\bibitem{Hernandez-Pinto:2015ysa}
R.~J. Hern\'andez-Pinto, G.~F.~R. Sborlini and G.~Rodrigo, \emph{{Towards gauge
  theories in four dimensions}},
  \href{http://dx.doi.org/10.1007/JHEP02(2016)044}{\emph{JHEP} {\bf 02} (2016)
  044}, [\href{http://arxiv.org/abs/1506.04617}{{\tt 1506.04617}}].

\bibitem{Catani:2008xa}
S.~Catani, T.~Gleisberg, F.~Krauss, G.~Rodrigo and J.-C. Winter, \emph{{From
  loops to trees by-passing Feynman's theorem}},
  \href{http://dx.doi.org/10.1088/1126-6708/2008/09/065}{\emph{JHEP} {\bf 09}
  (2008) 065}, [\href{http://arxiv.org/abs/0804.3170}{{\tt 0804.3170}}].

\bibitem{Rodrigo:2008fp}
G.~Rodrigo, S.~Catani, T.~Gleisberg, F.~Krauss and J.-C. Winter, \emph{{From
  multileg loops to trees (by-passing Feynman's Tree Theorem)}},
  \href{http://dx.doi.org/10.1016/j.nuclphysbps.2008.09.114}{\emph{Nucl. Phys.
  Proc. Suppl.} {\bf 183} (2008) 262--267},
  [\href{http://arxiv.org/abs/0807.0531}{{\tt 0807.0531}}].

\bibitem{Bierenbaum:2010cy}
I.~Bierenbaum, S.~Catani, P.~Draggiotis and G.~Rodrigo, \emph{{A Tree-Loop
  Duality Relation at Two Loops and Beyond}},
  \href{http://dx.doi.org/10.1007/JHEP10(2010)073}{\emph{JHEP} {\bf 10} (2010)
  073}, [\href{http://arxiv.org/abs/1007.0194}{{\tt 1007.0194}}].

\bibitem{Bierenbaum:2012th}
I.~Bierenbaum, S.~Buchta, P.~Draggiotis, I.~Malamos and G.~Rodrigo,
  \emph{{Tree-Loop Duality Relation beyond simple poles}},
  \href{http://dx.doi.org/10.1007/JHEP03(2013)025}{\emph{JHEP} {\bf 03} (2013)
  025}, [\href{http://arxiv.org/abs/1211.5048}{{\tt 1211.5048}}].

\bibitem{Buchta:2014dfa}
S.~Buchta, G.~Chachamis, P.~Draggiotis, I.~Malamos and G.~Rodrigo, \emph{{On
  the singular behaviour of scattering amplitudes in quantum field theory}},
  \href{http://dx.doi.org/10.1007/JHEP11(2014)014}{\emph{JHEP} {\bf 11} (2014)
  014}, [\href{http://arxiv.org/abs/1405.7850}{{\tt 1405.7850}}].

\bibitem{Buchta:2015xda}
S.~Buchta, \emph{{Theoretical foundations and applications of the Loop-Tree
  Duality in Quantum Field Theories}}.
\newblock PhD thesis, Valencia U., IFIC, 2015.
\newblock \href{http://arxiv.org/abs/1509.07167}{{\tt 1509.07167}}.

\bibitem{Buchta:2015wna}
S.~Buchta, G.~Chachamis, P.~Draggiotis and G.~Rodrigo, \emph{{Numerical
  implementation of the loop–tree duality method}},
  \href{http://dx.doi.org/10.1140/epjc/s10052-017-4833-6}{\emph{Eur. Phys. J.}
  {\bf C77} (2017) 274}, [\href{http://arxiv.org/abs/1510.00187}{{\tt
  1510.00187}}].

\bibitem{Hahn:2000kx}
T.~Hahn, \emph{{Generating Feynman diagrams and amplitudes with FeynArts 3}},
  \href{http://dx.doi.org/10.1016/S0010-4655(01)00290-9}{\emph{Comput. Phys.
  Commun.} {\bf 140} (2001) 418--431},
  [\href{http://arxiv.org/abs/hep-ph/0012260}{{\tt hep-ph/0012260}}].

\bibitem{Mertig:1990an}
R.~Mertig, M.~Bohm and A.~Denner, \emph{{FEYN CALC: Computer algebraic
  calculation of Feynman amplitudes}},
  \href{http://dx.doi.org/10.1016/0010-4655(91)90130-D}{\emph{Comput. Phys.
  Commun.} {\bf 64} (1991) 345--359}.

\bibitem{Shtabovenko:2016sxi}
V.~Shtabovenko, R.~Mertig and F.~Orellana, \emph{{New Developments in FeynCalc
  9.0}},  \href{http://arxiv.org/abs/1601.01167}{{\tt 1601.01167}}.

\bibitem{Collins:2016aya}
J.~C. Collins and J.~A.~M. Vermaseren, \emph{{Axodraw Version 2}},
  \href{http://arxiv.org/abs/1606.01177}{{\tt 1606.01177}}.

\bibitem{Peskin:1995ev}
M.~E. Peskin and D.~V. Schroeder, \emph{{An Introduction to Quantum Field
  Theory}}.
\newblock Addison-Wesley, 1995.

\bibitem{Ryder:1985wq}
L.~H. Ryder, \emph{{Quantum Field Theory}}.
\newblock Cambridge University Press, 1996.

\bibitem{Bollini:1972ui}
C.~G. Bollini and J.~J. Giambiagi, \emph{{Dimensional Renormalization: The
  Number of Dimensions as a Regularizing Parameter}},
  \href{http://dx.doi.org/10.1007/BF02895558}{\emph{Nuovo Cim.} {\bf B12}
  (1972) 20--26}.

\bibitem{tHooft:1972tcz}
G.~'t~Hooft and M.~J.~G. Veltman, \emph{{Regularization and Renormalization of
  Gauge Fields}},
  \href{http://dx.doi.org/10.1016/0550-3213(72)90279-9}{\emph{Nucl. Phys.} {\bf
  B44} (1972) 189--213}.

\bibitem{Collins:1986a}
J.~C. Collins, \emph{Renormalization -- An introduction to renormalization, the
  renormalization group and the operator-product expansion}.
\newblock Cambridge University Press, 1986.

\bibitem{Bern:1991aq}
Z.~Bern and D.~A. Kosower, \emph{{The Computation of loop amplitudes in gauge
  theories}}, \href{http://dx.doi.org/10.1016/0550-3213(92)90134-W}{\emph{Nucl.
  Phys.} {\bf B379} (1992) 451--561}.

\bibitem{Bern:2002zk}
Z.~Bern, A.~De~Freitas, L.~J. Dixon and H.~L. Wong, \emph{{Supersymmetric
  regularization, two loop QCD amplitudes and coupling shifts}},
  \href{http://dx.doi.org/10.1103/PhysRevD.66.085002}{\emph{Phys. Rev.} {\bf
  D66} (2002) 085002}, [\href{http://arxiv.org/abs/hep-ph/0202271}{{\tt
  hep-ph/0202271}}].

\bibitem{Stockinger:2005gx}
D.~Stockinger, \emph{{Regularization by dimensional reduction: consistency,
  quantum action principle, and supersymmetry}},
  \href{http://dx.doi.org/10.1088/1126-6708/2005/03/076}{\emph{JHEP} {\bf 03}
  (2005) 076}, [\href{http://arxiv.org/abs/hep-ph/0503129}{{\tt
  hep-ph/0503129}}].

\bibitem{Signer:2008va}
A.~Signer and D.~Stockinger, \emph{{Using Dimensional Reduction for Hadronic
  Collisions}},
  \href{http://dx.doi.org/10.1016/j.nuclphysb.2008.09.016}{\emph{Nucl. Phys.}
  {\bf B808} (2009) 88--120}, [\href{http://arxiv.org/abs/0807.4424}{{\tt
  0807.4424}}].

\bibitem{Wilson:1972cf}
K.~G. Wilson, \emph{{Quantum field theory models in less than
  four-dimensions}},
  \href{http://dx.doi.org/10.1103/PhysRevD.7.2911}{\emph{Phys. Rev.} {\bf D7}
  (1973) 2911--2926}.

\bibitem{Fazio:2014xea}
R.~A. Fazio, P.~Mastrolia, E.~Mirabella and W.~J. Torres~Bobadilla, \emph{{On
  the Four-Dimensional Formulation of Dimensionally Regulated Amplitudes}},
  \href{http://dx.doi.org/10.1140/epjc/s10052-014-3197-4}{\emph{Eur. Phys. J.}
  {\bf C74} (2014) 3197}, [\href{http://arxiv.org/abs/1404.4783}{{\tt
  1404.4783}}].

\bibitem{Cheung:2009a}
C.~Cheung and D.~O'Connell, \emph{Amplitudes and spinor-helicity in six
  dimensions},
  \href{http://dx.doi.org/10.1088/1126-6708/2009/07/075}{\emph{Journal of High
  Energy Physics} {\bf 2009} (2009) 075--075},
  [\href{http://arxiv.org/abs/0902.0981}{{\tt 0902.0981}}].

\bibitem{BaetaScarpelli:2000zs}
A.~P. Baeta~Scarpelli, M.~Sampaio and M.~C. Nemes, \emph{{Consistency relations
  for an implicit n-dimensional regularization scheme}},
  \href{http://dx.doi.org/10.1103/PhysRevD.63.046004}{\emph{Phys. Rev.} {\bf
  D63} (2001) 046004}, [\href{http://arxiv.org/abs/hep-th/0010285}{{\tt
  hep-th/0010285}}].

\bibitem{BaetaScarpelli:2001ix}
A.~P. Baeta~Scarpelli, M.~Sampaio, B.~Hiller and M.~C. Nemes, \emph{{Chiral
  anomaly and CPT invariance in an implicit momentum space regularization
  framework}}, \href{http://dx.doi.org/10.1103/PhysRevD.64.046013}{\emph{Phys.
  Rev.} {\bf D64} (2001) 046013},
  [\href{http://arxiv.org/abs/hep-th/0102108}{{\tt hep-th/0102108}}].

\bibitem{Pittau:2012zd}
R.~Pittau, \emph{{A four-dimensional approach to quantum field theories}},
  \href{http://dx.doi.org/10.1007/JHEP11(2012)151}{\emph{JHEP} {\bf 11} (2012)
  151}, [\href{http://arxiv.org/abs/1208.5457}{{\tt 1208.5457}}].

\bibitem{Donati:2013iya}
A.~M. Donati and R.~Pittau, \emph{{Gauge invariance at work in FDR:
  $H\to\gamma\gamma$}},
  \href{http://dx.doi.org/10.1007/JHEP04(2013)167}{\emph{JHEP} {\bf 04} (2013)
  167}, [\href{http://arxiv.org/abs/1302.5668}{{\tt 1302.5668}}].

\bibitem{tHooft:1973a}
G.~'t~Hooft, \emph{Dimensional regularization and the renormalization group},
  \href{http://dx.doi.org/https://doi.org/10.1016/0550-3213(73)90376-3}{\emph{Nuclear
  Physics B} {\bf 61} (1973) 455 -- 468}.

\bibitem{Weinberg:1973a}
S.~Weinberg, \emph{New approach to the renormalization group},
  \href{http://dx.doi.org/10.1103/PhysRevD.8.3497}{\emph{Phys. Rev. D} {\bf 8}
  (Nov, 1973) 3497--3509}.

\bibitem{Field:1989uq}
R.~D. Field, \emph{{Applications of Perturbative QCD}}, {\emph{Front. Phys.}
  {\bf 77} (1989) 1--366}.

\bibitem{Kinoshita:1962ur}
T.~Kinoshita, \emph{{Mass singularities of Feynman amplitudes}},
  \href{http://dx.doi.org/10.1063/1.1724268}{\emph{J. Math. Phys.} {\bf 3}
  (1962) 650--677}.

\bibitem{Lee:1964is}
T.~D. Lee and M.~Nauenberg, \emph{{Degenerate Systems and Mass Singularities}},
  \href{http://dx.doi.org/10.1103/PhysRev.133.B1549}{\emph{Phys. Rev.} {\bf
  133} (1964) B1549--B1562}.

\bibitem{Kunszt:1992tn}
Z.~Kunszt and D.~E. Soper, \emph{{Calculation of jet cross-sections in hadron
  collisions at order $\alpha_S^3$}},
  \href{http://dx.doi.org/10.1103/PhysRevD.46.192}{\emph{Phys. Rev.} {\bf D46}
  (1992) 192--221}.

\bibitem{Frixione:1995ms}
S.~Frixione, Z.~Kunszt and A.~Signer, \emph{{Three jet cross-sections to
  next-to-leading order}},
  \href{http://dx.doi.org/10.1016/0550-3213(96)00110-1}{\emph{Nucl. Phys.} {\bf
  B467} (1996) 399--442}, [\href{http://arxiv.org/abs/hep-ph/9512328}{{\tt
  hep-ph/9512328}}].

\bibitem{Catani:1996jh}
S.~Catani and M.~H. Seymour, \emph{{The Dipole formalism for the calculation of
  QCD jet cross-sections at next-to-leading order}},
  \href{http://dx.doi.org/10.1016/0370-2693(96)00425-X}{\emph{Phys. Lett.} {\bf
  B378} (1996) 287--301}, [\href{http://arxiv.org/abs/hep-ph/9602277}{{\tt
  hep-ph/9602277}}].

\bibitem{Catani:1996vz}
S.~Catani and M.~H. Seymour, \emph{{A General algorithm for calculating jet
  cross-sections in NLO QCD}},
  \href{http://dx.doi.org/10.1016/S0550-3213(96)00589-5,
  10.1016/S0550-3213(98)81022-5}{\emph{Nucl. Phys.} {\bf B485} (1997)
  291--419}, [\href{http://arxiv.org/abs/hep-ph/9605323}{{\tt
  hep-ph/9605323}}].

\bibitem{GehrmannDeRidder:2005cm}
A.~G.-D. Ridder, T.~Gehrmann and E.~W.~N. Glover, \emph{{Antenna subtraction at
  NNLO}}, \href{http://dx.doi.org/10.1088/1126-6708/2005/09/056}{\emph{JHEP}
  {\bf 09} (2005) 056}, [\href{http://arxiv.org/abs/hep-ph/0505111}{{\tt
  hep-ph/0505111}}].

\bibitem{Seth:2016hmv}
S.~Seth and S.~Weinzierl, \emph{{Numerical integration of subtraction terms}},
  \href{http://dx.doi.org/10.1103/PhysRevD.93.114031}{\emph{Phys. Rev.} {\bf
  D93} (2016) 114031}, [\href{http://arxiv.org/abs/1605.06646}{{\tt
  1605.06646}}].

\bibitem{Catani:2007vq}
S.~Catani and M.~Grazzini, \emph{{An NNLO subtraction formalism in hadron
  collisions and its application to Higgs boson production at the LHC}},
  \href{http://dx.doi.org/10.1103/PhysRevLett.98.222002}{\emph{Phys. Rev.
  Lett.} {\bf 98} (2007) 222002},
  [\href{http://arxiv.org/abs/hep-ph/0703012}{{\tt hep-ph/0703012}}].

\bibitem{Catani:2009sm}
S.~Catani, L.~Cieri, G.~Ferrera, D.~de~Florian and M.~Grazzini, \emph{{Vector
  boson production at hadron colliders: a fully exclusive QCD calculation at
  NNLO}}, \href{http://dx.doi.org/10.1103/PhysRevLett.103.082001}{\emph{Phys.
  Rev. Lett.} {\bf 103} (2009) 082001},
  [\href{http://arxiv.org/abs/0903.2120}{{\tt 0903.2120}}].

\bibitem{Czakon:2010td}
M.~Czakon, \emph{{A novel subtraction scheme for double-real radiation at
  NNLO}}, \href{http://dx.doi.org/10.1016/j.physletb.2010.08.036}{\emph{Phys.
  Lett.} {\bf B693} (2010) 259--268},
  [\href{http://arxiv.org/abs/1005.0274}{{\tt 1005.0274}}].

\bibitem{Bolzoni:2010bt}
P.~Bolzoni, G.~Somogyi and Z.~Trocsanyi, \emph{{A subtraction scheme for
  computing QCD jet cross sections at NNLO: integrating the iterated
  singly-unresolved subtraction terms}},
  \href{http://dx.doi.org/10.1007/JHEP01(2011)059}{\emph{JHEP} {\bf 01} (2011)
  059}, [\href{http://arxiv.org/abs/1011.1909}{{\tt 1011.1909}}].

\bibitem{DelDuca:2015zqa}
V.~Del~Duca, C.~Duhr, G.~Somogyi, F.~Tramontano and Z.~Trócsányi,
  \emph{{Higgs boson decay into b-quarks at NNLO accuracy}},
  \href{http://dx.doi.org/10.1007/JHEP04(2015)036}{\emph{JHEP} {\bf 04} (2015)
  036}, [\href{http://arxiv.org/abs/1501.07226}{{\tt 1501.07226}}].

\bibitem{Boughezal:2015dva}
R.~Boughezal, C.~Focke, X.~Liu and F.~Petriello, \emph{{${W}$-boson production
  in association with a jet at next-to-next-to-leading order in perturbative
  QCD}}, \href{http://dx.doi.org/10.1103/PhysRevLett.115.062002}{\emph{Phys.
  Rev. Lett.} {\bf 115} (2015) 062002},
  [\href{http://arxiv.org/abs/1504.02131}{{\tt 1504.02131}}].

\bibitem{Gaunt:2015pea}
J.~Gaunt, M.~Stahlhofen, F.~J. Tackmann and J.~R. Walsh, \emph{{N-jettiness
  Subtractions for NNLO QCD Calculations}},
  \href{http://dx.doi.org/10.1007/JHEP09(2015)058}{\emph{JHEP} {\bf 09} (2015)
  058}, [\href{http://arxiv.org/abs/1505.04794}{{\tt 1505.04794}}].

\bibitem{DelDuca:2016ily}
V.~Del~Duca, C.~Duhr, A.~Kardos, G.~Somogyi, Z.~Szőr, Z.~Trócsányi et~al.,
  \emph{{Jet production in the CoLoRFulNNLO method: event shapes in
  electron-positron collisions}},
  \href{http://dx.doi.org/10.1103/PhysRevD.94.074019}{\emph{Phys. Rev.} {\bf
  D94} (2016) 074019}, [\href{http://arxiv.org/abs/1606.03453}{{\tt
  1606.03453}}].

\bibitem{DelDuca:2016csb}
V.~Del~Duca, C.~Duhr, A.~Kardos, G.~Somogyi and Z.~Trócsányi,
  \emph{{Three-Jet Production in Electron-Positron Collisions at
  Next-to-Next-to-Leading Order Accuracy}},
  \href{http://dx.doi.org/10.1103/PhysRevLett.117.152004}{\emph{Phys. Rev.
  Lett.} {\bf 117} (2016) 152004}, [\href{http://arxiv.org/abs/1603.08927}{{\tt
  1603.08927}}].

\bibitem{Cutkosky:1960sp}
R.~E. Cutkosky, \emph{{Singularities and discontinuities of Feynman
  amplitudes}}, \href{http://dx.doi.org/10.1063/1.1703676}{\emph{J. Math.
  Phys.} {\bf 1} (1960) 429--433}.

\bibitem{Cutkosky:1961a}
R.~E. Cutkosky, \emph{Anomalous thresholds},
  \href{http://dx.doi.org/10.1103/RevModPhys.33.448}{\emph{Rev. Mod. Phys.}
  {\bf 33} (Jul, 1961) 448--455}.

\bibitem{Mandelstam:1960zz}
S.~Mandelstam, \emph{{Unitarity Condition Below Physical Thresholds in the
  Normal and Anomalous Cases}},
  \href{http://dx.doi.org/10.1103/PhysRevLett.4.84}{\emph{Phys. Rev. Lett.}
  {\bf 4} (1960) 84--87}.

\bibitem{Feynman:1963ax}
R.~P. Feynman, \emph{{Quantum theory of gravitation}}, {\emph{Acta Phys.
  Polon.} {\bf 24} (1963) 697--722}.

\bibitem{Collins:1989gx}
J.~C. Collins, D.~E. Soper and G.~F. Sterman, \emph{{Factorization of Hard
  Processes in QCD}},
  \href{http://dx.doi.org/10.1142/9789814503266_0001}{\emph{Adv. Ser. Direct.
  High Energy Phys.} {\bf 5} (1989) 1--91},
  [\href{http://arxiv.org/abs/hep-ph/0409313}{{\tt hep-ph/0409313}}].

\bibitem{Catani:2011st}
S.~Catani, D.~de~Florian and G.~Rodrigo, \emph{{Space-like (versus time-like)
  collinear limits in QCD: Is factorization violated?}},
  \href{http://dx.doi.org/10.1007/JHEP07(2012)026}{\emph{JHEP} {\bf 07} (2012)
  026}, [\href{http://arxiv.org/abs/1112.4405}{{\tt 1112.4405}}].

\bibitem{Soper:1999rd}
D.~E. Soper, \emph{{QCD calculations by numerical integration}},
  \href{http://dx.doi.org/10.1016/S0920-5632(99)00748-3}{\emph{Nucl. Phys.
  Proc. Suppl.} {\bf 79} (1999) 444--446}.

\bibitem{Soper:2000a}
D.~E. Soper, \emph{{Techniques for QCD calculations by numerical integration}},
  \href{http://dx.doi.org/10.1103/PhysRevD.62.014009}{\emph{Phys. Rev. D} {\bf
  62} (2000) 014009}.

\bibitem{Soper:2001hu}
D.~E. Soper, \emph{{Choosing integration points for QCD calculations by
  numerical integration}},
  \href{http://dx.doi.org/10.1103/PhysRevD.64.034018}{\emph{Phys. Rev.} {\bf
  D64} (2001) 034018}, [\href{http://arxiv.org/abs/hep-ph/0103262}{{\tt
  hep-ph/0103262}}].

\bibitem{Kramer:2002cd}
M.~Krämer and D.~E. Soper, \emph{{Next-to-leading order numerical calculations
  in Coulomb gauge}},
  \href{http://dx.doi.org/10.1103/PhysRevD.66.054017}{\emph{Phys. Rev.} {\bf
  D66} (2002) 054017}, [\href{http://arxiv.org/abs/hep-ph/0204113}{{\tt
  hep-ph/0204113}}].

\bibitem{Becker:2010ng}
S.~Becker, C.~Reuschle and S.~Weinzierl, \emph{{Numerical NLO QCD
  calculations}}, \href{http://dx.doi.org/10.1007/JHEP12(2010)013}{\emph{JHEP}
  {\bf 12} (2010) 013}, [\href{http://arxiv.org/abs/1010.4187}{{\tt
  1010.4187}}].

\bibitem{Becker:2012aqa}
S.~Becker, C.~Reuschle and S.~Weinzierl, \emph{{Efficiency Improvements for the
  Numerical Computation of NLO Corrections}},
  \href{http://dx.doi.org/10.1007/JHEP07(2012)090}{\emph{JHEP} {\bf 07} (2012)
  090}, [\href{http://arxiv.org/abs/1205.2096}{{\tt 1205.2096}}].

\bibitem{Passarino:2001a}
G.~Passarino, \emph{An approach toward the numerical evaluation of multi-loop
  feynman diagrams},
  \href{http://dx.doi.org/10.1016/S0550-3213(01)00528-4}{\emph{Nuclear Physics
  B} {\bf 619} (2001) 257 -- 312}.

\bibitem{Ferroglia:2002mz}
A.~Ferroglia, M.~Passera, G.~Passarino and S.~Uccirati, \emph{{All purpose
  numerical evaluation of one loop multileg Feynman diagrams}},
  \href{http://dx.doi.org/10.1016/S0550-3213(02)01070-2}{\emph{Nucl. Phys.}
  {\bf B650} (2003) 162--228}, [\href{http://arxiv.org/abs/hep-ph/0209219}{{\tt
  hep-ph/0209219}}].

\bibitem{Nagy:2003qn}
Z.~Nagy and D.~E. Soper, \emph{{General subtraction method for numerical
  calculation of one loop QCD matrix elements}},
  \href{http://dx.doi.org/10.1088/1126-6708/2003/09/055}{\emph{JHEP} {\bf 09}
  (2003) 055}, [\href{http://arxiv.org/abs/hep-ph/0308127}{{\tt
  hep-ph/0308127}}].

\bibitem{Nagy:2006a}
Z.~Nagy and D.~E. Soper, \emph{Numerical integration of one-loop feynman
  diagrams for ${N}$-photon amplitudes},
  \href{http://dx.doi.org/10.1103/PhysRevD.74.093006}{\emph{Phys. Rev. D} {\bf
  74} (Nov, 2006) 093006}.

\bibitem{Anastasiou:2007qb}
C.~Anastasiou, S.~Beerli and A.~Daleo, \emph{{Evaluating multi-loop Feynman
  diagrams with infrared and threshold singularities numerically}},
  \href{http://dx.doi.org/10.1088/1126-6708/2007/05/071}{\emph{JHEP} {\bf 05}
  (2007) 071}, [\href{http://arxiv.org/abs/hep-ph/0703282}{{\tt
  hep-ph/0703282}}].

\bibitem{Moretti:2008jj}
M.~Moretti, F.~Piccinini and A.~D. Polosa, \emph{{A Fully Numerical Approach to
  One-Loop Amplitudes}},  \href{http://arxiv.org/abs/0802.4171}{{\tt
  0802.4171}}.

\bibitem{Gong:2008ww}
W.~Gong, Z.~Nagy and D.~E. Soper, \emph{{Direct numerical integration of
  one-loop Feynman diagrams for ${N}$-photon amplitudes}},
  \href{http://dx.doi.org/10.1103/PhysRevD.79.033005}{\emph{Phys. Rev.} {\bf
  D79} (2009) 033005}, [\href{http://arxiv.org/abs/0812.3686}{{\tt
  0812.3686}}].

\bibitem{Kilian:2009wy}
W.~Kilian and T.~Kleinschmidt, \emph{{Numerical Evaluation of Feynman Loop
  Integrals by Reduction to Tree Graphs}},
  \href{http://arxiv.org/abs/0912.3495}{{\tt 0912.3495}}.

\bibitem{Becker:2012nk}
S.~Becker and S.~Weinzierl, \emph{{Direct contour deformation with arbitrary
  masses in the loop}},
  \href{http://dx.doi.org/10.1103/PhysRevD.86.074009}{\emph{Phys. Rev.} {\bf
  D86} (2012) 074009}, [\href{http://arxiv.org/abs/1208.4088}{{\tt
  1208.4088}}].

\bibitem{Freitas:2016sty}
A.~Freitas, \emph{{Numerical multi-loop integrals and applications}},
  \href{http://dx.doi.org/10.1016/j.ppnp.2016.06.004}{\emph{Prog. Part. Nucl.
  Phys.} {\bf 90} (2016) 201--240},
  [\href{http://arxiv.org/abs/1604.00406}{{\tt 1604.00406}}].

\bibitem{Ellis:2007qk}
R.~K. Ellis and G.~Zanderighi, \emph{{Scalar one-loop integrals for QCD}},
  \href{http://dx.doi.org/10.1088/1126-6708/2008/02/002}{\emph{JHEP} {\bf 02}
  (2008) 002}, [\href{http://arxiv.org/abs/0712.1851}{{\tt 0712.1851}}].

\bibitem{Sborlini:2013jba}
G.~F.~R. Sborlini, D.~de~Florian and G.~Rodrigo, \emph{{Double collinear
  splitting amplitudes at next-to-leading order}},
  \href{http://dx.doi.org/10.1007/JHEP01(2014)018}{\emph{JHEP} {\bf 01} (2014)
  018}, [\href{http://arxiv.org/abs/1310.6841}{{\tt 1310.6841}}].

\bibitem{Sborlini:2016fcj}
G.~F.~R. Sborlini, \emph{{Loop-tree duality and quantum field theory in four
  dimensions}}, \href{http://dx.doi.org/10.22323/1.235.0082}{\emph{PoS} {\bf
  RADCOR2015} (2016) 082}, [\href{http://arxiv.org/abs/1601.04634}{{\tt
  1601.04634}}].

\bibitem{Altarelli:1977zs}
G.~Altarelli and G.~Parisi, \emph{{Asymptotic Freedom in Parton Language}},
  \href{http://dx.doi.org/10.1016/0550-3213(77)90384-4}{\emph{Nucl. Phys.} {\bf
  B126} (1977) 298--318}.

\bibitem{deFlorian:2015ujt}
D.~de~Florian, G.~F.~R. Sborlini and G.~Rodrigo, \emph{{QED corrections to the
  Altarelli–Parisi splitting functions}},
  \href{http://dx.doi.org/10.1140/epjc/s10052-016-4131-8}{\emph{Eur. Phys. J.}
  {\bf C76} (2016) 282}, [\href{http://arxiv.org/abs/1512.00612}{{\tt
  1512.00612}}].

\bibitem{Sborlini:2014kla}
G.~F.~R. Sborlini, D.~de~Florian and G.~Rodrigo, \emph{{Polarized
  triple-collinear splitting functions at NLO for processes with photons}},
  \href{http://dx.doi.org/10.1007/JHEP03(2015)021}{\emph{JHEP} {\bf 03} (2015)
  021}, [\href{http://arxiv.org/abs/1409.6137}{{\tt 1409.6137}}].

\bibitem{Sborlini:2014mpa}
G.~F.~R. Sborlini, D.~de~Florian and G.~Rodrigo, \emph{{Triple collinear
  splitting functions at NLO for scattering processes with photons}},
  \href{http://dx.doi.org/10.1007/JHEP10(2014)161}{\emph{JHEP} {\bf 10} (2014)
  161}, [\href{http://arxiv.org/abs/1408.4821}{{\tt 1408.4821}}].

\bibitem{Catani:2003vu}
S.~Catani, D.~de~Florian and G.~Rodrigo, \emph{{The Triple collinear limit of
  one loop QCD amplitudes}},
  \href{http://dx.doi.org/10.1016/j.physletb.2004.02.039}{\emph{Phys. Lett.}
  {\bf B586} (2004) 323--331}, [\href{http://arxiv.org/abs/hep-ph/0312067}{{\tt
  hep-ph/0312067}}].

\bibitem{Catani:2002hc}
S.~Catani, S.~Dittmaier, M.~H. Seymour and Z.~Trocsanyi, \emph{{The Dipole
  formalism for next-to-leading order QCD calculations with massive partons}},
  \href{http://dx.doi.org/10.1016/S0550-3213(02)00098-6}{\emph{Nucl. Phys.}
  {\bf B627} (2002) 189--265}, [\href{http://arxiv.org/abs/hep-ph/0201036}{{\tt
  hep-ph/0201036}}].

\bibitem{GehrmannDeRidder:2009fz}
A.~Gehrmann-De~Ridder and M.~Ritzmann, \emph{{NLO Antenna Subtraction with
  Massive Fermions}},
  \href{http://dx.doi.org/10.1088/1126-6708/2009/07/041}{\emph{JHEP} {\bf 07}
  (2009) 041}, [\href{http://arxiv.org/abs/0904.3297}{{\tt 0904.3297}}].

\bibitem{Abelof:2012he}
G.~Abelof, O.~Dekkers and A.~Gehrmann-De~Ridder, \emph{{Antenna subtraction
  with massive fermions at NNLO: Double real initial-final configurations}},
  \href{http://dx.doi.org/10.1007/JHEP12(2012)107}{\emph{JHEP} {\bf 12} (2012)
  107}, [\href{http://arxiv.org/abs/1210.5059}{{\tt 1210.5059}}].

\bibitem{Abelof:2014fza}
G.~Abelof, A.~Gehrmann-De~Ridder, P.~Maierhofer and S.~Pozzorini, \emph{{NNLO
  QCD subtraction for top-antitop production in the $ q\bar{q} $ channel}},
  \href{http://dx.doi.org/10.1007/JHEP08(2014)035}{\emph{JHEP} {\bf 08} (2014)
  035}, [\href{http://arxiv.org/abs/1404.6493}{{\tt 1404.6493}}].

\bibitem{Bonciani:2015sha}
R.~Bonciani, S.~Catani, M.~Grazzini, H.~Sargsyan and A.~Torre, \emph{{The $q_T$
  subtraction method for top quark production at hadron colliders}},
  \href{http://dx.doi.org/10.1140/epjc/s10052-015-3793-y}{\emph{Eur. Phys. J.}
  {\bf C75} (2015) 581}, [\href{http://arxiv.org/abs/1508.03585}{{\tt
  1508.03585}}].

\bibitem{Rodrigo:1999qg}
G.~Rodrigo, M.~S. Bilenky and A.~Santamaria, \emph{{Quark mass effects for jet
  production in e+ e- collisions at the next-to-leading order: Results and
  applications}},
  \href{http://dx.doi.org/10.1016/S0550-3213(99)00293-X}{\emph{Nucl. Phys.}
  {\bf B554} (1999) 257--297}, [\href{http://arxiv.org/abs/hep-ph/9905276}{{\tt
  hep-ph/9905276}}].

\bibitem{Wilczek:1977zn}
F.~Wilczek, \emph{{Decays of Heavy Vector Mesons Into Higgs Particles}},
  \href{http://dx.doi.org/10.1103/PhysRevLett.39.1304}{\emph{Phys. Rev. Lett.}
  {\bf 39} (1977) 1304}.

\bibitem{Georgi:1977gs}
H.~M. Georgi, S.~L. Glashow, M.~E. Machacek and D.~V. Nanopoulos, \emph{{Higgs
  Bosons from Two Gluon Annihilation in Proton Proton Collisions}},
  \href{http://dx.doi.org/10.1103/PhysRevLett.40.692}{\emph{Phys. Rev. Lett.}
  {\bf 40} (1978) 692}.

\bibitem{Rizzo:1980a}
T.~G. Rizzo, \emph{Gluon final states in higgs-boson decay},
  \href{http://dx.doi.org/10.1103/PhysRevD.22.178}{\emph{Phys. Rev. D} {\bf 22}
  (Jul, 1980) 178--183}.

\bibitem{Ellis:1975ap}
J.~R. Ellis, M.~K. Gaillard and D.~V. Nanopoulos, \emph{{A Phenomenological
  Profile of the Higgs Boson}},
  \href{http://dx.doi.org/10.1016/0550-3213(76)90382-5}{\emph{Nucl. Phys.} {\bf
  B106} (1976) 292}.

\bibitem{Ioffe:1976sd}
B.~L. Ioffe and V.~A. Khoze, \emph{{What Can Be Expected from Experiments on
  Colliding e+ e- Beams with e Approximately Equal to 100-GeV?}}, {\emph{Sov.
  J. Part. Nucl.} {\bf 9} (1978) 50}.

\bibitem{Shifman:1979eb}
M.~A. Shifman, A.~I. Vainshtein, M.~B. Voloshin and V.~I. Zakharov,
  \emph{{Low-Energy Theorems for Higgs Boson Couplings to Photons}},
  {\emph{Sov. J. Nucl. Phys.} {\bf 30} (1979) 711--716}.

\bibitem{Gastmans:2011wh}
R.~Gastmans, S.~L. Wu and T.~T. Wu, \emph{{Higgs Decay into Two Photons,
  Revisited}},  \href{http://arxiv.org/abs/1108.5872}{{\tt 1108.5872}}.

\bibitem{Spira:1995rr}
M.~Spira, A.~Djouadi, D.~Graudenz and P.~M. Zerwas, \emph{{Higgs boson
  production at the LHC}},
  \href{http://dx.doi.org/10.1016/0550-3213(95)00379-7}{\emph{Nucl. Phys.} {\bf
  B453} (1995) 17--82}, [\href{http://arxiv.org/abs/hep-ph/9504378}{{\tt
  hep-ph/9504378}}].

\bibitem{Harlander:2005rq}
R.~Harlander and P.~Kant, \emph{{Higgs production and decay: Analytic results
  at next-to-leading order QCD}},
  \href{http://dx.doi.org/10.1088/1126-6708/2005/12/015}{\emph{JHEP} {\bf 12}
  (2005) 015}, [\href{http://arxiv.org/abs/hep-ph/0509189}{{\tt
  hep-ph/0509189}}].

\bibitem{Aglietti:2006tp}
U.~Aglietti, R.~Bonciani, G.~Degrassi and A.~Vicini, \emph{{Analytic Results
  for Virtual QCD Corrections to Higgs Production and Decay}},
  \href{http://dx.doi.org/10.1088/1126-6708/2007/01/021}{\emph{JHEP} {\bf 01}
  (2007) 021}, [\href{http://arxiv.org/abs/hep-ph/0611266}{{\tt
  hep-ph/0611266}}].

\bibitem{Schwinn:2006ca}
C.~Schwinn and S.~Weinzierl, \emph{{SUSY ward identities for multi-gluon
  helicity amplitudes with massive quarks}},
  \href{http://dx.doi.org/10.1088/1126-6708/2006/03/030}{\emph{JHEP} {\bf 03}
  (2006) 030}, [\href{http://arxiv.org/abs/hep-th/0602012}{{\tt
  hep-th/0602012}}].

\bibitem{Ferrario:2006np}
P.~Ferrario, G.~Rodrigo and P.~Talavera, \emph{{Compact multigluonic scattering
  amplitudes with heavy scalars and fermions}},
  \href{http://dx.doi.org/10.1103/PhysRevLett.96.182001}{\emph{Phys. Rev.
  Lett.} {\bf 96} (2006) 182001},
  [\href{http://arxiv.org/abs/hep-th/0602043}{{\tt hep-th/0602043}}].

\bibitem{Beneke:1997zp}
M.~Beneke and V.~A. Smirnov, \emph{{Asymptotic expansion of Feynman integrals
  near threshold}},
  \href{http://dx.doi.org/10.1016/S0550-3213(98)00138-2}{\emph{Nucl. Phys.}
  {\bf B522} (1998) 321--344}, [\href{http://arxiv.org/abs/hep-ph/9711391}{{\tt
  hep-ph/9711391}}].

\bibitem{Smirnov:2002pj}
V.~A. Smirnov, \emph{{Applied asymptotic expansions in momenta and masses}},
  {\emph{Springer Tracts Mod. Phys.} {\bf 177} (2002) 1--262}.

\bibitem{Zheng:1990qa}
H.-Q. Zheng and D.-D. Wu, \emph{{First order QCD corrections to the decay of
  the Higgs boson into two photons}},
  \href{http://dx.doi.org/10.1103/PhysRevD.42.3760}{\emph{Phys. Rev.} {\bf D42}
  (1990) 3760--3763}.

\bibitem{Djouadi:1990aj}
A.~Djouadi, M.~Spira, J.~J. van~der Bij and P.~M. Zerwas, \emph{{QCD
  corrections to gamma gamma decays of Higgs particles in the intermediate mass
  range}}, \href{http://dx.doi.org/10.1016/0370-2693(91)90879-U}{\emph{Phys.
  Lett.} {\bf B257} (1991) 187--190}.

\bibitem{Dawson:1992cy}
S.~Dawson and R.~P. Kauffman, \emph{{QCD corrections to $H\to\gamma\gamma$}},
  \href{http://dx.doi.org/10.1103/PhysRevD.47.1264}{\emph{Phys. Rev.} {\bf D47}
  (1993) 1264--1267}.

\bibitem{Fleischer:2004vb}
J.~Fleischer, O.~V. Tarasov and V.~O. Tarasov, \emph{{Analytical result for the
  two loop QCD correction to the decay $H\to\gamma\gamma$}},
  \href{http://dx.doi.org/10.1016/j.physletb.2004.01.063}{\emph{Phys. Lett.}
  {\bf B584} (2004) 294--297}, [\href{http://arxiv.org/abs/hep-ph/0401090}{{\tt
  hep-ph/0401090}}].

\bibitem{Aglietti:2004nj}
U.~Aglietti, R.~Bonciani, G.~Degrassi and A.~Vicini, \emph{{Two loop light
  fermion contribution to Higgs production and decays}},
  \href{http://dx.doi.org/10.1016/j.physletb.2004.06.063}{\emph{Phys.Lett.}
  {\bf B595} (2004) 432--441}, [\href{http://arxiv.org/abs/hep-ph/0404071}{{\tt
  hep-ph/0404071}}].

\bibitem{Actis:2008ts}
S.~Actis, G.~Passarino, C.~Sturm and S.~Uccirati, \emph{{NNLO Computational
  Techniques: The Cases $H\to\gamma\gamma$ and $H\to gg$}},
  \href{http://dx.doi.org/10.1016/j.nuclphysb.2008.11.024}{\emph{Nucl. Phys.}
  {\bf B811} (2009) 182--273}, [\href{http://arxiv.org/abs/0809.3667}{{\tt
  0809.3667}}].

\bibitem{Passarino:2007fp}
G.~Passarino, C.~Sturm and S.~Uccirati, \emph{{Complete Two-Loop Corrections to
  $H\to\gamma\gamma$}},
  \href{http://dx.doi.org/10.1016/j.physletb.2007.09.002}{\emph{Phys. Lett.}
  {\bf B655} (2007) 298--306}, [\href{http://arxiv.org/abs/0707.1401}{{\tt
  0707.1401}}].

\bibitem{Degrassi:2005mc}
G.~Degrassi and F.~Maltoni, \emph{{Two-loop electroweak corrections to the
  Higgs-boson decay $H\to\gamma\gamma$}},
  \href{http://dx.doi.org/10.1016/j.nuclphysb.2005.06.027}{\emph{Nucl. Phys.}
  {\bf B724} (2005) 183--196}, [\href{http://arxiv.org/abs/hep-ph/0504137}{{\tt
  hep-ph/0504137}}].

\bibitem{Fugel:2004ug}
F.~Fugel, B.~A. Kniehl and M.~Steinhauser, \emph{{Two loop electroweak
  correction of $\mathcal{O}(G(F)\,M(t)^2)$ to the Higgs-boson decay into
  photons}},
  \href{http://dx.doi.org/10.1016/j.nuclphysb.2004.09.018}{\emph{Nucl. Phys.}
  {\bf B702} (2004) 333--345}, [\href{http://arxiv.org/abs/hep-ph/0405232}{{\tt
  hep-ph/0405232}}].

\bibitem{Maierhofer:2012vv}
P.~Maierh{\"o}fer and P.~Marquard, \emph{{Complete three-loop QCD corrections
  to the decay $H\to\gamma \gamma$}},
  \href{http://dx.doi.org/10.1016/j.physletb.2013.02.040}{\emph{Phys. Lett.}
  {\bf B721} (2013) 131--135}, [\href{http://arxiv.org/abs/1212.6233}{{\tt
  1212.6233}}].

\bibitem{Steinhauser:1996wy}
M.~Steinhauser, \emph{Corrections of $\mathcal{O}(\alpha_s^2)$ to the decay of
  an intermediate mass higgs boson into two photons},  in \emph{{The Higgs
  puzzle - what can we learn from LEP-2, LHC, NLC and FMC? Proceedings,
  Ringberg Workshop, Tegernsee, Germany, December 8-13, 1996}}, pp.~177--185,
  1996.
\newblock \href{http://arxiv.org/abs/hep-ph/9612395}{{\tt hep-ph/9612395}}.

\bibitem{Anastasiou:2016cez}
C.~Anastasiou, C.~Duhr, F.~Dulat, E.~Furlan, T.~Gehrmann, F.~Herzog et~al.,
  \emph{{High precision determination of the gluon fusion Higgs boson
  cross-section at the LHC}},
  \href{http://dx.doi.org/10.1007/JHEP05(2016)058}{\emph{JHEP} {\bf 05} (2016)
  058}, [\href{http://arxiv.org/abs/1602.00695}{{\tt 1602.00695}}].

\bibitem{Bonetti:2018ukf}
M.~Bonetti, K.~Melnikov and L.~Tancredi, \emph{{Higher order corrections to
  mixed QCD-EW contributions to Higgs boson production in gluon fusion}},
  \href{http://dx.doi.org/10.1103/PhysRevD.97.056017,
  10.1103/PhysRevD.97.099906}{\emph{Phys. Rev.} {\bf D97} (2018) 056017},
  [\href{http://arxiv.org/abs/1801.10403}{{\tt 1801.10403}}].

\bibitem{Chetyrkin:1981qh}
K.~G. Chetyrkin and F.~V. Tkachov, \emph{{Integration by Parts: The Algorithm
  to Calculate beta Functions in 4 Loops}},
  \href{http://dx.doi.org/10.1016/0550-3213(81)90199-1}{\emph{Nucl. Phys.} {\bf
  B192} (1981) 159--204}.

\bibitem{Laporta:2001dd}
S.~Laporta, \emph{{High precision calculation of multiloop Feynman integrals by
  difference equations}},
  \href{http://dx.doi.org/10.1016/S0217-751X(00)00215-7,
  10.1142/S0217751X00002157}{\emph{Int. J. Mod. Phys.} {\bf A15} (2000)
  5087--5159}, [\href{http://arxiv.org/abs/hep-ph/0102033}{{\tt
  hep-ph/0102033}}].

\bibitem{Caffo:1998du}
M.~Caffo, H.~Czyz, S.~Laporta and E.~Remiddi, \emph{{The Master differential
  equations for the two loop sunrise selfmass amplitudes}}, {\emph{Nuovo Cim.}
  {\bf A111} (1998) 365--389}, [\href{http://arxiv.org/abs/hep-th/9805118}{{\tt
  hep-th/9805118}}].

\bibitem{Laporta:2004rb}
S.~Laporta and E.~Remiddi, \emph{{Analytic treatment of the two loop equal mass
  sunrise graph}},
  \href{http://dx.doi.org/10.1016/j.nuclphysb.2004.10.044}{\emph{Nucl. Phys.}
  {\bf B704} (2005) 349--386}, [\href{http://arxiv.org/abs/hep-ph/0406160}{{\tt
  hep-ph/0406160}}].

\bibitem{Carter:2010hi}
J.~Carter and G.~Heinrich, \emph{{SecDec: A general program for sector
  decomposition}},
  \href{http://dx.doi.org/10.1016/j.cpc.2011.03.026}{\emph{Comput. Phys.
  Commun.} {\bf 182} (2011) 1566--1581},
  [\href{http://arxiv.org/abs/1011.5493}{{\tt 1011.5493}}].

\bibitem{Borowka:2015mxa}
S.~Borowka, G.~Heinrich, S.~P. Jones, M.~Kerner, J.~Schlenk and T.~Zirke,
  \emph{{SecDec-3.0: numerical evaluation of multi-scale integrals beyond one
  loop}}, \href{http://dx.doi.org/10.1016/j.cpc.2015.05.022}{\emph{Comput.
  Phys. Commun.} {\bf 196} (2015) 470--491},
  [\href{http://arxiv.org/abs/1502.06595}{{\tt 1502.06595}}].

\bibitem{Bilenky:1994ad}
M.~S. Bilenky, G.~Rodrigo and A.~Santamaria, \emph{{Three jet production at LEP
  and the bottom quark mass}},
  \href{http://dx.doi.org/10.1016/0550-3213(94)00586-4}{\emph{Nucl. Phys.} {\bf
  B439} (1995) 505--535}, [\href{http://arxiv.org/abs/hep-ph/9410258}{{\tt
  hep-ph/9410258}}].

\bibitem{Remiddi:1999ew}
E.~Remiddi and J.~A.~M. Vermaseren, \emph{{Harmonic polylogarithms}},
  \href{http://dx.doi.org/10.1142/S0217751X00000367}{\emph{Int. J. Mod. Phys.}
  {\bf A15} (2000) 725--754}, [\href{http://arxiv.org/abs/hep-ph/9905237}{{\tt
  hep-ph/9905237}}].

\end{thebibliography}\endgroup

\end{document}